\documentclass{ws-ijmpe}

\usepackage{mathrsfs,amssymb,amsmath,amsopn}
\usepackage{graphicx}
\usepackage{color}
\usepackage[dvips,bookmarks,pdfhighlight=/O,pdfstartview=FitH]{hyperref}

\newcommand{\lsim}{\lesssim}
\newcommand{\gsim}{\gtrsim}
\newcommand{\tildeint}[1]{\ensuremath{\int \frac{\mathrm{d}^{3} #1}{
      (2\pi)^{3} 2 \omega_{#1}}}}
\newcommand{\ii}{\ensuremath{\mathrm{i}}}
\newcommand{\dd}{\ensuremath{\mathrm{d}}}
\newcommand{\erw}[1]{\ensuremath { %
    \left \langle {#1} \right \rangle}}
\newcommand{\Lag}{\ensuremath{\mathscr{L}}}
\newcommand{\MeV}{\ensuremath{\mathrm{MeV}}}
\newcommand{\GeV}{\ensuremath{\mathrm{GeV}}}
\newcommand{\fm}{\ensuremath{\mathrm{fm}}}
\newcommand{\Tr}{\ensuremath{\mathrm{Tr}}}
\newcommand{\comm}[2]{\ensuremath{ \left[ {#1}, {#2} \right] }}
\newcommand{\bvec}[1]{\boldsymbol{#1}}

\DeclareMathOperator{\im}{Im}
\DeclareMathOperator{\re}{Re}

\makeatletter
  \newcommand{\fslash}[2][0mu]{%
    \mathchoice
      {\fsl@sh\displaystyle{#1}{#2}}%
      {\fsl@sh\textstyle{#1}{#2}}%
      {\fsl@sh\scriptstyle{#1}{#2}}%
      {\fsl@sh\scriptscriptstyle{#1}{#2}}}
  \newcommand{\fsl@sh}[3]{%
    \m@th\ooalign{$\hfil#1\mkern#2/\hfil$\crcr$#1#3$}}
\makeatother

\begin{document}

\markboth{R. Rapp, H. van Hees}{Heavy quarks in the quark-gluon plasma}

\catchline{}{}{}{}{}

\title{HEAVY QUARKS IN THE QUARK-GLUON PLASMA}

\author{RALF RAPP}

\address{Cyclotron Institute and Physics Department\\ 
Texas A\&M University, College Station, Texas 77843-3366, U.S.A.\\
rapp@comp.tamu.edu}

\author{HENDRIK VAN HEES}

\address{Institut f{\"u}r Theoretische Physik,
  Justus-Liebig-Universit{\"a}t Giessen, D-35392 Giessen, Germany\\
Hendrik.vanHees@theo.physik.uni-giessen.de}

\maketitle

\begin{history}
\received{July 20, 2009}
\revised{July 20, 2009}
\end{history}

\begin{abstract}
  Heavy-flavor particles are believed to provide valuable probes of the
  medium produced in ultrarelativistic collisions of heavy nuclei. In
  this article we review recent progress in our understanding of the
  interactions of charm and bottom quarks in the Quark-Gluon Plasma
  (QGP).  For individual heavy quarks, we focus on elastic interactions
  for which the large quark mass enables a Brownian motion
  treatment. This opens a unique access to thermalization mechanisms for
  heavy quarks at low momentum, and thus to their transport coefficients
  in the quark-gluon fluid. Different approaches to evaluate heavy-quark
  diffusion are discussed and compared, including perturbative QCD,
  effective potential models utilizing input from lattice QCD and
  string-theoretic estimates in conformal field theories.  Applications
  to heavy-quark observables in heavy-ion collisions are realized via
  relativistic Langevin simulations, where we illustrate the important
  role of a realistic medium evolution to quantitatively extract the
  heavy-quark diffusion constant. In the heavy quarkonium sector, we
  briefly review the current status in potential-model based
  interpretations of correlation functions computed in lattice QCD,
  followed by an evaluation of quarkonium dissociation reactions in the
  QGP. The discussion of the phenomenology in heavy-ion reactions
  focuses on thermal model frameworks paralleling the open heavy-flavor
  sector. We also emphasize connections to the heavy-quark diffusion
  problem in both potential models and quarkonium regeneration
  processes.
\end{abstract}


\section{Introduction}
\label{sec_intro}

The investigation of strongly interacting matter constitutes a major
challenge in modern nuclear and particle physics. Of particular interest
are phase changes between hadronic and quark-gluon matter, similar to
the one which is believed to have occurred in the early Universe at a
few microseconds after its birth. While the theory of the strong force
is by now well established in terms of Quantum Chromodynamics
(QCD)\cite{Fritzsch:1973pi,Gross:1973id,Politzer:1973fx}, two of its
major manifestations in the world around us - the confinement of quarks
and gluons and the generation of hadronic masses - are subject of
vigorous contemporary research. Both phenomena occur at energy-momentum
scales of $Q^2\lsim 1 \, \mathrm{GeV}^2$ where the QCD coupling constant
is rather large, $\alpha_s \gsim 0.3$, and therefore perturbation theory
is not reliable and/or applicable. In a hot and dense medium at
sufficiently large temperature ($T$) and/or quark chemical potential
($\mu_q$), one expects the finite-size hadrons to be dissolved into a
deconfined Quark-Gluon Plasma (QGP) where the condensates underlying
hadronic-mass generation have melted. Numerical simulations of
lattice-discretized QCD (lQCD) at finite temperature predict the phase
change from hadronic to quark-gluon matter to occur at a
``pseudo-critical'' temperature of $T_c\simeq 200 \,
\MeV$\cite{Cheng:2007jq}. This appears to be a rather small scale for a
``perturbative QGP'' (pQGP) of weakly interacting quarks and gluons to
be realized, even though the computed energy density matches that of an
ideal (non-interacting and massless) QGP within 20\,\% or so for $T\gsim
1.2 \, T_c$.

In the laboratory, one hopes to create a QGP by colliding heavy atomic
nuclei at ultrarelativistic energies, with a center-of-mass energy per
colliding nucleon pair well above the nucleon rest mass, $\sqrt{s}/A \gg
M_N$. If the energy deposition in the reaction zone is large
enough, and if the interactions of the produced particles are strong
enough, the notion of an interacting medium may be justified, despite
its transient nature. This notion has been convincingly verified in 
nuclear collision experiments over the last $\sim$$25$ years at the 
Super-Proton-Synchrotron (SPS) at the European Organization for Nuclear
Research (CERN)\cite{Heinz:2000bk} and at the Relativistic Heavy-Ion
Collider (RHIC) at Brookhaven National Laboratory
(BNL)\cite{Arsene:2004fa}.
Transverse-momentum ($p_T$) spectra of different hadron species in the 
low-$p_T$ regime ($p_T\lsim 2$-$3 \, \mathrm{GeV}$) reveal that 
the produced medium explodes collectively reaching expansion velocities 
in excess of half the speed of light. In the high-$p_T$ regime 
($p_T\gsim 5 \, \mathrm{GeV}$), which in the heavy-ion environment 
became available at RHIC for the first time, hadron spectra are 
suppressed by up to a factor of $\sim$$5$ relative to $p$-$p$ 
collisions, indicative for a strong absorption of high-energy partons 
traversing the medium\cite{Gyulassy:2003mc}. The inclusive production 
of charm-quark bound states ($J/\psi$ mesons) is suppressed by a factor 
of 3-5 at both SPS and RHIC, indicative for their dissolution in the 
medium (possibly related to
deconfinement)\cite{Rapp:2008tf,BraunMunzinger:2009ih,Kluberg:2009yd}.
A large excess of electromagnetic radiation (photons and dileptons) is 
observed, indicative for medium temperatures around $200 \,
\mathrm{MeV}$ and a ``melting'' of the $\rho$-meson resonance (possibly
related to hadronic mass de-generation)\cite{Gale:2003iz,Rapp:2009yu}.
A more differential analysis of hadron spectra in non-central Au-Au
collisions at RHIC reveals a large elliptic asymmetry of the collective
flow (``elliptic flow''): the spatial asymmetry of the initial nuclear 
overlap zone is converted into an opposite asymmetry in the final
hadron $p_T$ spectra. Within a hydrodynamic modeling of the exploding 
fireball this observation requires a rapid thermalization and a very 
small viscosity of the interacting 
medium\cite{Kolb:2003dz,Shuryak:2004cy,Huovinen:2006jp,Hirano:2008aj}.
Only then can spatial pressure gradients build up fast enough to 
facilitate an effective conversion into azimuthal asymmetries in the 
energy-momentum tensor of the system. The agreement of hydrodynamic
predictions with elliptic-flow data at RHIC led to the notion of an
``almost perfect liquid'', with a ratio of viscosity to
entropy density close to a conjectured lower bound of any quantum
mechanical system\cite{Kovtun:2004de}. The microscopic mechanisms 
underlying these rather
remarkable transport properties are yet to be determined. In this
context, heavy quarks (charm and bottom, $Q$=$c$ and $b$) and their
bound states (charmonia and bottomonia) are recognized as particularly
suitable probes of the medium produced in ultrarelativistic heavy-ion
collisions (URHICs)\cite{Frawley:2008kk}\footnote{The (weak-decay)
  lifetime of the top quark of $\sim$$0.1\,\fm/c$ is too short to render
  it a viable probe in URHICs; thus, heavy quarks will exclusively refer
  to charm and bottom in this article. Strange quarks are in between the
  heavy- and light-quark limit, forming their own complex of valuable
  observables\cite{Muller:2008ae}.}. In the present article we will
attempt to review the current status of the theory and phenomenology of
this promise.

Let us first focus on the sector of individual heavy quarks (open heavy
flavor). The fact that their masses are well above the typical
temperature of the system, $m_Q\gg T$, has at least three important
implications:
\begin{itemize}
\item[(1)] The (hard) $Q\bar{Q}$ production process is essentially
  restricted to primordial $N$-$N$ collisions\cite{Levai:1994dx}, i.e.,
  re-interactions in the subsequently evolving medium are not expected
  to change the number of heavy quarks (reminiscent of the
  ``factorization theorem'' of perturbative QCD\cite{Collins:1985gm});
  this is borne out experimentally by a scaling of $c\bar{c}$
  production, $N_{c\bar c}$, with the number of binary $N$-$N$
  collisions, $N_{\rm coll}$, at different collision
  centralities\cite{Adare:2006nq}.
\item[(2)] The thermal relaxation time of heavy quarks ought to be
  larger than for light quarks, parameterically by a factor 
  $\sim$$m_Q/T\approx$~5-20. 
  With a light-quark and gluon thermalization time
  of $\tau_{q,g}\simeq$0.3-1\,fm/$c$ (as indirectly inferred
  from hydrodynamic modeling at RHIC) and an estimated QGP lifetime of
  $\tau_{\rm QGP}\simeq$~5\,fm/$c$ in central Au-Au collisions,
  one expects $\tau_c$ ($\tau_b$) to be on the same order as
  (significantly larger than) $\tau_{\rm QGP}$. Thus, charm (and
  especially bottom) quarks are not expected to reach thermal
  equilibrium, but their re-interactions should impart noticeable
  modifications on the initial momentum spectrum (less pronounced for
  bottom). The final heavy-quark (HQ) spectra may therefore encode a
  ``memory'' of the interaction history throughout the evolving
  fireball, by operating in between the limits of thermalization and
  free streaming.
\item[(3)] The theoretical task of describing HQ interactions is
  amenable to a diffusion treatment, i.e., Brownian motion of a heavy
  test particle in a bath of a light-particle fluid.
  Nonrelativistically, the typical thermal momentum of a heavy quark is
  $p_{\mathrm{th}}^2 \simeq 3m_Q T \gg T^2$, and therefore much larger
  than the typical momentum transfer from the medium, $Q^2\sim
  T^2$. This allows to expand the Boltzmann equation in momentum
  transfer to arrive at a Fokker-Planck description of HQ diffusion in
  the QGP, which directly yields the pertinent transport coefficients as
  well.
\end{itemize}
The above three points provide a well-defined framework to construct
in-medium HQ interactions in QCD matter and test them against
observables in URHICs (quantitative comparisons additionally require to
account for effects of hadronization of the quarks, as well as
reinteractions in the hadronic medium). The Fokker-Planck approach is
readily implemented for the case of \emph{elastic} $p+Q \to p+Q$
scattering off partons in the medium ($p=q,\bar
q,g$)\cite{Svetitsky:1987gq,GolamMustafa:1997id,vanHees:2004gq,Moore:2004tg,Mustafa:2004dr,vanHees:2005wb}.
In the light-hadron sector, however, the large suppression of high-$p_T$
spectra is believed to be largely caused by \emph{radiative} energy loss
of high-energy partons traversing the QGP, i.e., medium-induced gluon
radiation of type $q+g \to
q+g+g$\cite{Gyulassy:2003mc,Eskola:2004cr,Qin:2007rn}. Even in the
low-$p_T$ regime, perturbative $2,3\leftrightarrow 3$ scattering
processes have been suggested to facilitate the rapid thermalization
required by phenomenology (albeit in connection with rather large
coupling constants of $\alpha_s\simeq 0.5$)\cite{Xu:2007jv}. The
situation could be quite different in the HQ sector. In the low-momentum
limit, gluon-Bremsstrahlung of a heavy quark is
suppressed\cite{Dokshitzer:2001zm} and the dominant momentum-transfer
reaction is elastic scattering\cite{Moore:2004tg}. As is well known from
classical electrodynamics, the radiative energy loss of a muon is
suppressed relative to an electron by a mass ratio $(m_\mu/m_e)^4$. In
perturbative QCD (pQCD) it is currently an open question at what
momentum scale radiative energy loss of a heavy quark takes over from
the collisional one (which, most likely, will depend on additional
parameters such as temperature, path length, etc.). In fact, this may
not even be a well-defined question since nonperturbative processes at
moderate momentum transfers may supersede perturbative ones before the
elastic part of the latter dominates over the radiative one. 
The relations between perturbative and nonperturbative interactions 
is one of the key issues to be addressed in this review.

Experimental signatures for the modifications of HQ spectra in URHICs
are currently encoded in single-electron ($e^\pm$) spectra
associated with the semileptonic decays of charm and bottom hadrons, $D,
B, \Lambda_c, \ldots \to e \nu X$. These measurements require a careful
subtraction of all possible ``photonic'' sources of electrons, such as
photon conversions in the detector material, Dalitz decays of $\pi$ and
$\eta$, vector-meson decays, and others. The modifications of the
``non-photonic'' electron spectra (associated with heavy-flavor decays)
in Au-Au collisions are then quantified by the standard nuclear
modification factor, $R_{AA}^e$, and elliptic flow coefficient, $v_2^e$.
The available RHIC data in semicentral and central Au-Au collisions at
$\sqrt{s_{NN}} =200\,\mathrm{GeV}$ exhibit a substantial elliptic flow
of up to $v_2^e\simeq 10\%$ and a large high-$p_T$ suppression down to
$R_{AA}^e\simeq 0.25$,
respectively\cite{Adler:2005xv,Adare:2006nq,Abelev:2006db,Awes:2008qi}.
Both values are quite comparable to those for light hadrons (the pion
$v_2$ reaches somewhat higher, to about 15\%).  Radiative energy-loss
models\cite{Armesto:2005mz} based on perturbative QCD cannot explain the
$e^\pm$ data. These data were, in fact, instrumental\cite{Akiba:2005bs}
in reconsidering elastic scattering as a significant source of parton
energy loss in the
QGP\cite{vanHees:2004gq,Moore:2004tg,Mustafa:2004dr,Wicks:2005gt}.
The combination of pQCD elastic and radiative scattering does not 
suffice either to reproduce the observed suppression once a realistic
bottom component is accounted for in the electron
spectra\cite{Wicks:2005gt}. Elastic scattering based on nonperturbative
interactions, as proposed in Refs.\cite{vanHees:2004gq,vanHees:2005wb},
simultaneously accounts for the $e^\pm$ elliptic flow and suppression
reasonably well\cite{Adare:2006nq}.  This has reinforced the hope that
HQ observables provide the promised precision tool to characterize
transport properties of the ``strongly coupled QGP'' (sQGP). E.g., if a
clear mass hierarchy in thermal relaxation times, $\tau_Q \propto
\frac{m_Q}{T} \tau_q$ (as well as $\tau_b/\tau_c = m_b/m_c$) emerges
from a quantitative analysis of URHIC data, it would be suggestive for a
universal behavior of light- and heavy-quark transport in the
QGP. However, there is still a substantial way to go before such a
program can be realized, as discussed below\footnote{A recent review
  article\cite{Rapp:2008qc} addresses similar topics but from a more
  elementary perspective; see also Ref.\cite{Linnyk:2008hp}}.

It should not be surprising if in-medium properties of open heavy
flavor, especially at low momentum, are closely related 
to medium modifications of heavy quarkonia. The latter have a long
history as ``probes'' of the QGP in heavy-ion collisions, especially 
as potential indicators of the deconfinement transition,
cf.~Refs.\cite{Rapp:2008tf,BraunMunzinger:2009ih,Kluberg:2009yd} for a
broad up-to-date coverage of this topic. In particular, progress in
finite-temperature lattice
QCD\cite{Asakawa:2003re,Karsch:2003jg,Kaczmarek:2004gv} has triggered
vigorous reconsideration of the question whether quarkonia, especially
their ground states, can survive in the QGP significantly above the
critical temperature.  These developments include the application of
potential models at finite temperature, coupled with the hope that
heavy-quark free energies as computed in thermal lQCD can serve as a
model-independent input for the low-energy heavy-quark interaction. If
charmonium binding indeed remains sufficiently strong in the QGP to
support bound states up to rather high temperatures, it is conceivable
that the underlying interaction is of a more general relevance and
therefore also operative in heavy-light\cite{vanHees:2007me} and
maybe even light-light\cite{Shuryak:2004tx} systems. Especially in the 
former case, from the point of view of elastic (on-shell) scattering of 
a heavy quark in the medium, the conditions for momentum transfer are 
comparable to the heavy-heavy interaction governing quarkonium
properties. Since low-momentum HQ interactions determine their transport
properties, one immediately recognizes an intimate relation between HQ
transport and in-medium quarkonia. These connections are also being
exploited in the analysis of thermal lQCD computations of quarkonium
correlation functions\cite{Petreczky:2008px}.  In addition to the
binding properties, the inelastic reaction rates of quarkonia with
surrounding partons or hadrons are a key ingredient for a quantitative
description of their spectral function in QCD matter (also here
``quasi-elastic" scattering of thermal partons with a heavy quark
inside the quarkonium bound states may play an important role,
especially if the binding energy becomes small\cite{Grandchamp:2001pf}).
A good control over all of these aspects is mandatory to utilize
quarkonium properties as diagnostic tool in heavy-ion collisions and
eventually deduce more general properties of the medium produced in
these reactions. As in the open heavy-flavor sector, this has to be
built on a solid knowledge of the space-time history of nuclear
collisions, as well as of the initial conditions on quarkonium
spectra. The latter aspect could be more involved than for single 
heavy quarks, since (a) measurements in $p$-$A$ collisions show that 
cold-nuclear-matter (CNM) effects from the incoming nuclei (e.g., the
so-called nuclear absorption) affect the primordial charmonium
number significantly (e.g., with up to 60\% suppression for $J/\psi$ at
SPS energies when extrapolated to central Pb-Pb collisions); (b) the
bound-state formation time introduces another rather long time scale
(soft energy scale) which is easily of the order of (or longer than) 
the thermalization time of the medium (at least for charmonia and 
excited bottomonia at RHIC energies and higher).

Charmonium suppression beyond the level of CNM effects has been 
established in semi-/central Pb-Pb and Au-Au
collisions at SPS\cite{Alessandro:2004ap} and RHIC\cite{Adare:2006ns},
respectively. An intriguing finding is that the
observed suppression pattern and magnitude is very comparable at SPS and
RHIC, despite the different collision energies which lead to substantial
variations in, e.g., light-hadron observables (most notably a factor
$\sim$$2$ larger charged particle rapidity density and stronger
collective phenomena at RHIC). However, this ``degeneracy'' was
predicted\cite{Grandchamp:2001pf} as a consequence of charmonium 
regeneration
mechanisms\cite{Thews:2000rj,Braun-Munzinger:2000px,Gorenstein:2000ck}:
a stronger suppression in the hotter and denser medium at RHIC is 
compensated by the coalescence of $c$ and $\bar c$ quarks in the QGP 
and/or at hadronization (the $c\bar c$ production cross section at RHIC 
is about a factor of $\sim$100 larger than at SPS energies). While an
``extra'' source of charmonia increases the complexity of pertinent
observables in heavy-ion reactions, it also provides another, rather
direct, connection between the open and hidden heavy-flavor sectors. 
Obviously, the secondary yield from $c$-$\bar c$ coalescence
necessarily carries imprints of the charm-quark distributions, both
in its magnitude (softer $c$-quark spectra are expected to result in
larger coalescence probabilities) and in its momentum spectra (including
elliptic flow). A comprehensive theoretical and phenomenological
analysis of open and hidden heavy flavor is thus becoming an
increasingly pressing and challenging issue.  As a final remark on
quarkonia in this introduction, we point out that bottomonium production
in heavy-ion reactions is less likely to receive regeneration 
contributions (at least at RHIC and possibly neither at LHC). In 
addition, the increase in bottomonium binding energies (compared to 
charmonia) render them rather sensitive probes of color screening which 
strongly influences its dissociation rates\cite{Grandchamp:2005yw}. 
Bottomonia thus remain a promising observable to realize the originally 
envisaged ``spectral analysis of strongly interacting
matter''\cite{Karsch:1990wi}.

Our review is organized as follows: In Sec.~\ref{sec_hq-int} we outline
the theoretical framework of evaluating HQ diffusion in equilibrium QCD
matter. We first recall basic steps in setting up the HQ diffusion
equation (Sec.~\ref{ssec_diff}) which determines the time evolution of
the HQ distribution function in terms of pertinent transport
coefficients based on elastic scattering amplitudes. This is followed by
a discussion of several microscopic approaches to calculate the HQ
friction and diffusion coefficients in the QGP: perturbative QCD
(Sec.~\ref{ssec_pqcd}) at leading (\ref{sssec_lo-0}, \ref{sssec_lo-htl},
\ref{sssec_lo-run}) and next-to-leading order (\ref{sssec_nlo}) as well
as for three-body scattering (\ref{sssec_lo-3}); nonperturbative
calculations (Sec.~\ref{ssec_non-pert}) implementing resonance-like
correlations in the QGP using HQ effective theory
(Sec.~\ref{sssec_reso}), in-medium $T$-matrices with HQ potentials
estimated from thermal lattice QCD (Sec.~\ref{sssec_tmat}), or
collisional-dissociation mechanisms of heavy mesons
(Sec.~\ref{sssec_coll-diss}); and string-theoretic evaluations based on
the conjectured correspondence to conformal field theories
(Sec.~\ref{ssec_string}). The variety of the proposed approaches calls
for an attempt to reconcile the underlying assumptions and basic
interactions~(Sec.~\ref{ssec_recon}). This is followed by a discussion
of inelastic (radiative) energy-loss calculations and their relation to
elastic ones (Sec.~\ref{ssec_rad}).  We briefly consider interactions of
open heavy-flavor hadrons in hadronic matter (Sec.~\ref{ssec_hadronic}).
In Sec.~\ref{sec_hq-obs} we discuss applications of HQ diffusion to
URHICs using relativistic Langevin simulations of the Fokker-Planck
equation within an expanding finite-size thermal medium
(Sec.~\ref{ssec_langevin}). A realistic description of the latter
(utilizing hydrodynamics, transport models or suitable parameterizations
thereof) is an essential prerequisite to enable a quantitative
extraction of transport properties of the QCD medium
(Sec.~\ref{ssec_medium}). Further ingredients are reliable initial
conditions (possibly modified by nuclear effects) and the conversion of
quarks to hadrons (Sec.~\ref{ssec_initial}). Implementations of
different HQ diffusion coefficients in various space-time models are
quantitatively analyzed in terms of the resulting HQ spectra at RHIC, in
particular their nuclear modification factor and elliptic flow
(Sec.~\ref{ssec_hq-spectra}).  Including effects of hadronization (as
well as semileptonic electron decays), a quantitative comparison of
these calculations to single-electron spectra at RHIC is conducted
(Sec.~\ref{ssec_rhic-obs}).  We emphasize the importance of a consistent
(simultaneous) description of $p_t$ spectra and elliptic flow. Only then
can these observables be converted into a meaningful (albeit
preliminary) estimate of charm- and bottom-quark diffusion coefficients
in the QGP. We finish the discussion on open heavy flavor with an
attempt to utilize these coefficients for a schematic estimate of the
ratio of shear viscosity to entropy density in the QGP
(Sec.~\ref{ssec_viscosity}). In Sec.~\ref{sec_onia} we elaborate on
theoretical and phenomenological analyses of quarkonia in medium and
their production in heavy-ion collisions. We first address spectral
properties of quarkonia in equilibrium matter (Sec.~\ref{ssec_spec});
Euclidean correlation functions computed in lattice QCD with good
precision have been analyzed in terms of potential models based on
screened HQ potentials (Sec.~\ref{sssec_lat}). The interplay of color
screening and parton-induced dissociation reactions has important
consequences for the evaluation of quarkonium dissociation widths
(Sec.~\ref{sssec_diss}). In light of the charmonium equilibrium
properties the current status of the phenomenology in heavy-ion
collisions is discussed (Sec.~\ref{ssec_prod}). First, quarkonium
transport equations are introduced along with their main ingredients,
i.e., dissociation widths and equilibrium numbers using relative
chemical equilibrium at fixed HQ number (Sec.~\ref{sssec_trans}); this
is followed by model comparisons to $J/\psi$ data at SPS and RHIC,
scrutinizing suppression vs.  regeneration mechanisms and their
transverse-momentum dependencies (Sec.~\ref{sssec_phen}), and a brief
illustration of predictions for $\Upsilon$ production at RHIC. In
Sec.~\ref{sec_concl} we recollect the main points of this article and
conclude.


\section{Heavy-Quark Interactions in  QCD Matter}
\label{sec_hq-int}

At an energy scale of the (pseudo-) critical QCD transition
temperature, the large charm- and bottom-quark masses imply that the HQ
diffusion problem is a nonrelativistic one (unless initial conditions
bring in an additional large scale). In the weak-coupling regime this
further implies that the dominant interactions of the heavy quark are
elastic scattering (gluon radiation is suppressed by an extra power in
$\alpha_s$ and cannot be compensated by a large momentum transfer as
could be the case for a fast quark; see, e.g., the discussion in
Ref.\cite{Moore:2004tg}). It turns out, however, that the perturbative
expansion of the charm-quark diffusion coefficient, evaluated using
thermal field theory, is not well convergent even for a strong coupling
constant as low as $\alpha_s=0.1$\cite{CaronHuot:2007gq}. Thus,
non-perturbative methods, e.g., resummations of large contributions or
interactions beyond perturbation theory, are necessary to improve the
estimates of HQ diffusion. This is not surprising since transport
coefficients usually involve the zero-momentum limit of correlation
functions rendering them susceptible to threshold effects which may
increase with the mass of the particles. A simple example of such kind
are Coulomb-like bound states (e.g., heavy quarkonia), where the binding
energy increases with increasing HQ mass, 
$\epsilon_B \propto \alpha_s^2 m_Q$, to be compared to thermal
effects, e.g., at a scale $\sim$$gT$ for Debye screening (to
leading order in $g$) or at $\sim$$T$ for inelastic dissociation
reactions with thermal partons. An
interesting question in this context is whether potential models are a
viable means to evaluate HQ interactions in the QGP. If a suitable
formulation of a potential at finite temperature can be established, a
promising opportunity arises by extracting these from first principle
lattice computations of the HQ free energy. In the heavy-quarkonium
sector such a program has been initiated a few years
ago\cite{Mocsy:2005qw,Wong:2006bx,Cabrera:2006wh,Alberico:2006vw,Laine:2007gj}
with fair success, although several open questions
remain\cite{Mocsy:2007jz,Mocsy:2007yj,Alberico:2007rg,Cabrera:2006wh}. If
applicable, potential models have the great benefit of allowing for
nonperturbative solutions utilizing Schr\"odinger or Lippmann-Schwinger
equations; the calculated scattering amplitudes can then be
straightforwardly related to transport coefficients. A key issue in this
discussion is the transition to the (ultra-) relativistic regime, which
becomes inevitable in applications to experiment toward high
momentum. While relativistic kinematics can be readily accounted for,
the opening of inelastic (radiative) channels poses major
problems. However, here the contact to perturbative calculations may be
possible and provide a valuable interface to match the different
regimes, at least parametrically (e.g., in the limit of a small coupling
constant and/or high temperature). This reiterates the importance of
identifying the common grounds of seemingly different calculations for
HQ properties in medium.

We start the discussion in this Section by setting up the Brownian 
Motion framework for heavy
quarks in the QGP (Sec.~\ref{ssec_diff}).  The main part of this Section
is devoted to the evaluation of the Fokker-Planck transport
coefficients.  We focus on elastic interactions, classified into
(various levels of) perturbative (Sec.~\ref{ssec_pqcd}) and
nonperturbative approaches (Secs.~\ref{ssec_non-pert} and
\ref{ssec_string}). As we will see, there is considerable conceptual 
overlap in the calculations available in the literature, the
main difference being that they are carried out in different
approximation schemes (Sec.~\ref{ssec_recon}). Our presentation also
encompasses inelastic reactions with an additional gluon in the final
and/or initial state, i.e., radiative energy-loss calculations
within perturbative QCD (Sec.~\ref{ssec_rad}). This raises the issue 
of their relative magnitude compared to
elastic interactions which has recently received considerable
re-consideration even for light quarks and gluons.  Finally, we 
address interactions of hadrons carrying charm or bottom 
in hadronic matter (Sec.~\ref{ssec_hadronic}). Bottom-up
extrapolations in temperature (or density) in the hadronic world 
are useful complements to top-down ones in the QGP, to reveal 
qualitative trends of, e.g., the HQ diffusion coefficient toward $T_c$.


\subsection{Heavy-Quark Diffusion in the Quark-Gluon Plasma}
\label{ssec_diff}
As emphasized in the Introduction, an attractive feature in analyzing
HQ motion in a QGP is the ensuing simplification to a Brownian
motion framework\cite{Svetitsky:1987gq}. The latter 
is characterized by a Fokker-Planck equation where HQ interactions  
are conveniently encoded in transport coefficients. These, in turn, 
are readily related to underlying (elastic) scattering matrix elements 
on light partons in the QGP which allow for direct comparisons of 
microscopic models of the HQ interaction (as elaborated in subsequent 
sections).  

Starting point for the derivation of the Fokker-Planck
equation\cite{Svetitsky:1987gq} is the Boltzmann equation for the HQ
phase-space distribution, $f_Q$, 
\begin{equation}
\label{boltz}
\left [ \frac{\partial}{\partial t} + \frac{\bvec{p}}{\omega_{\bvec{p}}}
  \frac{\partial}{\partial \bvec{x}} + \bvec{F}
  \frac{\partial}{\partial{\bvec{p}}} \right ]
f_{\mathrm{Q}}(t,\bvec{x},\bvec{p}) = C[f_{\mathrm{Q}}],
\end{equation}
where $\omega_{\bvec{p}}=\sqrt{m_Q^2+\bvec{p}^2}$ denotes the energy of
a heavy quark with three-momentum $\bvec{p}$, $\bvec{F}$ is the
mean-field force, and $C[f_Q]$ summarizes the collision integral which 
will be analyzed in more detail below. In the
following, mean-field effects will be neglected, and by integration over
the fireball volume, Eq.~(\ref{boltz}) simplifies to an equation for
the momentum distribution,
\begin{equation}
\label{boltz2}
\frac{\partial}{\partial t} f_Q(t,\bvec{p})=C[f_Q],
\end{equation}
where
\begin{equation}
\label{2.1.2b}
f_Q(t,\bvec{p})=\int \dd^3 \bvec{x} f_Q(t,\bvec{x},\bvec{p}).
\end{equation}
The collision integral on the right-hand side of Eq.~(\ref{boltz2})
encodes the transition rate of heavy quarks due to collisions into and
out of a small momentum cell $\dd^3 \bvec{p}$ around the HQ momentum
$\bvec{p}$,
\begin{equation}
\label{2.1.3}
C[f]=\int \dd^3 \bvec{k} [w(\bvec{p}+\bvec{k},\bvec{k})
f_Q(\bvec{p}+\bvec{k})-w(\bvec{p},\bvec{k}) f_Q(\bvec{p}) ] \ .
\end{equation}
Here $w(\bvec{p},\bvec{k})$ is the transition rate for collisions 
of a heavy quark with heat-bath particles with momentum transfer
$\bvec{k}$, changing the HQ momentum from $\bvec{p}$ to
$\bvec{p}-\bvec{k}$. Accordingly the first (gain) term in the integral
describes the transition rate for HQ scattering from 
a state with momentum $\bvec{p}+\bvec{k}$, into a state with
momentum $\bvec{p}$, while the second (loss) term  
the scattering out of the momentum state $\bvec{p}$.

The transition rate, $w$, can be expressed through the cross section of
the collision processes in the heat bath. For \emph{elastic} scattering
of a heavy quark with momentum $\bvec{p}$ on a light quark in the heat
bath with momentum $\bvec{q}$, one finds
\begin{equation}
\label{2.1.4}
w(\bvec{p},\bvec{k})=\gamma_{q,g} \int \frac{\dd^3 \bvec{q}}{(2 \pi)^3}
f_{q,g}(\bvec{q}) v_{\mathrm{rel}} \, 
\frac{\dd \sigma}{\dd \Omega}(\bvec{p},\bvec{q}
\rightarrow \bvec{p}-\bvec{k},\bvec{q}+\bvec{k}),
\end{equation} 
where $f_{q,g}$ are the Fermi or Bose distributions for thermal light quarks 
or gluons, and $\gamma_q=6$ or $\gamma_g=16$ the respective spin-color 
degeneracy factors. The relative velocity is defined as
\begin{equation}
\label{2.1.4b}
v_{\mathrm{rel}}=\frac{\sqrt{(p \cdot q)^2-(m_Q m_q)^2}}{\omega_Q \omega_q},
\end{equation}
where $p=(\omega_{\bvec{p}},\bvec{p})$ and
$q=(\omega_{\bvec{q}},\bvec{q})$ are the four momenta of the incoming
heavy and light quark, respectively. Upon expressing the invariant
differential cross section, $\dd \sigma/\dd \Omega$, in
Eq.~(\ref{2.1.4}) in terms of the spin-color summed matrix element,
$\sum |\mathcal{M}|^2$, the collision term, Eq.~(\ref{2.1.3}), takes the
form
\begin{equation}
\begin{split}
\label{2.1.3b}
C[f_Q]= & \frac{1}{2 \omega_{\bvec{p}}} \tildeint{\bvec{q}}
\tildeint{\bvec{p}'}
\tildeint{\bvec{q}'} \frac{1}{\gamma_Q} \sum |\mathcal{M}|^2 \\
& \times (2 \pi)^4 \delta^{(4)}(p+q-p'-q') [f_Q(\bvec{p}')
f_{q,g}(\bvec{q}') - f_Q(\bvec{p}) f_{q,g}(\bvec{q}) ] \ 
\end{split}
\end{equation}
with $\bvec{k}=\bvec{p}-\bvec{p}'=\bvec{q}'-\bvec{q}$.

The key approximation is now that the relevant momentum transfers 
to the heavy quark obey $|\bvec{k}| \ll |\bvec{p}|$. This enables 
to expand the HQ momentum distribution function,
$f_Q$, and the first argument of the transition rate, $w$, 
in the collision integral, Eq.~(\ref{2.1.3}), with respect to 
$\bvec{k}$ up to second
order,\footnote{According to the Pawula theorem\cite{pawula1967alb} any
  truncation of the collision integral at finite order is only
  consistent with fundamental properties of Markov processes if the
  truncation is made at the 2$^{\mathrm{nd}}$-order term.}
\begin{equation}
\begin{split}
\label{2.1.6}
  w(\bvec{p}+\bvec{k},\bvec{k}) f_Q (\bvec{p}+\bvec{k},\bvec{k}) \simeq &
  w(\bvec{p},\bvec{k}) f_Q(\bvec{p}) \\
& + \bvec{k}
  \frac{\partial}{\partial{\bvec{p}}}[w(\bvec{p},\bvec{k}) f_Q(\bvec{p})] +
  \frac{1}{2} k_i k_j \frac{\partial^2}{\partial p_i \partial p_j}
  [w(\bvec{p},\bvec{k}) f_Q(\bvec{p})] \ 
\end{split}
\end{equation}
($i,j$=1,2,3 denote the spatial components of the 3-vectors, with
standard summation convention for repeated indices). 
The collision integral then simplifies to
\begin{equation}
\label{2.1.7}
C[f_Q] \simeq \int \dd^3 \bvec{k} \left [k_i \frac{\partial}{\partial p_i}
  + \frac{1}{2} k_i k_j \frac{\partial^2}{p_i p_j} \right ]
[w(\bvec{p},\bvec{k}) f_Q(\bvec{p})] \ ,
\end{equation}
i.e., the Boltzmann equation (\ref{boltz2}) is approximated by the
Fokker-Planck equation,
\begin{equation}
\label{2.1.8}
\frac{\partial}{\partial t} f_Q(t,\bvec{p}) = \frac{\partial}{\partial p_i}
\left \{ A_{i}(\bvec{p}) f_Q(t,\bvec{p})+\frac{\partial}{\partial
    p_j}[B_{ij}(\bvec{p}) f_Q(t,\bvec{p})] \right \} \ .
\end{equation}
The drag and diffusion coefficients are given according to
Eq.~(\ref{2.1.7}) by
\begin{equation}
\begin{split}
\label{2.1.9}
A_i(\bvec{p})&=\int \dd^3 \bvec{k} w(\bvec{p},\bvec{k}) k_i \ , \\
B_{ij}(\bvec{p}) &= \frac{1}{2} \int \dd^3 \bvec{k} w(\bvec{p},
\bvec{k}) k_i k_j \ .
\end{split}
\end{equation}
For an isotropic background medium, especially in the case of (local)
equilibrium (implying that the coefficients are defined in the local
rest frame of the heat bath), rotational symmetry enables to simplify
the coefficients to
\begin{equation}
\begin{split}
\label{2.1.10}
A_i(\bvec{p}) &= A(\bvec{p}) p_i \ , \\
B_{ij}(\bvec{p}) &= B_0(\bvec{p}) P_{ij}^{\parallel}(\bvec{p}) +
B_1(\bvec{p}) P_{ij}^{\perp}(\bvec{p})\ ,
\end{split}
\end{equation}
where the projection operators on the longitudinal and transverse
momentum components read
\begin{equation}
\label{2.1.11}
P_{ij}^{\parallel}(\bvec{p}) = \frac{p_i p_j}{\bvec{p}^2} \ , \quad
P_{ij}^{\perp}(\bvec{p}) = \delta_{ij} - \frac{p_i p_j}{\bvec{p}^2} \ .
\end{equation}
Implementing these simplifications into the collision integral,
Eq.~(\ref{2.1.3b}), the scalar drag and diffusion coefficients in
Eq.~(\ref{2.1.10}) are given by integrals of the form
\begin{equation}
\begin{split}
\label{2.1.10b}
\erw{X(\bvec{p}')}= &\frac{1}{2 \omega_{\bvec{p}}} \tildeint{\bvec{q}}
\tildeint{\bvec{p}'} \tildeint{\bvec{q}'} \frac{1}{\gamma_Q} \sum_{g,q} 
|\mathcal{M}|^2 \\
& \times (2 \pi)^4 \delta^{(4)}(p+q-p'-q') f_{q,g}(\bvec{q}) X(\bvec{p}') \ .
\end{split}
\end{equation}
In this notation, the coefficients can be written as
\begin{equation}
\begin{split}
\label{2.1.10c}
A(\bvec{p}) &= \erw{1-\frac{\bvec{p} \bvec{p}'}{\bvec{p}^2}} \ , \\
B_{0}(\bvec{p}) &= \frac{1}{4} \erw{\bvec{p}'{}^2-\frac{(\bvec{p}'
    \bvec{p})^2}{\bvec{p}^2}} \ , \\
B_{1}(\bvec{p}) &= \frac{1}{2} \erw{\frac{(\bvec{p'}
    \bvec{p})^2}{\bvec{p}^2} - 2 \bvec{p}' \bvec{p} + \bvec{p}^2} \ . 
\end{split}
\end{equation}
Note that Eq.~(\ref{2.1.10b}) includes the sum over gluons and light
quarks ($u$, $d$, $s$).

The physical meaning of the coefficients becomes clear in the
non-relativistic approximation of constant coefficients,
$\gamma\equiv A(\bvec{p})=\text{const}$ and
$D\equiv B_0(\bvec{p})=B_1(\bvec{p})=\text{const}$, in which case
the Fokker-Planck equation further simplifies to
\begin{equation}
\label{2.1.12}
\frac{\partial}{\partial t} f_Q(t,\bvec{p}) = \gamma
\frac{\partial}{\partial p_i} [p_i f_Q(t,\bvec{p})] 
              + D \Delta_{\bvec{p}} f_Q(t,\bvec{p}) \ .
\end{equation}
E.g., for an  initial condition 
\begin{equation}
\label{2.1.13}
f_Q(t=0,\bvec{p})=\delta^{(3)}(\bvec{p}-\bvec{p}_0) \ , 
\end{equation}
the solution takes the form of a Gaussian distribution,
\begin{equation}
\label{2.1.14}
f_Q(t,\bvec{p})= \left \{\frac{\gamma}{2 \pi D} \left[1-\exp(-2 \gamma
    t)\right] \right \}^{-3/2} \exp \left[-\frac{\gamma}{2 D}
  \frac{[\bvec{p}-\bvec{p}_0 \exp(-\gamma t)]^2}{1-\exp(-2 \gamma t)} 
  \right ] \ .
\end{equation}
From the equation for the mean momentum,
\begin{equation}
\label{2.1.15}
\erw{\bvec{p}}=\bvec{p_0} \exp(-\gamma t) \ ,
\end{equation}
one sees that $\gamma$ determines the relaxation rate of the average
momentum to its equilibrium value, i.e., it is a drag or friction
coefficient. The standard deviation of the momentum evolves according to
\begin{equation}
\label{2.1.16}
\erw{\bvec{p}^2}-\erw{\bvec{p}}^2=\frac{3 D}{\gamma} [1-\exp(-2 \gamma t)] \ ,
\end{equation}
i.e., $D$ is the momentum-diffusion constant, describing the momentum
fluctuations.

In the limit $t \rightarrow \infty$, Eq.~(\ref{2.1.14}) approaches the
(non-relativistic) Boltzmann distribution,
\begin{equation}
\label{2.1.17}
f_Q(t,\bvec{p}) = \left (\frac{2 \pi D}{\gamma} \right )^{3/2} \exp\left
  (-\frac{\gamma \bvec{p}^2}{2 D} \right ) \ .
\end{equation}
Since in thermal equilibrium the heavy quarks have to obey  
an equilibrium distribution with the temperature, $T$, of the heat bath,
the drag and diffusion coefficients should satisfy the \emph{Einstein
  dissipation-fluctuation relation},
\begin{equation}
\label{einstein}
D=m_Q \gamma T \ .
\end{equation}
The relativistic Fokker-Planck equation will be discussed in
Sec.~\ref{ssec_langevin} in connection with its formulation in terms
of stochastic Langevin equations.

We note that the spatial diffusion coefficient, $D_s$, which describes
the broadening of the spatial distribution with time,  
\begin{equation}
\label{spat-diff-coeff}
\erw{\bvec{x}^2(t)}-\erw{\bvec{x}(t)}^2 \simeq 6 D_s t \ ,
\end{equation}
is related to the drag and momentum-diffusion coefficient through
\begin{equation}
\label{spat-diff-mom-diff}
D_s=\frac{T}{m_Q \gamma}=\frac{T^2}{D}.
\end{equation}


\subsection{Perturbative QCD Approaches}
\label{ssec_pqcd}
In a first step to evaluate HQ diffusion in a QGP perturbation theory
has been applied, thereby approximating the medium as a weakly
interacting system of quark and gluon quasiparticles. Such a treatment
is expected to be reliable if the temperature is large enough
for the typical momentum transfers, $Q^2\sim T^2$, to be in the
perturbative regime, $Q^2 \ge 2 \; \GeV^2$ or so. This is most
likely not satisfied for matter conditions realized at 
SPS and RHIC. For more realistic applications to experiment several
amendments of the perturbative approach have been suggested which
are discussed subsequently (focusing again on elastic 
HQ scattering on light partons).

\subsubsection{Schematic Leading Order}
\label{sssec_lo-0}
\begin{figure}[!t]
\begin{center}
\begin{minipage}{0.27\linewidth}{\includegraphics[width=\textwidth]{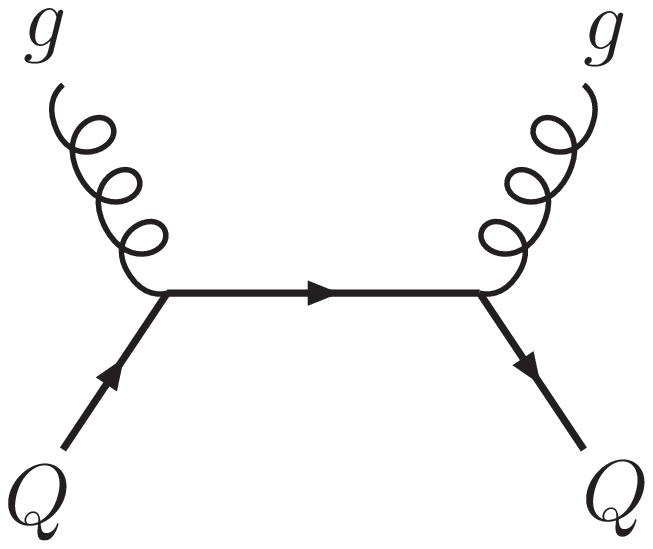}}
\end{minipage}
\begin{minipage}{0.27\linewidth}{\includegraphics[width=\textwidth]{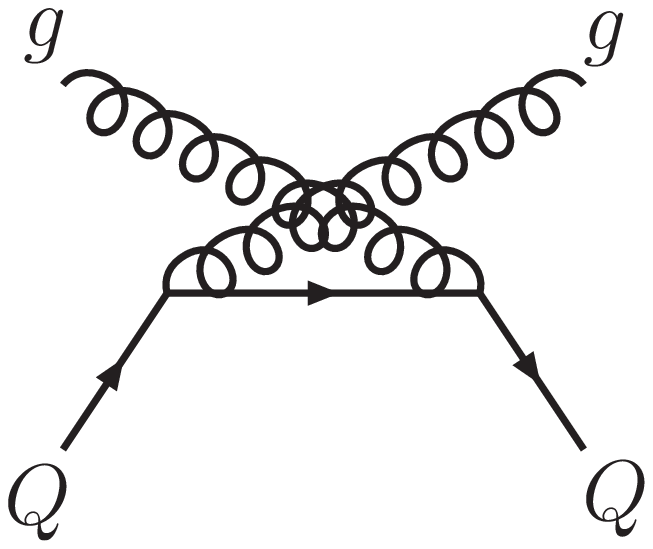}}
\end{minipage}
\begin{minipage}{0.2\linewidth}{\includegraphics[width=\textwidth]{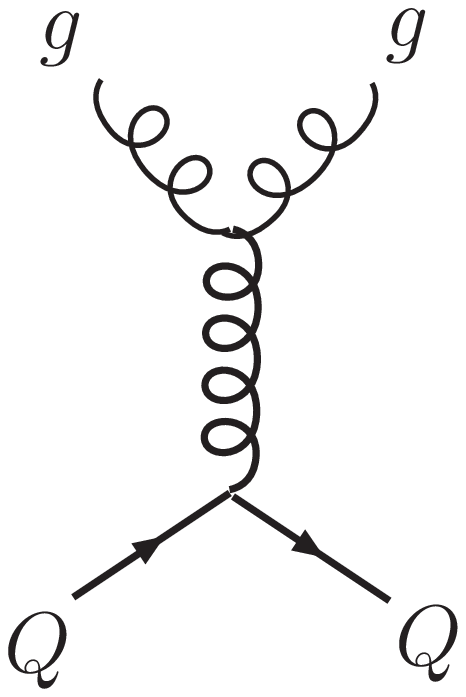}}
\end{minipage}
\begin{minipage}{0.2\linewidth}{\includegraphics[width=\textwidth]{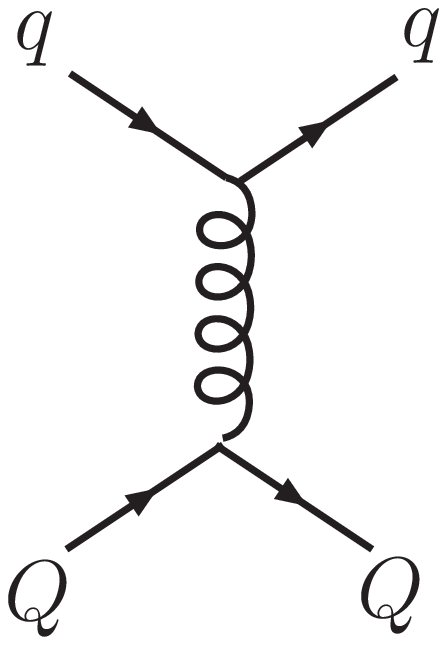}}
\end{minipage}
\end{center}
\caption{Feynman diagrams for leading-order perturbative HQ scattering
  off light partons.}
\label{fig_dia-pqcd}
\end{figure}
The initial estimates of equilibration times and energy loss of heavy
quarks in the QGP\cite{Svetitsky:1987gq} have started from the
leading-order (LO) perturbative diagrams involving the minimum of two
strong-interaction vertices, as displayed in Fig.~\ref{fig_dia-pqcd}.
Pertinent matrix elements\cite{Combridge:1978kx} figuring into
Eq.~(\ref{2.1.10b}) in the vacuum have been computed in
Ref.\cite{Combridge:1978kx}. The dominant contribution arises from gluon
$t$-channel exchange, i.e., the $3^{\mathrm{rd}}$ and $4^{\mathrm{th}}$
diagram in Fig.~\ref{fig_dia-pqcd}. For forward scattering, the gluon
propagator develops the well-known infrared singularity which has been
regularized by introducing a Debye-screening mass,
\begin{equation}
G(t) = \frac{1}{t} \to \frac{1}{t-\mu_D} \ , \quad \mu_D = g T \ ,
\label{Gdebye}
\end{equation}
where $g=\sqrt{4 \pi \alpha_s}$ denotes the strong coupling constant.
Even for a value as large as $\alpha_s=0.4$, and at a temperature of
$T=300\; \MeV$ (typical for the early stages in heavy-ion collisions at
RHIC), the thermal relaxation time, $\tau_{\mathrm{eq}} = 1/\gamma$, for
charm (bottom) quarks turns out around $\sim$$15(40)\,\fm/c$ (and 
therefore much larger than a typical QGP lifetime of
$\sim$5\,fm/$c$ at RHIC), see, e.g., right panel of
Fig.~\ref{fig_tau-reso} (in Ref.\cite{GolamMustafa:1997id} the
corrections due to quantum-equilibrium distributions (Bose/Fermi) 
have been investigated and found to be small). Note that
with the above gluon propagator, the pertinent total HQ-parton cross
section is parametrically given by $\sigma_{Qp} \propto
\alpha_s^2/\mu_D^2$, i.e., it essentially increases only linearly in
$\alpha_s$ ($p=q,\bar q,g$).

\subsubsection{Leading Order with Hard Thermal Loop Resummation}
\label{sssec_lo-htl}

\begin{figure}[!t]
\begin{minipage}{0.45\linewidth}
\includegraphics[width=\textwidth]{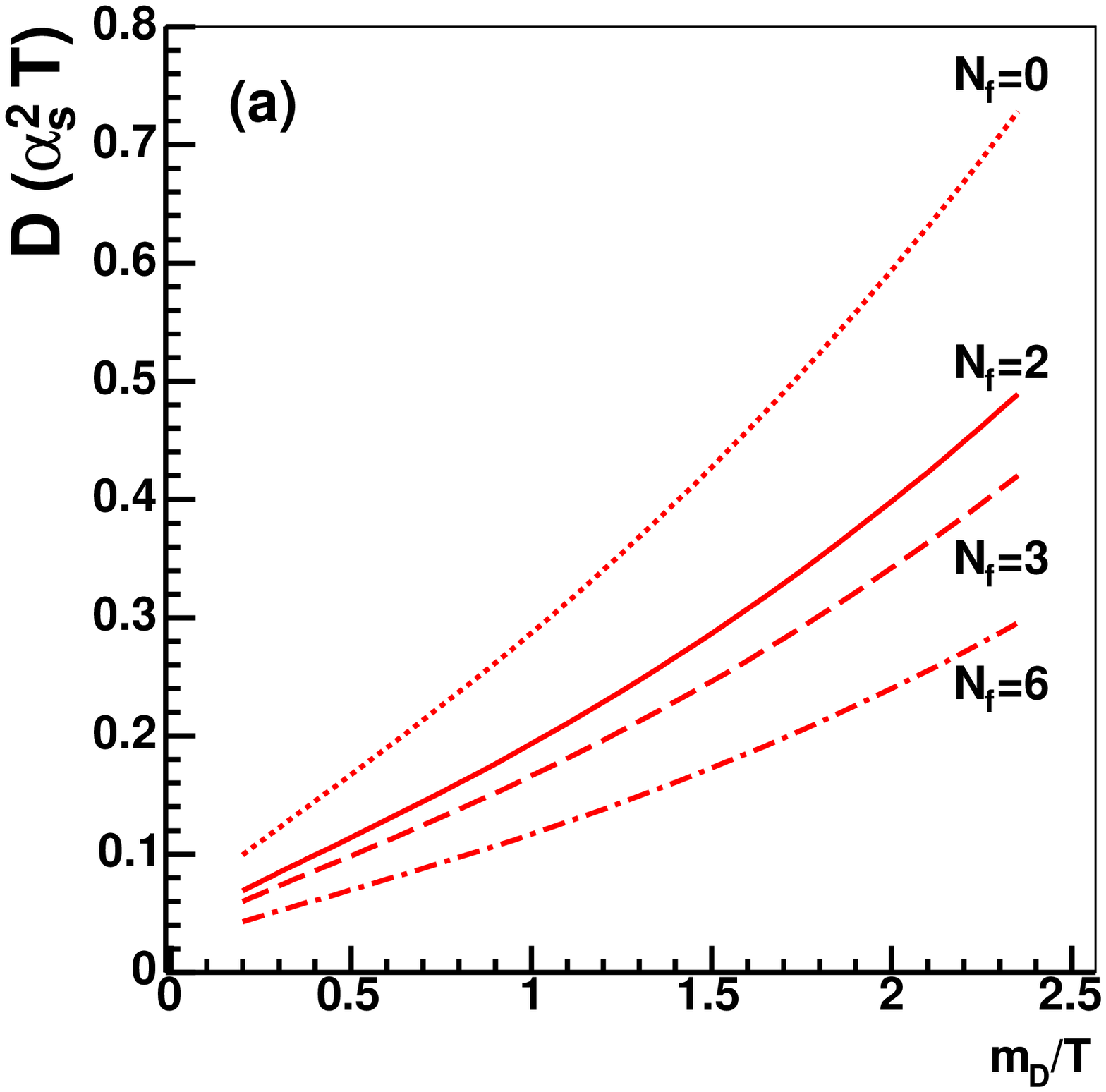}
\end{minipage} \hfill
\begin{minipage}{0.45\linewidth}
\includegraphics[width=\textwidth]{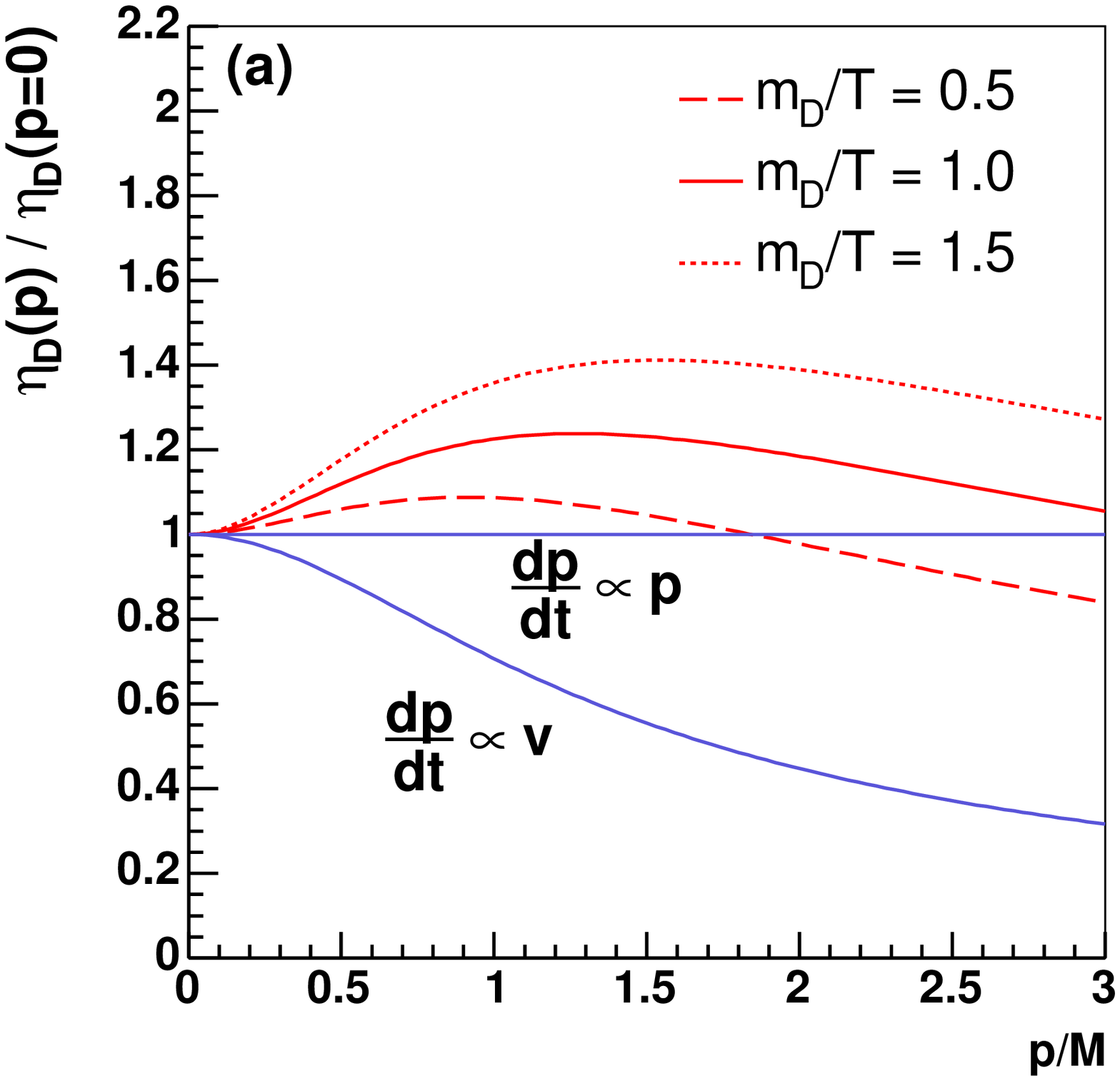}
\end{minipage}
\caption{(Color online) HQ transport coefficients in HTL 
  improved perturbation theory\protect\cite{Moore:2004tg}. Left panel:
  spatial diffusion coefficient at $p=0$ as a function of an
  independently varied Debye mass, $m_D\equiv\mu_D$, figuring into 
  the $t$-channel gluon exchange propagator, for different quark-flavor
  content of the (Q)GP. Right panel: momentum dependence of
  the drag coefficient, $\eta_D(p)\equiv A(p)$, for three values of
  $\mu_D$ in $t$-channel gluon exchange; the lower curve, 
  with $\dd p/\dd t\propto v$, resembles a calculation in the 
  non-relativistic limit ($M\equiv m_c =1.4\,\GeV$
  in the $x$-axis label denotes the charm-quark mass).}
\label{fig_MT-coeffs}
\end{figure}

In Ref.\cite{Moore:2004tg}, the schematic introduction of the Debye mass
into the $t$-channel gluon-exchange propagator has been extended by a LO
hard-thermal loop (HTL) calculation of the charm-quark drag and
diffusion coefficients in the QGP. In this approach, the screening of
the gluon propagator in the $t$-channel diagrams
(Fig.~\ref{fig_dia-pqcd}) is realized by inserting the HTL gluon
propagator for the region of small momentum exchange. In Coulomb gauge,
with $q=|\bvec{q}|$, this propagator is given by
\begin{equation}
\label{htl.1}
G_{\mu \nu}(\omega,q)=-\frac{\delta_{\mu 0} \delta_{\nu 0}}{q^2+\Pi_{00}} 
+ \frac{\delta_{ij} - q_i q_j/q^2}{q^2-\omega^2+\Pi_T} \ ,
\end{equation}
where the $i,j\in \{1,2,3\}$ denote the spatial components of $\mu,\nu
\in \{0,1,2,3\}$. The HTL self-energies read
\begin{equation}
\begin{split}
\label{Ghtl}
\Pi_T(\omega,\bvec{q}) &= \mu_D^2 \left \{\frac{\omega^2}{2\bvec{q}^2} +
  \frac{\omega(\bvec{q}^2-\omega^2)}{4 q^3} \left [\ln \left
      (\frac{q+\omega}{q-\omega} \right ) - \ii \pi \right ] \right \} \
,
\\
\Pi_{00}(\omega,\bvec{q}) &= \mu_D^2 \left \{ 1 - \frac{\omega}{2 q} \left [
    \ln \left (\frac{q+\omega}{q-\omega} \right ) - \ii \pi \right ]
\right \} \ .
\end{split}
\end{equation}
For small energy transfers, $\omega$, and a slowly moving heavy quark,
$v\ll 1$, only the time component of the propagator contributes to the
squared matrix elements which in this limit reduces to the
Debye-screened Coulomb-like propagator,
Eq.~(\ref{Gdebye}). Fig.~\ref{fig_MT-coeffs} shows the spatial diffusion
coefficient and the momentum dependence of the drag coefficient
resulting from this calculation. Compared to the screening description
with a constant Debye mass, the drag coefficient shows a slight increase
for an intermediate range of momenta (cf., e.g., the pQCD curves in
Fig.~\ref{fig_gamma-tmat}).

\subsubsection{Leading Order with Running Coupling}
\label{sssec_lo-run}
As indicated in the Introduction, the current data situation at RHIC
does not allow for an understanding of the electron data in terms of LO
pQCD with reasonably small coupling constant (say, $\alpha_s\le 0.4$).
This was a motivation for more recent 
studies\cite{Peshier:2008bg,Gossiaux:2008jv},
augmenting the LO pQCD framework in search for stronger effects. Two
basic amendments have been introduced. First, the idea of
Ref.\cite{Moore:2004tg} of introducing a reduced screening mass in the
gluon propagator was made more quantitative.  Starting from an ansatz
for the screened gluon propagator,
\begin{equation}
\label{gos.1}
G_r(t) \propto \frac{1}{t- r \mu_D^2} \ .
\end{equation}
the objective is to obtain an estimate for the constant $r$ (it was
denoted $\kappa$ in Refs.\cite{Peshier:2008bg,Gossiaux:2008jv}; we 
changed the notation to
avoid conflicts in what follows below). This has been done in analogy to
a corresponding QED calculation\cite{Braaten:1991jj,Braaten:1991we}, by
requiring that the energy loss of a high-energy quark obtained in a
LO-pQCD calculation with the screened propagator, Eq.~(\ref{gos.1}),
matches a calculation where for low momentum transfers, $|t|<|t^*|$, the
HTL propagator, Eq.~(\ref{htl.1}), and for $|t|>|t^*|$ the perturbative
gluon propagator, Eq.~(\ref{Gdebye}), is used; $|t^*|$ is a
momentum-transfer scale between $g^2 T^2$ and $T^2$. The QED
calculation\cite{Braaten:1991jj,Braaten:1991we} yields an energy loss
which is independent of the matching scale $|t^*|$, while this is not
the case in QCD. This problem is treated by introducing an
infrared-regulator mass into the hard part of the energy-loss integrals
involving the $t$-channel exchange-matrix elements, chosen such that the
dependence on $|t^*|$ is weak for $|t^*|<T^2$ (the validity range of the
HTL approximation). This translates into effective values for the $r$
coefficient in Eq.~(\ref{gos.1}) of $r \simeq 0.15$-0.2.

Second, a running strong coupling constant is introduced well into the
nonperturbative regime but with an infrared-finite limit. The
justification for such a procedure\cite{Dokshitzer:1995qm} is that it
can account for (low-energy) physical observables (e.g., in $e^+e^-$
annihilation\cite{Mattingly:1993ej}) in an effective way. The
parameterization adopted in Refs.\cite{Peshier:2008bg,Gossiaux:2008jv} 
is based on an
extrapolation of Ref.\cite{Dokshitzer:1995qm} into the spacelike regime,
\begin{equation}
\label{gos.2}
\alpha_{\mathrm{eff}}(Q^2)=\frac{4 \pi}{\beta_0} \begin{cases}
L_-^{-1} & \text{for $Q^2 \leq 0$} \\
1/2-\pi^{-1} \arctan(L_+/\pi) & \text{for $Q^2>0$},
\end{cases}
\end{equation}
where $\beta_0=11-2 N_f/3$, $N_f=3$, and $L_{\pm}=\ln(\pm
Q^2/\Lambda^2)$. The pertinent substitution in the $t$-channel
gluon-exchange matrix elements amounts to
\begin{equation}
\label{gos.3}
\frac{\alpha}{t} \rightarrow \frac{\alpha_{\mathrm{eff}}(t)}{t-\tilde{\mu}^2}
 \ ,
\end{equation}
where the regulator mass is chosen as $\tilde{\mu}^2 \in [1/2,2]
\tilde{\mu}_D^2$, while the Debye-screening mass is determined
self-consistently from the equation
\begin{equation}
\label{gos.4}
\tilde{\mu}_D^2=\left(\frac{N_c}{3} +\frac{N_f}{6} \right) 4 \pi
\alpha(-\tilde{\mu}_D^2) T^2 \ .
\end{equation}
\begin{figure}[!t]
\begin{minipage}{0.45\linewidth}
\includegraphics[width=\textwidth]{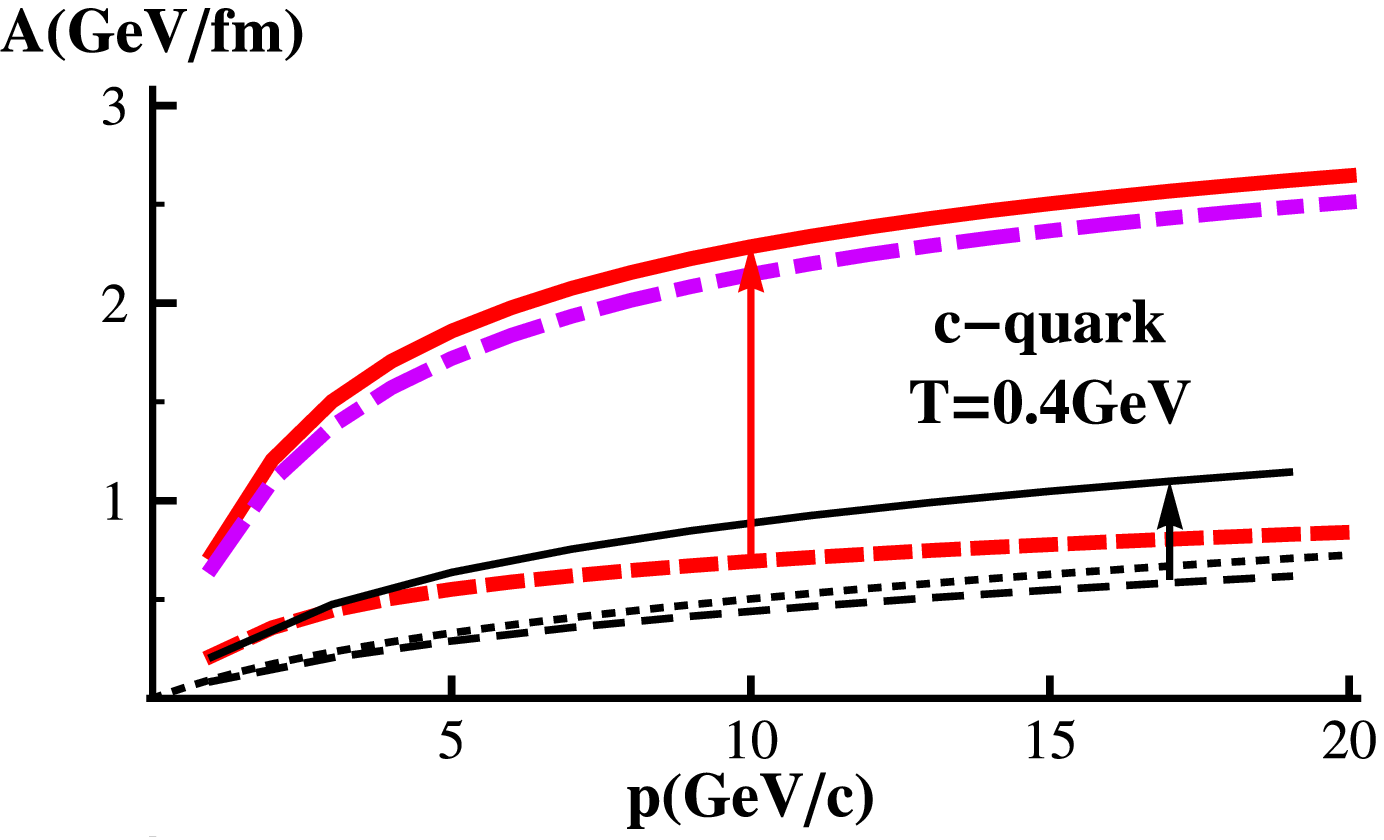}
\end{minipage} \hfill
\begin{minipage}{0.5\linewidth}
\includegraphics[width=\textwidth]{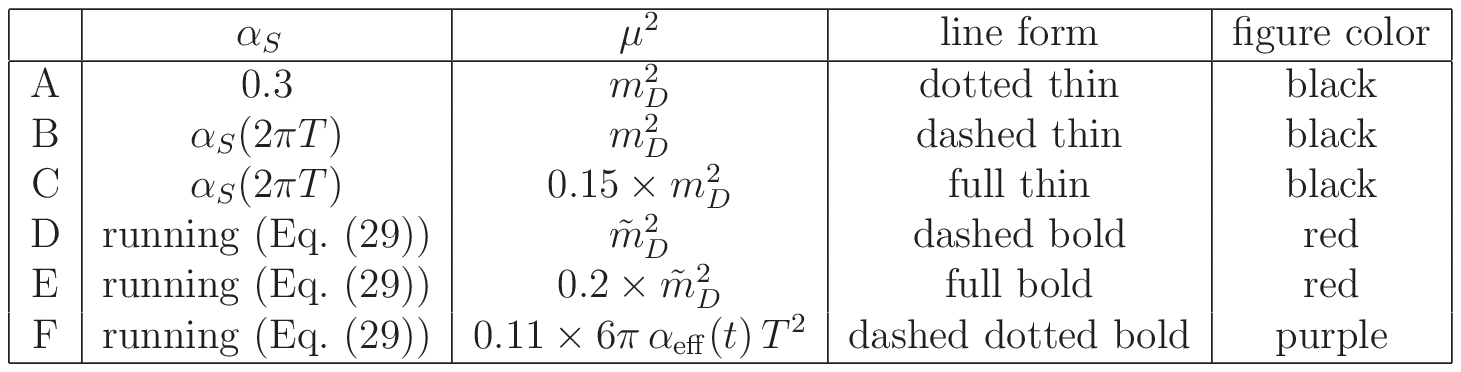}
\end{minipage}
\caption{(Color online) The drag coefficient as a function of HQ
  three-momentum in the amended pQCD scheme with reduced infrared
  regulator and running coupling constant (left
  panel)\protect\cite{Peshier:2008bg,Gossiaux:2008jv}. 
  The corresponding legend (right panel) details the different 
  parameter choices in the calculation.}
\label{fig.gos-coeffs}
\end{figure}

To find the optimal value for the regulator mass a similar strategy of
matching the energy loss with a Born approximation has been employed,
using the substitution, Eq.~(\ref{gos.3}), in the $t$-channel diagrams,
with a HTL calculation along the same lines as summarized above for the
calculation with non-running $\alpha_s$. The results for the drag
coefficients for charm quarks under the various model assumptions
described above are depicted in Fig.~\ref{fig.gos-coeffs}. Changing the
screening mass from the standard Debye mass, $\mu_D$, to that
reproducing the HTL energy loss, with $r=0.15$ in Eq.~(\ref{gos.1}),
increases the drag coefficient by a factor of $2$. In view of the large
reduction in $r$ this appears to be a rather moderate effect. This is
simply due to the fact that the change mostly enhances forward
scattering which is little effective in thermalizing (isotropizing) a
given momentum distribution. Implementing the running-coupling scheme
with a small screening mass yields a substantial enhancement by a factor
of $\sim$$5$.

\subsubsection{Next-to-Leading Order}
\label{sssec_nlo}
The rather large values of the coupling constant employed in the
calculations discussed in the previous sections imminently raise
questions on the convergence of the perturbative series.  This problem
has been addressed in a rigorous next-to-leading-order (NLO) calculation
for the HQ momentum-diffusion coefficient, $\kappa=2 D$, in
Refs.\cite{CaronHuot:2007gq,CaronHuot:2008uh}. This work starts from the
definition of $\kappa$ as the mean squared momentum transfer per unit
time, which in gauge theories is given by the time-integrated correlator
of color-electric-field operators connected by fundamental Wilson lines:
\begin{equation}
\label{kappa-gauge}
\kappa=\frac{g^2}{3 d_H} \int \dd t \Tr_H \erw{W^{\dagger}(t,0) E_i^a
  (t) T_H^a W(t,0) E_i^b(0) T_H^b} \ ;
\end{equation}
$W(t;0)$ denotes a fundamental Wilson line running from $t'=0$ to $t$
along the static trajectory of the heavy quark, $T_H^a$ are the
generators of the gauge group in the representation of the heavy quark
and $d_H$ its dimension. In leading order this reduces to a
Wightman-two-point function of $A^0$ fields at zero frequency, i.e., in
the usual real-time propagator notation,
\begin{equation}
\label{kappa-lo}
\kappa \simeq \frac{C_H g^2}{3} \int \frac{\dd^3 \bvec{p}}{(2 \pi)^3}
p^2 G^{>00}(\omega=0,\bvec{p}),
\end{equation}
with $C_H=4/3$ the Casimir operator of the HQ representation. The
integral is IR regulated by HTL corrections, i.e., a Debye mass,
$\mu_D^2=g^2 T^2 (N_c+N_f/2)/3$.
\begin{figure}[!t]
\begin{minipage}{0.4\linewidth}
\includegraphics[width=\textwidth]{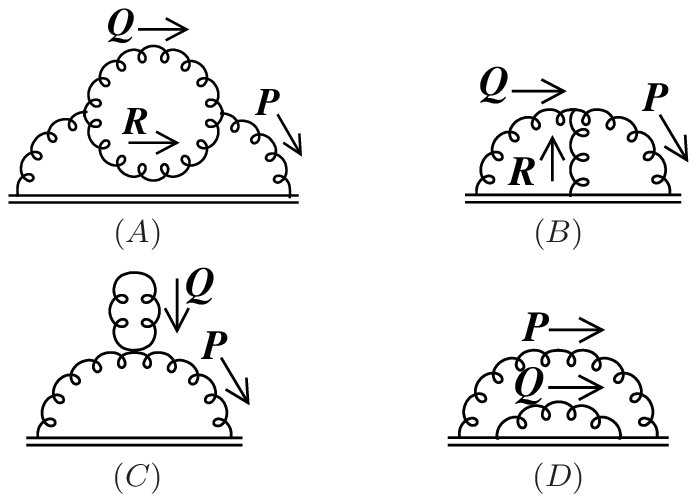}
\end{minipage} \hfill
\begin{minipage}{0.55\linewidth}
\includegraphics[width=\textwidth]{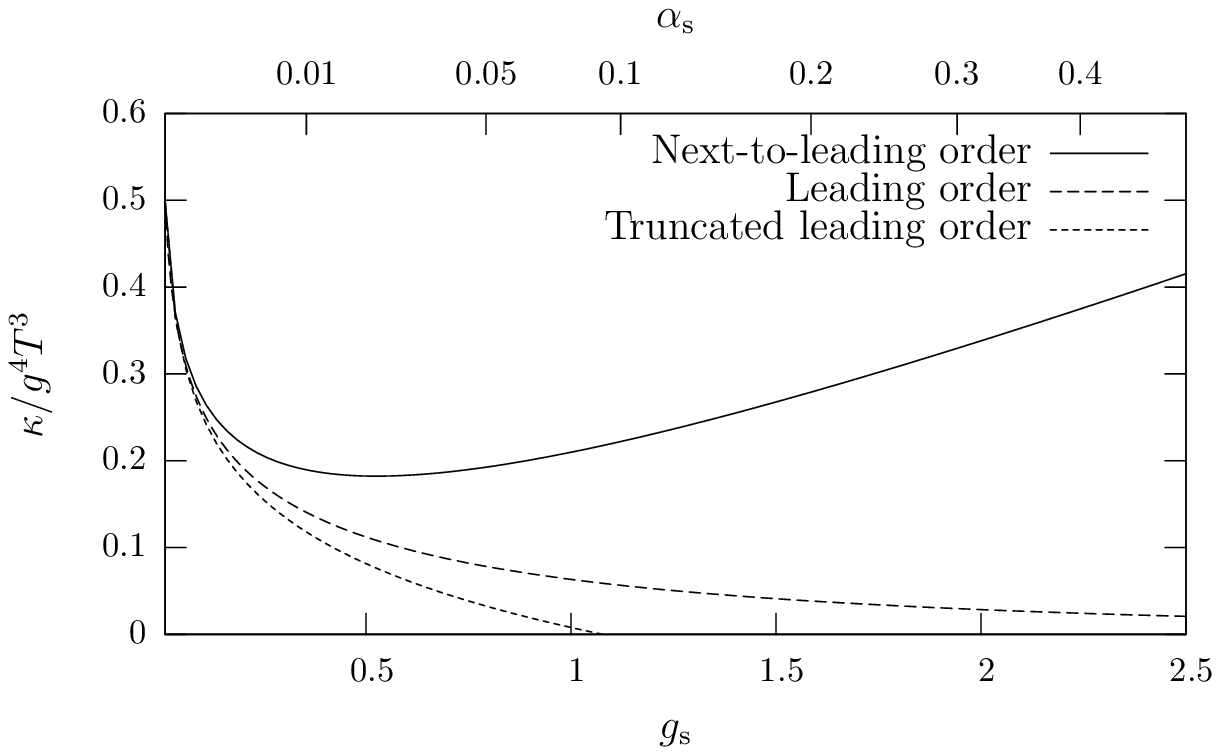}
\end{minipage}
\caption{NLO calculations for HQ diffusion in the 
  QGP\protect\cite{CaronHuot:2007gq}. 
  Left panel: NLO diagrams for the momentum-diffusion coefficient, 
  $\kappa$; the double line represents the heavy quark, all propagators 
  are soft and HTL resummed, and all vertices include HTL vertices. 
  Right panel: comparison of LO to NLO result for $\kappa$ as 
  a function of the strong coupling, $\alpha_s$.}
\label{fig.nlo-kappa-CHM}
\end{figure}
In the left panel of Fig.~\ref{fig.nlo-kappa-CHM} the NLO corrections to
the LO result, Eq.~(\ref{kappa-lo}), are depicted in terms of Feynman
diagrams. The double line represents the heavy quark, and all
propagators and vertices include HTL corrections, leading to a gauge
invariant expression as it should be the case for an observable quantity
like $\kappa$. The diagrams are evaluated in Coulomb gauge within the
closed-time path (real-time) Keldysh formalism of thermal quantum-field
theory (TQFT). 
The real part of diagram (A) provides a correction to the Debye mass. 
Diagrams (C) and (D) take into account real and virtual corrections 
by additional soft scattering or plasmon emission/absorption of the 
light or heavy scatterer, respectively. Diagram (B) represents
interference between scattering events occurring on the light
scatterer's and on the heavy quark's side. Contrary to naive power
counting, the NLO calculation provides $\mathcal{O}(g)$ corrections due
to scattering with soft gluons with momentum, $q \simeq \mu_D$, and due
to overlapping scattering events, dominated by $t$-channel Coulombic
scatterings involving soft momentum transfers, $\simeq \mu_D \propto
g T$. The right panel of Fig.~\ref{fig.nlo-kappa-CHM} shows that the
NLO correction to $\kappa$ is positive, i.e., the momentum-diffusion
coefficient becomes larger compared to the LO calculation. The
convergence is poor even for rather small coupling constants. A 
rigorous resummation scheme to cure this behavior is not known 
to date, especially to establish convergence in the typical range of
coupling constants under conditions in relativistic heavy-ion
collisions, $\alpha_s \simeq 0.3$-$0.4$. In Ref.\cite{CaronHuot:2008uh} 
the investigation of NLO corrections is
extended to the weak-coupling limit of $\mathcal{N}=4$ supersymmetric
Yang-Mills (SYM) theory. Also in this case the perturbative series turns
out to be poorly convergent, even for low couplings.

\subsubsection{Three-body elastic scattering}
\label{sssec_lo-3}

\begin{figure}
\begin{center}
\includegraphics[width=0.6\textwidth]{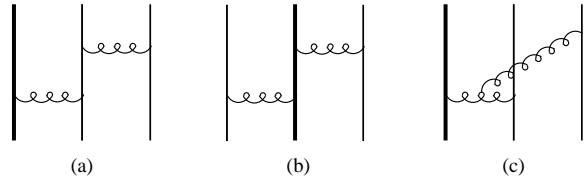}
\end{center}
\caption{Different topological classes of diagrams for three-body 
 elastic scattering of a heavy quark (thick lines) off light
  quarks and antiquarks (thin lines)\protect\cite{Liu:2006vi}.}
\label{fig_3-body}
\end{figure}

Another step in the (would-be) perturbative hierarchy are three-body
collisions, which are expected to become increasingly important at high
parton density. An attempt to assess the effects of three-body elastic
scattering for HQ diffusion has been conducted in Ref.\cite{Liu:2006vi},
with pertinent Feynman diagrams as depicted in Fig.~\ref{fig_3-body}.
Special care has to be taken in regularizing contributions from diagrams
with intermediate particles going on-shell; these can lead to divergent
real parts in the scattering amplitude and represent successive two-body
scatterings (rather than genuine three-body scattering).  Therefore, in
Ref.\cite{Liu:2006vi} the intermediate quark lines in diagram (a) and
(b) are supplemented with an in-medium collisional width, and only the
real part of their propagator is kept in the evaluation of the
diagrams. For three-body elastic processes involving one or two gluons,
it has been assumed that the dominant contributions arise from diagrams
with similar topology as diagram (b) in Fig.~\ref{fig_3-body} for $Qqq$
scattering; all other contributions are neglected. To compare with
two-body gluo-radiative inelastic scattering the LO diagrams have been
used to evaluate matrix elements for $Qq \to Qqg$, $Q\bar{q} \to Q
\bar{q} g$ and $Qg \to Qgg$ processes. Within this scheme, at
temperatures $T=200$-$300\,\MeV$, three-body elastic scattering
processes are estimated to contribute to the $c$- and $b$-quark friction
coefficients with a magnitude comparable to two-body elastic scattering.
Again, this raises the question of how to control the perturbative
series for HQ diffusion.  As a by-product, the friction coefficient for
radiative scattering, $Qp\to Qpg$, was estimated to exceed the one from
elastic two-body scattering for HQ momenta $p\gsim 12 \, \GeV$ (for both
charm and bottom).

\subsection{Non-Perturbative Interactions}
\label{ssec_non-pert}

The evidence for the formation of a strongly coupled QGP (sQGP) at RHIC
has motivated vigorous theoretical studies of the possible origin of the
interaction strength (see, e.g., Ref.\cite{Shuryak:2008eq} for a recent
review). In particular, several lattice QCD computations of hadronic
correlation functions at finite temperature have found indications that
hadronic resonances (or bound states) survive up to temperatures of
twice the critical one or more (for both a gluon plasma (GP) and a
QGP)\cite{Asakawa:2003re,Karsch:2003jg,Asakawa:2003nw,Aarts:2007pk},
cf.~also Sec.~\ref{sssec_lat} of this article. Pertinent spectral
functions (extracted from Euclidean correlators using probabilistic
methods, i.e., the maximum entropy method) exhibit resonance peaks in
both $Q\bar{Q}$ and $q\bar{q}$ channels. The consequences of hadronic
resonances in the QGP for HQ transport have been elaborated in
Refs.\cite{vanHees:2004gq,vanHees:2005wb,Adil:2006ra,vanHees:2007me}.
The starting point in
Refs.\cite{vanHees:2004gq,vanHees:2005wb,Adil:2006ra} is the postulate
that heavy-light quark ($Q\bar{q}$) resonances, i.e., ``$D$'' and
``$B$'' mesons, persist in the QGP. In
Refs.\cite{vanHees:2004gq,vanHees:2005wb} this has been realized within
an effective resonance model for $Q$-$\bar q$ scattering
(Sec.~\ref{sssec_reso}) while in Ref.\cite{Adil:2006ra} HQ fragmentation
into mesons and their subsequent momentum broadening was considered
(Sec.~\ref{sssec_coll-diss}). The phenomenological success of these
models (cf., e.g., the right panel of Fig.~\ref{fig_elec-reso}) called
for a more microscopic evaluation of the heavy-light quark
correlations. This was realized in Ref.\cite{vanHees:2007me} where
in-medium heavy-light quark $T$-matrices were computed with interaction
potentials estimated from HQ free energies in lattice QCD
(cf.~Sec.~\ref{sssec_tmat}). This approach is the direct analog to the
potential models used in the heavy quarkonium context
(cf.~Sec.~\ref{sssec_lat}). We finish this section with a brief
discussion of a recent suggestion to extract information on HQ diffusion
more directly from thermal lattice QCD (Sec.~\ref{sssec_hq-lat}), which
would constitute a valuable benchmark for both perturbative and
non-perturbative calculations.

\subsubsection{Effective $Q\bar{q}$-resonance model}
\label{sssec_reso}
The heavy-light quark resonance model\cite{vanHees:2004gq} has been set
up by combining HQ effective theory (HQET) with chiral symmetry in the
light-quark sector, $q=(u,d)$, based on the Lagrangian,
\begin{equation}
\begin{split}
\Lag_{Dcq} =& \Lag_D^0 + \Lag_{c,q}^0 - \ii G_S \left( \bar q \Phi_0^*
\frac{1+\fslash{v}}{2} c - \bar q \gamma^5 \Phi \frac{1+\fslash{v}}{2}
 c + h.c. \right)
\\
& - G_V \left( \bar q \gamma^{\mu} \Phi_{\mu}^* \frac{1+\fslash{v}}{2} c -
  \bar q \gamma^5 \gamma^{\mu} \Phi_{1\mu} \frac{1+\fslash{v}}{2} c + h.c.
\right) \ ,
\end{split}
\label{L_hq-eff}
\end{equation}
written in the charm sector (an equivalent one in the bottom sector
follows via the replacements $c\to b$ and $D\to B$ for the HQ and
resonance fields, respectively; $v$: HQ four-velocity). The pertinent
free Lagrangians read
\begin{equation}
\begin{split}
\Lag_{c,q}^0 &= \bar{c}(\ii \fslash{\partial}-m_c) c+\bar{q} \, \ii
\fslash{\partial} q,\\
\Lag_D^0 & =  (\partial_{\mu} \Phi^{\dagger})(\partial^{\mu} \Phi) +
(\partial_{\mu} {\Phi_0}^{*\dagger})(\partial^{\mu} \Phi_0^*)
-m_S^2(\Phi^{\dagger} \Phi+\Phi_0^{*\dagger} \Phi_0^*) \\
& \quad -\frac{1}{2} (\Phi_{\mu \nu}^{*\dagger} \Phi^{*\mu \nu}
+ \Phi_{1 \mu \nu}^{\dagger}
\Phi_1^{\mu \nu}) + m_V^2 (\Phi_{\mu}^{*\dagger} \Phi^{*\mu} +
\Phi_{1 \mu}^{\dagger} \Phi_1^{\mu}) \ .
\end{split}
\label{L_0-eff}
\end{equation}
$\Phi$ and $\Phi_0^*$ denote the pseudoscalar and scalar meson fields
(corresponding to $D$ and $D_0^*$ mesons) which are assumed to be
degenerate chiral partners (mass $m_S$) as a consequence of chiral
restoration in the QGP. The same reasoning applies to the vector and
axialvector states (mass $m_V$), $\Phi_{\mu}^*$ and $\Phi_{1 \mu}$
(corresponding to $D^*$ and $D_1^*$). HQ spin symmetry furthermore
asserts the degeneracy of spin-0 and -1 states with identical angular
momentum, implying $m_S=m_V$ and the equality of the coupling constants,
$G_S=G_V$. In the strange-quark sector only the pseudoscalar ($D_s$) and
vector ($D_s^*$) resonance states are considered (i.e., chiral symmetry
is not imposed).

\begin{figure}[!t]
\begin{minipage}{0.41\linewidth}
\includegraphics[width=\textwidth]{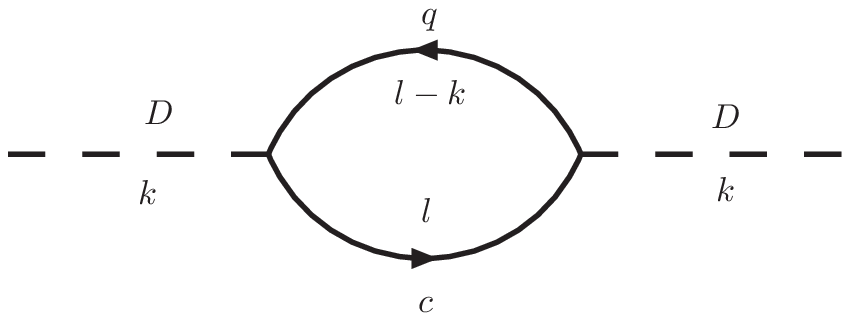}
\end{minipage}\hfill
\begin{minipage}{0.35\linewidth}
\includegraphics[width=\textwidth]{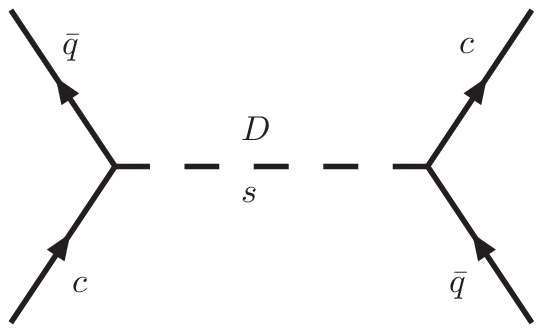}
\end{minipage}
\begin{minipage}{0.22\linewidth}
\includegraphics[width=\textwidth]{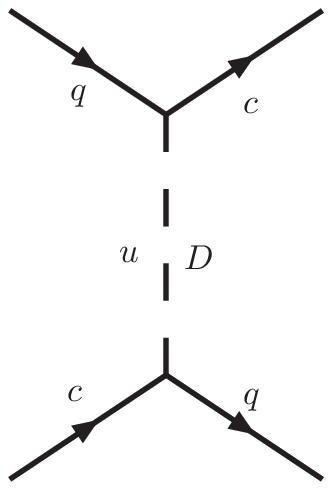}
\end{minipage}
\caption{Left panel: one-loop diagram representing the $D$-meson
  self-energy in the QGP within the effective resonance
  model\protect\cite{vanHees:2004gq}. Right and middle panels: elastic
  $Q\bar{q}$ and $Q q$ scattering diagrams for $s$- and $u$-channel
  resonance exchange, respectively.}
\label{fig_dia-reso}
\end{figure}
The boson-resonance propagators are dressed with heavy-light quark
self-energies at the one-loop level (cf.~left panel of
Fig.~\ref{fig_dia-reso}). To leading order in HQET, in accordance with
spin symmetry, the self-energies for the vector/axialvector resonances
are given by
\begin{equation}
\Pi_{D^*,\mu \nu}=(v_{\mu} v_{\nu}-g_{\mu \nu}) \Pi_D(s),
\end{equation}
where $s=p^2$ denotes the meson's four momentum, and $\Pi_D$ is the
self-energy of the pseudoscalar/scalar resonances. Its imaginary part
reads
\begin{equation}
\im \Pi_D(s)=-\frac{3 G^2}{8 \pi} \frac{(s-m_c^2)^2}{s} \Theta(s-m_c^2) \ ,
\end{equation}
while the real part is calculated from a twice-subtracted dispersion
relation with the wave-function and mass counter terms adjusted such
that the following renormalization conditions hold,
\begin{equation}
\label{2.1.37}
\partial_s \Pi_{D}^{(\text{ren})}(s)|_{s=0}=0 \ , \quad 
\re \Pi_{D}^{(\text{ren})}(s)|_{s=m_D^2}=0 \ . 
\end{equation}
As an alternative regularization scheme, dipole form factors,
\begin{equation}
\label{2.1.38}
F(|\bvec{q}|)=
\left ( \frac{2 \Lambda^2}{2 \Lambda^2+\bvec{q}^2} \right )^2 \ ,
\end{equation}
have been supplemented to simulate finite-size vertices of the resonance
model,
\begin{equation}
\label{2.1.39}
\im \Pi_{D}^{(\text{ff})}(s)= \im \Pi_{D}(s) F^2(|\bvec{q}|) \ ,
\end{equation}
with $|\bvec{q}|=(s-m_c^2)/(2\sqrt{s})$. In this scheme, the real part
is calculated from an unsubtracted dispersion relation, while the bare
resonance mass is adjusted to obey the second renormalization condition
in Eq.~(\ref{2.1.37}).

\begin{figure}[!t]
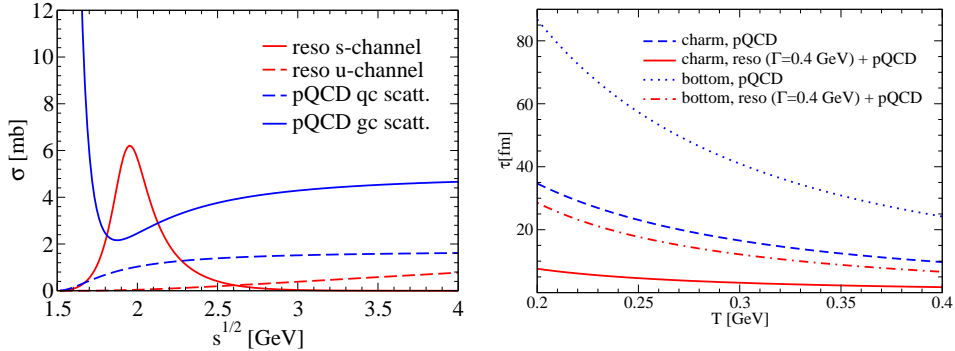

\begin{minipage}{0.48\textwidth}
\includegraphics[width=\linewidth]{charm-xsec.eps}
\end{minipage}
\hspace{0.2cm}
\begin{minipage}{0.48\textwidth}
\vspace{-0.2cm}
\includegraphics[width=\linewidth]{tau-charm-bottom.eps}
\end{minipage}
\caption{(Color online) Left panel: total HQ scattering cross sections
  off light partons in LO pQCD (blue lines) and within the effective
  resonance model (red lines). Right panel: thermalization times,
  $\tau=1/A(\bvec{p}=0)$, for charm and bottom quarks in LO pQCD with
  $\alpha_s=0.4$ and Debye-screening mass $\mu_D=g T$, compared to the
  results from the resonance+pQCD model, as a function of QGP
  temperature.}
\label{fig_tau-reso}
\end{figure}

With charm- and bottom-quark masses of $m_c=1.5\;\mathrm{GeV}$ and
$m_b=4.5 \; \mathrm{GeV}$, the physical resonance masses are adjusted to
$m_D=2\;\mathrm{GeV}$ and $m_B=5 \;\mathrm{GeV}$, respectively. This is
in approximate accordance with earlier $T$-matrix models of heavy-light
quark interactions\cite{Blaschke:2002ws,Blaschke:2003ji}. Likewise, the
coupling constant, $G$, is adjusted such that the resonance widths vary
as $\Gamma_{D,B}=0.4 \ldots 0.75 \;\mathrm{GeV}$.  The resulting
heavy-light quark scattering matrix elements (cf.~middle and right
panels of Fig.~\ref{fig_dia-reso}) have been injected into
Eq.~(\ref{2.1.10c}) to calculate HQ drag and diffusion coefficients. In
the left and right panel of Fig.~\ref{fig_tau-reso} we compare the total
HQ elastic scattering cross sections and resulting thermal relaxation
times, $\tau_{\text{eq}}=1/A(\bvec{p}=0)$, of the resonance model with
LO pQCD (cf.~the diagrams in Fig.~\ref{fig_dia-pqcd}).  Although the
total cross sections are not very different in magnitude, the
thermalization times decrease by around a factor of $\sim$$3$-$4$ when
adding resonant scattering, for all temperatures $T=$~1-2\,$T_c$.
The main reason for this behavior is that $s$-channel $Q \bar{q}$
scattering is isotropic in the rest frame of the resonance, while the
pQCD cross section is largely forward-peaked ($t$-channel gluon
exchange), and thus produces a much less efficient transport cross
section (which encodes an extra angular weight). The charm-quark
equilibration times in the resonance+pQCD model, $\tau_{\mathrm{eq}}^c =
2$-$10\;\mathrm{fm}/c$, are comparable to the expected QGP
lifetime at RHIC of around $\tau_{\mathrm{QGP}} \simeq 5 \;
\mathrm{fm}/c$. Thus, at least for charm quarks, substantial
modifications of their $p_t$ spectra towards local equilibrium in the
flowing medium can be expected.

\begin{figure}[!t]
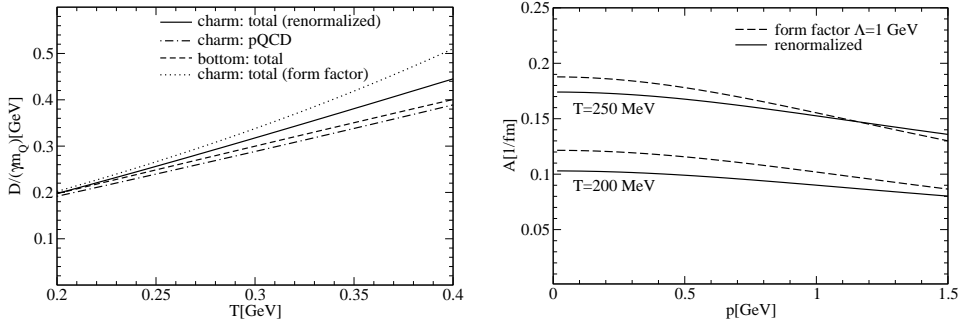

\begin{center}
\includegraphics[width=0.48\linewidth]{temperature-consist}\hfill
\includegraphics[width=0.48\linewidth]{reso-model-form-A}
\end{center}
\caption{Left panel: check of the dissipation-fluctuation relation at
  $\bvec{p}=0$ for $c$ and $b$ quarks with and without resonance
  interactions and in the renormalization and form-factor cutoff
  schemes. Right panel: momentum dependence of the drag coefficient
  in the renormalization and the formfactor-cutoff scheme, where the
  coupling constants have been chosen in both schemes as to obtain
  a resonance width $\Gamma=0.4\,\GeV$.}
\label{fig.2.1.6}
\end{figure}

The consistency of the Fokker-Planck approach can be checked with the
dissipation-fluctuation relation, Eq.~(\ref{2.1.30}), at $\bvec{p}=0$,
cf.~left panel of Fig.~\ref{fig.2.1.6}. For the forward-peaked
pQCD-matrix elements, the relation is fulfilled within $3\%$, while with
the isotropic resonance scattering deviations reach up to $11\%$ in the
renormalization scheme and up to $26\%$ in the formfactor-cutoff scheme
at the highest temperatures considered ($T=400\; \MeV$). Note however,
that for a typical thermal evolution at RHIC, average fireball
temperatures above $T=250\; \MeV$ are only present within the first
$\fm/c$\cite{Rapp:2008qc}; below this temperature, the deviations are
less than $5\%$ for all cases. The right panel of Fig.~\ref{fig.2.1.6}
illustrates that (for identical resonance widths) the formfactor
regularization scheme leads to somewhat larger (smaller) friction
coefficients at low (high) momentum than the renormalization scheme.

\begin{figure}[!t]
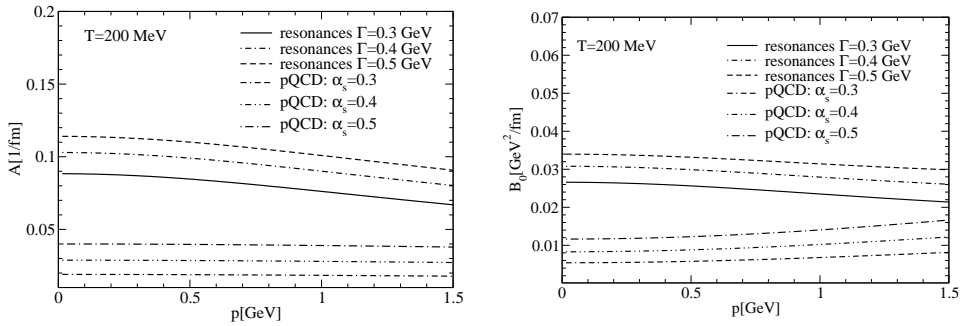

\begin{minipage}{0.48\textwidth}
\includegraphics[width=\textwidth]{reso-model-A}
\end{minipage}\hfill
\begin{minipage}{0.48\textwidth}
\includegraphics[width=\textwidth]{reso-model-B0}
\end{minipage}\vspace*{0.5mm}
\caption{Drag (left panel) and transverse-diffusion coefficient (right
  panel) for pQCD and resonance model with varying interactions
  strengths as a function of the HQ momentum at a temperature of
  $T=200\;\mathrm {MeV}$.}
\label{fig_AB-reso}
\end{figure}

In Fig.~\ref{fig_AB-reso} the momentum dependence of the drag and
transverse diffusion coefficients is depicted using either
resonance-scattering or pQCD-matrix elements. Resonance scattering
becomes relatively less efficient for higher HQ momenta since the
center-of-mass energy in collisions with thermal light antiquarks
increasingly exceeds the resonance pole. The variations of the
coefficients with the strong coupling constant in the pQCD
scattering-matrix elements or the resonance-coupling constant in the
effective resonance-scattering model are rather moderate. This is due to
compensating effects of an increase of the matrix elements with
$\alpha_s^2$ or $G^4$, on the one hand, and the accordingly increased
Debye-screening mass for pQCD scattering or the broadening of the
resonances widths, on the other hand.

\subsubsection{In-Medium $T$-matrix with lQCD-based Potentials}
\label{sssec_tmat}
The idea of utilizing HQ free energies computed in lattice QCD to
extract a driving kernel for heavy-light quark interactions in the QGP
has been carried out in Ref.\cite{vanHees:2007me}, with the specific
goal of evaluating HQ diffusion. Since the latter is, in principle,
determined by low-energy HQ interactions, the potential-model framework
appears to be suitable for this task. Moreover, with a potential
extracted from lQCD, the calculation could be essentially
parameter-free. Currently, however, such an approach bears significant
uncertainty, both from principle and practical points of view, e.g.,
whether a well-defined potential description can be constructed in
medium\cite{Laine:2006ns,Laine:2007qy,Brambilla:2008cx} and, if so, how
to extract this information from, say, the HQ free energy. In the
vacuum, both questions have been answered
positively\cite{Bali:2000gf,Brambilla:2004wf}, thus validating the 30
year-old phenomenological approaches using Cornell potentials for 
heavy quarkonia, which
provide a very successful spectroscopy\cite{Eichten:1979ms}. The
potential approach has been extended to heavy-light mesons in
Refs.\cite{Godfrey:1985xj,Avila:1994vi}.

\begin{figure}[!t]
\centerline{\includegraphics[width=0.8\linewidth]{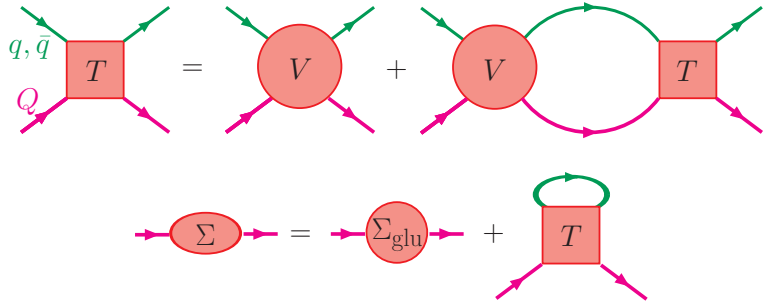}}
\caption{(Color online) Diagrammatic representation of the Brueckner
  many-body calculation for the coupled system of the $T$-matrix based
  on the lQCD static internal potential energy as the interaction kernel
  and the HQ self-energy.}
\label{fig_dia-tmat}
\end{figure}

A Brueckner-like in-medium
$T$-matrix approach for heavy-light quark scattering in the QGP has been
applied in Ref.\cite{vanHees:2007me}, diagrammatically represented in
Fig.~\ref{fig_dia-tmat}. The underlying (static) two-body potential has
been identified with the internal energy
\begin{equation}
\label{2.1.40}
U_1(r,T)=F_1(r,T)+T S_1(r,T)=F_1(r,T)-T \frac{\partial F_1(r,T)}{\partial
  T} \ , 
\end{equation} 
extracted from two lQCD computations of the color-singlet HQ free energy
above $T_c$, for quenched\cite{Kaczmarek:2003dp} and 
two-flavor\cite{Kaczmarek:2003ph} QCD (pertinent parameterizations are 
given in Refs.\cite{Wong:2004zr} and \cite{Shuryak:2004tx}, hereafter
referred to as [Wo] and [SZ], respectively).  This choice (rather than,
e.g., the free energy) provides an upper limit for the interaction
strength\cite{Shuryak:2004tx,Wong:2004zr,Mannarelli:2005pz,Cabrera:2006wh}.
To use Eq.~(\ref{2.1.40}) as a potential in a $T$-matrix calculation,
the internal energy has to be subtracted such that it vanishes for $r
\rightarrow \infty$,
\begin{equation}
\label{2.1.41}
V_1(r,T) \equiv U_1(r,T)-U_1(r \rightarrow \infty,T) \ ,  
\end{equation}
which is dictated by the convergence of the $T$-matrix integral in
momentum space. It is suggestive to interpret the asymptotic value
$U_1^\infty\equiv U_1(r \rightarrow \infty,T)$ as an in-medium HQ-mass,
\begin{equation}
\label{2.1.42}
m_Q(T)=m_{Q}^0+\frac{1}{2} U_1^\infty \ , 
\end{equation}
where $m_{Q}^0$ denotes the bare HQ mass (e.g., $m_c^0 \simeq 1.25 \;
\GeV$\cite{Amsler:2008zzb} for the bare $c$-quark mass). However, close
to $T_c$, the values for $U_1^\infty(T)$ extracted from lQCD
calculations develop a rather pronounced peak
structure\cite{Petreczky:2004pz,Kaczmarek:2005gi}, which renders a mass
interpretation problematic. Progress in understanding these properties
is closely connected with the proper identification of the
potential. First lQCD estimates of the in-medium HQ mass (extracted by
relating zero-mode contributions to quarkonium correlators to the HQ
susceptibility) indicate a moderate increase when approaching
$T_c$ from above\cite{Petreczky:2008px}. In Ref.\cite{vanHees:2007me}
constant (average) in-medium charm- and bottom-quark masses of
$m_c=1.5\,\GeV$ and $m_b=4.5\,\GeV$, respectively,
have been employed.

The $T$-matrix approach is readily generalized to color-configurations
other than the singlet channel of the $qQ$ pair. The complete set of
color states for $Q \bar{q}$ (singlet and octet) and $Qq$ (anti-triplet
and sextet) pairs has been taken into account assuming Casimir scaling
as in pQCD,
\begin{equation}
\label{2.1.43}
V_8=-\frac{1}{8} V_1, \quad V_{\bar{3}}=\frac{1}{2} V_1, \quad
V_6=-\frac{1}{4} V_1 \ .
\end{equation}
which is, in fact, supported by finite-$T$
lQCD\cite{Nakamura:2005hk,Doring:2007uh}. To augment the static
(color-electric) potentials with a minimal relativistic (magnetic)
correction for moving quarks\cite{Brown:2003km}, the so-called Breit
correction as known from electrodynamics\cite{Brown:1952} has been
implemented via the substitution
\begin{equation}
\label{2.1.43b}
V_a \rightarrow V_a(1-\hat{\alpha}_1 \cdot \hat{\alpha}_2),
\end{equation}
where $\hat{\alpha}_{1,2}$ are quasiparticle velocity operators.

\begin{figure}[!t]
\begin{minipage}{0.5\linewidth}
\includegraphics[width=\textwidth]{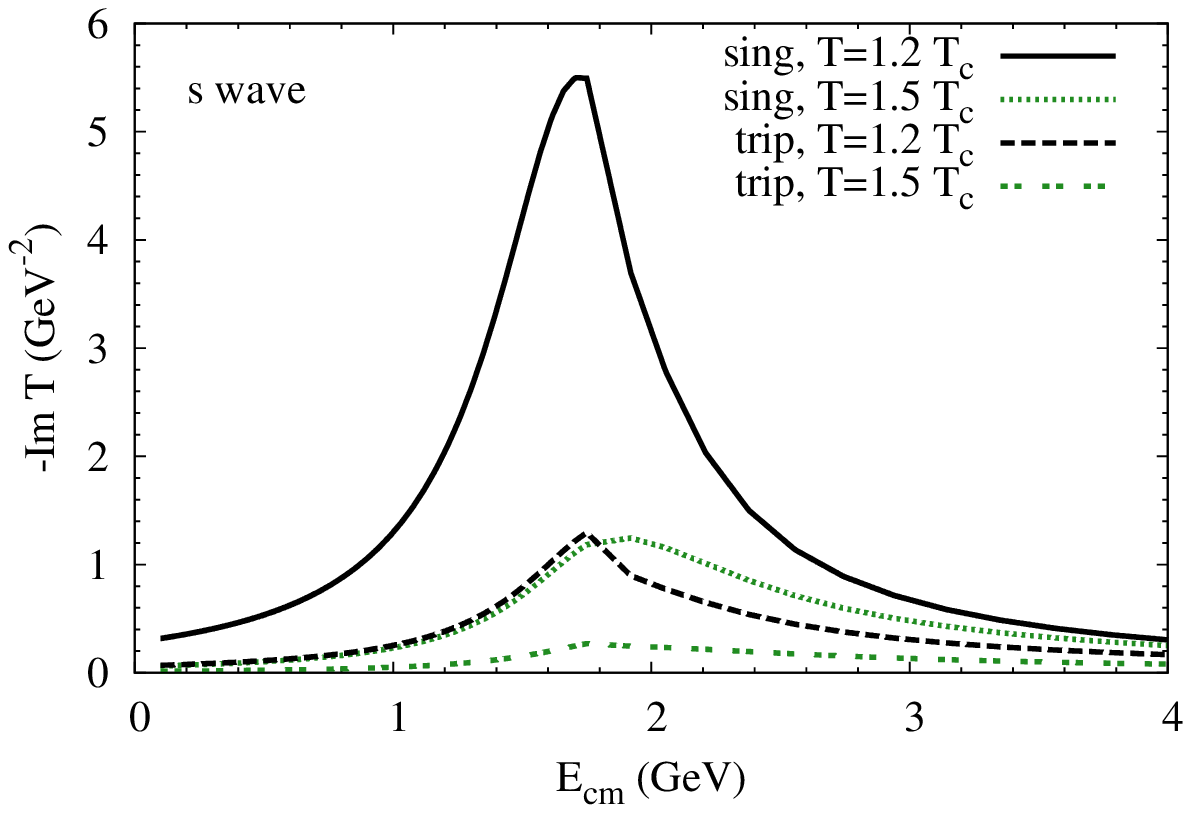}
\end{minipage}\hfill
\begin{minipage}{0.47\linewidth}
\includegraphics[width=\textwidth]{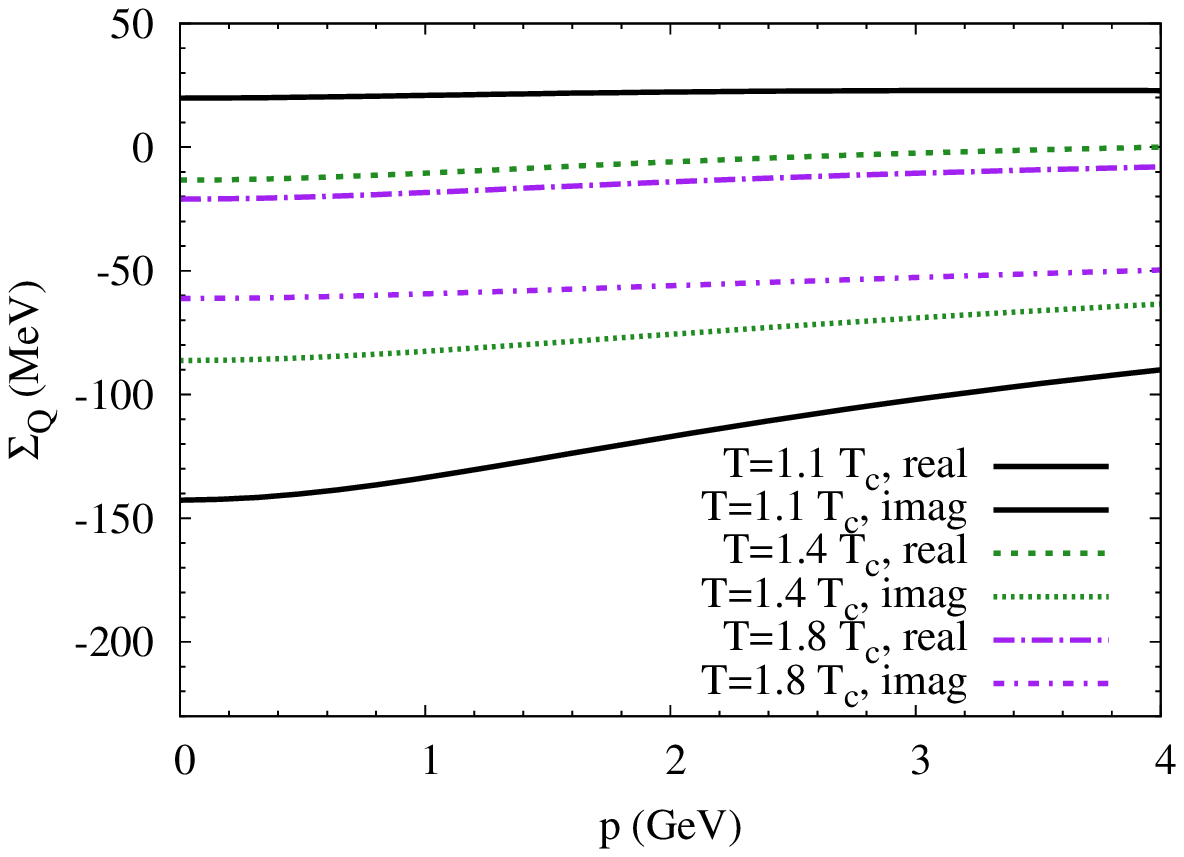}
\end{minipage}
\caption{(Color online) Results of a Brueckner-type approach for
  $c$-quarks in a QGP\protect\cite{vanHees:2007me} based on a potential
  corresponding to the internal energy extracted from quenched lattice
  QCD\protect\cite{Kaczmarek:2003dp,Wong:2004zr}. Left panel: imaginary
  part of the in-medium $T$ matrices for $S$-wave $c$-$q$ scattering in
  color-singlet and -triplet channels at two different temperatures;
  right panel: real and imaginary parts of $c$-quark selfenergies based
  on the $T$-matrices in the left panel.}
\label{fig_ImT-Sig}
\end{figure}

The above constructed heavy-light potentials can now be resummed in a
two-body scattering equation. In accordance with the static nature of
the potentials, it is appropriate to use a three-dimensional reduction
of the full four-dimensional Bethe-Salpeter equation. This leads to the
well-known ladder series which is resummed by the Lippmann-Schwinger
(LS) integral equation for the $T$-matrix In the $q$-$Q$
center-of-mass (CM) frame it takes the form
\begin{equation}
\begin{split}
\label{Tmat}
T_a(E;\bvec{q}',\bvec{q}) = &V_a(\bvec{q}',\bvec{q})-\int \frac{\dd^3
  \bvec{k}}{(2 \pi)^3} V_a(\bvec{q}',\bvec{k}) \ G_{qQ}(E;k) \\
& \times T_a(E;\bvec{k},\bvec{q})\ [1-f_q(\omega_k^q)-f_Q(\omega_k^Q)] \ .
\end{split}
\end{equation}
The driving kernel (potential) can now be identified with the Fourier 
transform of the coordinate-space potential extracted from lQCD, 
\begin{equation}
\label{2.1.46}
V_a(\bvec{q}',\bvec{q})=\int \dd^3 \bvec{r} \ V_a(r) \ 
\exp[\ii(\bvec{q}-\bvec{q}')\cdot\bvec{r}]
\end{equation}
in a given color channel, $a \in \{1,\bar{3},6,8 \}$. The concrete form
of the intermediate $q$-$Q$ (or $\bar{q}$-$Q$) propagator,
$G_{qQ}(E,k)$, depends on the reduction scheme of the underlying
Bethe-Salpeter equation. It has been
verified\cite{Mannarelli:2005pz,vanHees:2007me} that, e.g., the
Thompson\cite{Thompson:1970wt} and
Blancenbecler-Sugar\cite{Blankenbecler:1965gx} scheme lead to very
similar results in the present context (as was found for nucleon-nucleon
scattering). In the former, the two-particle propagator is given by
\begin{equation}
\label{G_qQ}
G_{qQ}(E,k)=\frac{1}{E-(\omega_k^q+\ii \Sigma_I^q)-(\omega_k^Q+\ii
  \Sigma_I^Q)} \ ,
\end{equation}
where $E$ and $k$ denote the CM energy and relative momentum of the 
$qQ$ pair, respectively. The quasi-particle widths are chosen as
$\Gamma_{I}^{q,Q}=2\Sigma_I^{q,Q}=200\; \mathrm{MeV}$, and the light
quark masses as constant at $m_q=0.25$~GeV, with on-shell energies
\begin{equation}
\label{2.1.44b}
\omega_{k}^{q,Q}=\sqrt{m_{q,Q}^2+k^2} \ . 
\end{equation}
The latter figure into the Pauli blocking factor with Fermi-Dirac
distributions,
\begin{equation}
\label{2.1.47}
f_{q,Q}(\omega^{q,Q})=\frac{1}{\exp(\omega^{q,Q}/T)+1} \ 
\end{equation}
(at the considered temperatures their impact is negligible). The
solution of the $T$-matrix Eq.~(\ref{Tmat}) is facilitated by a an
expansion into partial waves, $l$,
\begin{equation}
\begin{split}
\label{2.1.48}
V_a(\bvec{q}',\bvec{q})=4 \pi \sum_{l} (2l+1) \ V_{a,l}(q',q) \ 
P_l[\cos\angle(\bvec{q},\bvec{q}')] \ , \\
T_a(E;\bvec{q}',\bvec{q})=4 \pi \sum_{l} (2l+1) \ T_{a,l}(E;q',q) \ 
P_l[\cos\angle(\bvec{q},\bvec{q}')] \ ,
\end{split}
\end{equation}
which yields a one-dimensional LS equation,
\begin{equation}
\begin{split}
\label{Tmat-1}
T_{a,l}(E;q',q)=V_{a,l}&(q',q) + \frac{2}{\pi} \int \dd k k^2
V_{a,l}(q',k) \ G_{Qq}(E;k) \\
&\times T_{a,l}(E;k,q)\ [1-f_F(\omega_k^Q)-f_F(\omega_k^q)] \ ,
\end{split}
\end{equation}
for the partial-wave components, $T_{a,l}$, of the $T$-matrix.
Eq.~(\ref{Tmat-1}) can be solved numerically by discretization and
subsequent matrix-inversion with the algorithm of Haftel and
Tabakin\cite{Haftel1970}. The resulting $S$-wave ($l=0$) $T$-matrices
indeed show resonance structures in a QGP in the channels where the
potential is attractive, i.e., in the meson (color-singlet) and diquark
(color-antitriplet) channels. The pertinent peaks in the imaginary part
of the $T$-matrix develop close to the $Q$-$q$ threshold, and melt with
increasing temperature at around $1.7 \, T_c$ and $1.4\,T_c$,
respectively (cf.~left panel of Fig.~\ref{fig_ImT-Sig}). In the
repulsive channels, as well as for $P$-waves, the $T$-matrices carry
much reduced (non-resonant) strength. The increasing strength in the
meson and diquark channels (the latter relevant for baryon binding) 
when approaching $T_c$ from above is suggestive for ``pre-hadronic'' 
correlations toward the hadronization transition.

The next step is to use the $T$-matrices to compute the light-quark
contribution to HQ self-energies, i.e., the last diagram in the
second line of the Brueckner scheme illustrated in
Fig.~\ref{fig_dia-tmat}. In a given color-channel, $a$, the 
$T$-matrix induced self-energy is given by 
\begin{equation}
\label{2.1.50}
\Sigma_a^Q(\omega,p)=\frac{d_{\text{SI}} d_a}{6} \int \frac{k^2 \dd k 
\dd x}{4\pi^2} [f_F(\omega_k)+f_B(\omega+\omega_k)] \ 
T_a(E;\bvec{p},\bvec{k}) \ ,
\end{equation}
where $d_{\mathrm{SI}}=4(2l+1)N_f$ denotes the spin-isospin and angular
momentum degeneracy of all $Qq$ (or $Q\bar q$) configurations (assuming
spin and light-flavor symmetry) and $d_a$ the color degeneracy of
channel $a$; the factor $1/6$ averages over the incoming HQ color-spin
degrees of freedom. The resulting charm-quark selfenergies (summed over
all light quarks and antiquarks) are displayed in the right panel of
Fig.~\ref{fig_ImT-Sig}. One finds rather small corrections to
the HQ mass (presented by the real part of $\Sigma$), but the imaginary 
parts are substantial, $\Gamma_c=-2 \im \Sigma_c \simeq$~100-300\,MeV 
for temperatures $T=1.1$-$1.8\,T_c$ (with the
largest values attained close to $T_c$).  These values were the
motivation for the choice of input widths in the propagator,
Eq.~(\ref{G_qQ}), of the $T$-matrix equation, thus providing a rough
self-consistency.  The in-medium mass corrections, on the other hand,
are associated with the gluonic contribution to the HQ self-energy
(corresponding to the first term in the lower line
Fig.~\ref{fig_dia-tmat}), which have not been calculated explicitely in
Ref.~\cite{vanHees:2007me}, but represent a rough (average) 
representation of the asymptotic values of the HQ potential, 
$U^{(1)}_\infty$ (as discussed above).

\begin{figure}[!t]
\begin{minipage}{0.48\linewidth}
\includegraphics[width=\textwidth]{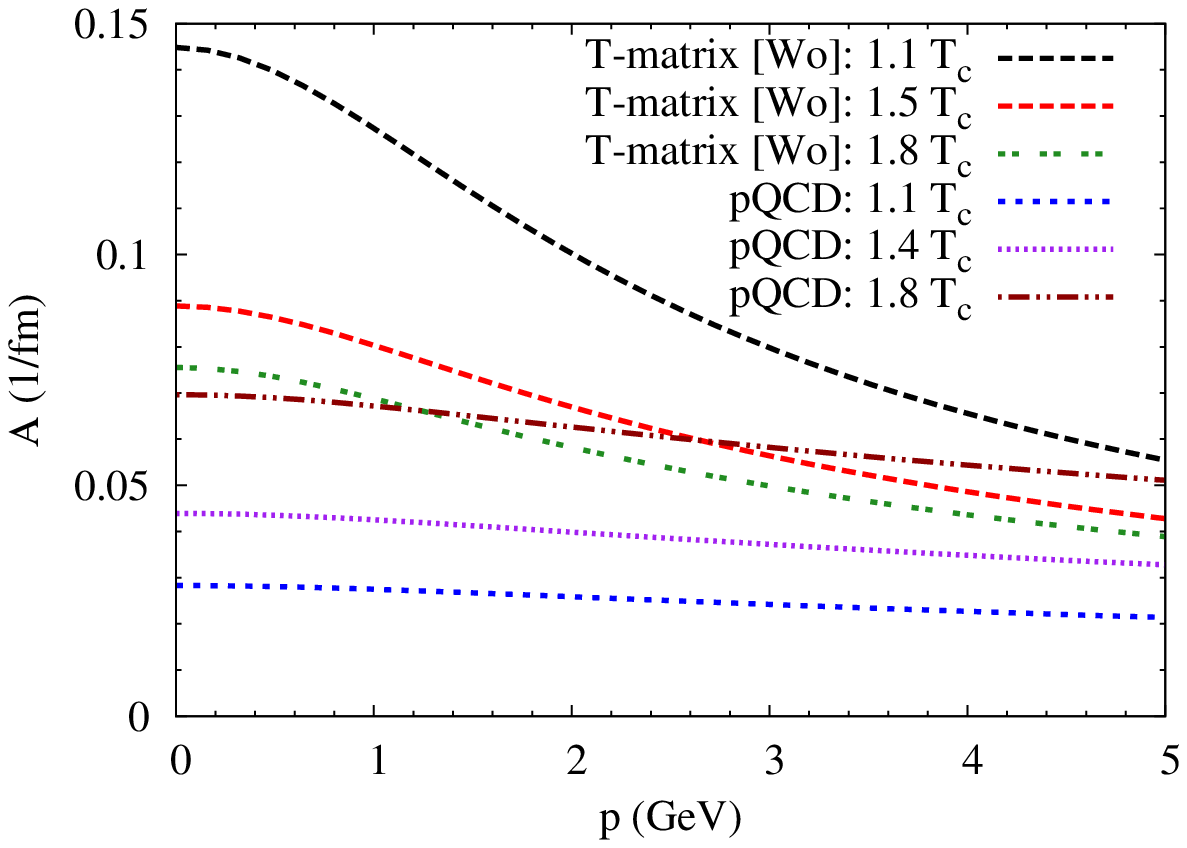}
\end{minipage}\hfill
\begin{minipage}{0.48\linewidth}
\includegraphics[width=\textwidth]{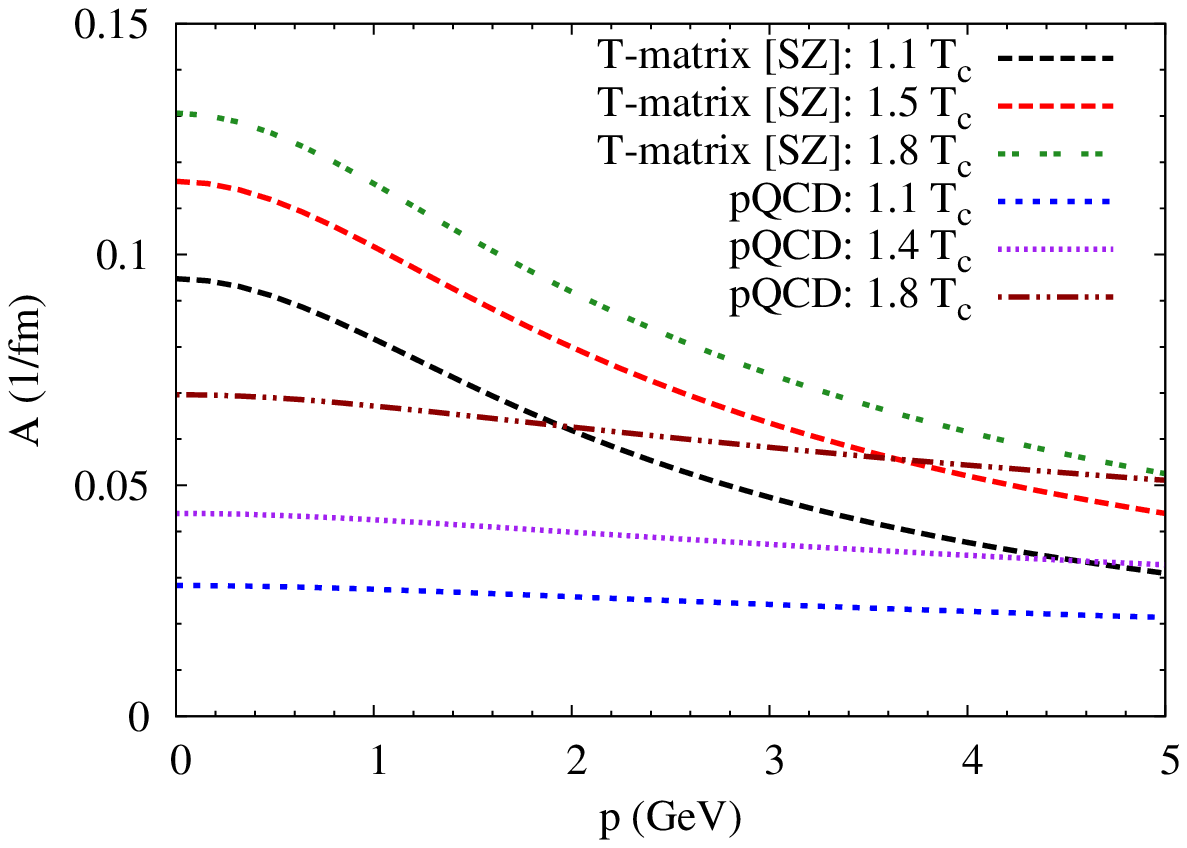}
\end{minipage}
\caption{(Color online) The drag coefficients at different temperatures,
  using the parameterization of the HQ potential from [Wo] (left panel)
  and [SZ] (right panel) compared to LO pQCD with $\alpha_s=0.4$ and
  $\mu_D=g T$.}
\label{fig_gamma-tmat}
\end{figure}

The final step is to implement the $T$-matrix elements into a
calculation of HQ drag and diffusion coefficients via
Eq.~(\ref{2.1.10c}); one finds
\begin{equation}
\begin{split}
\label{2.1.51}
\sum & |\mathcal{M}|^2=\frac{64\pi}{s^2} (s-m_q^2+m_Q^2)^2(s-m_Q^2+m_q^2)^2 \\
 &\times N_f\sum_{a} d_a (|T_{a,l=0}(s)|^2 +3 |T_{a,l=1} (s)
\cos(\theta_{\text{cm}})|^2) \ .
\end{split}
\end{equation}
The resulting friction coefficients are summarized in
Fig.~\ref{fig_gamma-tmat} as a function of momentum for three
temperatures and for two potential extractions from
lQCD~\cite{Shuryak:2004tx,Wong:2004zr}.
Generally, the $Qq$ $T$-matrix based coefficients are largest at low HQ
momentum, as to be expected from the resonance formation close to 
threshold. The values exceed the LO-pQCD coefficients at small 
temperatures and for both potentials by a factor of $\sim$$3$-$5$. At 
higher temperatures the enhancement reduces considerably, to a factor of 
less than $2$ in the [SZ] potential and to essentially equal strength 
for the [Wo] potential\footnote{Recall that the LO-pQCD calculations 
  employ a rather large coupling ($\alpha_s$=0.4), and are dominated by 
  scattering off gluons in the heat bath; thus a minimal merging of the 
  gluon sector with the $T$-matrix calculations
  would consist of adding the gluonic part of LO-pQCD; this procedure is
  adopted below whenever combined results are shown or utilized. In
  principle, a non-perturbative treatment should also be applied to
  HQ-gluon scattering.}.  In fact, the coefficients computed with the
[SZ] potential have a slightly ($30\%$) increasing trend with $T$, 
while the [Wo] potential leads to a decreasing trend. This difference 
needs to be scrutinized by future systematic comparisons of lQCD input 
potentials.  Compared to the resonance model (cf., e.g., left panel of 
Fig.~\ref{fig_AB-reso}), the $T$-matrix calculations yield
quantitatively similar results at temperatures not too far above $T_c$
but become smaller at higher $T$ due to resonance melting (which is 
presumably a more realistic feature of a non-perturbative 
interaction strength).

\subsubsection{Collisional Dissociation of Heavy Mesons in the QGP}
\label{sssec_coll-diss}

In Ref.\cite{Adil:2006ra}, a so-called reaction-operator (GLV) approach
has been applied to resum multiple elastic scatterings of a fast
$Q\bar{q}$ pair. The quenching of heavy quarks in a QGP is calculated 
by solving coupled rate equations for the fragmentation of $c$ and $b$
quarks into $D$ and $B$ mesons and their dissociation in the QGP. 
The main mechanisms for HQ energy loss are
collisional broadening of the meson's transverse momentum and the
distortion of its intrinsic light-cone wave function. The latter is
modeled in a Cornell-type potential ansatz\cite{Avila:1994vi} (note that
this bears some similarity to the $T$-matrix approach discussed in
Sec.~\ref{sssec_tmat}). This in-medium HQ/heavy-meson
fragmentation/dissociation mechanism leads to comparable high-$p_t$
suppression for $B$ and $D$ mesons, which is quite contrary to
perturbative calculations for both collisional and radiative energy loss
(where the suppression of $b$ quarks is significantly less pronounced
than for $c$ quarks). This feature largely results from the much smaller
formation times of $B$-mesons compared to $D$- mesons, leading to a
faster fragmentation-dissociation cycle for $b$ quarks/$B$ mesons.

\subsubsection{Estimates of HQ Diffusion in Lattice QCD}
\label{sssec_hq-lat}
It has recently been suggested that, unlike in the case of other
transport coefficients (e.g., the shear viscosity), the HQ diffusion
coefficient might be amenable to a determination within lQCD, based on
an analytic continuation of the color-electric-field correlator along a
Polyakov loop\cite{CaronHuot:2009uh} (see also 
Ref.~\cite{Petreczky:2005nh} for earlier related work). The starting 
point of these considerations is the spectral function of the HQ 
current correlator,
\begin{equation}
\label{hq-jj-corr}
\rho_V^{\mu \nu}(\omega) = \int \dd t \exp(\ii \omega t) \int \dd^3
\bvec{x} \erw {\frac{1}{2} \comm{J_Q^{\mu}(t,\bvec{x})}{J_Q^{\nu}(0,0)}}
 \ ,
\end{equation}
where $J_Q^{\mu}$ denotes the HQ current operator in the Heisenberg
picture. The spatial diffusion coefficient, $D_s$, can be extracted from
this spectral function by the pole position at $\omega=-\ii D_s
\bvec{k}^2$, where $\bvec{k}$ is the HQ momentum. The condition for a
pole leads to the Kubo relation
\begin{equation}
\label{kubo-Ds}
D_s=\frac{1}{3 \chi^{00}} \lim_{\omega \rightarrow 0} \sum_{i=1}^{3}
\frac{\rho_V^{ii}(\omega)}{\omega} \ ,
\end{equation}
where $\chi_{00}$ is the conserved-charge susceptibility
\begin{equation}
\label{chi00}
\chi^{00}=\frac{1}{T} \int \dd^3 \bvec{x} \erw{J_Q^{0}(t,\bvec{x}) 
J_Q^{0}(0,0)} \ .
\end{equation}
For a heavy quark, the spectral function, Eq.~(\ref{hq-jj-corr}), is
expected to develop a sharp peak around $\omega=0$ which can be
described by a Lorentzian function close to this point. The width of
this function is given by the drag coefficient, which obeys the
fluctuation-dissipation relations, discussed in
Sec.~\ref{ssec_diff}. Using HQ effective theory techniques it is shown
that in the static limit the momentum-diffusion coefficient, $\kappa=2
D$, is given by a correlator involving color-electric fields and $J_Q^0$
operators whose Euclidean analogue can be mapped to an expectation value
involving Wilson lines and color-electric fields, similar to
Eq.~(\ref{kappa-gauge}). This purely gluonic correlation function can in
principle be evaluated in lQCD.

Another lattice-based approach to assess HQ diffusion has been 
suggested in Ref.\cite{Laine:2009dd} in terms of (discretized) 
classical gauge theory. The limitation of this approach is set by the 
thermal (hard) scale $\sim$$\pi T$ where quantum theory suppresses 
excitations. Since the HQ thermalization rate is expected to be 
governed by the electric screening scale $\sim$$gT$, this limitation 
may not be severe for small and moderate coupling ($gT\ll T$), and 
thus allow for valuable insights. First, it has been verified that, in 
the weak coupling limit, the discretized (nonperturbative) classical 
computation indeed agrees well with pQCD. Upon increasing the coupling 
strength, the classical lattice results for the HQ momentum diffusion 
coefficient increasingly exceed the LO perturbative result, by about 
an order of magnitude for a moderate coupling strength corresponding 
to $\alpha_s\simeq0.2$. Next, the NLO term (with slightly increased 
strength to account for HTL effects), as calculated in 
Ref.\cite{CaronHuot:2007gq}, has been added to the LO calculation
which extends the agreement of pQCD with the classical lattice results 
to larger (but still weak) coupling. For$\alpha_s\simeq0.2$ the 
increase over LO amounts to a factor $\sim$2, which means that the 
classical lattice result remains substantially larger (by a factor
of $\sim$5) than the NLO value. Besides reconfirming the poor
convergence of pQCD, this also suggests that the perturbative series
is not alternating but that higher order terms keep increasing
the value of the HQ momentum diffusion coefficient. 
Semi-quantitatively, such an enhancement is in the ball park of the 
factor $\sim$3-4 found in the effective resonance model 
(Sec.~\ref{sssec_reso}) or $T$-matrix approach (Sec.~\ref{sssec_tmat}).


\subsection{String Theoretical Evaluations of Heavy-Quark Diffusion}
\label{ssec_string}

The conjectured correspondence between certain classes of string
theories, formulated in five-dimensional Anti-de-Sitter space
($\mathrm{AdS}_5$), and gauge theories with conformal invariance
(conformal field theory, CFT) has opened interesting possibilities to
address nonperturbative aspects of QCD. The so-called AdS/CFT
correspondence implies a ``duality'' of a weakly coupled gravity to a
strongly coupled supersymmetric (and conformal) gauge theory,
specifically ${\cal N}=4$ $\mathrm{SU}(N_c)$ super-Yang-Mills (SYM)
theory. This connection has been exploited to formulate the problem of
HQ diffusion at finite temperature and extract an ``exact''
nonperturbative result for HQ transport coefficients in the SYM
plasma\cite{Herzog:2006gh,Gubser:2006bz,CasalderreySolana:2006rq}.  The
translation to QCD matter is beset with several
caveats\cite{Gubser:2006qh}, e.g., the particle content of the SYM
medium is quite different compared to the QGP. While this may be
corrected for by a suitable rescaling of the temperature by matching,
e.g., the energy densities\footnote{This procedure works quite well when
  comparing quantities in quenched and unquenched lattice QCD
  computations, e.g.~for the critical temperature.}, a more problematic
difference is the absence of a scale (other than temperature) in
conformal SYM. Thus, the latter does not possess a breaking of scale
invariance, a running coupling constant, confinement nor spontaneous
chiral symmetry breaking, and consequently no notion of a critical
temperature, either. Thus SYM is quite different from QCD in the zero-
and low-temperature regimes. However, at sufficiently high $T$, where
the QCD medium deconfines its fundamental charges, the resemblance to
SYM might be much closer. E.g., the pressure in SYM in the strong
coupling limit amounts to about $75\%$ of the Stefan-Boltzmann limit,
close to what is found in thermal lattice QCD for a
wide range above $T_c$. In addition, the finding of an extremely small
shear viscosity in strongly coupled SYM, $\eta/s=1/4\pi$ (conjectured to
be a universal lower bound)\cite{Kovtun:2004de}, and the apparently
low-viscosity QCD medium deduced from the success of hydrodynamic models
at RHIC, is another good reason to further pursue exact nonperturbative
calculations in SYM for quantities that are relevant for RHIC
phenomenology. If nonperturbative effects in the strongly coupled QGP at
moderate temperatures, $T=1$-$2~T_c$, are ultimately connected to the
presence of the phase change(s) (and thus inherently to the critical
temperature as a relevant scale), the CFT-QCD connection would not be a
rigorous one. But even in this case, the nonperturbative computation
of transport coefficients of a strongly coupled system at a given
reference temperature ``not too close'' to $T_c$ should provide useful
insights.

The first step in computing HQ diffusion for CFT is the introduction of
a heavy quark into the conformal field theory. This can be achieved by
either introducing a heavy charge via breaking the gauge group from
$N_c+1$ to $N_c$ (which, strictly speaking, generates $(2N_c+1)$
``Higgsed'' ``$W$'' bosons)\cite{Maldacena:1998im}, or by adding a
finite-mass ${\mathcal{N}}=2$ hypermultiplet with charges in the
fundamental representation as a ``probe'' of the CFT medium. In either
case, the pertinent object on the 4-dimensional boundary of the 
5-dimensional AdS space represents a fundamental charge.  
In Refs.\cite{Herzog:2006gh,Gubser:2006bz}, the HQ drag has
been evaluated by computing its momentum degradation, $\dd p/\dd
t=-\gamma p$, through the force on the trailing string, resulting in a
friction (or drag) coefficient,
\begin{equation}
\gamma_{\rm AdS/CFT}=\frac{\pi \sqrt{\lambda} T_{\rm SYM}^2}{2m_Q} \ ,
\label{ads-cft}
\end{equation}
where $\lambda =g_{\mathrm{SYM}}^2 N_c$ denotes the 't~Hooft coupling
constant. Alternatively, in Ref.\cite{CasalderreySolana:2006rq} the
problem was formulated focusing on the diffusion term. For time scales
longer than the thermal relaxation time of the medium, but short
compared to the HQ relaxation time, the fluctuation term in the Langevin
equation~(\ref{langevin}) dominates over the drag term. The evaluation
of the noise (or force) correlator is then carried out via the
fluctuations of the string, resulting in a noise coefficient which is
directly related to the diffusion coefficient
(cf.~Eqs.~(\ref{force-corr}) and (\ref{Cjk-D}) below).  Furthermore, the
latter can be related to the friction coefficient using the Einstein
relation, Eq.~(\ref{einstein}); it turns out that the result is
identical to Eq.~(\ref{ads-cft}), which also verifies that a Langevin
process consistent with the fluctuation-dissipation theorem applies in
the SYM theory (see, however,
Refs.\cite{Gubser:2006nz,CasalderreySolana:2007qw}, where the
applicability of the Langevin framework in AdS/CFT for high-momentum
quarks is discussed). The square-root dependence of $\gamma_{\rm
  AdS/CFT}$ on the coupling constant $\lambda$ clearly characterizes its
nonperturbative nature; in this sense it is parametrically large for
comparatively small coupling constants. The temperature dependence is
rather ``conventional", as to be expected since there are no additional
scales in the problem (the HQ mass in the denominator implies the
standard suppression of the HQ relaxation rate by $\sim$$T/m_Q$).

The next question is how to convert the result into a (semi-)
quantitative estimate for the QCD plasma. Naively, one may just insert
the values of the strong coupling constant, $g_s$, and QGP temperature,
$T$, for $g_{\rm SYM}$ and $T_{\rm SYM}$, respectively. A more suitable 
identification
probably consists of matching physical quantities which leads to
somewhat different parameter values. E.g., in Ref.\cite{Gubser:2006qh},
comparable temperatures were identified by matching the energy densities
($\varepsilon$) of the QGP and SYM-plasma. Since the latter has a factor
$\sim$$3$ larger particle content (degeneracy factor), one has a smaller
temperature at the same $\varepsilon$, $T_{\rm SYM}\simeq T /
3^{1/4}$. For the coupling constant, one can exploit the fact that in
AdS/CFT the potential between a heavy charge and anticharge is
essentially of Coulomb-type, both at zero\cite{Maldacena:1998im} and
finite temperature\cite{Rey:1998bq,Brandhuber:1998bs}. In the latter
case, the potential goes to zero at some finite range, characteristic
for Debye-screening behavior. This range can be used to identify the
length scale in comparison to typical screening radii of heavy-quark
free energies as computed in thermal lattice QCD (although some
ambiguity remains)\cite{Gubser:2006qh}. Matching the magnitude of the
potentials at the screening radius then allows for a matching of the
coupling constants. This leads to significantly smaller values for
$\lambda$ (by a factor of 3-6) than the naive identification with
$\alpha_s=0.5$. In connection with the redefined temperature, the
improved AdS/CFT-based estimate for the HQ friction coefficient in QCD
amounts to $\gamma\simeq 0.3$-$0.9 \; c/\mathrm{fm}$ at $T = 250 \;
\mathrm{MeV}$, which is significantly smaller than the ``naive''
estimate of $\sim$$2~c/\mathrm{fm}$.


\subsection{Comparison of Elastic Diffusion Approaches}
\label{ssec_recon}

In view of the recent proliferation of seemingly different approaches to
evaluate HQ transport coefficients in the QGP it becomes mandatory to
ask to what extent they are related and encode similar 
microscopic mechanisms\cite{Rapp:2008zq}. It turns out that all of the
approaches discussed above incorporate a color-Coulomb-type
interaction. This is rather obvious for the $T$-matrix approach, where
the input potentials from lattice QCD clearly exhibit the Coulomb part
at sufficiently small distance (including effects of color
screening). The one-gluon exchange in pQCD (which is the dominant
contribution to HQ rescattering, recall the two right diagrams in
Fig.~\ref{fig_dia-pqcd}), also recovers the Coulomb potential in the
static limit (color screening enters via the Debye mass in the spacelike
gluon-exchange propagator). The collisional dissociation mechanism
involves the Cornell potential for the $D$- and $B$-meson wave functions
and thus incorporates a Coulomb interaction as well; the emphasis in
this approach is on formation-time effects essentially caused by the
different (free) binding energies of $D$ and $B$ mesons. In addition,
the confining part of the Cornell potential may play a role (as in the
$T$-matrix approach). Finally, in conformal field theory (AdS/CFT), the
absence of any scale promotes the Coulomb potential to the unique form
of a potential, $V(r) \propto 1/r$ (this is the only way of generating a
quantity with units of energy). On the other hand, scale-breaking
effects are present in the QCD-based approaches in terms of a running
coupling constant (pQCD), while the Cornell and lQCD-based potentials
additionally feature linear terms $\propto \sigma r$ where the string
tension introduces a further (nonperturbative) scale. In fact, in
Ref.\cite{Kaczmarek:2005gi} it has been argued, based on an analysis of
lQCD results for the heavy-quark free (and internal) energy, that
``remnants'' of the confining force play a prominent role for 
temperatures not too far above $T_c$ (e.g., for heavy quarkonium 
binding).

If one assumes the prevalence of the Coulomb interaction, the obvious
first question is with what strength (coupling constant) it figures into
the different approaches, which should be fairly straightforward to
determine. A more involved issue is to scrutinize the underlying
approximation schemes and their applicability. E.g., perturbative
approaches with large (running) coupling constants have poor (if any)
control over higher-order corrections. As usual in such situations,
diagrams with large contributions should be identified and resummed
(which is, of course, a non-trivial task, e.g., maintaining gauge
invariance); it would be illuminating to extract a static
gluon-exchange (Coulomb) potential for a given set of parameters. The
$T$-matrix approach performs a resummation of the ladder series of a
static (color-electric) potential; magnetic interactions are implemented
in a simplified manner using the Breit current-current interaction from 
electrodynamics. It has been verified that for large center-of-mass 
energies, the $qQ$ $T$-matrix recovers the result for perturbative
scattering. However, a number of effects are neglected and need to be
scrutinized, including the interactions with gluons beyond pQCD,
retardation, extra gluon or particle/antiparticle emission (e.g., in a
coupled channel treatment) and the validity (and/or accuracy) of a
potential approach at finite temperature (this issue will reappear in
the context of heavy quarkonia in Sec.~\ref{sec_onia}). In the
collisional dissociation approach, it would be interesting to explore
medium effects in the employed potential (i.e.,
on the mesonic wave function). Ideally, by improving on specific
assumptions in a given approach, an agreement would emerge establishing
a common result. Explicit connections with the AdS/CFT results are more
difficult to identify. Maybe it is possible to push the $T$-matrix
approach into a regime of ``large'' coupling, or study the existence and
properties of ($D$ and $B$) bound states in the string theory setting.
\begin{figure}[!t]
\begin{minipage}{0.495\textwidth}
\includegraphics[width=0.96\textwidth]{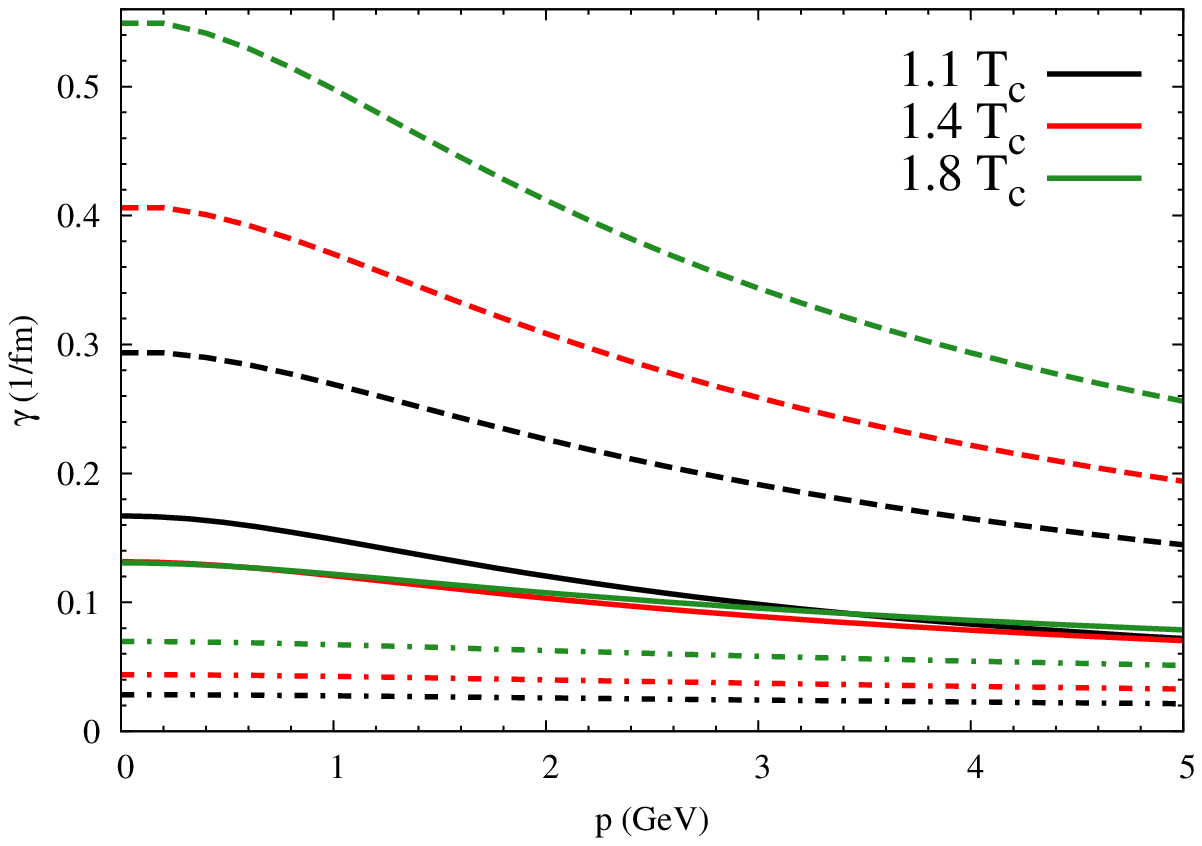}
\end{minipage}
\begin{minipage}{0.495\textwidth}
\includegraphics[width=0.98\textwidth]{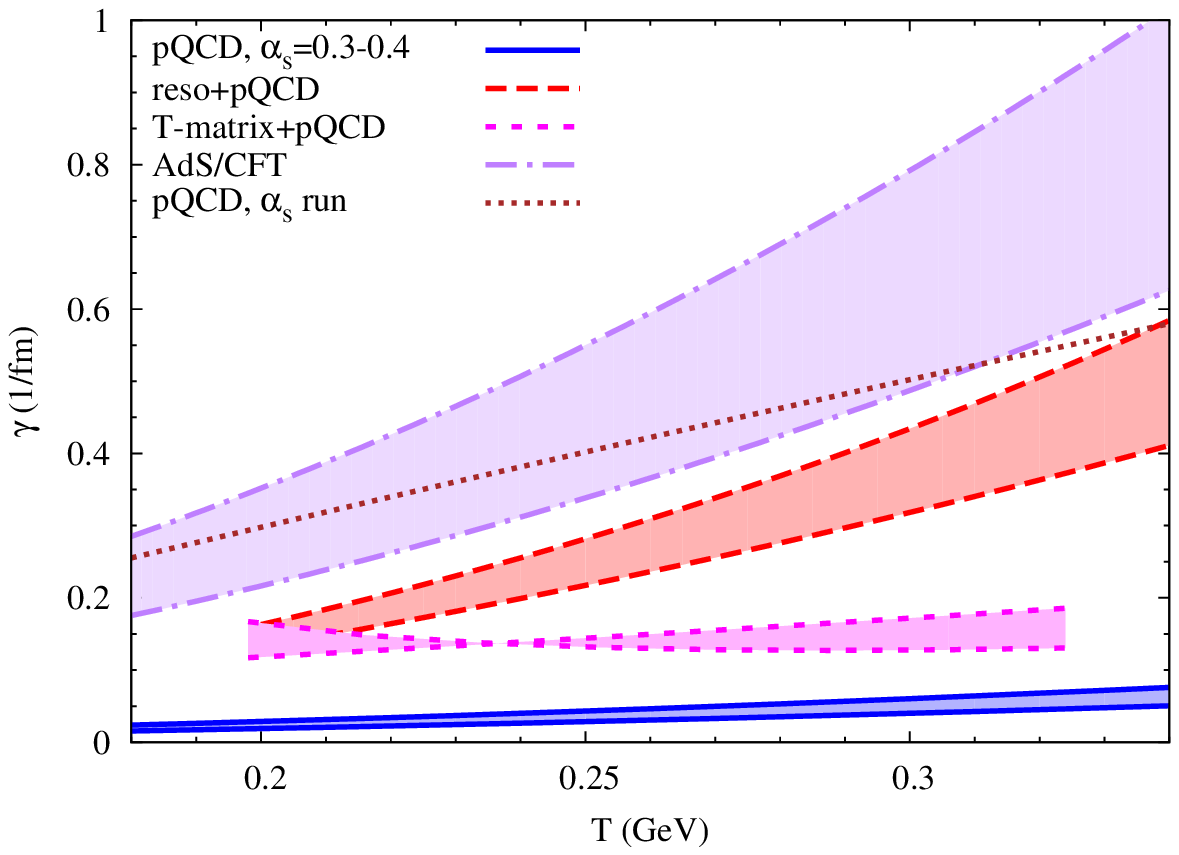}
\end{minipage}
\caption{(Color online) Charm-quark friction coefficients, $\gamma$, in
  the QGP.  Left panel: three-momentum dependence at three temperatures
  (color code) for: LO-pQCD with fixed $\alpha_s=0.4$ and $\mu_D=gT$
  (dash-dotted lines), heavy-light quark $T$-matrix plus LO-pQCD for
  gluons (solid lines)\protect\cite{vanHees:2007me}, and pQCD
  with running $\alpha_s$ and reduced infrared regulator (dashed
  lines)\protect\cite{Peshier:2008bg,Gossiaux:2008jv}. 
  Right panel: temperature
  dependence of $\gamma$ for LO-pQCD, $T$-matrix plus LO-pQCD (gluons
  only), pQCD with running $\alpha_s$, and from AdS/CFT correspondence
  matched to QCD{\protect\cite{Gubser:2006qh}} with
  $C=1.5$-$2.6${\protect\cite{Akamatsu:2008ge}}.}
\label{fig_gam}
\end{figure}
In Fig.~\ref{fig_gam} we summarize the drag coefficients as function of
momentum (for three temperatures, left panel) and temperature (for $p=0$, 
right panel) resulting from the approaches discussed above, i.e.,
\begin{enumerate}
\item[(i)] leading-order pQCD calculations with fixed $\alpha_s=0.4$ and
  Debye-screening mass, $\mu_D=g T$, in the $t$-channel gluon-exchange
  contributions to the matrix elements for elastic $gQ$ and $qQ$
  scattering,
\item[(ii)] in-medium $T$-matrix calculations using lQCD-based $qQ$
  potentials, augmented by the leading-order pQCD matrix elements for
  elastic $gQ$ scattering\cite{vanHees:2007me},
\item[(iii)] pQCD calculations with running $\alpha_s$ and reduced
  screening mass\cite{Peshier:2008bg,Gossiaux:2008jv}, and
\item[(iv)] the AdS/CFT correspondence matched to
  QCD\cite{Gubser:2006qh} with $\gamma_{\rm QCD} = C T^2/m_Q$ for
  $C=1.5$-2.6\cite{Akamatsu:2008ge} .
\end{enumerate}
At all temperatures, the $T$-matrix approach, (ii), produces significantly 
more HQ interaction strength than LO pQCD, (i), while for $T$$>$~0.2\,GeV 
the thermalization rate for the $T$-matrix is a factor of $\sim$2-4 less 
than for AdS/CFT, (iv), or for LO-pQCD with running coupling and reduced
infrared regulator, (iii). Close to $T_c$$\simeq$~180\,MeV, however,  
the three approaches (ii), (iii) and (iv) are not much different and
share overlap around $\gamma$$\simeq$~0.2\,$c$/fm. 
The spread in the numerical results reiterates the
necessity for systematic checks as indicated above.

\begin{figure}[!t]
\begin{minipage}{0.495\textwidth}
\includegraphics[width=0.98\textwidth]{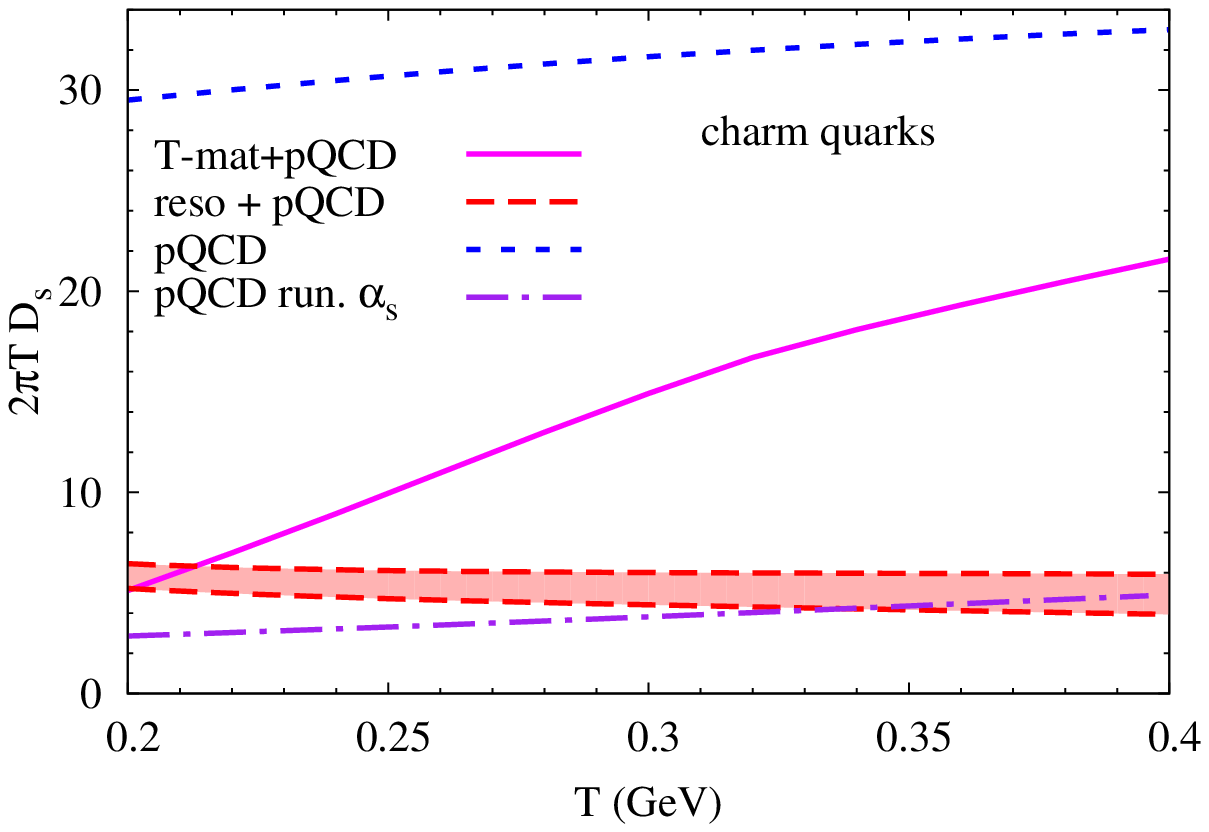}
\end{minipage}
\begin{minipage}{0.495\textwidth}
\includegraphics[width=0.98\textwidth]{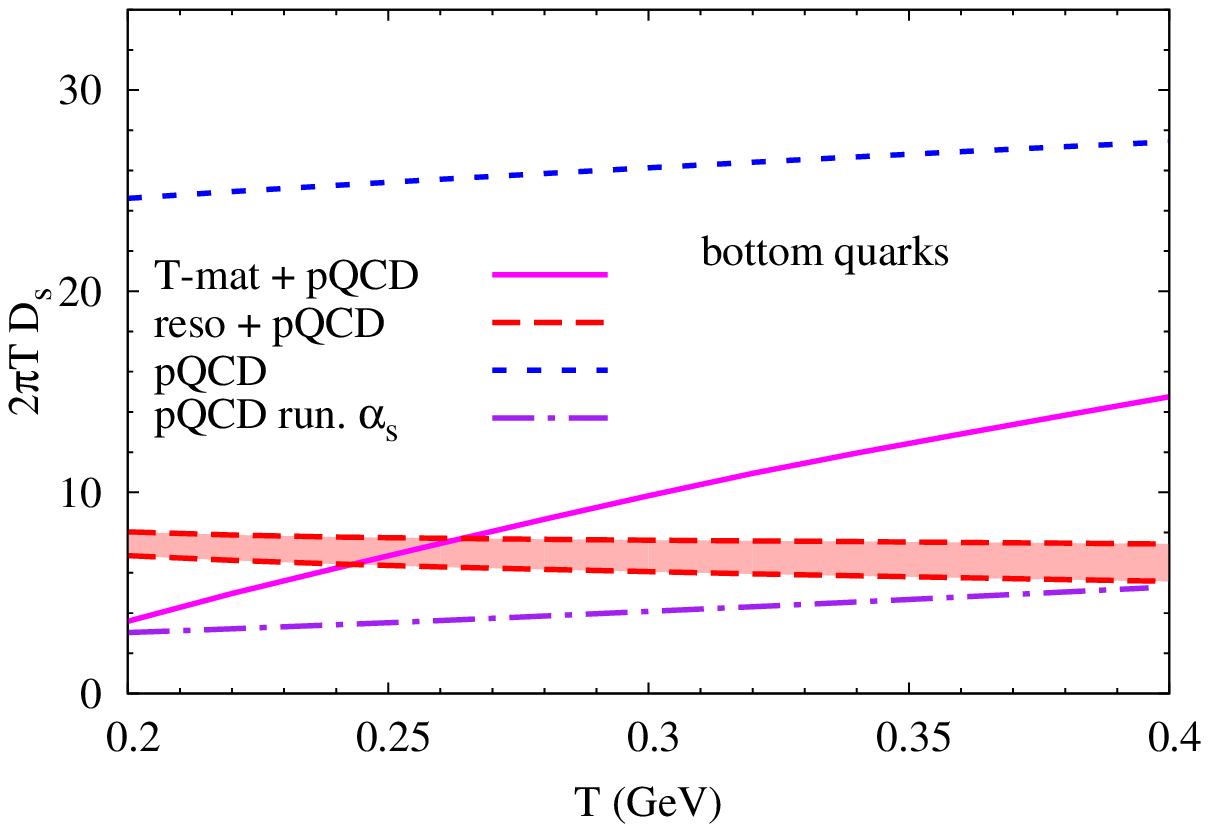}
\end{minipage}
\caption{(Color online) Spatial diffusion coefficient, $D_s=T/(\gamma
  m_Q)$, for $c$ (left) and $b$ quarks (right) in a QGP for: LO-pQCD
  with fixed $\alpha_s=0.4$ (dashed lines), effective resonance model +
  LO-pQCD (bands for
  $\Gamma_{D,B}=0.4$-$0.75\;\mathrm{GeV}$){\protect\cite{vanHees:2004gq}},
  $T$-matrix approach + LO-pQCD (gluons
  only){\protect\cite{vanHees:2007me}} and pQCD with running $\alpha_s$
  (dash-dotted line)\protect\cite{Peshier:2008bg,Gossiaux:2008jv}. 
  The AdS/CFT result
  corresponds to $2\pi T D_s=2\pi/C \simeq 1.5-4$ (not shown in the
  plots).}
\label{fig_Ds}
\end{figure}

Finally, one can convert the drag coefficients into estimates of other
HQ transport coefficients of the QGP. Within the Fokker-Planck approach
the spatial diffusion coefficient, $D_s$, is directly related to the
drag coefficient, $\gamma$, as given by Eq.~(\ref{spat-diff-mom-diff}).
Fig.~\ref{fig_Ds} shows the dimensionless quantity $2 \pi T D_s$ for
charm (left panel) and bottom quarks (right panel) as a function of
temperature for LO pQCD, LO pQCD with running coupling and reduced
infrared regulator, effective resonance model and $T$-matrix approach.
The former three are fairly constant as a function of temperature while
the $T$-matrix approach exhibits a significant increase with
temperature, indicating maximal interaction strength close to
$T_c$. This originates from the increasing potential strength (decrease
in color-screening) with decreasing temperature, enhancing resonance
correlations at lower temperature. It is tempting to interpret this
feature as a precursor phenomenon of hadronization. 
However, its robustness needs to be checked with a broader range of 
lattice potentials.
We recall that the internal-energy based potentials probably
provide an upper estimate for the strength of the interaction. It is
interesting to note that for all approaches the results for $b$
quarks coincide with the ones for $c$ quarks within
$\sim$$20$-$30\%$. The largest deviation is seen in the $T$-matrix
approach, where the (spatial) diffusion coefficient is smaller for $b$
quarks than for $c$ quarks ($B$-meson resonances survive until higher
temperatures than $D$ resonances). This is qualitatively similar to what
has been found for the collisional dissociation mechanism, where the
relative enhancement of the $b$-quark energy loss (compared to charm) 
is due to smaller $B$-meson formation times.  Since the latter are
related to larger $B$-meson binding energies, the dynamical origin of
the smaller $D_s$ for $b$ quarks appears to be of similar origin as in
the $T$-matrix approach. The relative magnitudes of the various
approaches reflect what we discussed before for the drag coefficient.


\subsection{Collisional vs. Radiative Energy Loss}
\label{ssec_rad}

For slowly moving heavy quarks in the QGP, the
parametrically dominant interaction is elastic scattering. However, at
high $p_T$, radiative scattering is believed to eventually become the
prevailing energy-loss mechanism. It is currently not known at which
$p_T$ this transition occurs. Therefore, it is important to assess the
relative importance between elastic and inelastic scattering processes
in the medium, even at the level of perturbative scattering only. Toward
this purpose, we first recollect basic results on the
gluon-Bremsstrahlung mechanism for light-parton, and then HQ, energy
loss in the QGP, followed by a direct comparison to collisional energy
loss for heavy quarks.

A seminal perturbative treatment of gluo-radiative energy loss (E-loss) 
of high-energy partons in the QGP has been given in
Refs.\cite{Baier:1994bd,Baier:1996kr} (BDMPS). The medium is modeled as
static scattering centers which implies that the E-loss is purely
radiative. The key finding is that the E-loss due to multiple
in-medium scattering of a high-energy parton grows as $L^2$, where $L$
is the path length of the parton traversing the medium. The static
scattering centers, at positions $\bvec{x}_i$, are described by screened
Coulomb potentials,
\begin{equation}
\label{bdmps-scat-cent}
V_i(\bvec{q})=\frac{g}{\bvec{q}^2+\mu_D^2} \exp(-\ii \bvec{q} \bvec{x}_i) \ .
\end{equation}
The range of the potentials is assumed to be small compared to the mean
free path, $\lambda$, of the scattered parton, i.e., 
$1/\mu_D \ll \lambda$. In this case successive scatterings can be 
considered as independent, thus enabling an eikonal approximation for 
the elastic scattering on static centers, i.e., a classical propagation 
of the particle with energy $E \gg \mu_D$, undergoing independent kicks, 
thereby radiating Bremsstrahlung gluons. In analogy to the QED case an 
important ingredient is the coherent resummation of the 
multiple-scattering Bremsstrahlung amplitudes 
(``Landau-Pomeranchuk-Migdal effect'') which can be formulated as a 
diffusion equation for the effective
scattering amplitudes (or pertinent currents). The total radiative
E-loss of a high-energy parton traversing a medium of path length
$L$ is then given by
\begin{equation}
\label{qhat-e-loss}
\Delta E=\frac{\alpha_s}{2} \hat{q} L^2  \ ,
\end{equation}
where $\hat{q}$ is the diffusion coefficient for transverse-momentum
broadening in scattering off the
static scattering centers, $\erw{q_T^2}=\hat{q} L$. Perturbative
calculations of the transport coefficient result in a value of about
$\hat{q}\simeq 1 \; \GeV/\fm^2$ at typical energy densities of $\epsilon
\simeq 10 \;\GeV/\fm^3$ (translating into $T\simeq 250 \; \MeV$)
relevant for the QGP at RHIC\cite{Baier:2002tc}. It turns out, however,
that the description of high-momentum pion suppression at RHIC in
the BDMPS formalism requires an approximate ten-fold increase of the
perturbative value for $\hat{q}$\cite{Eskola:2004cr}. Recent 
calculations of perturbative E-loss including both
elastic and radiative contributions within a thermal-field-theory
framework indicate that collisional E-loss may be significant even
for high-$p_T$ light partons\cite{Qin:2007rn}. This would imply a
reduction of the value required for $\hat{q}$ from RHIC phenomenology.
                                                   
An early calculation\cite{Mustafa:1997pm} of radiative charm-quark
E-loss, $-\dd E/\dd x$, in the QGP has found that it dominates 
over the elastic one down to rather small momenta, 
$p\le 2$\,GeV\cite{GolamMustafa:1997id}. 
In Ref.\cite{Dokshitzer:2001zm} it has been pointed out that the
application of radiative E-loss to heavy quarks leads to the
appearance of the so-called ``dead cone'', i.e., the suppression of
forward gluon radiation for $\Theta<m_Q/E$, where $\Theta$ denotes the
direction of motion of the gluon with energy $E$, relative to the
direction of the HQ momentum\cite{Dokshitzer:2001zm}. It has been 
predicted that the reduced HQ E-loss leads to a 
heavy-to-light hadron ratio above one in the high-$p_T$ regime
accessible at RHIC.  Within the BDMPS model, extended to heavy quarks,
it has been argued\cite{Armesto:2003jh}, however, that medium-induced
gluon radiation tends to fill the dead cone. As will be discussed in
Sec.~\ref{ssec_rhic-obs}, a similar value for $\hat{q}$ as in the
light-hadron sector is necessary to come near the observed suppression
of high-$p_T$ electrons from HQ decays in terms of radiative E-loss
alone\cite{Armesto:2005mz}.

The BDMPS formalism for light partons has been generalized to resum an
expansion of gluo-radiative parton E-loss in the GP with opacity,
$\bar{n}=L/\lambda$, employing a so-called reaction operator
approach\cite{Gyulassy:2000er} (GLV).  A reaction operator $\hat{R}_n$
is constructed that relates the $n^{\mathrm{th}}$ power in a
opacity-inclusive radiation probability distribution to classes of
diagrams of order $n-1$. This results in a recursion relation for the
radiation probability distribution, corresponding to a certain
resummation to all orders in opacity, which can be implemented in
Monte-Carlo simulations for jet quenching. The GLV reaction-operator
method for light-parton radiative E-loss in the QGP has been
extended to heavy quarks in Ref.\cite{Djordjevic:2003zk} (DGLV), 
implementing the kinematical suppression of gluon radiation by the HQ 
mass in the ``dead cone''.
 
A direct study of the relative magnitude of collisional (elastic) and
radiative pQCD HQ E-loss in the GP has been undertaken in
Ref.\cite{Wicks:2005gt}. For the elastic E-loss of a parton with
color Casimir constant, $C_R$, the leading logarithm expression in an
ideal QGP with $N_f$ effective quark flavors at temperature $T$,
\begin{equation}
\label{el-loss-leading-log}
\frac{\dd E^{\text{el}}}{\dd x} = C_R \pi \alpha_s^2 T^2 \left
  (1+\frac{N_f}{6} \right) f(v) \ln(B_c),
\end{equation}
has been used. In an ultrarelativistic gas of massless partons the
jet-velocity function is given by
\begin{equation}
\label{jet-vel-f}
f(v)=\frac{1}{v^2} \left [v+\frac{1}{2}(v^2-1) \ln \left
    (\frac{1+v}{1-v} \right ) \right ],
\end{equation}
while estimates for $B_c$ are taken from Refs.~[Bj]\cite{Bjorken:1982tu},
[TG]\cite{Thoma:1990fm}, and
[BT]\cite{Braaten:1991jj,Braaten:1991we}. The different
values for $B_c$ obtained in these models are considered as reflecting
theoretical uncertainties. The radiative E-loss within the DGLV
reaction-operator approach is calculated in Ref.\cite{Djordjevic:2003zk}
based on Refs.\cite{Gyulassy:2000er,Gyulassy:2001nm}.
\begin{figure}[!t]
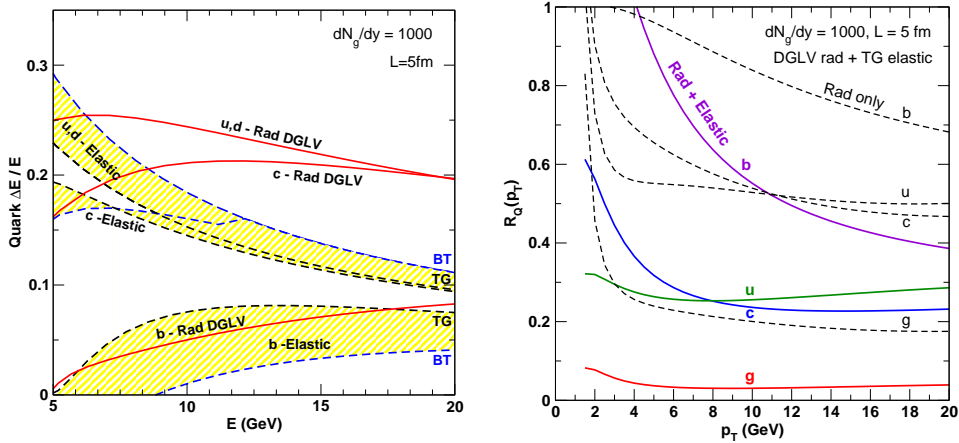

\begin{minipage}{0.48\linewidth}
\includegraphics[width=\textwidth]{de-over-e-coll-vs-rad-Wicks.eps}
\end{minipage}\hfill
\begin{minipage}{0.48\linewidth}
\includegraphics[width=\textwidth]{raa_parton_l5.eps}
\end{minipage}
\caption{(Color online) Left panel: average relative E-loss, 
 $\Delta E/E$, for $u$, $c$ and $,b$ quarks as a function of jet energy, 
 $E$, in a longitudinally (Bjorken) expanding QGP, with fixed path 
 length $L=5$\,fm, initial gluon rapiditiy density $\dd N_g/\dd y=1000$ 
 and fixed $\alpha_s=0.3$; the gluon (light-quark) mass is set to 
 $\mu_D/\sqrt{2}$ ($\mu_D/2$), the $c$($b$)-quark mass to 
 $m_c=1.2(4.75)$\,GeV (solid lines: radiatve E-loss, dashed bands: 
 elastic E-loss in two schemes as discussed in the text). 
 Right panel: parton nuclear modification factor, $R_{AA}$$\equiv$$R_Q$, 
 for gluons, $u$-, $c$-, and $b$-quarks as a function of $p_T$ for a 
 fixed path length and $\dd N_g/\dd y=1000$ (dashed lines: radiative 
 E-loss, solid lines: radiative+elastic E-loss).}
\label{fig.rad-coll-eloss-el-raa-Wicks}
\end{figure}
The left panel of Fig.~\ref{fig.rad-coll-eloss-el-raa-Wicks} compares
pQCD radiative and collisional E-loss for various quark flavors
(masses) at high $p_T>5 \, \GeV$ in a gluon plasma (GP) with $T\simeq
240\,\MeV$. For light and charm (bottom) quarks the elastic E-loss
is comparable to the radiative one up to $p_T\simeq 10(20)\,\GeV$, and
still significant above. The right panel of
Fig.~\ref{fig.rad-coll-eloss-el-raa-Wicks} reiterates that, within pQCD,
collisional E-loss is an essential component in calculating the
suppression of light-parton and especially HQ spectra at RHIC. Note that
the relative importance of collisional E-loss is expected to
increase if non-perturbative effects become relevant (which
predominantly figure toward lower $p_T$), or if the GP is replaced by a
QGP.


\subsection{$D$ Mesons in the Hadronic Phase}
\label{ssec_hadronic}

To complete the discussion of open charm in QCD matter we briefly
address medium modifications of charm hadrons in hadronic matter.
Pertinent studies may be divided into calculations for cold nuclear
matter as well as for hot meson matter.

\begin{figure}[!t]
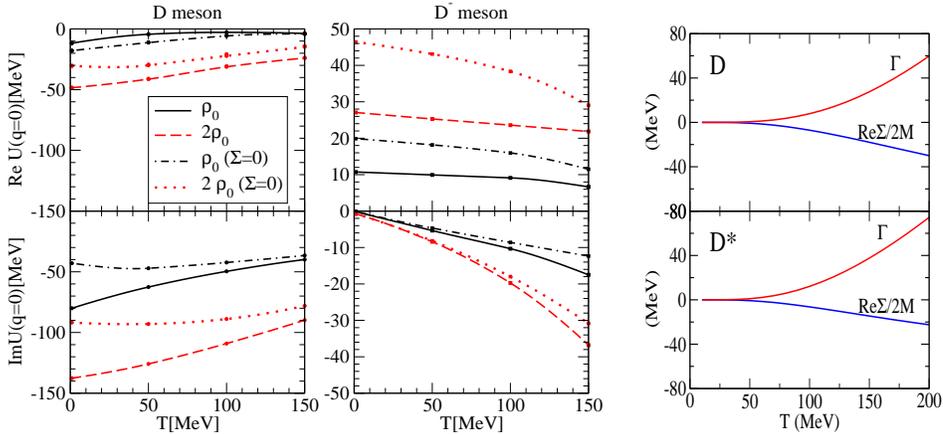

\begin{minipage}{0.66\textwidth}
\includegraphics[width=0.95\textwidth]{D-med-nuc.eps}
\end{minipage}
\begin{minipage}{0.33\textwidth}
\vspace{0.3cm}
\includegraphics[width=0.95\textwidth,height=1.3\textwidth]{D-med-pi.eps}
\end{minipage}
\caption{(Color online) Real and imaginary parts of $D$-meson
  selfenergies in dense and/or hot hadronic matter. Left and middle
  panels: selfconsistent coupled-channel calculations for in-medium
  $D^+$ (left) and $D^-$ (right) potentials based on $DN$ scattering in
  a hot nucleon gas as a function of
  temperature\protect\cite{Tolos:2007vh} for two different nuclear
  densities (and for two different inputs for the $\bar DN$ scattering
  length); the real (upper panels) and imaginary (lower panels) part of
  the potentials are defined in terms of the on-shell selfenergy via
  $U(q)=\Sigma_D(\omega_D^{\rm on}(q),q)/2\omega_D^{\rm on}$ with 
  $\omega_D^{\rm on}$ the
  quasiparticle energy; note that the width is given by $\Gamma=-2 \im
  U$.  Right panel: width (upper lines) and mass shift (lower lines) of
  $D$ (top) and $D^*(2010)$ (bottom) mesons in a hot pion gas based on
  resonant scattering via chiral partners\protect\cite{Fuchs:2004fh};
  note that the isospin symmetric pion gas implies equal effects on
  $D$-mesons ($D^+$, $D^0$) and anti-$D$ mesons ($D^-$, $\bar D^0$).  }
\label{fig_D-med-hg}
\end{figure}

Early studies of $D$-mesons in cold nuclear matter focused on possible
mass shifts due to scalar and vector mean fields acting on the
light-quark content of the meson\cite{Sibirtsev:1999js}. At normal
nuclear matter density $\varrho_N\equiv\varrho=0.16$\,fm$^{-3}$, 
attractive mass shifts of up to $-100$\,MeV have been reported for $D^+$
and $D^0$ mesons (where both mean fields contribute with the same sign)
while the mass change of the $D^-$ and $\bar D^0$ turned out to be small
due to a cancellation of the mean fields. Similar findings have been
reported in QCD sum rule calculations\cite{Hayashigaki:2000es} where the
(isospin-averaged) $D$-meson mass is reduced by about $-50$\,MeV, mostly 
as a consequence of the reduction in the light-quark condensate. 
Rather different results are obtained in
microscopic calculations of $D$-meson selfenergies (or spectral
functions) based on coupled channel $T$-matrices for $DN$ scattering 
in nuclear matter\cite{Tolos:2004yg,Lutz:2005vx}. These calculations
incorporate hadronic many-body effects, most notably $DN$ excitations
into charm-baryon resonances not too far from the $DN$ threshold, e.g.,
$\Lambda_c$(2593) and $\Sigma_c$(2625), as well as charm exchange into
$\pi\Lambda_c$ and $\pi\Sigma_c$ channels.  In Ref.\cite{Tolos:2004yg}
separable meson-baryon interactions have been employed with parameters
constrained to dynamically generate the $\Lambda_c(2593)$ state. Since
the in-medium $D$-meson spectral function figures back into the
$T$-matrix, one is facing a selfconsistency problem (much like for the
heavy-light quark $T$-matrix discussed in
Sec.~\ref{sssec_tmat}). Selfconsistent calculations including nucleon
Pauli blocking and dressing of intermediate pion and nucleon propagators
result in $D$-meson spectral functions with a significant broadening of
up to $\Gamma_D\simeq 100$\,MeV but a rather small shift of the peak
position of about $-10 \; \mathrm{MeV}$ (for 
$\varrho_N\equiv\varrho=0.16 \, \mathrm{fm}^{-3}$). In
Ref.\cite{Lutz:2005vx}, a somewhat stronger coupling of $DN$ to the
$\Lambda_c(2593)$ results in a stronger collective
$DN^{-1}\Lambda_c$(2593) mode (about $250 \; \mathrm{MeV}$ below the
free $D$-meson mass) and a pertinent level repulsion which pushes up the
``elementary'' $D$-peak by $\sim$$30 \, \mathrm{MeV}$. Also in this
calculation the broadening is significant, by about $\sim$$50 \,
\mathrm{MeV}$. The $D^-$ was found to be rather little affected, neither
in mass nor in width. Investigations in the selfconsistent
coupled-channel framework have been extended to a nucleon gas at finite
temperature\cite{Tolos:2007vh} with a more complete treatment of $DN$
scattering, cf. left and middle panels of Fig.~\ref{fig_D-med-hg}.
The thermal motion of nucleons implies that a larger kinematic regime in
the center-of-mass (cm) energies in the scattering amplitude is probed
(compared to $T$=0). For the real parts this leads to a further 
averaging of the positive and negative parts of the amplitude, while the
imaginary parts are negative definite (some loss of interaction strength
may occur in channels with resonances close to threshold). More
quantitatively, at normal nuclear matter density, the resulting mass
shifts are $10$-$20 \, \mathrm{MeV}$ at $T=0$ (attractive for $D$ and
repulsive for $\bar{D}$), decreasing to about half (or less) at $T=150\,
\mathrm{MeV}$. On the other hand, the $D$-meson width is around $100\,
\mathrm{MeV}$ at both zero and finite temperature, while the $\bar D$
width is small at $T=0$ but increases to about $30\; \mathrm{MeV}$ at
$T=150 \, \mathrm{MeV}$.

Medium modifications of $D$-mesons in a hot pion gas have been studied
in Ref.\cite{Fuchs:2004fh}. The main idea in this work is to implement
the recently discovered scalar $D_0^*(2310)$ and axialvector
$D_1'(2430)$ states as chiral partners of the pseudoscalar $D$ and
vector $D^*(2010)$ mesons, respectively. Their large widths of
200-400\,MeV are primarily attributed to $S$-wave pion decays 
into $D$ and $D^*$. In a thermal pion gas, $D_0^*(2310)$ and
$D_1'(2430)$ therefore act as strong resonances in $D\pi$ and $D^*\pi$
scattering, which have been treated in Breit-Wigner
approximation. In addition, $D$-wave resonances, $D_1(2420)$ and
$D_2^*(2460)$, have been accounted for. The resulting collisional widths
of $D$ and $D^*$ reach up to about $\sim$$40$-$60\,\MeV$ for
temperatures around $T=175\;\MeV$, while the mass shifts are attractive
up to $-20 \ \MeV$. It can be expected that (e.g., in a selfconsistent
calculation) the inclusion of medium effects on the resonances (e.g.,
chiral partners are expected to approach degeneracy towards chiral
restoration) will lead to a reduction in the mass shift (not so much for
the widths). Above the critical temperature, it is then natural to
switch to a quark-based description, i.e., $c+q$ scattering, much like
in the effective resonance model discussed in Sec.~\ref{sssec_reso}.

When combining the effects of pion and nucleon scattering on $D$-mesons
in hot hadronic matter, their total width at temperatures around $T_c$ 
adds up to $\Gamma_D^{\rm tot}(T=180\,{\rm MeV})\simeq 150$\,MeV. This 
is only by about a factor of $\sim$2 smaller than what was found for 
$c$-quarks at $T=1.1\,T_c$ in the $T$-matrix approach for $c$-$q$ 
scattering, cf.~right panel of Fig.~\ref{fig_ImT-Sig}. Since it can be 
expected that other excited hadrons contribute to $D$-meson rescattering 
(albeit with less strength), $D$-meson transport properties may not be 
much different from those of $c$-quarks in the QGP, at least at
temperatures close to $T_c$. It is therefore of considerable interest 
to employ in-medium $D$-meson $T$-matrices to evaluate heavy-flavor
transport coefficients in hadronic matter.


\section{Heavy-Quark Observables in Relativistic Heavy-Ion Collisions}
\label{sec_hq-obs}

One of the main motivations for the vigorous theoretical studies of HQ
diffusion in the QGP is the
possibility of utilizing HQ observables in ultrarelativistic heavy-ion
collisions as a quantitative probe of the matter produced in these
reactions. If the latter reaches approximate local thermal equilibrium,
such applications can be performed by solving the Fokker-Planck equation
for a heavy quark diffusing within a collectively expanding background
medium with space-time dependent temperature and flow field (applicable 
for ``sufficiently slow'' charm and bottom quarks). This is typically
realized with a Monte-Carlo simulation using a test-particle ansatz for
an equivalent stochastic Langevin equation. In such a formulation, a
direct relation between the input in terms of a (temperature-dependent)
HQ diffusion coefficient and the modifications of HQ spectra in the
evolution can be established. In addition, the Langevin formulation
admits an efficient implementation of the dissipation-fluctuation
relation for relativistic kinematics.

Alternatively, the modifications of HQ spectra in URHICs have been 
evaluated by implementing test particles into numerical transport
simulations of the background 
medium\cite{Bratkovskaya:2004ec,Zhang:2005ni,Molnar:2006ci}.
This, in principle, accounts for non-equilibrium effects in the medium
evolution (which could be particularly relevant for high-$p_t$ particles
in the bulk), but the connection to the diffusion concept becomes less
direct (HQ cross sections need to be evaluated in an equilibrium medium
to extract ``equivalent'' diffusion coefficients).  However, when
analyzing theoretical predictions of HQ spectra we also compare to
results of transport models for the bulk evolution.

We start our presentation in this Section by briefly outlining how the
Fokker-Planck equation is implemented into numerical simulations based
on a relativistic Langevin process (Sec.~\ref{ssec_langevin}). This is
followed by a discussion of different models for the background medium
in relativistic heavy-ion collisions, where we focus on thermal models
including hydrodynamics and expanding fireballs
(Sec.~\ref{ssec_medium}). The main task here is to provide 
realistic benchmarks for the conversion of transport coefficients
into modifications of HQ spectra. This furthermore requires the
definition of controlled initial conditions for the HQ spectra
(Sec.~\ref{ssec_initial}), usually taken from $p$-$p$ collisions,
possibly augmented by nuclear effects (in particular a ``Cronin''
$p_t$-broadening). Available Langevin simulations combining different
inputs are quantitatively compared at the level of the final-state HQ
spectra resulting from the QGP and ``mixed'' phases in central and
semicentral Au-Au collisions at RHIC (Sec.~\ref{ssec_hq-spectra}).  Even
though HQ spectra are not observable, they provide the cleanest
theoretical level of comparison, before further processing through
hadronization, electron decay and charm/bottom composition occurs. The
latter three steps are necessary to enable comparisons to currently
available electron data (Sec.~\ref{ssec_rhic-obs}), and thus arrive at
an empirical estimate of the HQ diffusion coefficient characterizing the
QCD medium produced at RHIC. In a more speculative step, the extracted
HQ transport coefficient may be used to schematically estimate the ratio
of shear viscosity to entropy density (Sec.~\ref{ssec_viscosity}), which
has recently received considerable attention in connection with viscous
hydrodynamic simulations at RHIC.


\subsection{Relativistic Langevin Simulations}
\label{ssec_langevin}
The Fokker-Planck equation, introduced in Sec.~\ref{ssec_diff}, is
equivalent to an ordinary stochastic differential equation. Neglecting
mean-field effects of the medium, the force acting on the heavy particle
is divided into a ``deterministic'' part, describing its average
interactions with the light particles in the medium (friction or drag),
and a ``stochastic'' part, taking into account fluctuations around the
average on the level of the standard deviation. Thus the relativistic
equations of motion for a heavy quark become a coupled set of stochastic
differential equations, which for an isotropic medium can be written in
the form
\begin{equation}
\begin{split}
\dd x_j &= \frac{p_j}{E} \dd t \ , \\
\dd p_j &= -\Gamma p_j \dd t+ \sqrt{\dd t} C_{jk} \rho_k \ , 
\end{split}
\label{langevin}
\end{equation}
where $E=(m_Q^2+\bvec{p}^2)^{1/2}$, and $\Gamma$ and $C_{jk}$ are 
functions of $(t,\bvec{x},\bvec{p})$
with $j,k$=1,2,3; they are related to the transport coefficients $A$ and
$B$ (discussed in the previous section) below.  $\Gamma$ and $C_{jk}$
describe the deterministic friction (drag) force and the stochastic
force in terms of independent Gaussian-normal distributed random
variables $\bvec{\rho}=(\rho_1,\rho_2,\rho_3)$,
\begin{equation}
\label{2.1.20}
P(\bvec{\rho})=\left (\frac{1}{2 \pi} \right )^3 \exp \left
  (-\frac{\bvec{\rho}^2}{2} \right )  \ ,
\end{equation}
In the limit $\dd t \rightarrow 0$, the covariance of the
fluctuating force is thus given by
\begin{equation}
\erw{F_j^{(\mathrm{fl})}(t) 
F_k^{(\mathrm{fl})}(t')}=C_{jl} C_{kl} \delta(t-t') \ . 
\label{force-corr}
\end{equation}
However, with these specifications the stochastic process is not yet
uniquely defined, but depends on the specific choice of the momentum
argument of the covariance matrix, $C_{jk}$, in
Eq.~(\ref{langevin})\cite{Dunkel-Haenggi:2008}, i.e., the definition of
the stochastic integral. Usual schemes are given by the pre-point Ito,
the mid-point Stratonovic-Fisk, and the post-point Ito (or
H{\"a}nggi-Klimontovich\cite{Haenggi:2005}) interpretation of the
stochastic integral. We can summarize all these realizations of the
stochastic process by specifying the actual momentum argument in the
covariance matrix by
\begin{equation}
\label{2.1.21}
C_{jk} \rightarrow C_{jk}(t,\bvec{x},\bvec{p}+\xi \dd \bvec{p}) \ ,
\end{equation}
where $\xi=0,1/2,1$ corresponds to the pre-point Ito, the mid-point
Stratonovic, and the post-point Ito realizations, respectively. The
equation for the corresponding phase-space distribution function can be
found by calculating the average change of an arbitrary phase-space
function, $g(\bvec{x},\bvec{p})$, with time. According to
Eq.~(\ref{langevin}), with the specification Eq.~(\ref{2.1.21}) of the
stochastic process, we find
\begin{equation}
\begin{split}
\label{2.1.22}
\erw{g(\bvec{x}+\dd \bvec{x},\bvec{p}+\dd \bvec{p}) -
  g(\bvec{x},\bvec{p})} = \Bigg \langle & \frac{\partial g}{\partial
  x_j} \frac{p_j}{E} + \frac{\partial g}{\partial p_j} \left ( -\Gamma p_j +
  \xi
  \frac{\partial C_{jk}}{\partial p_l} C_{lk} \right ) \\
& + \frac{1}{2} \frac{\partial^2 g}{\partial p_j \partial p_k} C_{jl}
C_{kl} \Bigg \rangle \dd t+ \mathcal{O}(\dd t^{3/2}).
\end{split}
\end{equation}
Here, the arguments of both, $\Gamma$ and $C_{jk}$, have to be taken at
$(t,\bvec{x},\bvec{p})$ since the corrections are of the neglected
order, $\mathcal{O}(\dd t^{3/2})$. In the derivation of this equation
the statistical properties of the random variables $\rho_i$, implied by
Eq.~(\ref{2.1.20}),
\begin{equation}
\label{2.1.23}
\erw{\rho_j}=0 \ , \quad \erw{\rho_j \rho_k}=\delta_{jk} \ ,
\end{equation}
have been used. It follows that the average of an arbitrary phase-space
function is by definition given by the phase-space distribution function
for the heavy particle (in our context a heavy quark),
$f_Q(t,\bvec{x},\bvec{p})$, e.g.,
\begin{equation}
\label{2.1.24}
\frac{\dd}{\dd t} \erw{g(\bvec{x},\bvec{p})} = \int \dd^3 \bvec{x} \int
\dd^3 \bvec{p} \, g(\bvec{x},\bvec{p}) \frac{\partial}{\partial t}
f_Q(t,\bvec{x},\bvec{p}) \ . 
\end{equation}
After integrations by parts, and since Eq.~(\ref{2.1.22}) holds for any
function $g$, one finally arrives at the Fokker-Planck equation,
\begin{equation}
\label{fp-2}
\frac{\partial f_Q}{\partial t} + \frac{p_j}{E} \frac{\partial f_Q}{\partial
  x_j} = \frac{\partial}{\partial p_j} \left [\left (\Gamma p_j-\xi C_{lk}
    \frac{\partial C_{jk}}{\partial p_l} \right ) f_Q \right ]
+\frac{1}{2} \frac{\partial^2}{\partial p_j \partial p_k} \left
  (C_{jl} C_{kl} f_Q \right) \ .
\end{equation}
The drag term, i.e., the first term on the right-hand side of this
equation, depends on the definition of the stochastic integral in terms
of the parameter $\xi$. Comparison with Eq.~(\ref{2.1.8}) shows that,
independent of the choice of $\xi$, the covariance matrix is related to
the diffusion matrix by
\begin{equation}
\label{Cjk-B}
C_{jk}=\sqrt{2 B_0} P_{jk}^{\perp}+\sqrt{2 B_1} P_{jk}^{\parallel} \ ,
\end{equation}
while the friction force is given by
\begin{equation}
\label{2.1.25b}
  \Gamma p_j=A p_j-\xi C_{lk}\frac{\partial C_{jk}}{\partial p_l} \ .
\end{equation}
Numerical investigations have shown that the drag and diffusion 
coefficients inferred from microscopic models according to 
Eqs.~(\ref{2.1.10c}) in general do not warrant a good agreement 
of the long-time limit of the solution to the Fokker-Planck evolution
with the relativistic equilibrium J{\"u}ttner-Boltzmann distribution
(where the temperature is given by the background medium).  
The problem is with the longitudinal diffusion coefficient, $B_1$, 
which induces an overestimate of the corresponding fluctuating forces. 
Thus, one typically \emph{adjusts} the drag coefficient by
choosing $B_1$ in Eq.~(\ref{fp-2}) to satisfy the asymptotic 
equilibration condition\cite{Arnold:2000a,Arnold:2000b,Moore:2004tg}.

To find the dissipation-fluctuation relation, imposed by the
equilibration condition, we first study the heavy quark's motion in a
heat bath in thermal equilibrium in its rest frame. Then the momentum
distribution of the heavy quarks should become a J{\"u}ttner-Boltzmann
distribution,
\begin{equation}
\label{2.1.26}
f_Q^{\mathrm{eq}}(\bvec{p}) \propto \exp \left(-E/T\right) 
\end{equation}
with the temperature, $T$, imposed by the heat bath. For a Langevin
process with $B_0=B_1=D$, i.e.,
\begin{equation}
\label{Cjk-D}
C_{jk}=\sqrt{2 D(E)} \delta_{jk} \ ,
\end{equation}
where the diffusion coefficient has been written as a function of the
heavy quark's energy, $E$, the equilibration condition for a given
parameter $\xi$ in Eq.~(\ref{2.1.21}) is obtained by using
Eqs.~(\ref{2.1.26}) and Eq.~(\ref{Cjk-D}) in Eq.~(\ref{fp-2}):
\begin{equation}
\label{2.1.28}
A(E) E T - D(E) + T(1-\xi) D'(E)=0 \ .
\end{equation}
Since the drag and diffusion coefficients are usually given numerically,
the most convenient update rule for the Langevin process is achieved by
setting $\xi=1$, i.e., using the post-point Ito
(H{\"a}nggi-Klimontovich) rule for the stochastic integral in
Eq.~(\ref{langevin}) and imposing the simple relativistic
dissipation-fluctuation relation,
\begin{equation}
\label{2.1.29}
D=A E T \ .
\end{equation}
This guarantees the proper approach of the heavy quark's phase-space
distribution to the appropriate equilibrium distribution with the
temperature imposed by the heat bath.

For the more general form of the covariance matrix, Eq.~(\ref{Cjk-B}),
the post-point Ito value, $\xi=1$, has been chosen in
Ref.\cite{vanHees:2005wb}, and the longitudinal diffusion coefficient is
set to
\begin{equation}
\label{2.1.30}
B_1=A E T \ , 
\end{equation}
while the drag coefficient $A$ as well as the transverse diffusion
coefficient, $B_0$, are used as given by Eq.~(\ref{2.1.10c}) for the
various microscopic models for HQ scattering in the QGP.
Comparing to Eq.~(\ref{2.1.28}), one finds that
this is equivalent to the strategy followed in Ref.\cite{Moore:2004tg}
of using the prepoint-Ito rule, $\xi=0$, but to adjust the drag
coefficient according to the dissipation-fluctuation relation
Eq.~(\ref{2.1.28}).


\subsection{Background Medium in Heavy-Ion Collisions}
\label{ssec_medium}
For HQ transport coefficients computed in an equilibrium QGP,
the natural and consistent framework to describe the evolving
medium in heavy-ion collisions are hydrodynamic simulations, 
formulated in the same (thermodynamic) variables. This choice is
further rendered attractive by the success of ideal hydrodynamics in
describing bulk observables at RHIC, in particular $p_T$ spectra and
elliptic flow of the most abundant species of
hadrons\cite{Kolb:2003dz,Shuryak:2004cy,Huovinen:2006jp,Hirano:2008aj}. 
The agreement with meson and baryon spectra
typically extends to $p_T\simeq 2$-$3\,\GeV$, respectively. At the
parton level, this converts into a momentum of $p_t\simeq 1\,\GeV$, 
which approximately coincides with the ``leveling-off''
of the experimentally observed $v_2(p_t)$ (at higher momenta
hydrodynamics overestimates the elliptic flow). On the one hand, this
appears as a rather small momentum in view of the ambition of describing
HQ spectra out to, say, $p_t\simeq 5\,\GeV$. However, one
should realize that (a) more than $90\%$ of the bulk matter is comprised
of light partons with momenta below $p_t\simeq1\,\GeV$, and
(b) the velocity of a $p_t=5\,\GeV$ charm quark (with $m_c=1.5\,\GeV$) 
is very similar to a $p_t=1\,\GeV$ light
quark (with $m_q=0.3\,\GeV$). This suggests that most of the
interactions of a $p_t=5\,\GeV$ charm quark actually occur with soft 
light partons (which are well described by a hydrodynamic bulk).
This has been verified by explicit
calculations\cite{vanHees:2009} and is, after all, a prerequisite for
the applicability of the Fokker-Planck approach (i.e., small momentum
transfer per collision). On the other hand, one may be concerned that
the overestimate of the experimentally observed elliptic flow at
intermediate and high $p_t$ within hydrodynamics may exaggerate the HQ
elliptic flow in Langevin simulations. This is, however, not necessarily
the case, since the transfer of $v_2$ from the bulk to the heavy quark
critically depends on the light-parton phase space density (the drag 
coefficient is proportional to it); since the hydrodynamic spectra fall 
significantly below the experimental ones at higher $p_T$, the phase 
space density of the hydrodynamic component is relatively small. It is 
therefore not clear whether the (small) fraction of thermalized 
particles at high $p_t$ implies an overestimate of the total $v_2$; this 
may be judged more quantitatively by comparing to transport calculations.

In parallel to hydrodynamic descriptions of the bulk medium, expanding
fireball models have been employed. These are simplified (and thus
convenient) parameterizations of a full hydrodynamic calculation in
terms of an expanding volume and spatial flow-velocity field, but
otherwise based on similar principles and variables. E.g., entropy 
conservation is used to convert a given volume into a temperature via
an underlying equation of state (EoS). The reliability of a fireball model
crucially hinges on a realistic choice of the parameters, in connection
with a proper description of the final state in terms of particle
production and collective flow. In principle, a fireball model offers
some additional flexibility in varying the evolution, which may provide
useful checks of the sensitivity to specific aspects of the expansion.

\begin{table}[!t]
\begin{center}
\begin{tabular}{|p{0.12\linewidth}||p{0.18\linewidth}|p{0.18\linewidth}|
p{0.18\linewidth}|p{0.18\linewidth}|}
\hline
\quad & MT & HGR & AHH & GA \\
\hline \hline
$\tau_0$ [fm/$c$] & 1.0 & 0.33 & 0.6 & 0.6 \\
\hline
$T_0$ [MeV] & 265 (max)  & 340 (avg)  & 250 (avg) & 330\;(max) 
\hbox{ } \hfill   260\;(avg) \\
\hline
$T_{c}$ [MeV] & 165 & 180 & 170 & 165 \\
\hline
initial spatial & wounded nucleon  & isotropic & lin. comb. of
$N_{\mathrm{coll}}$ and $N_{\mathrm{part}}$ & lin. comb. of
$N_{\mathrm{coll}}$ and $N_{\mathrm{part}}$   \\
\hline
$b$ [fm] & 6.5 & 7 & 5.5 & 7 \\
\hline
bulk-$v_2$ & 5  & 5.5\,\%  & 3\% & 4.75\,\% \\
\hline
$\tau_{\text{FB}}$\,[fm/$c$] & $\sim$9 & $\sim$5 & $\sim$9  & $\sim$7 \\
\hline
QGP-EoS & massless ($N_f = 3$) & massless ($N_f=2.5$) &
massless ($N_f=3$) & massless ($N_f = 3$) \\  
\hline
HQ Int.  & pQCD HTL & pQCD+reso  pQCD+T-mat &  AdS/CFT & pQCD run. $\alpha_s$ \\ 
\hline
$m_{c,b}$\,[GeV]  & 1.4 & 1.5, 4.5 & 1.5, 4.8 & 1.5, 5.1 \\
\hline
\end{tabular}
\end{center}
\caption{Survey of parameters figuring into hydrodynamic and fireball 
  evolutions employed in the Langevin simulations of HQ spectra for
  semicentral Au-Au collisions at RHIC, corresponding to 
  Refs.~[MT]\protect\cite{Moore:2004tg,Teaney:2000cw},
  [HGR]\protect\cite{vanHees:2005wb,vanHees:2007me,Kolb:2000sd}, 
  [AHH]\protect\cite{Akamatsu:2008ge,Hirano:2005xf}  and 
  [GA]\protect\cite{Gossiaux:2008jv,Kolb:2003dz}.}
\label{tab_evo}
\end{table}
Several key parameters of thermal medium evolution models employed in 
HQ Langevin simulations are compiled in Tab.~\ref{tab_evo}. 
The starting point of both hydro and fireball models are the initial
conditions of the thermal medium, characterized by a formation time 
when the medium is first assumed to be (locally) equilibrated. 
At RHIC, typical formation times are estimated to be in the range of 
$\tau_0 \simeq 0.3$-$1 \, \mathrm{fm}/c$. With a total entropy fixed to 
reproduce the measured rapidity density of hadrons at a given centrality, 
e.g.~at impact parameter $b\simeq 7$~fm/$c$, these formation times 
translate into average initial temperatures of
$T_0\simeq250$-$350\,\MeV$. If the entropy density scales as $s\propto T^3$,
one can roughly compare the initial conditions in different approaches
using $S=s_0 V_0$ and $V_0\propto \tau_0$. E.g., an initial 
$T_0^{\rm max}=$~265\,MeV based on $\tau_0=$~1\,fm/$c$ increases 
by a factor of $3^{1/3}$ upon decreasing $\tau_0=$~0.33\,fm/$c$, 
resulting in $T_0^{\rm max}\simeq$~382\,MeV; if the number of light 
flavors in the EoS is reduced, $T_0$ increases as well; e.g., 
$T_0^{\rm avg}=260$\,MeV based on $\tau_0=$~0.6\,fm/$c$ and $N_f$=3 
(as in Ref.~\cite{Gossiaux:2008jv}) increases to 
$T_0^{\rm avg}=260\,{\rm MeV}\,(0.6/0.33)^{1/3}\,(47.5/42.25)^{1/3} 
\simeq$~330\,MeV for $\tau_0=$~0.33\,fm/$c$ and $N_f$=2.5 (reasonably 
consistent with Ref.~\cite{vanHees:2005wb}, cf.~Tab.~\ref{tab_evo}).
The QGP-dominated evolution lasts for about 2-4\,fm/$c$,
followed by a mixed phase of similar duration at a
critical temperature $T_c\simeq 165$-$180\,\MeV$. The effects of a
continuous (cross-over) transition, as well as of the hadronic phase,
have received little attention thus far, but are not expected to leave a
large imprint on HQ observables. After all, the cross-over
transition found in lQCD exhibits a marked change in energy
density over a rather narrow temperature interval. A more important
aspect is the consistency between the EoS used to extract
the temperature of the bulk evolution and the corresponding degrees of
freedom figuring into the calculation of the HQ transport coefficients.
In hydrodynamical backgrounds used thus
far\cite{Moore:2004tg,Akamatsu:2008ge,Gossiaux:2008jv} the evolution is
described with a 2+1-dimensional boost-invariant simulation with an
ideal massless-gas EoS. The initial state is typically
determined by distributing the entropy in the transverse plane according
to the wounded nucleon density. Unfortunately, the impact parameters in
current Langevin simulations vary somewhat, which is particularly 
critical for the magnitude of the subsequently developed elliptic flow. 
The value of the critical temperature has some influence on the QGP
lifetime (lower temperatures leading to larger QGP duration),
as does the hadron-gas EoS (more hadronic states
imply a larger entropy density at $T_c$ and thus a reduced 
duration of the mixed phase). The termination point of the evolution 
(beginning, middle or end of mixed phase) is rather significant, 
especially for the HQ $v_2$ which needs about 5\,fm/$c$ to build
up most of its magnitude.  

In Refs.\cite{vanHees:2005wb,vanHees:2007me} the medium is 
parameterized, guided by the detailed hydrodynamic calculations of
Ref.\cite{Kolb:2000sd}, as a homogeneous thermal elliptic
``fire cylinder'' of volume $V(t)$. The QGP temperature is determined
via the QGP entropy density, $s$, under the assumption of
isentropic expansion (total $S=\mathrm{const}$),
\begin{equation}
s=\frac{S}{V(t)}=\frac{4 \pi^2}{90} T^3 (16+10.5 N_f) \ .
\label{entro}
\end{equation}
The thus inferred temperature is used in Eq.~(\ref{2.1.10c}) to 
compute the friction coefficients, $A$, and transverse diffusion 
coefficient, $B_0$, with the
longitudinal diffusion coefficient fixed by the dissipation-fluctuation
relation, Eq.~(\ref{2.1.30}). In the mixed phase at $T_c=180\,\MeV$
the QGP drag and diffusion coefficients are scaled by a factor $\propto
\varrho^{2/3}$ to account for the reduction in parton densities (rather
than using hadronic calculations).
Special care has to be taken in the parameterization of the elliptic
flow in noncentral Au-Au collisions: the contours of constant flow
velocity are taken as confocal ellipses in the transverse plane with the
pertinent transverse flow set consistently in perpendicular direction.
The time evolution of the surface velocity of the semi-axes of the
elliptic fire cylinder parameterizes the corresponding results of the
hydrodynamic calculations in Ref.\cite{Kolb:2000sd}, in particular the
time-dependence of the elliptic-flow parameter, $v_2$, for the light
quarks. The parameters are adjusted such that the average surface
velocity reaches $v_{\perp}^{(s)}=0.5c$ and the anisotropy parameter
$v_2=5.5\%$ at the end of the mixed phase. Finally, the velocity field
is specified by scaling the boundary velocity linearly with distance
from the center of the fireball, again in accordance with the
hydrodynamic calculation\cite{Kolb:2000sd}.


\subsection{Initial Conditions and Hadronization}
\label{ssec_initial}

The Langevin simulations of HQ diffusion in the QGP require initial
conditions for charm- and bottom-quark phase-space distributions. For
the spatial part of the initial distribution in the transverse plane all
calculations adopt binary-collision scaling following from a Glauber
model, reflecting a hard process for the primordial production mechanism.  
Furthermore, all calculations thus far focus on a limited rapidity 
window around midrapidity, where the longitudinal distribution is
assumed to be uniform in space-time rapidity. As for the initial HQ
$p_t$ spectra, Ref.\cite{Moore:2004tg} employs a fit to a leading-order
parton-model calculation from the CompHEP package\cite{Pukhov:1999gg},
\begin{equation}
\label{2.1.32}
\frac{\dd N}{\dd y \dd \eta \dd^2 p_t} \propto \delta(\eta-y)
\frac{1}{(p_t^2+\Lambda^2)^{\alpha}},
\end{equation}
with $\alpha=3.5$ and $\Lambda=1.849\,\GeV$.
\begin{figure}[!t]
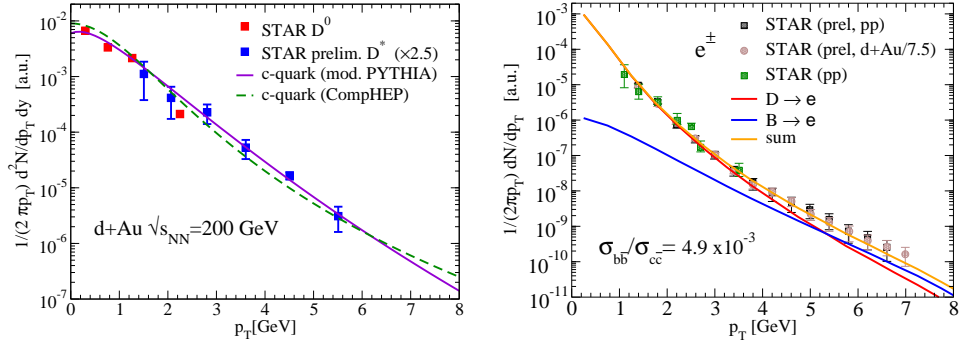

\begin{minipage}{0.48\linewidth}
\includegraphics[width=\textwidth]{D-spectra-vs-c-Spectra}
\end{minipage}\hfill
\begin{minipage}{0.48\linewidth}
\includegraphics[width=\textwidth]{spectra-elect-charm-bottom-pp-new3}
\end{minipage}
\caption{(Color online) Fits of $D$- and $D^*$-meson $p_T$ spectra in 
  $200 A\mathrm{GeV}$ d-Au collisions at RHIC with a modified PYTHIA
  simulation (left panel) and the corresponding non-photonic
  single-electron $p_t$ spectra in $p$-$p$ and d-Au
  collisions\protect\cite{Rapp:2005at}. The missing yield of high-$p_T$
  electrons is fitted with the analogous $B$-meson decay spectra, thus
  fixing the bottom-charm ratio at
  $\sigma_{b\bar{b}}/\sigma_{c\bar{c}}\simeq 4.9 \cdot 10^{-3}$.}
\label{fig_initial}
\end{figure}
In Refs.\cite{vanHees:2005wb,vanHees:2007me}, PYTHIA results for 
$c$-quark spectra have been tuned to reproduce available $D$-meson 
spectra in d-Au
collisions at RHIC (assuming $\delta$-function fragmentation, cf.~left
panel of Fig.~\ref{fig_initial}). The pertinent semileptonic
single-electron decay spectra approximately account for $p$-$p$ 
and d-Au spectra up to $p_T=4\,\GeV$; the missing part at higher $p_T$ 
is then supplemented by $B$-meson contributions. 
This procedure results in a crossing of the $D$- and $B$-meson 
decay electrons at $p_T \simeq5\,\GeV$ and a cross-section ratio of
$\sigma_{b\bar{b}}/\sigma_{c\bar{c}} \simeq 4.9 \cdot 10^{-3}$ (see
right panel of Fig.~\ref{fig_initial}), which is within the range of
pQCD predictions\cite{Cacciari:2005rk}.

With initial conditions and bulk medium evolution in place, one can
evolve HQ phase-space distributions through the QGP (and mixed phase) of
a heavy-ion collision. The final HQ spectra, however, require further
processing before comparisons to observables can be made. First, one has
to address the hadronization of the HQ spectra into charm and bottom
hadrons ($D$, $D^*$, $\Lambda_c$ etc.). Two basic mechanisms have been
widely considered in hadronic collisions, i.e., fragmentation of an
individual quark and recombination with an extra quark from the
environment. In general, the former is mostly applicable for high-energy
partons while the latter requires a sufficient overlap of the mesonic
wave function with the phase-space density of surrounding quarks and is
therefore more relevant toward lower momentum.

The fragmentation of a quark is implemented by applying the
factorization theorem of QCD\cite{Collins:1985gm}. At large transverse
momenta, the production process of a parton occurs on a short time 
scale, $\tau_{\mathrm{prod}} \simeq 1/p_t$, while hadronization occurs 
at the considerably larger time scale $\tau_{\mathrm{had}} \simeq
1/\Lambda_{\mathrm{QCD}}$. Thus the production cross section for a
hadron can be factorized into an elementary parton-production cross
section (hard process) and a phenomenological universal transition
probability distribution, $D_{h/i}(z)$, for a parton $i$ of momentum
$p_i$ to convert into a hadron with momentum fraction $z=p_h/p_i \leq
1$. For light quarks and gluons the fragmentation functions, $D_{h/i}$,
are rather broad distributions around $z \simeq 0.5$, but for heavy
quarks they become rather sharply peaked toward $z=1$ and are sometimes
even approximated by a $\delta$-function, $D(z)=\delta(1-z)$.

The mechanism of recombination of a produced quark with other quarks or
antiquarks in its environment (e.g., the valence quarks of the colliding
hadrons) has first been introduced in the late 1970's to explain flavor
asymmetries in $\pi$ and $K$ meson production in hadronic collisions at
forward rapidities\cite{Das:1977cp}. In particular, the recombination
idea has been rather
successful in describing flavor asymmetries in the charm
sector\cite{Braaten:2002yt,Rapp:2003wn}, even close to midrapidity. In
the context of heavy-ion collisions, quark coalescence models, applied at
the hadronization transition, provide a simple and intuitive explanation
for the observed constituent-quark number scaling (CQNS) of the elliptic
flow of light hadrons\cite{Adare:2006ti,Abelev:2007qg} and the
(unexpectedly) large baryon-to-meson ratios (e.g., $p/\pi \simeq 1$ or
$(\Lambda+\bar{\Lambda})/(4 K_S^0) \simeq 1.3$ in central 200\,$A$GeV 
Au-Au collisions at RHIC) at intermediate transverse momenta 
($2\,\GeV\lesssim p_T \lesssim 
5 \,\GeV$)\cite{Hwa:2002tu,Greco:2003mm,Fries:2003kq,Molnar:2003ff}.
CQNS refers to a scaling property of the hadronic elliptic flow,
$v_{2,h}(p_T)$, in terms of a universal function
$v_{2,q}(p_T/n)=v_{2,h}(p_T)/n$, where $n$ denotes the number of 
constituent quarks in a given hadron, $h$. CQNS naturally emerges from 
the recombination of approximately comoving quarks and antiquarks in a
collectively flowing medium. Thus, within this picture, $v_{2,q}(p_t)$
is interpreted as a universal  elliptic flow of the quarks with 
transverse momentum $p_t$ at the moment of hadronization (typically 
assumed to be the quark-hadron transition at $T_c$). It can be expected 
that the phenomenologically very successful coalescence concept also 
applies in the HQ sector of heavy-ion
collisions\cite{Greco:2003vf,Lin:2003jy}. Note that, unlike quark
fragmentation, quark recombination adds momentum and elliptic flow to
the produced hadron (through the quark picked up from the environment).

At this point it might be instructive to reiterate a conceptual
connection between the quark coalescence model and the idea of resonance
correlations in the QGP. The latter were found to be an efficient
mechanism for arriving at a small HQ diffusion constant, both within 
the effective resonance model (Sec.~\ref{sssec_reso}) and within the
$T$-matrix approach (Sec.~\ref{sssec_tmat}). Especially in the
$T$-matrix approach, the resonance correlations were found to strengthen
toward the expected hadronization transition, and thus provide a natural
emergence of heavy-light quark coalescence at $T_c$. These ideas have
recently been implemented in a resonance-based description of the
coalescence process in kinetic theory based on a Boltzmann 
equation\cite{Ravagli:2007xx}. This approach improves instantaneous
coalescence formulations in that it respects energy conservation and
establishes a well-defined equilibrium limit in the coalescence process
(i.e., the thermal distribution for the formed meson). Subsequently,
resonance-recombination has been combined with ``realistic'' quark 
phase-space distributions, as generated in relativistic Langevin 
simulations\cite{Ravagli:2008rt}. 
In particular, it was found that CQNS could be recovered under the
inclusion of space-momentum correlations in the quark phase-space 
distributions.

In Refs.\cite{vanHees:2005wb,vanHees:2007me} HQ spectra
at RHIC have been hadronized in a combined
coalescence plus fragmentation scheme. For the hadronization of, e.g.,
charm quarks into $D$ mesons one obtains the $D$-meson spectra as
\begin{equation}
  \frac{\dd N_D^{\mathrm{tot}}}{\dd y \, \dd^2p_T} =
  \frac{\dd N_D^{\mathrm{coal}}}{\dd y\,\dd^2p_T} +
  \frac{\dd N_c^{\mathrm{frag}}}{\dd y\,\dd^2p_T} \ .
\end{equation}
For the first term on the right-hand side, the quark-coalescence model
of Ref.\cite{Greco:2003vf} has been employed, where the $p_T$ spectrum
of a $D$ meson follows from a convolution of light anti-quark and
charm-quark phase-space distributions, $f_{\bar{q},c}$, as 
\begin{equation}
\label{q-coal} 
\frac{\dd N_D^{\mathrm{coal}}}{\dd y \dd^2 p_T}=g_D
\int \frac{p \cdot \dd \sigma}{(2\pi)^3} \int \dd^3 \bvec{q} f_D(\bvec{q},\bvec{x})
f_{\bar{q}}(\bvec{p}_{\bar{q}},\bvec{r}_{\bar{q}})
f_{c}(\bvec{p}_{c},\bvec{r}_{c}) \ .
\end{equation}
Here, $\bvec{p}=\bvec{p}_{\bar{q}}+\bvec{p}_{c}$ is the $D$-meson
momentum and $g_D$ a combinatorial factor accounting for 
color-neutrality and spin averaging. The $D$-meson Wigner function,
$f_D(q,x)$, is assumed as a double Gaussian in relative momentum
$\bvec{p}_c-\bvec{p}_{\bar{q}}$ and size,
$\bvec{r}_c-\bvec{r}_{\bar{q}}$, and $\dd \sigma$ is the hyper-surface
element 4-vector of the hadronization volume. The charm-quark distribution
corresponds to the Langevin output at the end of the mixed phase
of the fireball model, while the light-quark distributions are taken
from previous applications of the coalescence model to
light-hadron observables at RHIC\cite{Greco:2003mm}. This represents 
a parameter-free conversion of HQ distributions into heavy-meson spectra
(note that the final state of the expanding fireball
model\cite{vanHees:2005wb} has been matched to the parameterization of
collective velocity and elliptic flow for the light-quark distributions
in the coalescence model\cite{Greco:2003mm}). The coalescence mechanism
does not exhaust all heavy quarks in the hadronization process, especially
toward higher $p_t$ (where the light-quark phase-space density becomes
small). Therefore, the remaining heavy quarks are
hadronized using fragmentation, which for simplicity is treated in
$\delta$-function approximation (as has been done in connection with the
initial conditions). The formation of baryons containing heavy quarks
(e.g., $\Lambda_c$) has been neglected since it has been found to give
only small contributions, i.e., $\Lambda_c/D \ll 1$. Quantitative
refinements should, however, include these processes, see, e.g.,
Refs.\cite{MartinezGarcia:2007hf,Sorensen:2005sm}.

Finally, the comparison to electron spectra requires to compute
semileptonic decays of heavy-flavor hadrons. Thus far, these have been
approximated in three-body kinematics, e.g., $D\to e\nu K$. An important
finding in this context is that the resulting electron $v_2(p_T)$ traces
the one of the parent meson rather
accurately\cite{Greco:2003vf,Dong:2004ve}, implying that electron
spectra essentially carry the full information on the heavy-meson
$v_2$. In the $p_T$ spectra, the decay electrons appear at roughly half
of the momentum of the parent meson. It has also been pointed
out\cite{MartinezGarcia:2007hf,Sorensen:2005sm} that $\Lambda_c$ baryons
have a significantly smaller branching fraction into electrons (about
$4$-$5\%$) than $D$ mesons ($7\%$ and $17\%$ for neutral and charged
$D$'s, respectively).  Thus, in case of a large $\Lambda_c/D$
enhancement, a net electron ``loss'' could mimic a stronger suppression
than actually present at the HQ level. In fact, even variations in the
neutral to charged chemistry from $p$-$p$ to $A$-$A$
collisions\cite{Rapp:2003wn} could be quantitatively relevant.


\subsection{Model Comparisons of Heavy-Quark Spectra at RHIC}
\label{ssec_hq-spectra}
We are now in position to conduct quantitative comparisons of
diffusion calculations using transport simulations for HQ spectra 
in $200\,A\GeV$ Au-Au collisions at RHIC. We focus on Langevin
simulations but also allude to Boltzmann transport models. The
modifications of the initial spectra are routinely quantified in terms
of the nuclear modification factor, $R_{AA}$, and elliptic-flow
parameter, $v_2$, defined by
\begin{equation}
\begin{split}
  R_{AA}(p_t;b) &=\frac{\dd N_{\mathrm{Q}}^{AA}(b)/\dd p_t}
              {N_{\mathrm{coll}}(b) \ \dd N_Q^{pp}/\dd p_t} \ ,
\\
  v_2(p_t;b) &= \frac{\int \dd \phi \frac{\dd N_{\mathrm{Q}}^{AA}(b)}
            {\dd p_t \dd y \dd \phi} \cos(2 \phi)}{\int \dd \phi \frac{\dd
      N_{\mathrm{Q}}^{AA}(b)}{\dd p_t \dd y \dd \phi}} \ ,
\end{split}
\end{equation}
respectively; $\dd N_{\mathrm{Q}}^{AA}(b)/\dd p_t$ denotes the HQ $p_t$
spectrum in an $A$-$A$ collision at impact parameter, $b$, which is
scaled by the spectrum $\dd N_{\mathrm{Q}}^{pp}/\dd p_t$ from $p$-$p$
collisions times the number of binary nucleon-nucleon collisions,
$N_{\mathrm{coll}}(b)$ (to account for the same
number of heavy quarks). Thus, any deviation of $R_{AA}$ from one
indicates nuclear effects (from the QGP but possibly also in the
nuclear initial conditions or from the pre-equilibrium stages).  The
elliptic-flow parameter, $v_2(p_t)$, is the second Fourier coefficient
in the expansion of the (final) momentum distributions in the azimuthal
angle, $\phi$, relative to the reaction plane ($x$-$z$ plane) of the
nuclear collision. At midrapidity, where the ``directed'' flow ($v_1$) 
is expected to vanish, the $v_2$ coefficient is the leading source of
azimuthal asymmetries. A non-zero $v_2$ is only expected to occur in
noncentral $A$-$A$ collisions due to an ``almond''-shaped nuclear
overlap zone (with a long (short) axis in $y$ ($x$) direction). Typical
sources for a non-zero elliptic flow are a path-length difference for
absorption of particles traversing the reaction zone or an asymmetry in
the collective (hydrodynamic) flow due to stronger pressure gradients
across the short axis. Both effects convert the initial spatial
anisotropy, $v_2$, in a positive momentum anisotropy in the particle
$p_t$-spectra. While the former mechanism is usually associated with
high-$p_t$ particles (typical leading to a $v_2<5\%$), the latter
is driven by collective expansion due to thermal pressure mostly
applicable to low-$p_t$ particles (with significantly larger $v_2$
values, in excess of $5\%$). Since in the Langevin simulations heavy
quarks are assumed to be exclusively produced in primordial $N$-$N$
collisions (i.e., their number is conserved subsequently), the HQ
$R_{AA}$ can be simply calculated as the ratio of the HQ $p_t$
distribution function at the moment of hadronization to the initial 
distribution (taken from $p$-$p$ collisions),
\begin{equation}
\begin{split}
\label{2.1.33}
R_{AA} &= \frac{f_Q(t_{\mathrm{had}},p_t)}{f_Q(t_0,p_t)} \ ,  \\
v_2(p_t)&=\frac{\int \dd \phi f_Q(t_{\mathrm{had}},p_t,\phi)
  \cos(2 \phi)}{f_{\mathrm{Q}}(t_{\mathrm{had}},p_t)} \ ,
\end{split}
\end{equation}
while the $v_2$ is computed using its definition given above.  

\begin{figure}[!t]
\begin{center}
\includegraphics[width=0.45\linewidth]{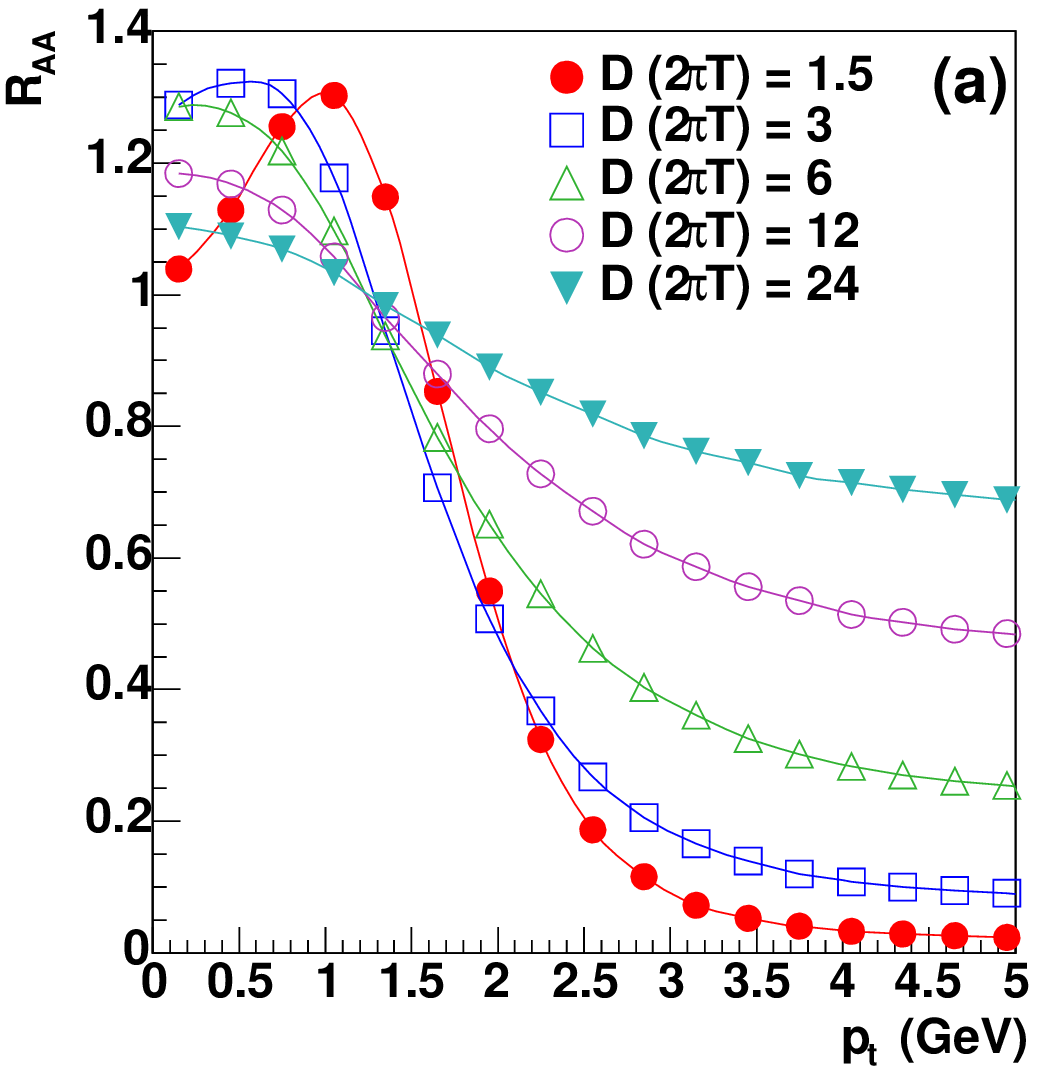}\hfill
\includegraphics[width=0.45\linewidth]{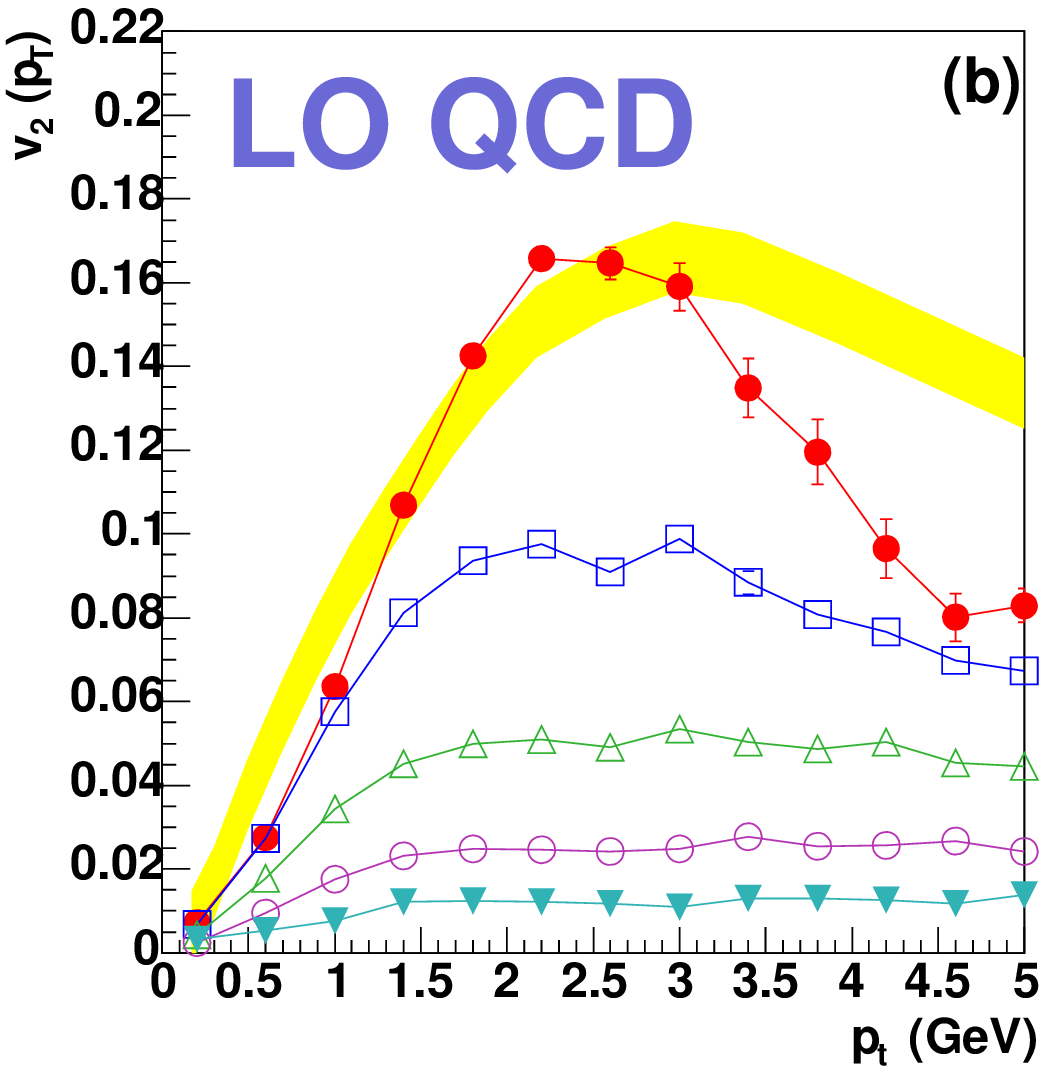}
\end{center}
\caption{(Color online) Nuclear modification factor (left panel) 
  and elliptic flow (right panel) of
  charm quarks as a function of transverse momentum in semicentral
  ($b=6.5\,\mathrm{fm}$) Au-Au collisions using a hydrodynamic
  evolution of the bulk medium at RHIC\protect\cite{Moore:2004tg}.
  The calculations are performed for HTL-improved LO-pQCD scattering 
  with variable strong coupling and fixed Debye-screening mass in 
  $t$-channel gluon-exchange scattering ($\mu_D=1.5\,T$). 
  The relation of the spatial diffusion coefficient, $D_s$ (denoted 
  $D$ in the figure legend), to the strong 
  coupling constant, $\alpha_s$, is given by the approximate relation 
  $2 \pi T D_s \approx 6 (0.5/\alpha_s)^2$.}
\label{fig_hq-MT}
\end{figure}
\begin{figure}[!t]
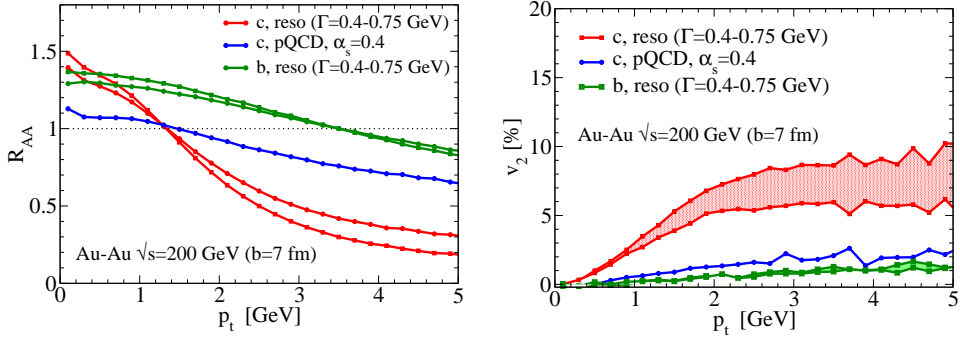

\begin{minipage}{0.48\linewidth}
\includegraphics[width=\textwidth]{quark-minbias-RAA.eps}
\end{minipage}\hfill
\begin{minipage}{0.48\linewidth}
\includegraphics[width=\textwidth]{quark-minbias-v2.eps}
\end{minipage}
\caption{(Color online) The HQ $R_{AA}$ (left panel) and $v_2$ (right
  panel) for semicentral ($b=7\,\fm$)  Au-Au collisions at RHIC within 
  the effective resonance + pQCD  model compared to results from 
   LO pQCD elastic scattering only with $\alpha_s=0.4$ and corresponding 
   Debye-screening mass $\mu_D=\sqrt{4\pi\alpha_s} T$.}
\label{fig_hq-reso}
\end{figure}
\begin{figure}[!t]
\begin{minipage}{0.49\linewidth}
\includegraphics[width=1.06\textwidth]{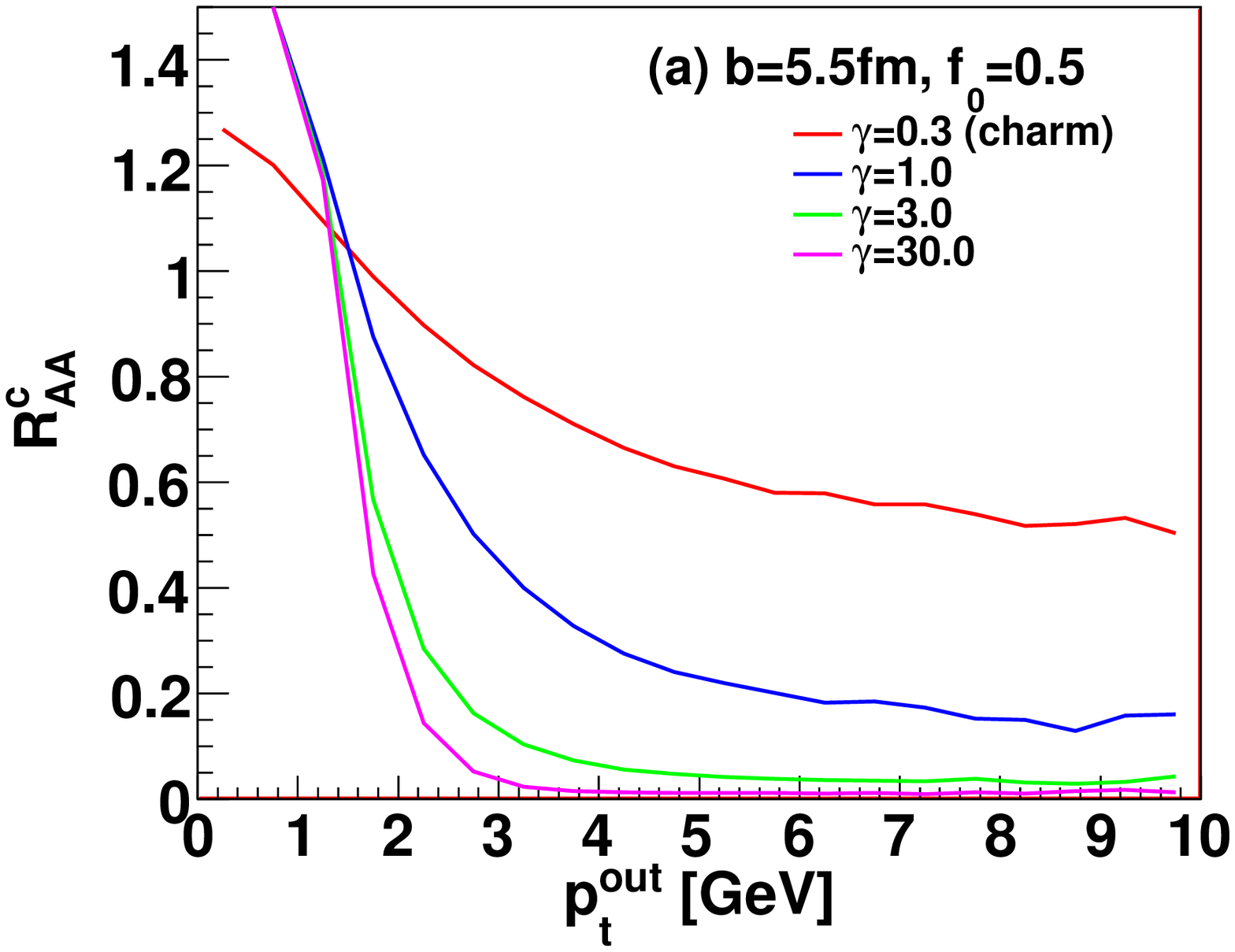}
\end{minipage}
\begin{minipage}{0.49\linewidth}
\includegraphics[width=1.06\textwidth]{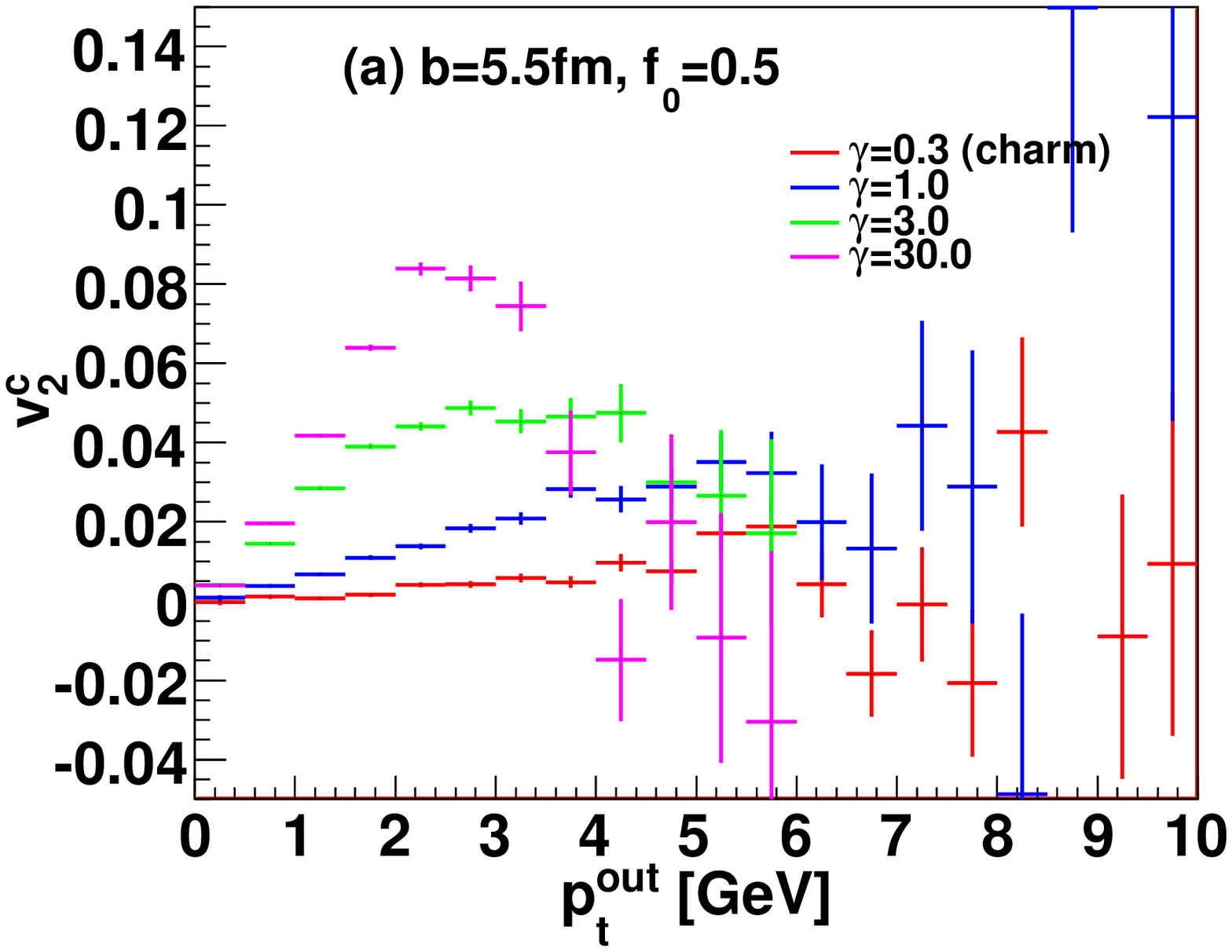}
\end{minipage}
\caption{(Color online) $R_{AA}$ (left) and $v_2$ (right) of charm
  quarks resulting from hydrodynamic simulations of $b=5.5$\,fm 
  Au-Au ($\sqrt{s_{NN}}=200$\,GeV) collisions
  using AdS/CFT-motivated charm-quark diffusion constants with variable
  strength parameter, $\gamma$\protect\cite{Akamatsu:2008ge}, which
  corresponds to the constant $C$ in Eq.~(\ref{ads-cft-par}).}
\label{fig_hq-AHH}
\end{figure}
\begin{figure}[!t]
\begin{minipage}{0.48\linewidth}
\includegraphics[width=\textwidth]{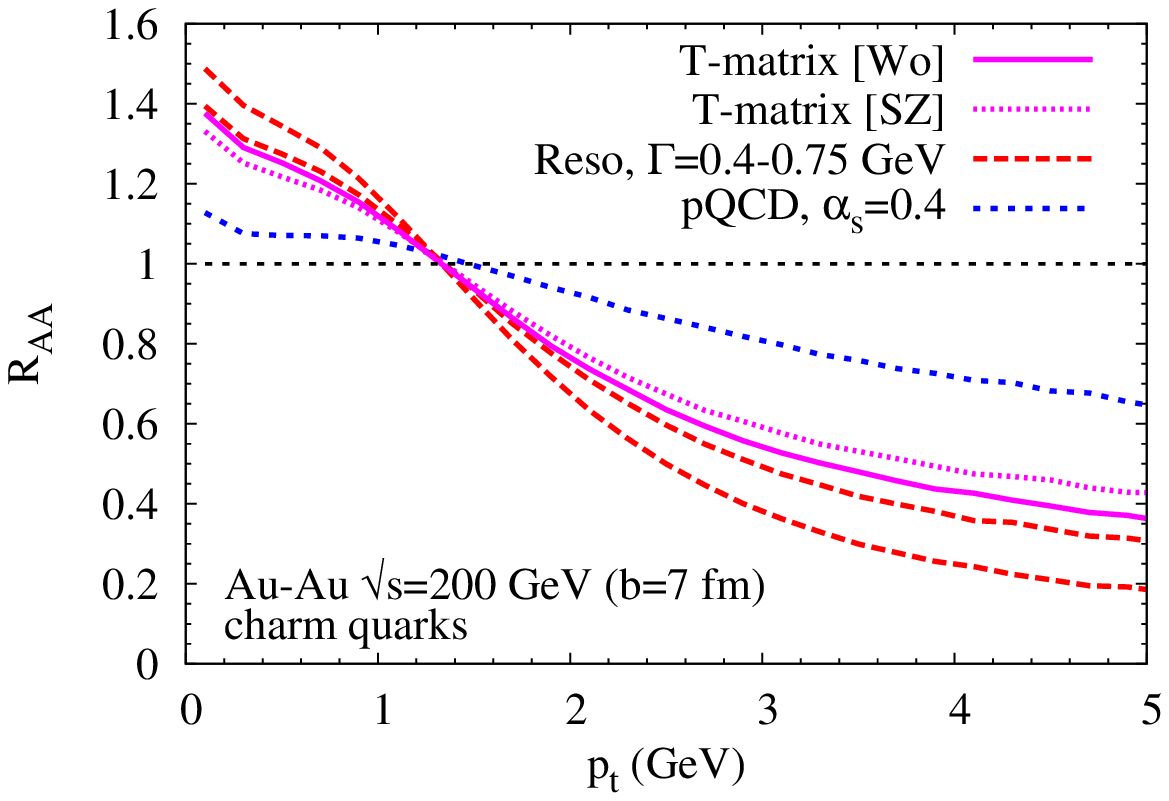}
\end{minipage}\hfill
\begin{minipage}{0.48\linewidth}
\includegraphics[width=\textwidth]{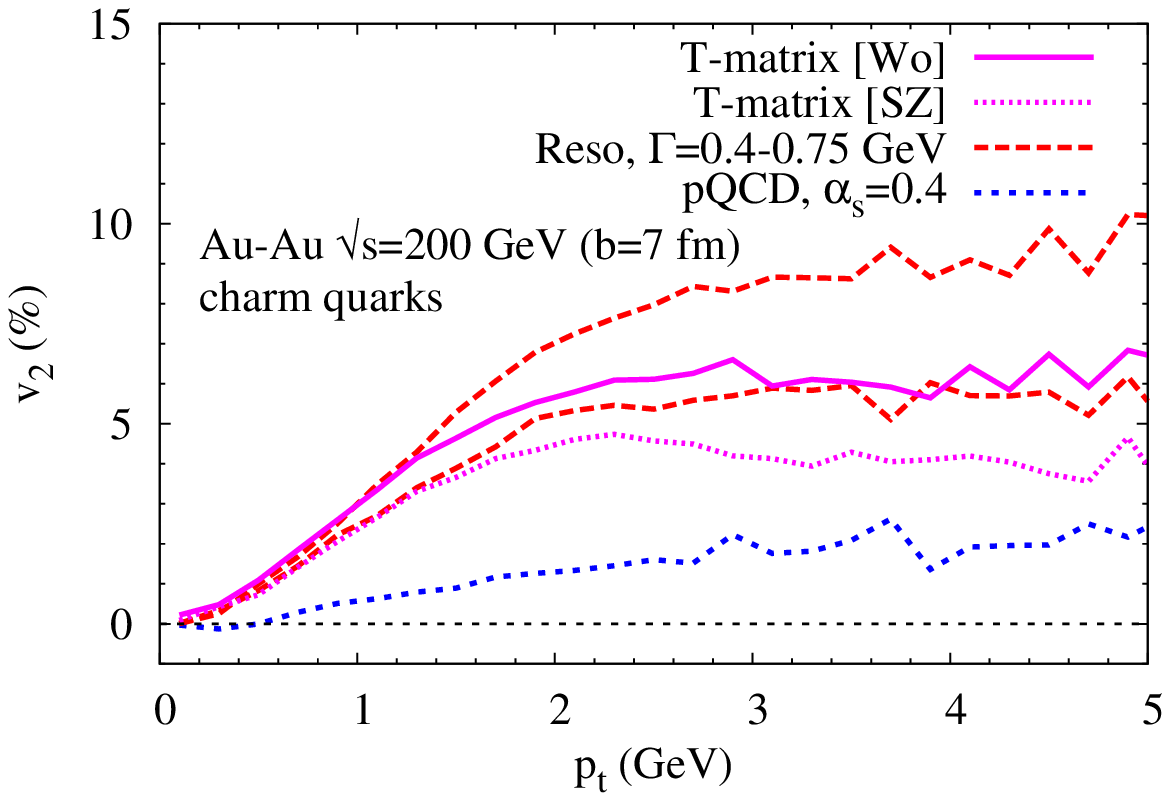}
\end{minipage}
\begin{minipage}{0.48\linewidth}
\includegraphics[width=\textwidth]{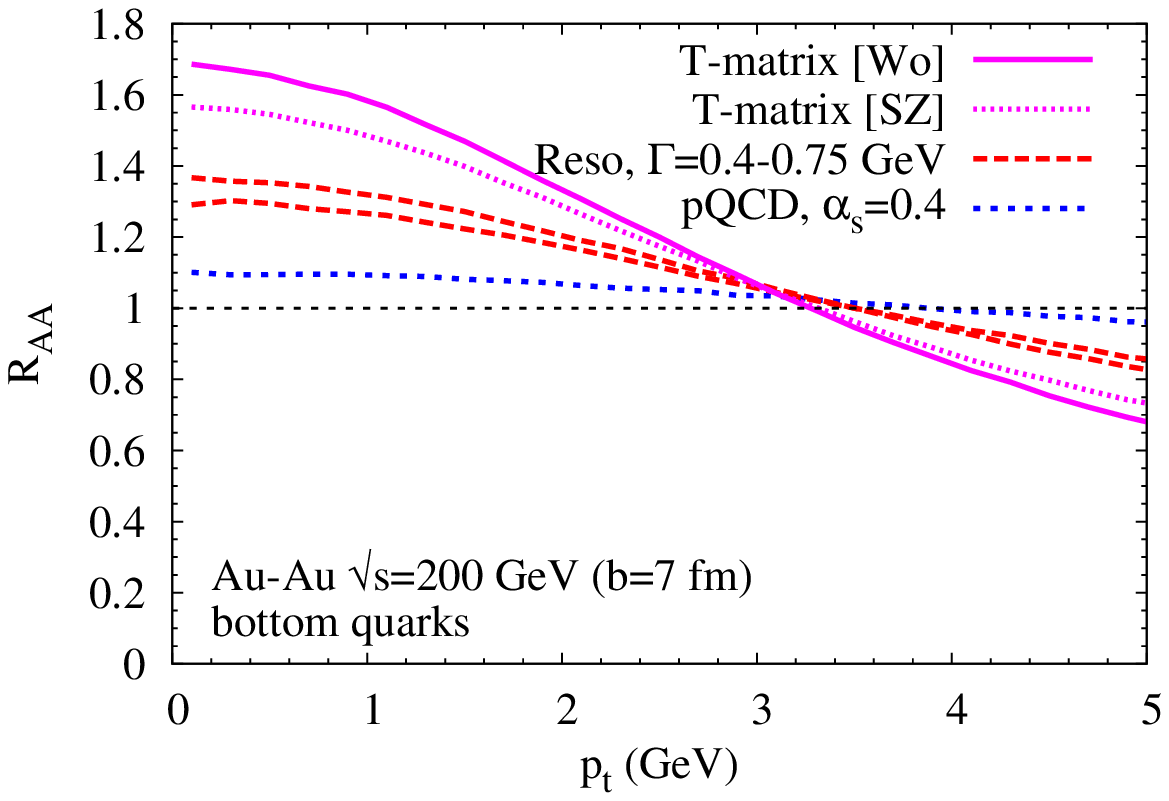}
\end{minipage}\hfill
\begin{minipage}{0.48\linewidth}
\includegraphics[width=\textwidth]{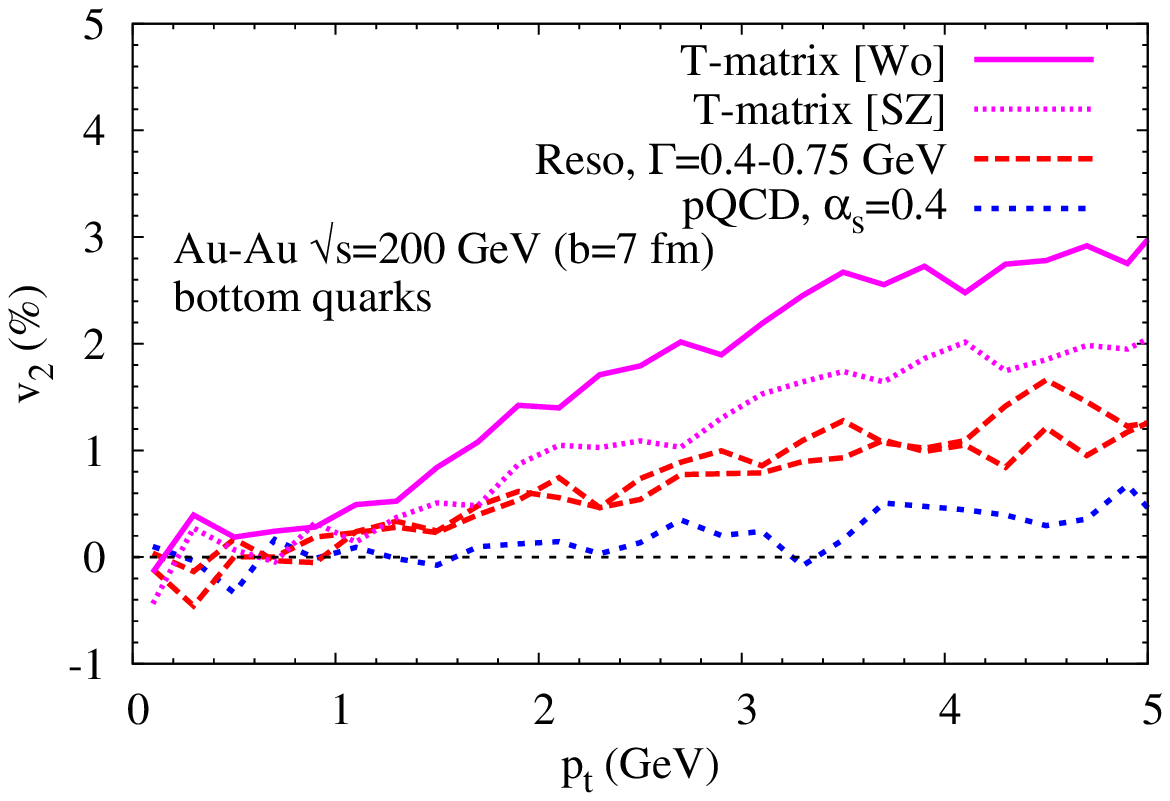}
\end{minipage}
\caption{(Color online) Charm- (top row) and bottom-quark (bottom row)
  spectra in Au-Au($\sqrt{s_{NN}}=200$\,GeV) collisions at
  RHIC using Langevin simulations for HQ diffusion in an expanding
  fireball model. The results using HQ diffusion based on LO-pQCD and
  resonance+pQCD approaches (see also Fig.~\ref{fig_hq-reso}) are
  compared to the $T$-matrix+pQCD
  calculations\protect\cite{vanHees:2007me} (the pertinent uncertainty
  band reflects different input potentials based on parameterizations of
  lQCD HQ free energies as given in
  Refs.\protect\cite{Wong:2004zr,Shuryak:2004tx}. The left (right)
  panels show the $R_{AA}$ ($v_2$) for central (semicentral)
  collisions.}
\label{fig_hq-tmat}
\end{figure}
\begin{figure}[!t]
\begin{minipage}{0.48\linewidth}
\includegraphics[width=\textwidth]{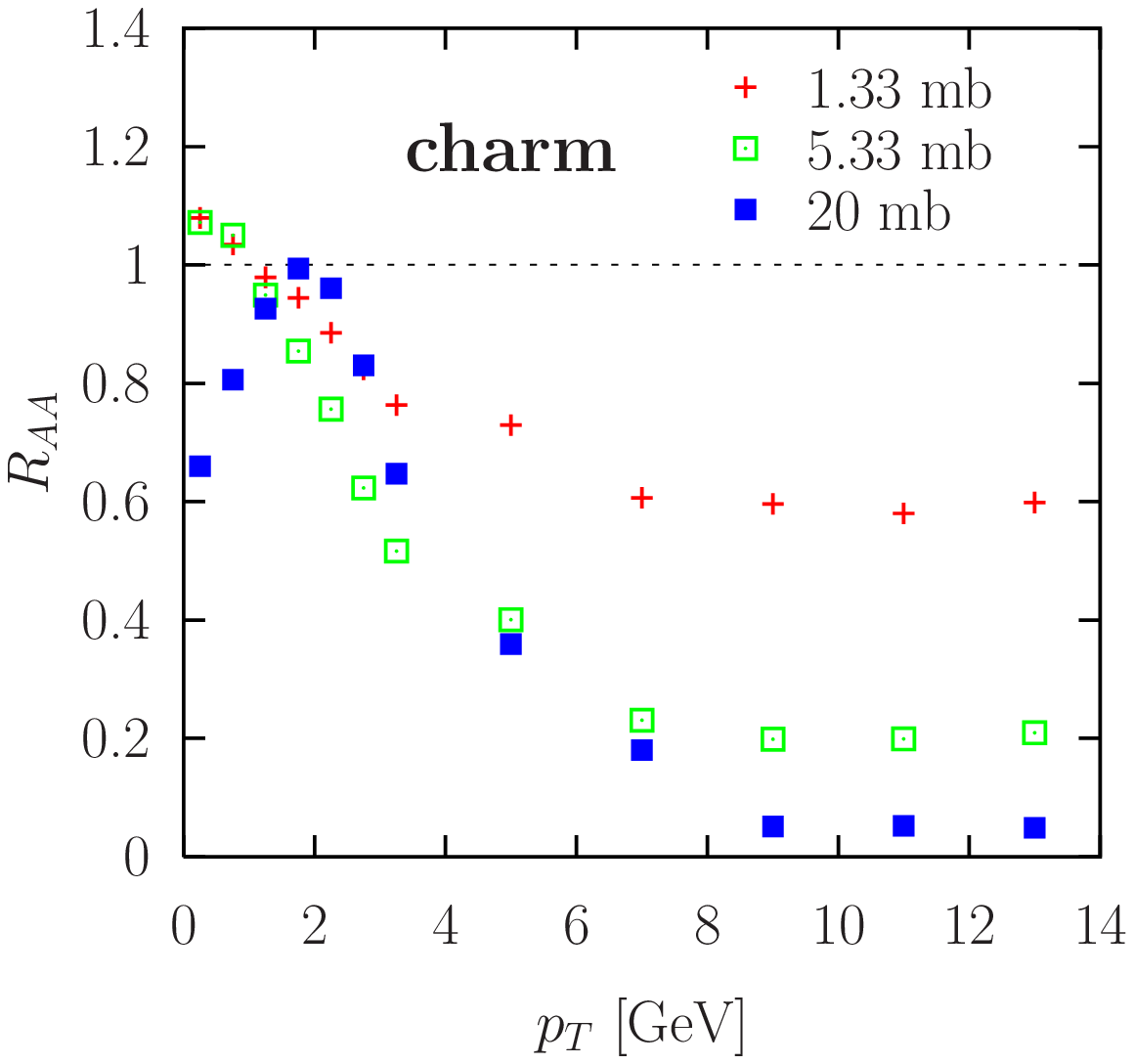}
\end{minipage}\hfill
\begin{minipage}{0.48\linewidth}
\includegraphics[width=\textwidth]{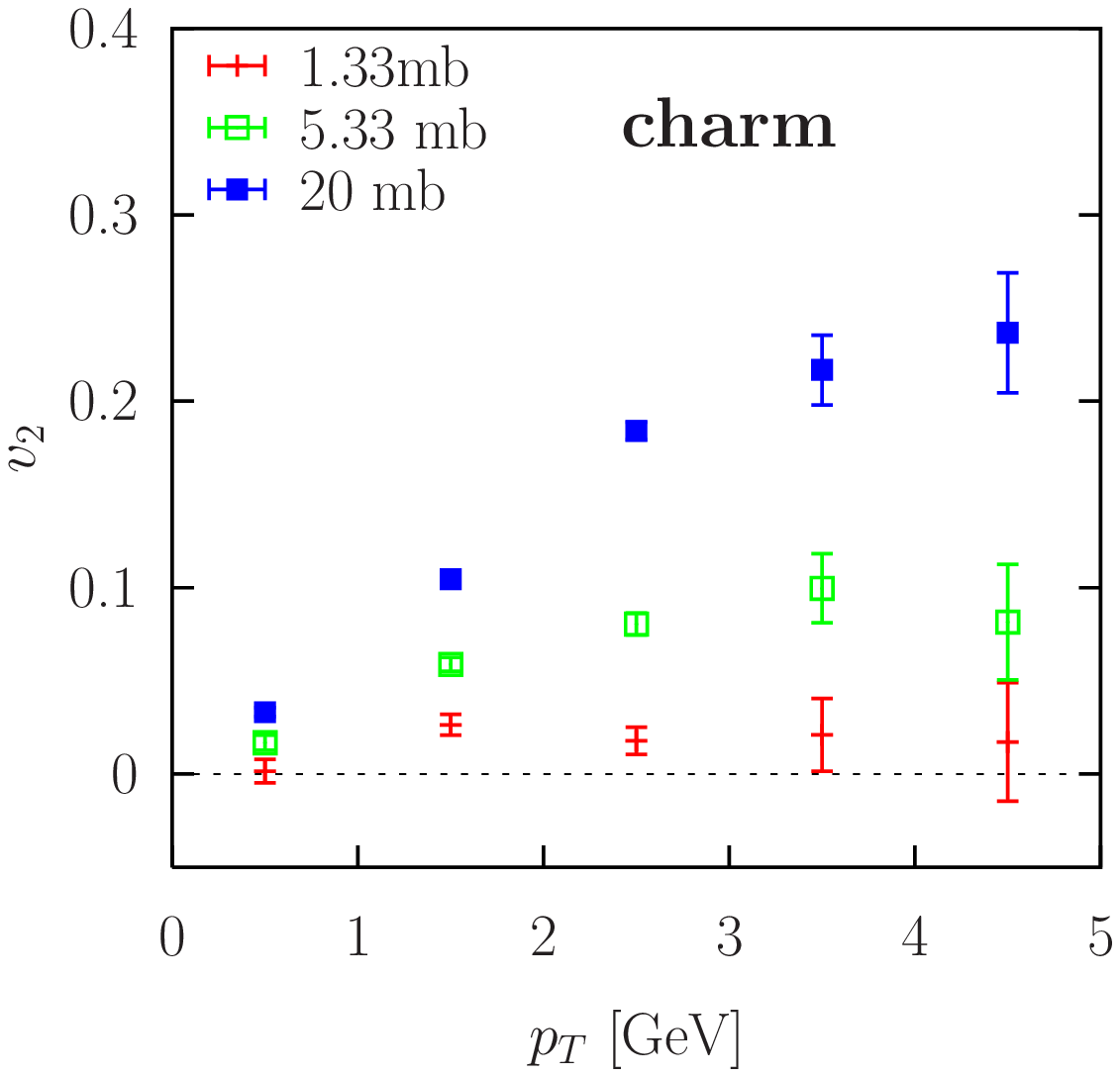}
\end{minipage}
\caption{(Color online) $R_{AA}$ (left) and $v_2$ (right) of charm
  quarks resulting from covariant transport simulations of $b=8 \; \fm$
  Au-Au ($\sqrt{s_{NN}}=200\; \GeV$)
  collisions\protect\cite{Molnar:2006ci}.}
\label{fig_hq-Mol}
\end{figure}

The next five figures (\ref{fig_hq-MT}-\ref{fig_hq-Mol}) encompass 
calculations of $R_{AA}$ and $v_2$ in
semicentral Au-Au collisions at RHIC for the following approaches:
\begin{itemize}
\item[(i)] Fig.~\ref{fig_hq-MT}~[MT]\cite{Moore:2004tg} displays
  Langevin simulations for $c$ quarks (with the pre-point Ito
  realization of the stochastic integral) using a hydrodynamic evolution
  for $b=6.5\,\fm$; the HQ drag and diffusion coefficients are based on
  LO hard-thermal loop scattering matrix elements with variable
  $\alpha_s$ but fixed Debye screening mass.
\item[(ii)] Fig.~\ref{fig_hq-reso}~[HGR]\cite{vanHees:2005wb} displays
  Langevin simulations for $c$ and $b$ quarks (with the post-point Ito
  (H{\"a}nggi-Klimontovich) realization) using a thermal fireball
  expansion for $b=7\,\fm$; the HQ drag and diffusion coefficients are
  based on the effective resonance+pQCD model\cite{vanHees:2004gq} for
  variable resonance width (coupling strength) and $\alpha_s=0.4$ in the
  pQCD part.
\item[(iii)] Fig.~\ref{fig_hq-AHH}~[AHH]\cite{Akamatsu:2008ge} displays
  Langevin simulations for $c$ quarks (with the pre-point Ito realization)
  using a hydrodynamic expansion for $b=5.5\,\fm$; the HQ drag and
  diffusion coefficients are based on the strong-coupling limit with
  AdS/CFT correspondence with a variable coupling strength estimated
  from matching to QCD\cite{Gubser:2006qh}.
\item[(iv)] Fig.~\ref{fig_hq-tmat}~[HMGR]\cite{vanHees:2007me} displays
  Langevin simulations as under (ii) but with HQ transport coefficients
  based on the $T$-matrix+pQCD approach for two lQCD-based input
  potentials.
\item[(v)] Fig.~\ref{fig_hq-Mol}~[Mol]\cite{Molnar:2006ci} displays
  Boltzmann transport simulations using a covariant transport model for
  $b=8\;\fm$; the HQ interactions are modeled by schematic LO pQCD cross
  sections, including upscaling by ``$K$ factors''.
\end{itemize}

Before going into details, let us try to extract generic features of the
calculations. In all cases there is a definite correlation between a
reduction in $R_{AA}(p_t>3\,\GeV)$ and an increase in $v_2(p_t)$,
i.e., both features are coupled to an increase in interaction strength
(decrease in the spatial HQ diffusion coefficient). Furthermore, the
$v_2(p_t)$ shows a typical, essentially linear, increase reminiscent of 
a quasi-thermal regime followed by a saturation characteristic for the
transition to a kinetic regime. In all Langevin calculations the 
saturation for charm quarks occurs at about $p_t=2$-$3\,\GeV$. For the
largest interaction strength considered ($D_s \simeq 1/(2\pi T)$), 
the left panels of
Figs.~\ref{fig_hq-MT} and \ref{fig_hq-AHH} even suggest a turnover of
$v_2$ (at this point one should recall that all calculations displayed
in this section utilize elastic scattering only which is expected to
receive appreciable corrections at high $p_t$ due to radiative
processes). On the other hand, for $p_t=2$-$3\; \GeV$ the nuclear
modification factor is still significantly falling, leveling off only 
at larger $p_t\simeq5$-$6\,\GeV$. As expected, bottom quarks exhibit much
reduced effects for comparable diffusion constants due to their
factor $\sim$3 larger mass (see Fig.~\ref{fig_hq-reso} and lower
panels in Fig.~\ref{fig_hq-tmat}).

Next, we attempt more quantitative comparisons. Some representative
numbers for the resulting $R_{AA}$ and $v_2$ values are compiled in
Tab.~\ref{tab_hq}\cite{Rapp:2008zq}.
\begin{table}[!b]
\begin{center}
\begin{tabular}{|c|c|c|c|c|}
\hline
Model [Ref.] & $D_s (2\pi T)$
& $b$ [fm] & $v_2^{\rm max}$ & $R_{AA}$($p_t$=5~GeV)
\\
\hline  \hline
hydro + LO-pQCD\cite{Moore:2004tg} & 24  & 6.5 & 1.5\,\% & 0.7 \\
\hline
hydro + LO-pQCD\cite{Moore:2004tg} & 6 & 6.5 & 5\,\% & 0.25 \\
\hline
fireball + LO-pQCD\cite{vanHees:2005wb} & $\sim$30  & 7 & 2\,\% & 0.65 \\
\hline
fireball + reso+LO-pQCD\cite{vanHees:2005wb} & $\sim$6 & 7 & 6\,\% & 0.3 \\
\hline
hydro + ``AdS/CFT" (\ref{ads-cft-par})\cite{Akamatsu:2008ge}
& 21 & 7.1 & 1.5-2\,\% & $\sim$0.7 \\
\hline
hydro + ``AdS/CFT" (\ref{ads-cft-par})\cite{Akamatsu:2008ge} & 2$\pi$ & 7.1 & 4\,\% & $\sim$0.3 \\
\hline
transport + LO pQCD\cite{Molnar:2006ci} & $\sim$30 & 8 & $\sim$$2\%$ &
$\sim$$0.65$ \\
\hline
transport + LO pQCD\cite{Molnar:2006ci} & $\sim$7 & 8 & 10\% & 
$\sim$$0.4$\\
\hline
\end{tabular}
\end{center}
\caption{Overview of model approaches ($1^{\mathrm{st}}$~column) 
  and input parameters
  ($2^{\mathrm{nd}}$~column: spatial charm-quark diffusion coefficient,
  $3^{\mathrm{rd}}$~column: nuclear impact parameter)
  for Langevin simulations of charm-quark
  spectra in Au-Au collisions at RHIC; selected values for the resulting
  elliptic flow ($v_2^{\rm max} \simeq v_2(p_t=5\; \GeV)$)
  and nuclear modification factor are quoted in columns 4 and 5.
  The last two rows represent charm-quark transport calculations in 
  a transport model for the bulk.}
\label{tab_hq}
\end{table}
First we compare the LO-pQCD calculations for HQ diffusion in the
hydrodynamic and fireball backgrounds corresponding to
Figs.~\ref{fig_hq-MT} and \ref{fig_hq-reso}, respectively; for a
comparable spatial diffusion coefficient, $D_s\simeq24$-$30/(2\pi T)$,
both calculations show a maximal $v_2$ of about $2\%$ and a
$R_{AA}$($p_t$=5\,GeV)~$\simeq$~0.7 (recall the smaller $b=$~6.5\,fm
in [MT] vs. 7\,fm in [HGR] which may lead to somewhat smaller $v_2$,
and the smaller $T_0=265\,\MeV$ [MT] vs. 340\,MeV in [HGR] which
entails somewhat less suppression). For the [AHH] hydro calculation 
with an AdS/CFT-motivated ansatz for the HQ friction constant,
\begin{equation}
  \gamma = C \, \frac{T^2}{m_Q} \ ,
\label{ads-cft-par}
\end{equation}
a diffusion constant of $D_s=21/(2\pi T)$ leads to similar results (note
that Tab.~\ref{tab_hq} contains results for
$b=$~7.1\,fm\cite{Akamatsu:2008pr} while Fig.~\ref{fig_hq-AHH} is
calculated for $b=$~5.5\,fm).
Let us now turn to stronger coupling, still focusing on the three
Langevin simulations in Figs.~\ref{fig_hq-MT}, \ref{fig_hq-reso} and
\ref{fig_hq-AHH} which all utilize friction coefficients with a similar
temperature dependence, essentially $\gamma\propto T^2$ (recall right
panel of Fig.~\ref{fig_gam}), corresponding to an approximately
constant spatial diffusion constant times temperature, 
$D_s (2\pi T)\simeq
\mathrm{const}$ (recall left panel of Fig.~\ref{fig_Ds}). 
For $D_s=6/(2\pi T)$, all calculations are again in
semi-/quantitative agreement, with a maximum $v_2$ of 4-6\,\% and
$R_{AA}$($p_t$=5\,GeV)~$\simeq$~0.25-0.3. The $4\%$ value for
Ref.\cite{Akamatsu:2008ge}~[AHH] will increase somewhat if the hydro
evolution is run to the end of the mixed phase rather than terminated in
the middle of the mixed phase (this is supported by the discussion in
Sec.~\ref{sssec_elec-langevin}). We also note that in the fireball model
of Ref.\cite{vanHees:2005wb}~[HGR] the inclusive ($p_t$-integrated)
$v_2$ at the end of the mixed phase was adjusted to the experimentally
observed light-particle $v_2\simeq$~5.5-6\,\% at an impact parameter of
$b=$~7\,fm, i.e., it presumably includes an extra $20\%$ of bulk-$v_2$
compared to the hydrodynamic calculations\footnote{This adjustment ensures
  compatibility of the fireball freezeout with the coalescence
  model\cite{Greco:2003mm}.}. Such an amount is typically built
up in the subsequent hadronic phase of hydrodynamic evolutions and thus
not present in pertinent HQ simulations within a QGP (+ mixed) phase.

We now make some comments specific to individual calculations. The
Langevin calculations using HQ $T$-matrix interactions (supplemented
with pQCD scattering off gluons) shown in Fig.~\ref{fig_hq-tmat}
are rather close to the effective resonance model, even
though they do not involve tunable parameters. However, they are still
beset with substantial uncertainty, as indicated by the use of two
different input potentials (in addition, the use of the free energy,
$F_{1}$, instead of the internal energy, $U_{1}$, as potential 
significantly reduces the effects). One also notices that the $v_2$ at 
low $p_t$ is very similar to the resonance model while the suppression 
at high $p_t$ is somewhat less pronounced. This is so since 
the $T$-matrix transport coefficients  (a) fall off stronger with 
three-momentum (the resonant correlations are close to the $Q$-$q$ 
threshold), and (b) decrease with increasing temperature (resonance 
melting). The latter combines with the facts that the suppression is 
primarily built up in the very early stages (where the $T$-matrix is 
less strong) while the bulk $v_2$ takes a few $\fm/c$ to build up (at 
which point the $T$-matrix has
become stronger).  Furthermore, the $T$-matrix calculations lead to
stronger medium effects on $b$ quarks than the effective
resonance model; this reflects the stronger binding due to the mass
effect in the $T$-matrix calculation.

A principal limitation of the Langevin approach is the treatment of
fluctuations which are by definition implemented in Gaussian
approximation. The latter arises due to enforcing the
dissipation-fluctuation relation (mandatory to ensure the HQ
distributions to approach equilibrium) which tends to underestimate the
momentum fluctuations especially at high momentum, compared to a full
transport calculation.  This leads to an overestimate of the quenching
effect at high $p_t$ even for the same \emph{average} energy loss. One
may assess these limitations more quantitatively by comparing to Boltzmann 
simulations including partonic phases\cite{Zhang:2005ni,Molnar:2006ci}, an 
example of which is displayed in Fig.~\ref{fig_hq-Mol} for charm quarks 
in $b=$~8\,fm Au-Au collisions at RHIC. The baseline LO-pQCD calculations
indicated by the crosses in Fig.~\ref{fig_hq-Mol}, labeled by ``$1.33\;
\mathrm{mb}$'', may be compared to the fireball-Langevin simulations
represented by the blue lines in Fig.~\ref{fig_hq-reso}. In both
calculations the underlying elastic parton-HQ cross sections correspond
to a strong coupling constant of $\alpha_s \simeq 0.4$. The quenching
and elliptic flow come out quite similar in both calculations at least
up to $p_t \simeq 5\; \GeV$, especially when accounting for the slightly
different centrality.  E.g., in the Boltzmann treatment, the $R_{AA}$
for $p_t=5\; \GeV$ charm quarks is about $0.6$-$0.7$ with a $v_2$ of a
few percent.  For a four-fold increase of the cross section (which would
roughly correspond to a reduction of $D_s (2 \pi T)$ from $\sim$$30$ to
$\sim$$7$), one finds $R_{AA}(p_t=5\;\GeV) \simeq 0.4$ and a maximum
$v_2$ of close to $10\%$. While the latter value is somewhat larger than
the Langevin predictions, the agreement is not too bad.

\begin{figure}[!tb]
\begin{minipage}{0.48\linewidth}
\includegraphics[width=\textwidth]{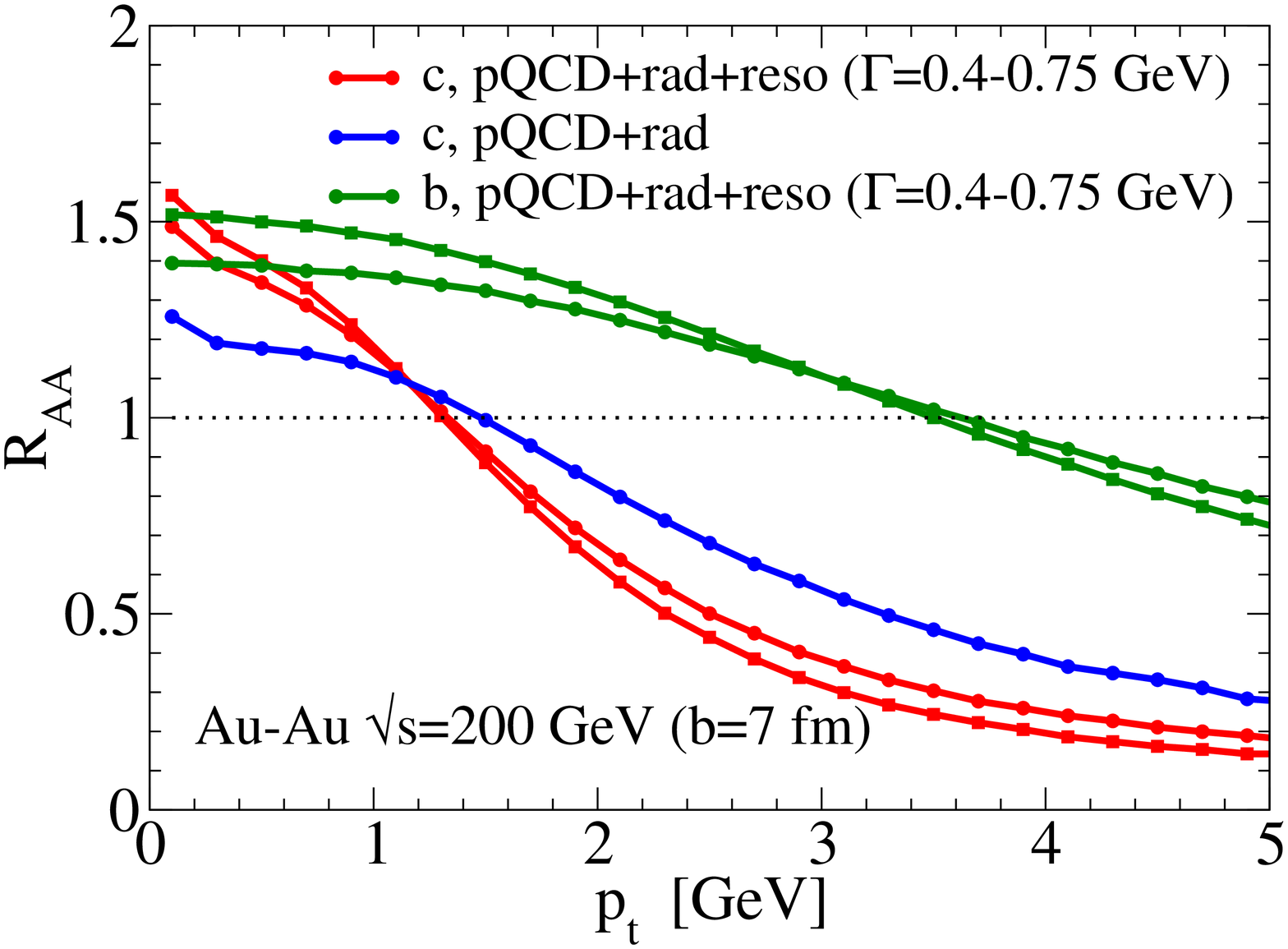}
\end{minipage}\hfill
\begin{minipage}{0.48\linewidth}
\includegraphics[width=\textwidth]{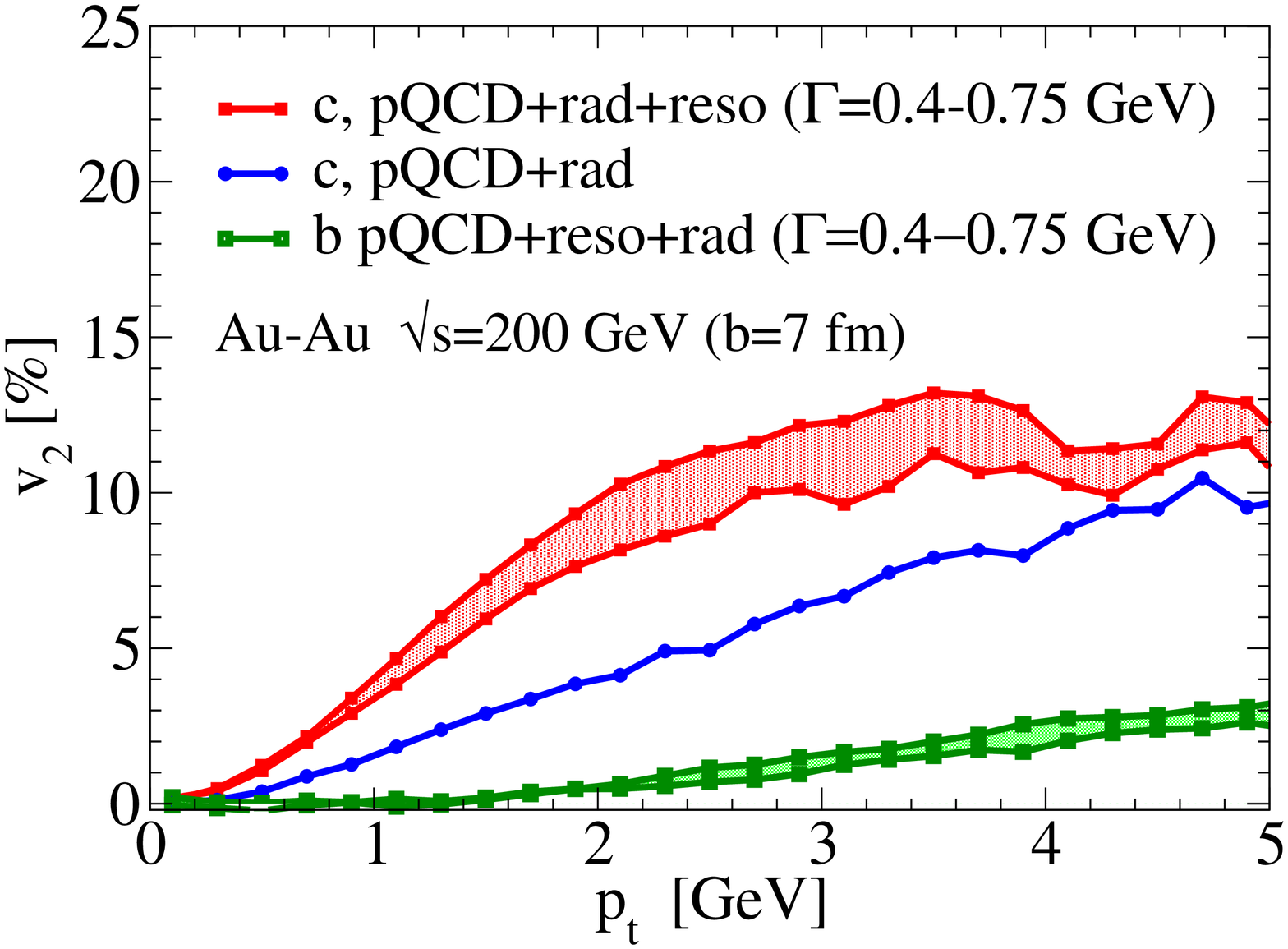}
\end{minipage}
\caption{(Color online) The nuclear modification factor, $R_{AA}$ (left
  panel), and elliptic flow, $v_2$ (right panel), for heavy quarks from
  collisional\protect\cite{vanHees:2005wb} and radiative 
  E-loss\protect\cite{Gyulassy:2000er},  
  cf.~Ref.\protect\cite{Vitev:2007jj}.}
\label{fig_hq-rad-coll-langevin}
\end{figure}
Finally, Fig.~\ref{fig_hq-rad-coll-langevin} shows results from an
exploratory calculation in the Langevin approach where HQ drag and
diffusion coefficients from elastic scattering in the effective
resonance model (cf.~Sec.~\ref{sssec_reso}) are combined with induced
gluon radiation in the DGLV E-loss formalism
(cf.~Sec.~\ref{ssec_rad})\cite{Vitev:2007jj}.  One of the
uncertainties in this calculation is the extrapolation of the 
radiative E-loss into the low-momentum regime, where it still
contributes rather substantially; e.g., the elliptic flow of charm
quarks is increased over elastic pQCD+resonance model by ca.~40\%, and
even more (ca.~100\%) relative to pQCD elastic scattering
only. Another limitation is the above mentioned caveat in Langevin
theory of underestimating the (E-loss) fluctuations implying an
overestimate of the quenching at high $p_t$. This can also be 
seen when comparing to the pQCD radiative E-loss
calculations\cite{Gyulassy:2003mc,Vitev:2005he,Wicks:2005gt}, where the
gluon radiation is treated microscopically within an opacity
expansion. A consistent merging of radiative and elastic processes in 
HQ transport thus remains a challenging task.


\subsection{Heavy-Meson and Electron Observables}
\label{ssec_rhic-obs}
To compare to observables, the HQ spectra discussed in the preceding
section need to be converted into spectra of color-neutral final-state
particles.  At the minimal level, this requires hadronization into charm
and bottom mesons and baryons. Thus, a measurement of identified HQ
hadrons constitutes the most direct way to make contact with theoretical
predictions. Currently, the richest source of information on HQ spectra
in Au-Au collisions at RHIC are single-electron ($e^{\pm}$) spectra,
which, after the subtraction of sources coupling to a photon
(``photonic sources''), are associated with semileptonic decays of HQ
hadrons. As discussed in Sec.~\ref{ssec_initial}, the decay electrons
largely preserve the modifications of the parent hadron spectra, albeit
shifted in $p_t$ (by roughly a factor of $\sim$2). The more severe
complication is the composition of the $e^{\pm}$ spectra, most notably
the partition into charm and bottom
parents\cite{Djordjevic:2005db}. Since the heavier bottom quarks are, in
general, less affected by the medium, their contribution significantly
influences the resulting $e^\pm$ spectra. Unless otherwise stated, the
calculations discussed below include a ``realistic'' input for the
charm/bottom partition, i.e., either in terms of pQCD predictions for
$p$-$p$ spectra or via empirical estimates from $D$-meson and electron
spectra in $p$-$p$ and $d$-Au. Within the current theoretical and
experimental uncertainties, both procedures agree, with an expected
crossing of charm ad bottom electrons at $p_t\simeq$~3-6\,GeV in $p$-$p$
collisions at RHIC energy.

Almost all of the approaches for computing HQ diffusion and/or energy
loss introduced in Sec.~\ref{sec_hq-int} have been applied to $e^{\pm}$
data at RHIC. We organize the following discussion into (mainly
perturbative) E-loss calculations (usually applied within a static
geometry of the nuclear reaction zone) as well as perturbative and
nonpertubative diffusion calculations using Langevin simulations for an
expanding medium.

\subsubsection{Energy-Loss Calculations}
\label{sssec_elec-eloss}
Radiative energy loss (E-loss) of high-energy partons in the QGP is 
believed to be the prevalent mechanism in the suppression of light 
hadrons with high $p_T\ge 6$\,GeV. It turns out that the application of 
this picture to the HQ sector (Sec.~\ref{ssec_rad}) cannot account for the
observed suppression in the non-photonic $e^{\pm}$ spectra.

\begin{figure}[!t]
\begin{minipage}{0.33\linewidth}
\includegraphics[width=\textwidth]{electron-Raa-Wicks.eps}
\end{minipage}
\hspace*{2mm}
\begin{minipage}{0.3\linewidth}
\includegraphics[width=\textwidth]{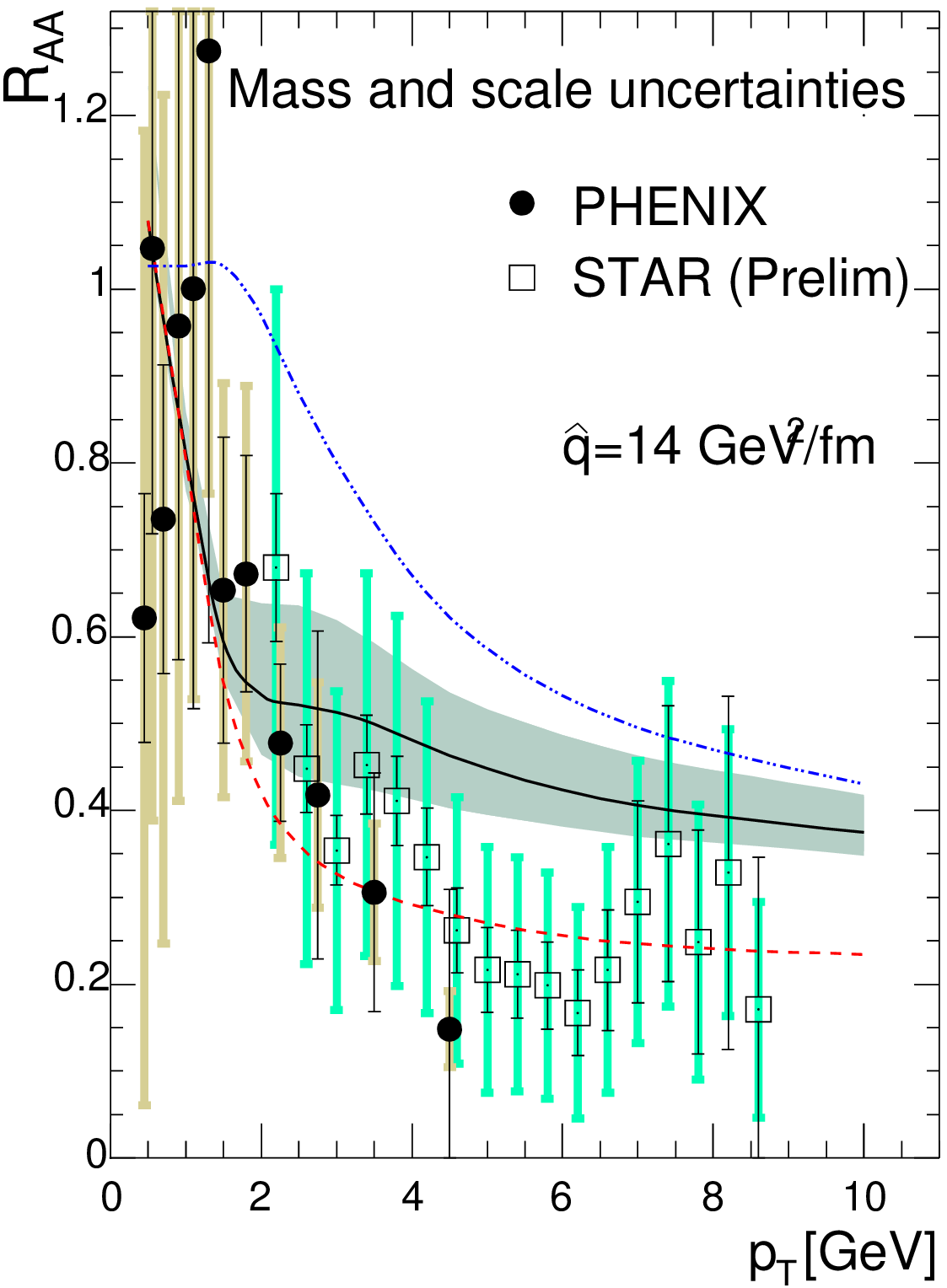}
\end{minipage}
\begin{minipage}{0.33\linewidth}
\includegraphics[width=\textwidth]{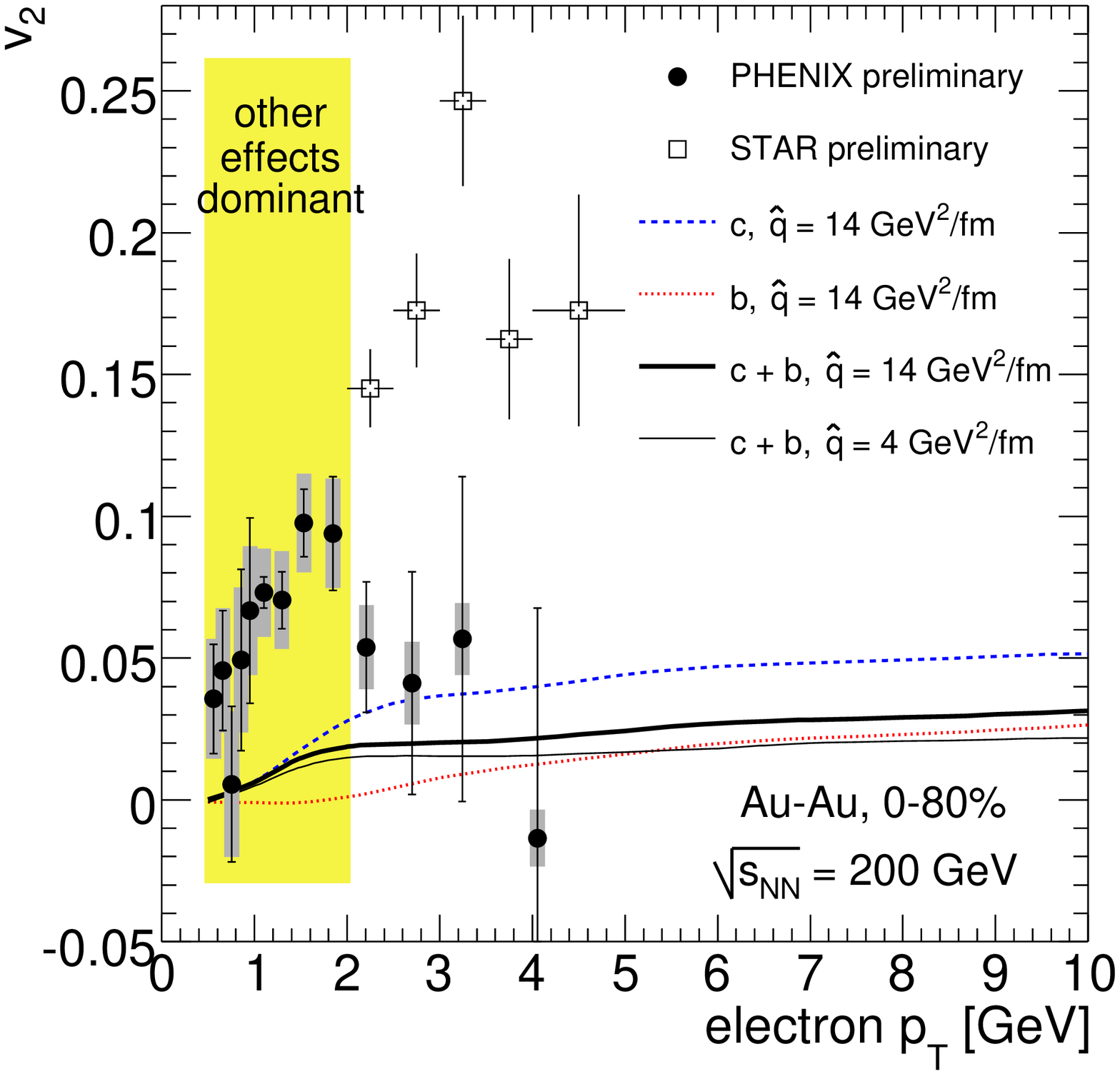}
\end{minipage}
\caption{(Color online) PQCD E-loss calculations for heavy quarks
  compared to single-electron observables in Au-Au collisions at
  RHIC\protect\cite{Adler:2005xv,Butsyk:2005qn,Abelev:2006db} Left
  panel: $e^\pm$ nuclear modification factor, $R_{AA}^e$, in central
  collisions in the DGLV approach\protect\cite{Wicks:2005gt} (upper
  band: radiative E-loss only; lower band: radiative plus elastic
  E-loss) for a gluon plasma with rapidity density $\dd N^{g}/\dd
  y$=1000; 
  the bands represent the uncertainty in the leading-logarithm 
  approximation of the elastic part (as described in Sec.~\ref{ssec_rad}).
  Middle and right panel: $R_{AA}^e$ in central and $v_2^e$ in 0-80\%
  central collisions within the BDMPS approach using a
  transport coefficient of
  $\hat{q}$=14~GeV$^2$/fm\protect\cite{Armesto:2005mz}; dashed and
  dash-dotted curves represent the results for $c$- and $b$-quark
  contributions separately, while the solid curve is the combined result
  with a band indicating the pQCD uncertainty in the charm/bottom
  partition\protect\cite{Cacciari:2005rk}.}
\label{fig_elec-rad}
\end{figure}

In the DGLV formalism, the high-$p_T$ $e^\pm$ suppression due to 
radiative E-loss of $c$ and $b$ quarks falls short of the data by
about a factor of $3$, cf.~left panel of
Fig.~\ref{fig_elec-rad}\cite{Wicks:2005gt}.
This led the authors to consider elastic E-loss (see also
Refs.\cite{Mustafa:2004dr,Djordjevic:2008iz}) which was found to be
comparable to the radiative one out to the highest electron $p_T$
measured thus far ($\sim$$10\,\GeV$). Their combined effect still
underestimates the measured suppression by about a factor of $\sim$$2$
for $p_t>4$\,GeV.  Similar findings were reported within the BDMPS
approach: for a transport coefficient of
$\hat{q}=14$\,GeV$^2$/fm\footnote{This value is a factor 5-10 larger
  than predicted by pQCD, and at the upper limit of being compatible
  with light hadron suppression\cite{Eskola:2004cr}.}, the $e^{\pm}$
spectra cannot be reproduced either, unless an unrealistic assumption of
neglecting the bottom contribution is made, cf.~middle panel of
Fig.~\ref{fig_elec-rad}\cite{Armesto:2005mz}.

Both E-loss calculations\cite{Wicks:2005gt,Armesto:2005mz} are
performed for a static (time-averaged) medium of gluons, with
fragmentation as the sole mechanism for hadronization. This is expected
to be a good approximation at high $p_T$. Processes leading to
an energy gain in the spectra, e.g., due to drag effects or
coalescence with a light quark, are not included. Such processes lead to
an increase in the $p_T$ of the final-state hadron and thus to an
increase in the electron $R_{AA}$, which would augment the discrepancy
with data. The neglect of the diffusive term becomes particularly
apparent in the elliptic flow. In the E-loss treatment the only
source of an azimuthal asymmetry in the $p_T$ spectra in non-central
Au-Au collisions is the spatial geometry of the overlap zone: particles
traveling along the short axis are less likely to be absorbed than those
moving along the long axis of the approximately elliptic reaction
zone. The positive $v_2$ generated by this mechanism amounts to up to a
few percent and significantly falls short of the observed electron
$v_2$, see right panel in Fig.~\ref{fig_elec-rad}.

\begin{figure}[!t]
\begin{minipage}{0.48\linewidth}
\includegraphics[width=\textwidth]{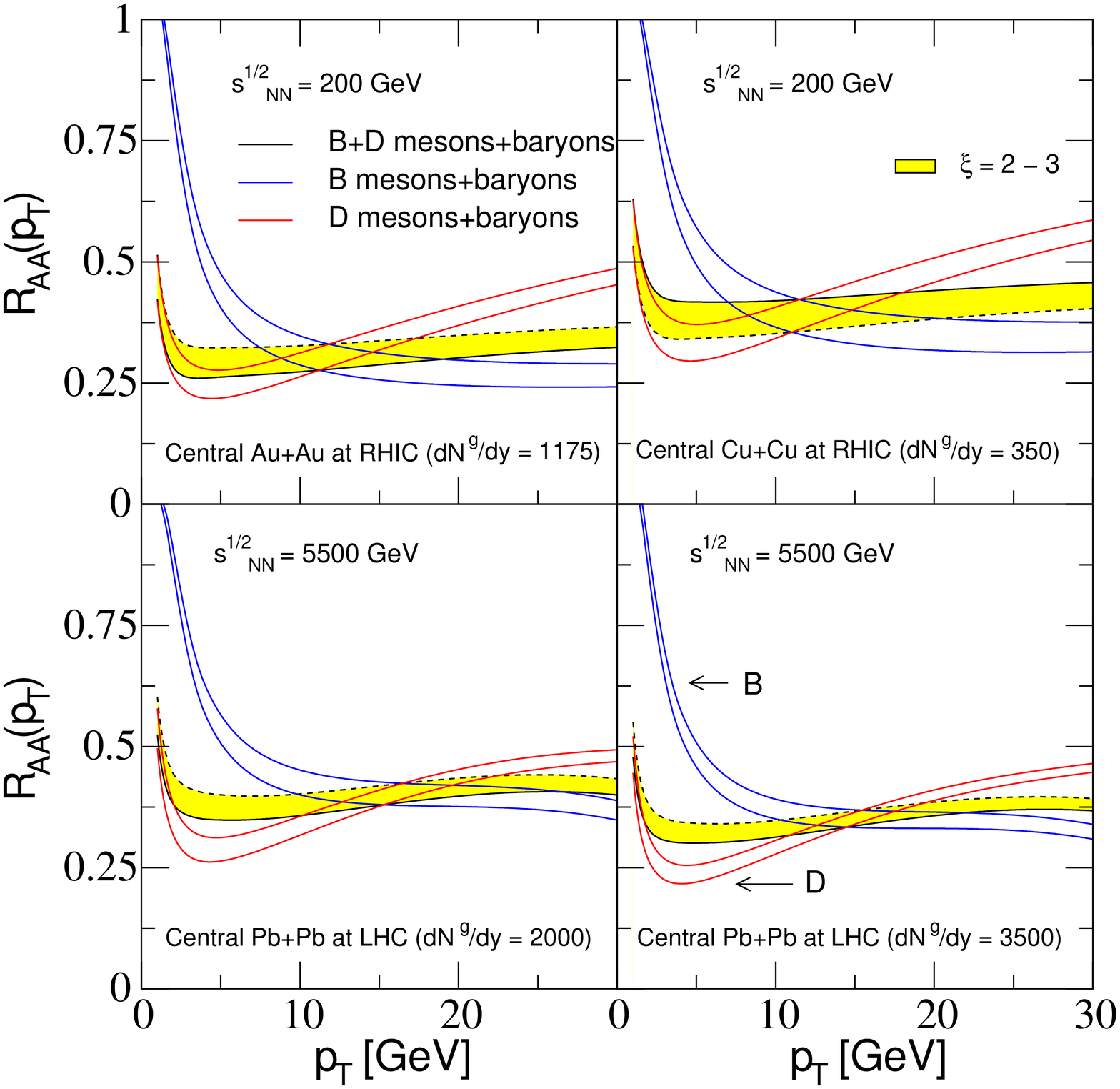}
\end{minipage}\hfill
\begin{minipage}{0.48\linewidth}
\includegraphics[width=\textwidth]{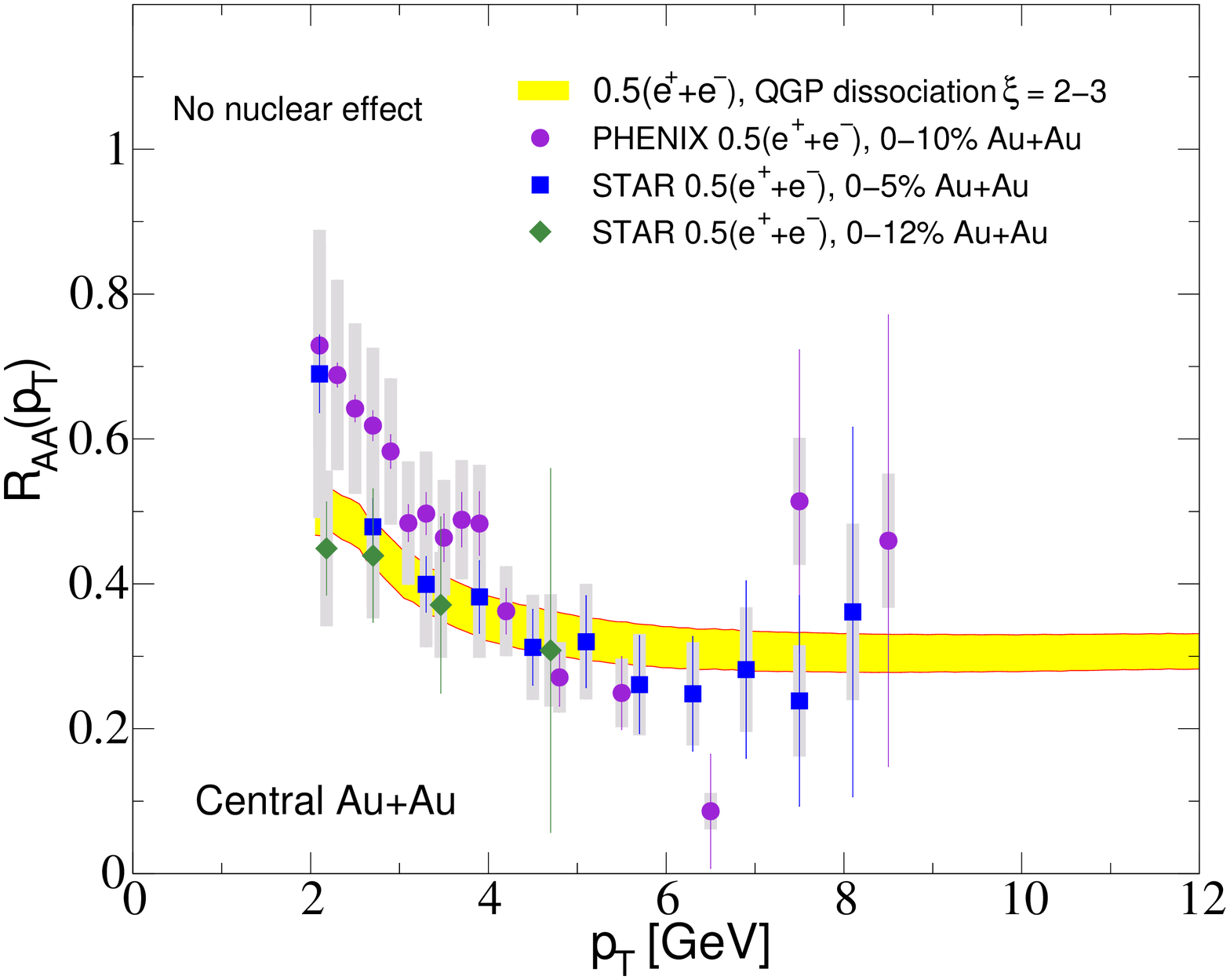}
\end{minipage}
\caption{(Color online) Energy-loss calculations employing the
  collisional dissociation mechanism for HQ fragmentation into heavy
  mesons in the QGP\protect\cite{Adil:2006ra}.  Left panels: nuclear
  modification factor for $D$ and $B$ mesons in central 200\,$A$GeV
  Au-Au and Cu-Cu at RHIC (upper panels) for gluon-rapidity densities of
  $\dd N^g/\dd y=1175$ and $350$, respectively, and in central $5.5 \, A
  \mathrm{TeV}$ Pb-Pb collisions at LHC for $\dd N^g/\dd y=2000$ and
  $3500$ (lower panels).  Right panel: $R_{AA}$ for $e^{\pm}$ in central
  200\,$A$GeV Au-Au collisions at RHIC compared to PHENIX and STAR
  data\protect\cite{Adare:2006nq,Abelev:2006db}; the yellow bands
  reflect theoretical uncertainties due to an impact parameter expansion
  of the heavy-light quark interaction.}
\label{fig_elec-coll-diss} 
\end{figure}
As an alternative mechanism, the collisional dissociation of $D$ and $B$
mesons from HQ fragmentation in the QGP (Sec.~\ref{sssec_coll-diss}) has
been implemented into an E-loss
calculation\cite{Adil:2006ra}\footnote{The compatibility with the
  radiative picture of high-$p_t$ light-hadron suppression is presumably
  little affected since the Lorentz-dilated formation times of light
  quarks largely result in hadronization beyond the QGP lifetime.}. A
rather striking prediction of this calculation is that the shorter
formation time of $B$ mesons leads to stronger suppression than for $D$
mesons above hadron momenta of $p_t\simeq 15\;\GeV$ at RHIC, cf.~left
panels in Fig.~\ref{fig_elec-coll-diss}.  This turns out to be an
important ingredient in the successful reproduction of the $e^\pm$
suppression data as shown in the right panel of
Fig.~\ref{fig_elec-coll-diss}.

\subsubsection{Langevin Simulations}
\label{sssec_elec-langevin}

The importance of elastic scattering for HQ diffusion \emph{and}
E-loss has been emphasized, prior to quantitative measurements of
$e^{\pm}$ spectra, in Refs.\cite{vanHees:2004gq,Moore:2004tg}, albeit
within rather different realizations of the underlying HQ interaction
(recall Secs.~\ref{sssec_reso} and \ref{sssec_lo-htl}, respectively).

\begin{figure}[!t]
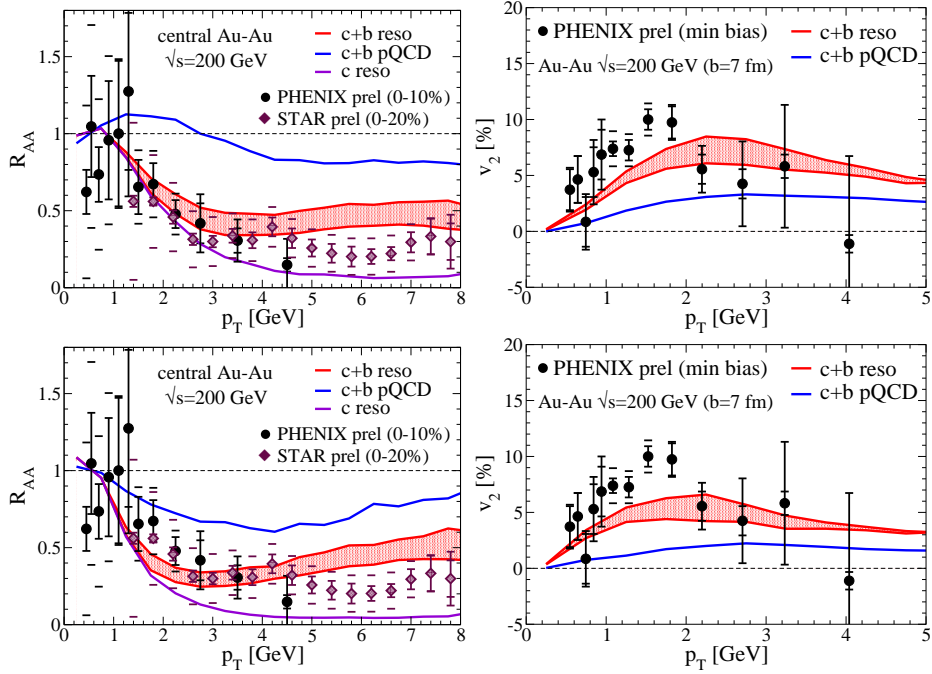

\begin{center}
\includegraphics[width=0.48\linewidth]{raa_e_cent-therm.eps}
\includegraphics[width=0.48\linewidth]{v2_e_MB_therm.eps}
\includegraphics[width=0.48\linewidth]{raa_e_cent_frag.eps}
\includegraphics[width=0.48\linewidth]{v2_e_MB_frag.eps}
\end{center}
\caption{(Color online) Nuclear modification factor (left panels) and
  elliptic flow (right panels) of $e^\pm$ spectra from heavy-flavor
  decays in Au-Au collisions at RHIC, as computed within HQ Langevin
  simulations in a thermal fireball employing the resonance+pQCD model
  for HQ transport in the QGP\protect\cite{vanHees:2005wb,Rapp:2006ta}.
  The upper panels include effects from heavy-light quark coalescence
  while the latter are switched off in the lower panels. The bands
  represent the full results for $D$- and $B$-meson resonance widths of
  $\Gamma=0.4$-$0.75 \; \GeV$; the blue lines are obtained for LO-pQCD
  interactions only, while the purple lines neglect bottom
  contributions. The data are from
  Refs.\protect\cite{Adler:2005xv,Butsyk:2005qn,Bielcik:2005wu}.}
\label{fig_elec-reso}
\end{figure}
The HQ spectra of the effective resonance + LO-pQCD
model\cite{vanHees:2004gq} (using Langevin simulations within an
expanding fireball, recall Fig.~\ref{fig_hq-reso}) have been converted
into $e^\pm$ spectra utilizing a combined coalescence/fragmentation
scheme at $T_c$ followed by heavy-meson three-body
decays\cite{vanHees:2005wb,Rapp:2006ta}. The predicted $e^{\pm}$ spectra
and elliptic flow show approximate agreement with 2005
PHENIX~\cite{Adler:2005xv,Butsyk:2005qn} and STAR\cite{Bielcik:2005wu}
data, see upper left and right panel of Fig.~\ref{fig_elec-reso},
respectively.  Compared to the results for LO pQCD interactions only
(blue lines), the resonance interactions (red bands) turn out to be
instrumental in generating the required suppression and elliptic flow
(see upper panels of Fig.~\ref{fig_elec-reso}). LO-pQCD scattering
alone, even with a strong coupling of $\alpha_s$=0.4, does not
produce sufficient coupling to the bulk medium to suppress the
primordial quark spectra, nor to drag the
heavy quarks along with the collective flow of the expanding
fireball. The effect of heavy-light quark coalescence is illustrated by
a calculation where only fragmentation is used as a hadronization
mechanism (lower panels in Fig.~\ref{fig_elec-reso}). In this case, the
shape of the $e^{\pm}$ $R_{AA}$ and the magnitude of the $v_2$ are not
well reproduced. Coalescence processes add both momentum and $v_2$ to
the meson (and thus to the $e^{\pm}$) spectra, i.e., the suppression
becomes smaller. It is furthermore instructive to compare the LO-pQCD
results with fragmentation only (lower left panel in
Fig.~\ref{fig_elec-reso}) to the pQCD E-loss calculations,
especially to the elastic DGLV results where $\alpha_s$=0.3 has been
used (left panel of Fig.~\ref{fig_elec-rad}). The suppression level in
the pertinent electron $R_{AA}$ is quite comparable for a rather large
range in $p_T$. The increasing trend in the Langevin calculations for
$p_T\gsim$~5\,GeV is presumably due to the dominant $b$-quark
contribution (which is barely suppressed even in the resonance model up
to $p_T\gsim$~5\,GeV, see left of Fig.~\ref{fig_hq-reso}). Let us also
estimate the impact of radiative contributions on the resonance
model. Within DGLV the electron suppression due to radiative E-loss
alone amounts to about 0.6-0.8 for $p_T\simeq$~4-10\,GeV.  Upon
multiplying the $R_{AA}$ for the resonance+pQCD model in the upper 
left panel of Fig.~\ref{fig_elec-reso} with this factor, the result 
would be compatible with current RHIC data.

\begin{figure}[!t]
\begin{minipage}{0.49\linewidth}
\vspace*{-3mm}
\includegraphics[width=0.97\linewidth]{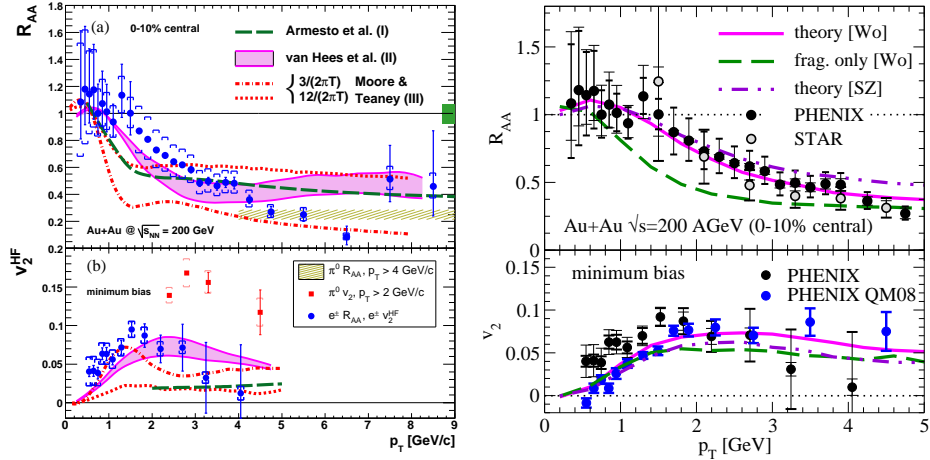}
\end{minipage}
\begin{minipage}{0.49\linewidth}
\includegraphics[width=0.96\linewidth]{RAA-v2-elec-tmat.eps}
\end{minipage}\hfill
\caption{(Color online) Left panel: Comparison of PHENIX $e^\pm$
  data\protect\cite{Adare:2006nq} to theoretical predictions based on
  (i) Langevin simulations with the
  resonance+LO-pQCD interactions plus quark coalescence
  (bands)\protect\cite{vanHees:2005wb}, (ii) Langevin simulations with
  (upscaled) HTL-improved LO-pQCD interactions (dash-dotted and dotted
  lines)\protect\cite{Moore:2004tg}, and (iii) radiative E-loss
  calculations (dashed lines)\protect\cite{Armesto:2005mz}. For further
  illustration, PHENIX data for $\pi^0$ suppression and elliptic flow
  are shown\protect\cite{Adler:2006hu}. Right panel: Comparison
  of PHENIX\protect\cite{Adare:2006nq,Awes:2008qi} and
  STAR\protect\cite{Abelev:2006db} $e^\pm$ data to Langevin-fireball
  simulations for HQ transport using $T$-matrices with lQCD-based 
  heavy-quark potentials\protect\cite{vanHees:2007me}; an uncertainty 
  due to different lQCD internal energies is indicated by the solid and
  dash-dotted curve; the dashed lines are obtained without the effects
  of heavy-light quark coalescence at $T_c$.}
\label{fig_elec-tmat}
\end{figure}
The PHENIX collaboration has conducted a comprehensive comparison of
their 2006 $e^\pm$ data\cite{Adare:2006nq} to theoretical calculations
predicting {\em both} $R_{AA}$ and
$v_2$\cite{vanHees:2005wb,Armesto:2005mz,Teaney:2006pr}, cf.~left panel
of Fig.~\ref{fig_elec-tmat}. The interpretation reiterates some of the
main points made above: (i) the missing drag in (radiative) E-loss
calculations entails a substantial underprediction of the $v_2$; (ii)
Langevin calculations using elastic scattering require rather small HQ
diffusion coefficients, $D_s (2\pi T)\simeq$~4-6, to be compatible with
the observed level of suppression and elliptic flow, and, (iii) quark
coalescence improves the simultaneous description of these two
observables.

The heavy-light quark $T$-matrix approach, based on input potentials
estimated from thermal lattice-QCD, has been applied within the same
Langevin-fireball + coalescence/fragmentation scheme as the effective
resonance model\cite{vanHees:2007me}. The pertinent $e^{\pm}$ spectra
(cf.~right panel of Fig~\ref{fig_elec-tmat}) exhibit a comparable level
of agreement with current RHIC
data\cite{Adare:2006nq,Abelev:2006db,Awes:2008qi} as in the resonance
model, with a similar uncertainty due to different extractions of the HQ
internal energy. Although the $T$-matrix calculations involve, in
principle, no tunable parameters, the inherent
theoretical uncertainties are appreciable (e.g., in the definition of
the in-medium potential in terms of internal or free energy). Let us,
however, recall a rather general feature of the $T$-matrix approach
which was visible already at the level of the HQ spectra in
Fig.~\ref{fig_hq-tmat}: the weak temperature dependence of, e.g., the
friction coefficient implies that the HQ coupling to the medium remains
rather strong in the later QGP and mixed phase stages of the evolution. 
Since the bulk $v_2$ is largest in these later stages, while the
suppression largely occurs in the first 1-2\,fm/$c$\cite{Rapp:2008qc}, 
the $T$-matrix
interactions generate relatively more $v_2$ than suppression compared
to, e.g., the resonance model (or, alternatively: for the same $v_2$,
the suppression in the $T$-matrix approach is smaller). This traces back
to the increasing color-Debye screening with increasing temperature,
which leads to a gradual melting of the resonance correlations and a
marked increase of the spatial diffusion constant, $D_s/(2\pi T)$, with
temperature (recall Fig.~\ref{fig_Ds}). Such a temperature dependence
appears to improve the consistency in the simultaneous
description of the $e^{\pm}$ $R_{AA}$ and $v_2$, but more precise data
are needed to scrutinize this feature (including $d$-A
collisions to quantify the Cronin effect, which could increase the
$R_{AA}$ without noticably affecting $v_2$).

\begin{figure}[!t]
\begin{center}
\includegraphics[width=0.45\linewidth]{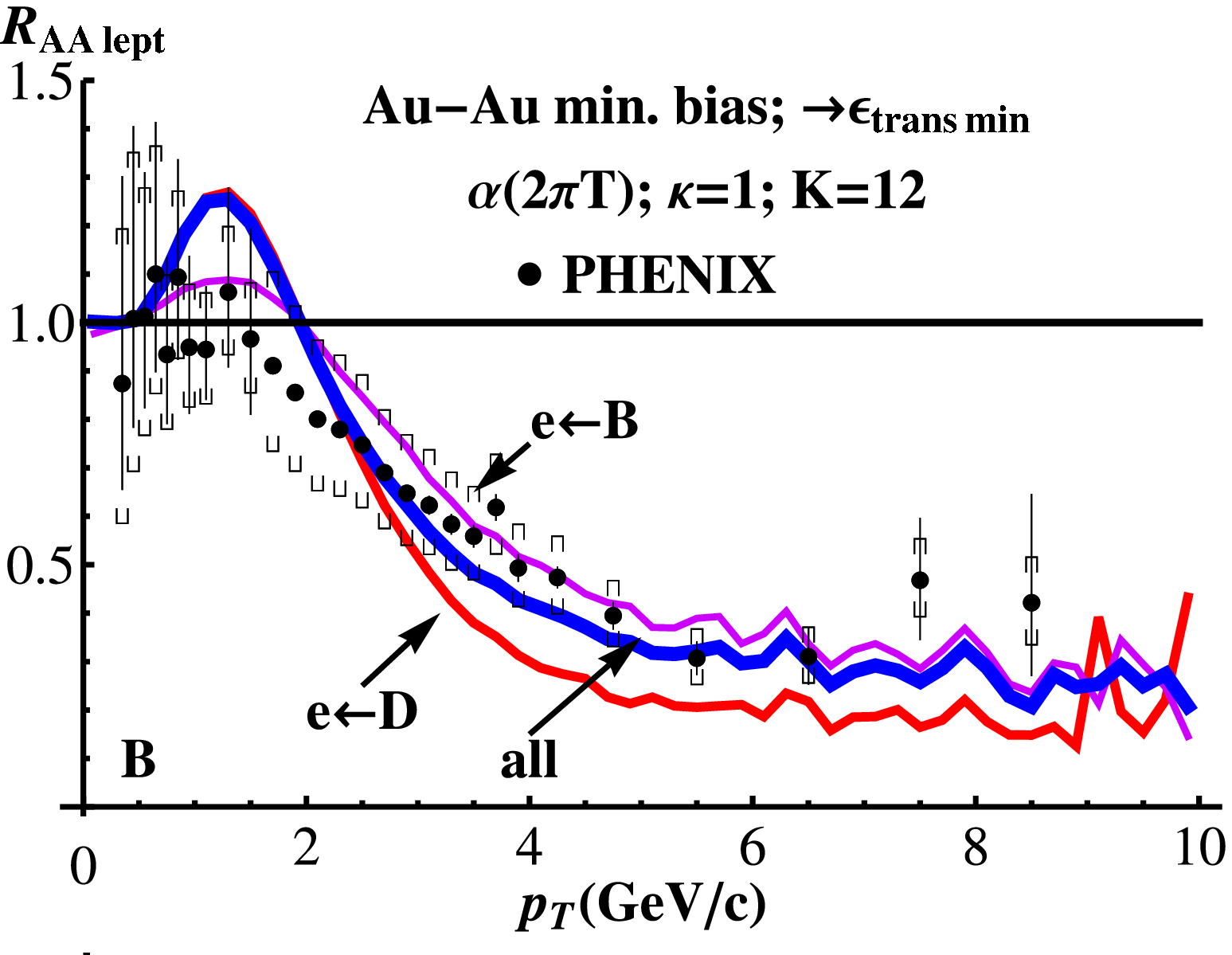}\hfill
\includegraphics[width=0.45\linewidth]{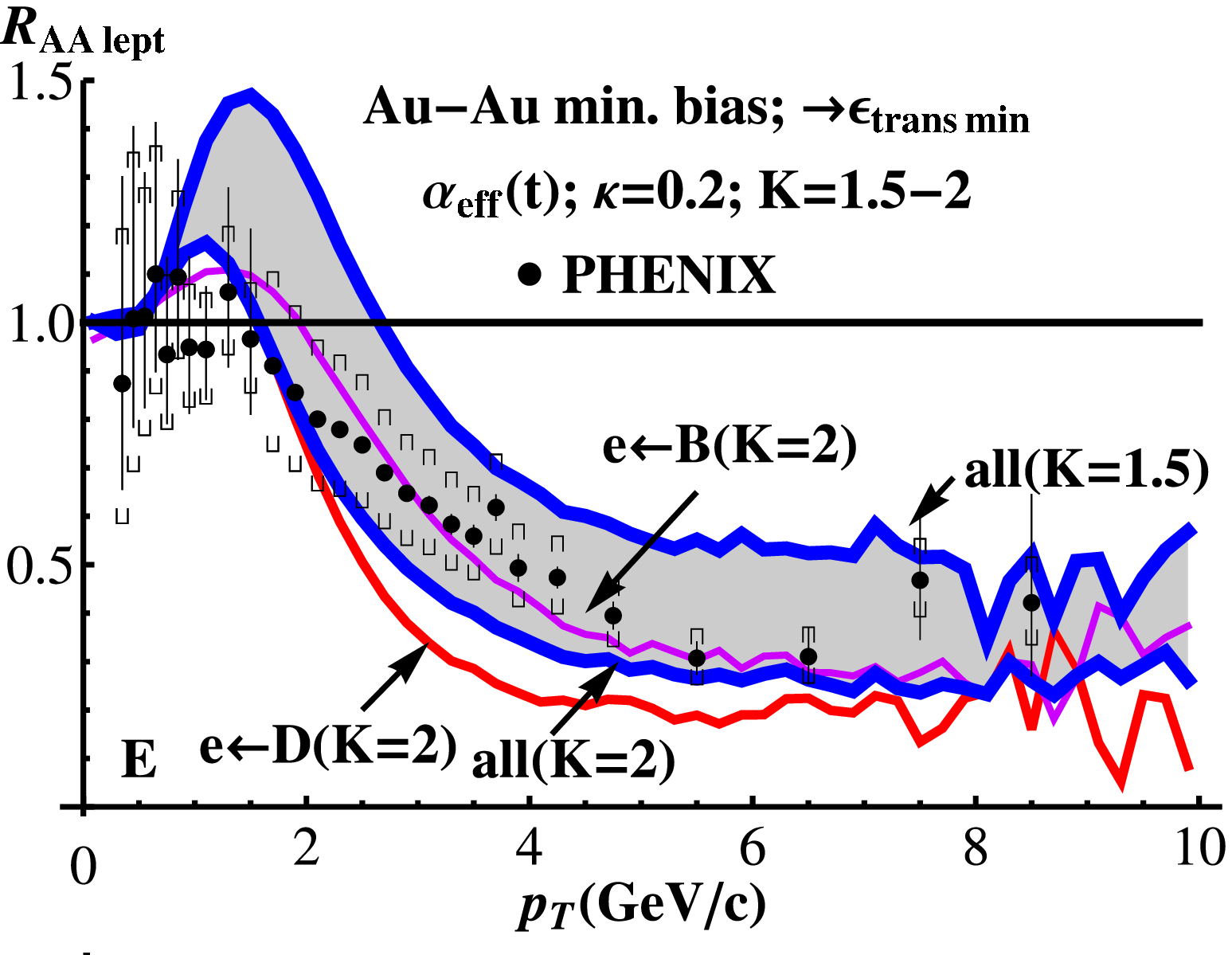}
\vspace*{3mm}
\includegraphics[width=0.45\linewidth]{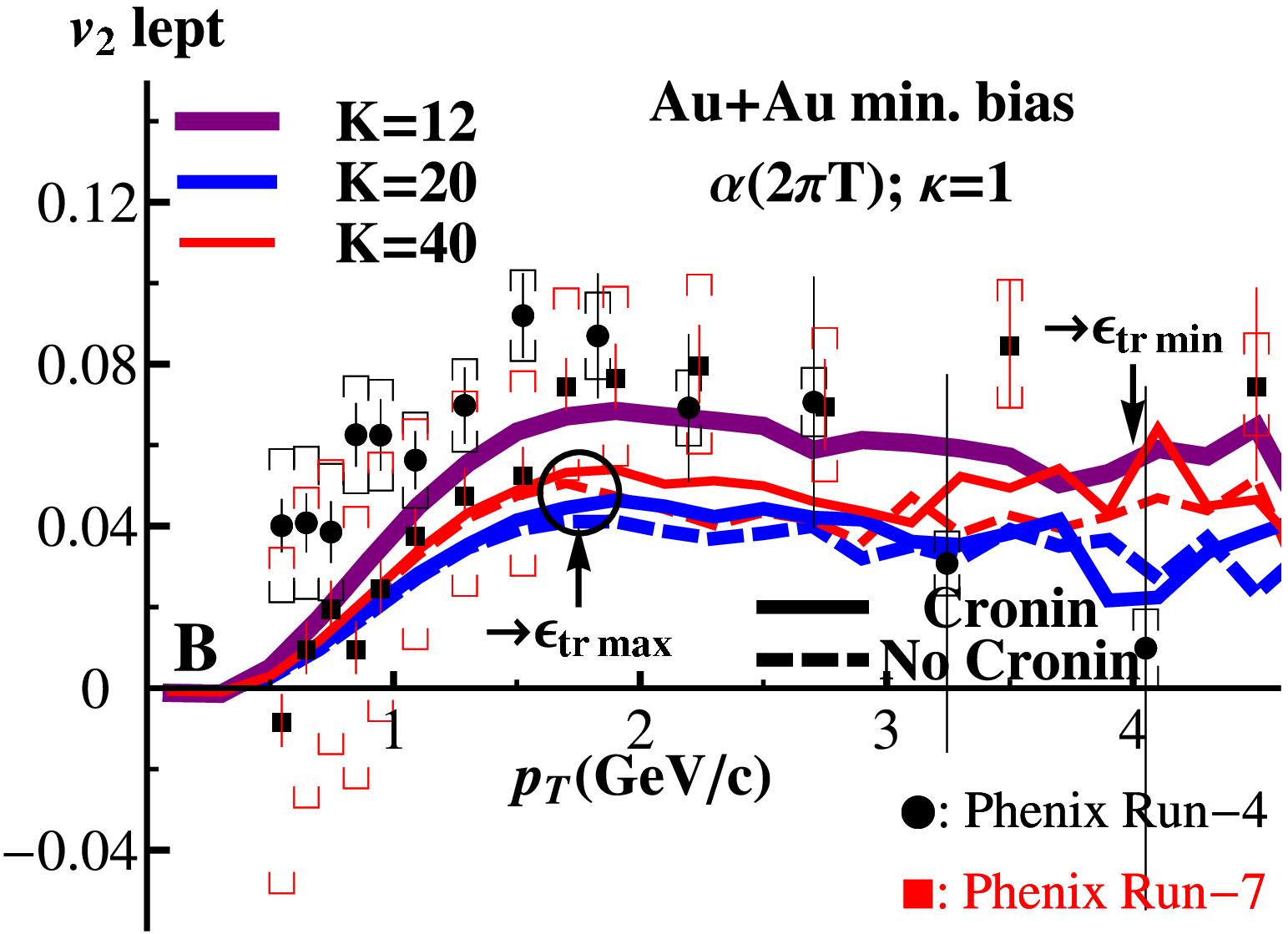}\hfill
\includegraphics[width=0.45\linewidth]{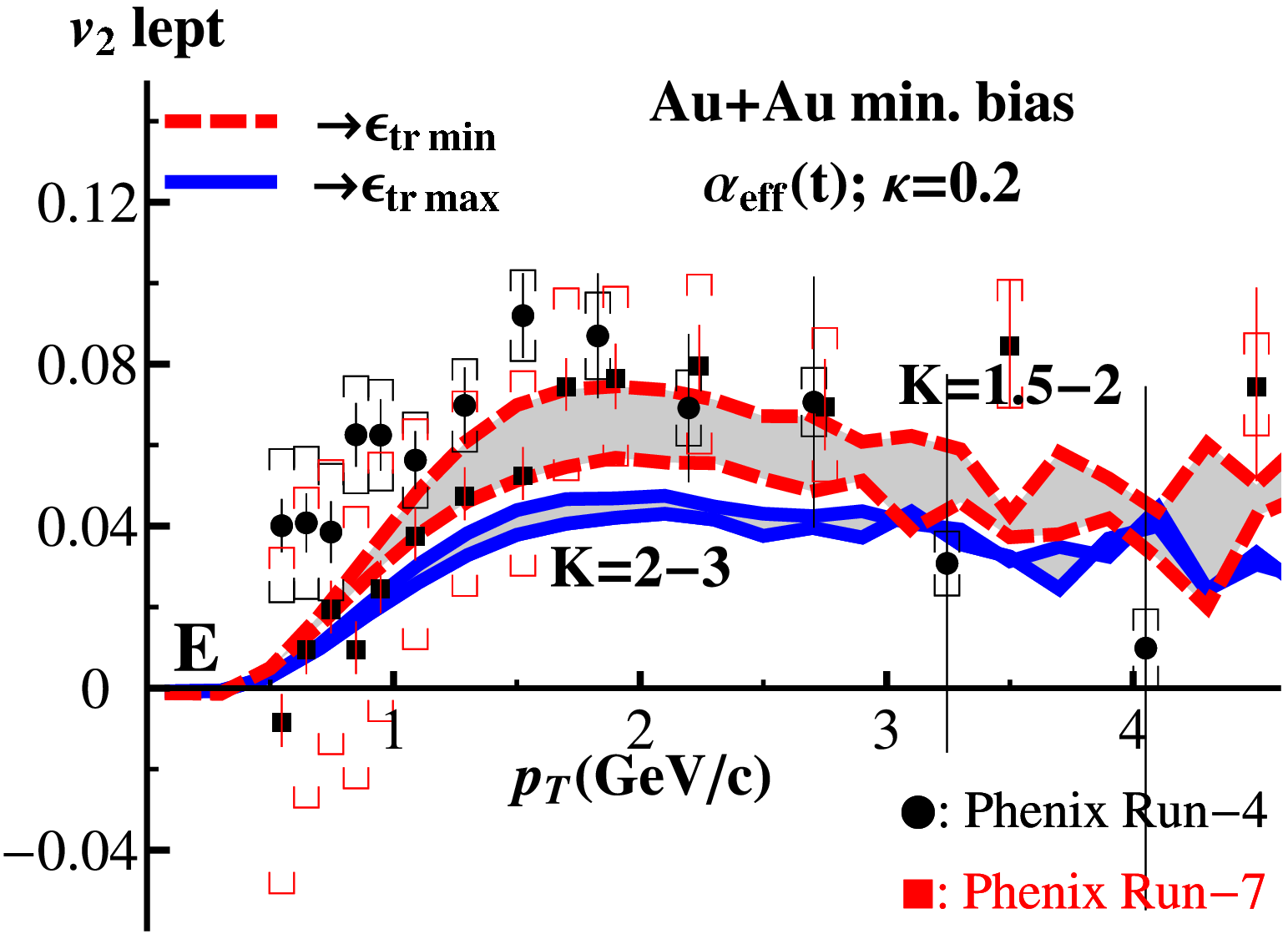}
\end{center}
\caption{(Color online) Boltzmann-transport model results for electron 
  $R_{AA}$ (upper panels) and $v_2$ (lower 
  panels)\protect\cite{Gossiaux:2008jv} in a hydrodynamic evolution
  using different versions of LO-pQCD HQ interactions, compared to 
  PHENIX data\protect\cite{Adare:2006nq,Awes:2008qi} in 
  200\,$A$GeV Au-Au collisions at RHIC. The curves in the left panels 
  are computed with fixed $\alpha_s(2\pi T)$ at given temperature, 
  conventional infrared regulator ($\tilde\mu_D^2 =r \mu_D^2$ with 
  $r$$\equiv$$\kappa$=1) and large $K$-factors; 
  the curves in the right panels are computed with a 
  running $\alpha_s$, reduced IR regulator ($r$=0.2) and reduced $K$ 
  factors of 1.5-2 (or 2-3), as represented by the bands. 
  $\epsilon_{tr max}$ or $\epsilon_{tr min}$ indicate freezeout at the 
  beginning or end of the mixed phase, respectively.}
\label{fig_v2-Raa-gossiaux}
\end{figure}
Finally, we reproduce in Fig.~\ref{fig_v2-Raa-gossiaux} selected results
of the Boltzmann-transport approach of HQ
diffusion\cite{Gossiaux:2008jv} with the background medium described by
the hydrodynamical model of Ref.\cite{Kolb:2003dz}; hadronization is
treated in a combined coalescence+fragmentation approach similar to the
one in Refs.\cite{vanHees:2005wb,vanHees:2007me}. The HQ interactions in
the QGP are implemented via the elastic pQCD scattering amplitudes
described in Sec.~\ref{sssec_lo-run}. The left panels in
Fig.~\ref{fig_v2-Raa-gossiaux} refer to a model with fixed coupling
constant, $\alpha_s(2 \pi T)$, at given temperature and standard 
screening mass ($r$=1). A large $K$ factor of $K=12$ is needed to
simultaneously reproduce the electron $R_{AA}$ and $v_2$ data. The final 
elliptic flow is found to be rather sensitive to the late QGP stages of 
the evolution, favoring hadronization at the end of the mixed phase (i.e., 
at a small transition energy-density, $\epsilon_{\text{trans}}$); this
is consistent with the findings of Ref.\cite{Rapp:2008qc} and the above
discussion of the $T$-matrix interaction. It thus corroborates that
anisotropic matter-flow can only be transferred to the heavy quarks if
the latter is sufficiently large, while E-loss (reflected in high-$p_T$
suppression) is mostly effective when the fireball density is high. The
simulations find little impact of the initial-state Cronin effect on the
final heavy-quark $v_2$, but the suppression is somewhat reduced
primarily for $e^\pm$ momenta of $p_T\simeq$~1-3\,GeV.  The right panel
of Fig.~\ref{fig_v2-Raa-gossiaux} illustrates that similar results can
be achieved with smaller $K$-factors, $K$=1.5-2, if the pQCD cross
sections are augmented by a running coupling, $\alpha_s(t)$ ($t$:
4-momentum transfer in the elastic scattering process), and a small
infrared regulator, $\tilde\mu_D^2=r \mu_D^2$ with $r$=0.2, in
$t$-channel gluon-exchange scattering.

\subsection{Viscosity?}
\label{ssec_viscosity}

In this section we utilize the quantitative estimates
for the HQ diffusion coefficient as extracted from current RHIC data to
obtain a rough estimate of the ratio of shear viscosity to entropy
density, $\eta/s$, in the QGP. This quantity has received considerable
attention recently since (a) it allows to quantify deviations from the
predictions of ideal fluid dynamics for observables like the elliptic
flow, and (b) conformal field theories in the strong coupling limit are
conjectured to set a universal lower bound for any liquid, given by
$\eta/s=1/(4 \pi)$\cite{Kovtun:2004de}, referred to as KSS bound. In the
following we will bracket the estimates derived from HQ observables by
using relations of $D_s$ and $\eta/s$ in the weak- and strong-coupling
limit (the latter assumed to be given by the AdS/CFT correspondence).

Following the discussion in Sec.~\ref{ssec_string}, the strong coupling
limit in the AdS/CFT framework results in a (spatial) HQ diffusion
constant of $D_s \simeq 1/(2 \pi T)$. Combining this with the lower
bound of the viscosity-to-entropy-density quoted above, one obtains
\begin{equation}
\frac{\eta}{s}=\frac{1}{2} T D_s \ .
\end{equation}
For a weakly coupled dilute gas, an estimate for a relation between
$D_s$ and $\eta$ can be inferred starting from the calculation of the
shear viscosity from kinetic theory for an ultrarelativistic
gas\cite{Israel:1970,Danielewicz:1984ww},
\begin{equation}
\label{eta-s-strong}
\eta \approx \frac{4}{15} n \erw{p} \lambda_{\mathrm{tr}} \ ,
\end{equation}
where $n$ denotes the particle density, $\erw{p}$ the average momentum
of the gas particles and $\lambda_{\mathrm{tr}}$ its transport-mean free
path. With the estimate, $n \erw{p} \simeq \epsilon$, for the energy
density and the corresponding equation of
state, $T s=\epsilon+P=4/3 \epsilon$, one arrives at
\begin{equation}
\frac{\eta}{s} \simeq \frac{1}{5} T \lambda_{\mathrm{tr}} \ .
\end{equation}
Finally, equating the transport mean-free path $\lambda_{\mathrm{tr}}$
to the mean-free time $\tau_{\mathrm{tr}}$ and taking into account the
delay due to the mass effect of the heavy quark on the thermalization 
time, $\tau_Q \approx \tau_{\mathrm{tr}} T/m_Q $, one finds
\begin{equation}
\label{eta-s-weak}
\frac{\eta}{s} \approx \frac{1}{5} T D_s  \ .
\end{equation}
In comparison to the ``strong-coupling'' estimate within AdS/CFT,
Eq.~(\ref{eta-s-strong}), the shear viscosity appears to be
underestimated when the kinetic theory for a dilute gas is applied to
liquids. 
\begin{figure}[!t]
\begin{minipage}{0.48\linewidth}
\includegraphics[width=0.96\textwidth]{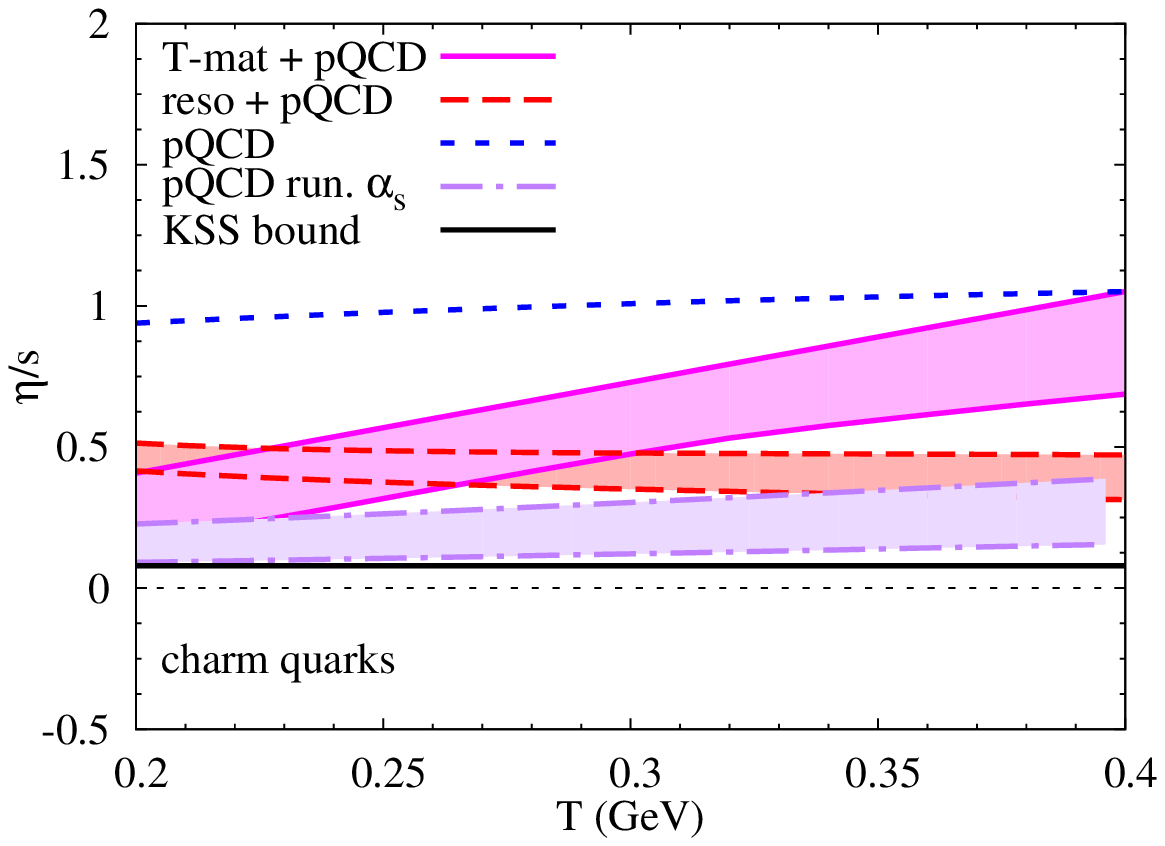}
\end{minipage}\hfill
\begin{minipage}{0.48\linewidth}
\vspace{-0.15cm}
\includegraphics[width=\textwidth]{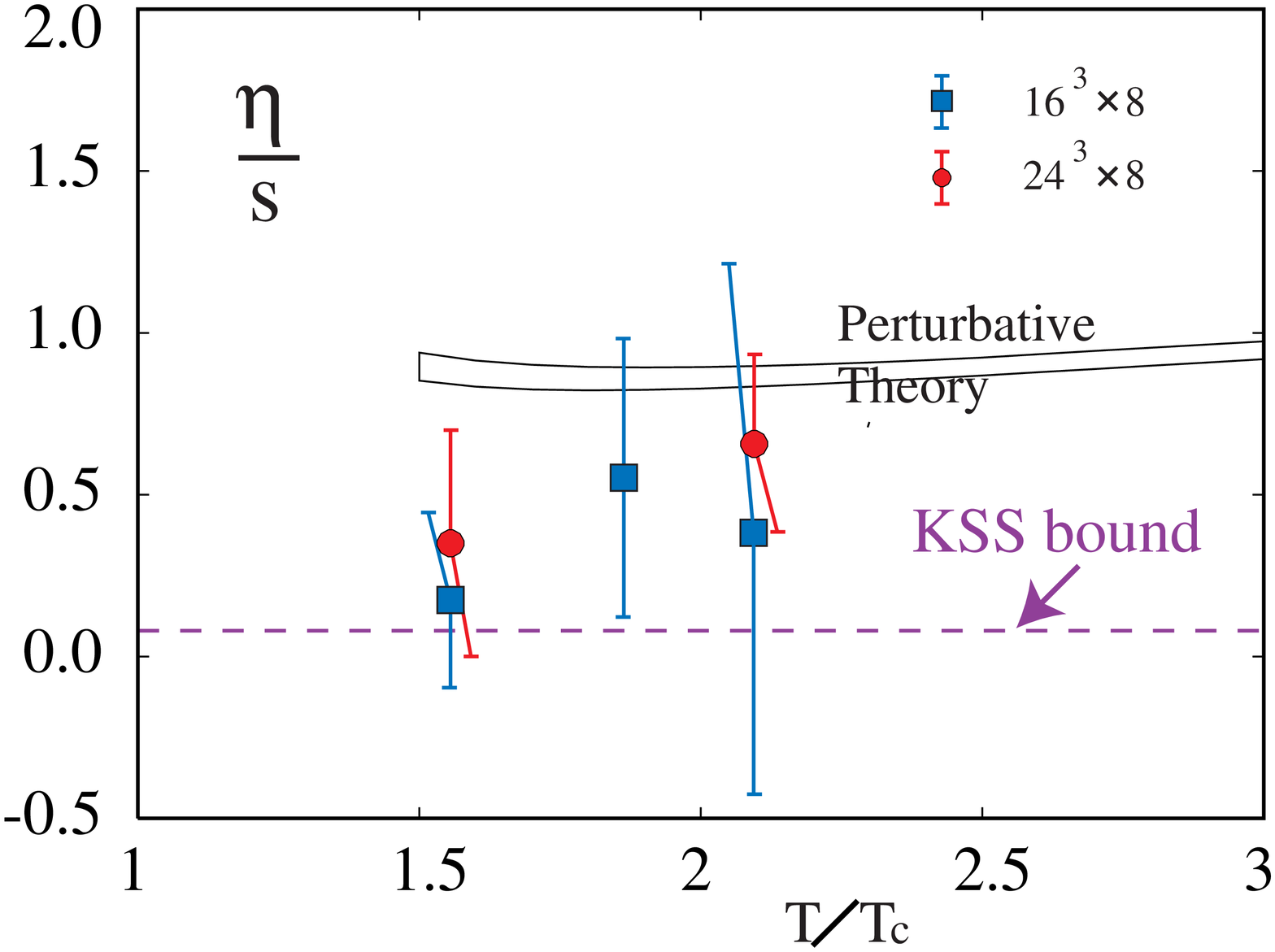}
\end{minipage}
\caption{(Color online) The ratio of shear viscosity to entropy density,
  $\eta/s$. Left panel: schematic estimates using charm-quark diffusion
  constants based on (a) schematic LO pQCD elastic scattering
  ($\alpha_s$=0.4) in the weakly interacting limit,
  Eq.~(\ref{eta-s-weak}) (dashed line), (b) pQCD elastic scattering with
  running coupling constant and small IR regulator (band enclosed
  by dash-dotted lines using the weak- and strong-coupling limits), 
  (c) the effective resonance + pQCD model in the strong-coupling limit,
  Eq.~(\ref{eta-s-strong}) (band enclosed by long-dashed lines for 
  $\Gamma$=0.4-0.75\,GeV), and (d) the
  lattice-QCD potential based $T$-matrix approach augmented by pQCD
  scattering off gluons (band enclosed by solid lines
  constructed from the weak- and strong-coupling limits). Right panel:
  lattice QCD computations in a gluon
  plasma\protect\cite{Nakamura:2004sy} compared to results inferred from
  perturbation theory\protect\cite{Blaizot:1999xk,Arnold:2003zc}.}
\label{fig_eta-s}
\end{figure}
These estimates are now applied to several of the HQ diffusion
calculations discussed above, see the left panel of
Fig.~\ref{fig_eta-s}.  Since $\eta/s\propto D_s (2\pi T)$, the main
features of Fig.~\ref{fig_Ds} are transmitted to $\eta/s$, in particular
the weak temperature dependence of the LO-pQCD calculations and the
effective resonance model. Of course, the absolute values of these
calculations differ considerably. A different behavior is only found for
the $T$-matrix+pQCD model, which suggests a transition from a strongly
coupled regime close to $T_c$ to relatively weak coupling above
$\sim$2$T_c$.  In fact, the uncertainty band has been constructed as
follows: for the lower limit, the weak-coupling estimate
Eq.~(\ref{eta-s-weak}) is used; for the upper limit, the strong-coupling
limit estimate, Eq.~(\ref{eta-s-strong}), at $T$=0.2\,GeV is linearly
interpolated with the LO-pQCD weak-coupling limit at $T$=0.4\,GeV (the
strong-coupling estimate for $T$-matrix+pQCD overshoots the LO-pQCD
result at this temperature). As for the spatial diffusion constant, the
increase of $\eta/s$ with temperature is related to color-Debye
screening of the lQCD-based potentials which entails a gradual melting
of the dynamically generated resonances in the heavy-light quark
$T$-matrix. It is tempting to interpret the decrease of $\eta/s$ when
approaching $T_c$ from above as a precursor-phenomenon of hadronization
and thus connected to the phase transition itself. It remains to be seen
whether a similar mechanism is operative in the light-quark and/or 
gluon sector (three-body interactions are unlikely to produce this
due to the decrease in particle density when approaching $T_c$ from
above). Such a behavior is rather general in that it has been observed
around phase-transition points for a large variety of substances, see,
e.g., the discussion in Refs.\cite{Lacey:2006pn,Csernai:2006zz}.

Finally we show in the right panel of Fig.~\ref{fig_eta-s} a quenched
lQCD computation of $\eta/s$\cite{Nakamura:2004sy}.  The error bars are
appreciable but the results tend to favor $\eta/s$ values which are
below LO-pQCD calculations.  The specific pQCD result included in this
plot employs a next-to-leading logarithm calculation for the shear
viscosity\cite{Arnold:2003zc} and a self-consistent hard-thermal-loop
calculation for the entropy density\cite{Blaizot:1999xk}. It is rather
close to the schematic LO calculation (using $\alpha_s$=0.4) in the left
panel of Fig.~\ref{fig_eta-s}.


\section{Heavy Quarkonia in Medium}
\label{sec_onia}

In recent years it has become increasingly evident that observables in
the heavy-quarkonium and open heavy-flavor sectors are intimately
connected. In the original picture of charmonium suppression as a probe
of color-screening in hot and dense QCD matter\cite{Matsui:1986dk} there
are no obvious such connections. Several recent developments have
changed this situation. Thermal lattice QCD calculations find that
charmonium correlation functions are remarkably stable up to
temperatures of $\sim$2\,$T_c$ or higher, suggestive for the survival of
the ground state ($\eta_c$, $J/\psi$) well into the QGP. This
interpretation is supported by probabilistic extractions of the
pertinent quarkonium spectral functions. It implies that quarkonia can
not only dissociate but also regenerate in the QGP\footnote{Note that
  higher dissociation rates also imply higher formation rates.}.  It
immediately follows that the yield and spectra of regenerated quarkonia
are, in principle, sensitive to the abundance and momentum spectra of
open-charm states in the system.  E.g., for a fixed total charm number
in the system, a softening of the heavy-quark spectra is expected to
increase $c$-$\bar c$ overlap in phase space and thus enhance the
probability for charmonium formation. At the same time, elliptic flow of
charm quarks will imprint itself on regenerated charmonia. Furthermore,
HQ interactions with light quarks (and possibly gluons) may be closely
related to the interaction (or potential) between two heavy
quarks. E.g., the $T$-matrix approach discussed in the previous section
is directly based on potentials which are extracted from the HQ free
energy computed in lattice QCD. As we argued there, this approach to
evaluate HQ diffusion has several attractive features, both
theoretically (it may provide maximal interaction strength in the
vicinity of $T_c$) and phenomenologically (it describes current HQ
observables at RHIC fairly well).

In the remainder of this section we address several aspects of quarkonia
in medium and in heavy-ion collisions with a focus on connections to the
open heavy-flavor sector. More extensive reviews on quarkonia in medium
have recently been given in
Refs.\cite{Rapp:2008tf,BraunMunzinger:2009ih,Kluberg:2009yd}, which we
do not attempt to reproduce here. In Sec.~\ref{ssec_spec} we give a
brief review of theoretical issues in the understanding of in-medium
quarkonium spectral properties, in terms of thermal lattice QCD results
for correlation and spectral functions and their interpretation using
effective potential models (Sec.~\ref{sssec_lat}). The latter are 
employing input potentials extracted from heavy-quark free energies
computed in lattice QCD, thus enabling, in principle, an internal
consistency check, provided a suitable potential can be defined. While
color screening is a key medium effect in the potentials (governing the
binding energy of the bound states), a quantitative assessment of
spectral functions requires the inclusion of finite-width effects
induced by dissociation reactions and possibly elastic scattering
(Sec.~\ref{sssec_diss}). In Sec.~\ref{ssec_prod} we elaborate on recent
developments in describing heavy-quarkonium production in heavy-ion
collisions. The main focus is on transport models which track the
dissociation and regeneration of charmonia (and bottomonia) through the
QGP, mixed and hadronic phases (Sec.~\ref{sssec_trans}), complemented by
a brief discussion of initial conditions as affected by 
cold-nuclear-matter (CNM) effects (shadowing, Cronin effect and 
nuclear absorption). This
is followed by an assessment of the current status of charmonium
phenomenology at SPS and RHIC.

\subsection{Spectral Properties of Quarkonia in the QGP}
\label{ssec_spec}
\subsubsection{Lattice QCD and Potential Models}
\label{sssec_lat}
The phenomenological Cornell potential\cite{Eichten:1979ms} for the
interaction between two heavy-quark (color-) charges in the
color-singlet channel,
\begin{equation}
V_{\bar QQ}(r;T=0) = -\frac{4}{3} \frac{\alpha_s}{r} + \sigma r \ ,
\label{Fqq-vac}
\end{equation}
has been very successful in reproducing the vacuum spectroscopy of
charmonium and bottomonium bound states. It consists of a (color-)
Coulomb plus a (linear) confining part with a very limited number of
parameters, i.e., a strong coupling constant, $\alpha_s$, and string
tension, $\sigma$ (in addition, an effective HQ mass, $m_Q$, needs to be
specified). Subsequent developments have put this framework on a more
rigorous footing by showing that (a) the potential description can be
recovered as a low-energy effective theory of QCD with heavy
quarks\cite{Bali:2000gf,Brambilla:2004wf}, and (b) lattice QCD
computations of the color-singlet heavy-quark free energy, $F_{1}$, 
have found excellent agreement with the functional form
(and parameters) of the Cornell potential\cite{Kaczmarek:2005ui}.
  
Early calculations\cite{Karsch:1987pv} of $Q$-$\bar Q$ bound-state
properties in the QGP have supplemented the Cornell potential by a
phenomenological ansatz for color screening of both the Coulomb and
confining parts,
\begin{equation}
V_{\bar QQ}(r;T)=\frac{\sigma}{\mu_D(T)} 
\left( 1-{\rm e}^{-\mu_D(T)r}\right) -\frac{4\alpha_s}{3r} \ 
{\rm e}^{-\mu_D(T)r}  \ . 
\label{Cornell-T}
\end{equation}
The key quantity carrying the temperature dependence is the Debye
screening mass, $\mu_D$ ($\propto gT$ in thermal pQCD). Already at 
that time the possibility was established that
ground-state charmonia (and even more so bottomonia) can survive until
temperatures (well) above $T_c$.
\begin{figure}
\begin{minipage}{0.48\linewidth}
\includegraphics[width=1.05\linewidth]{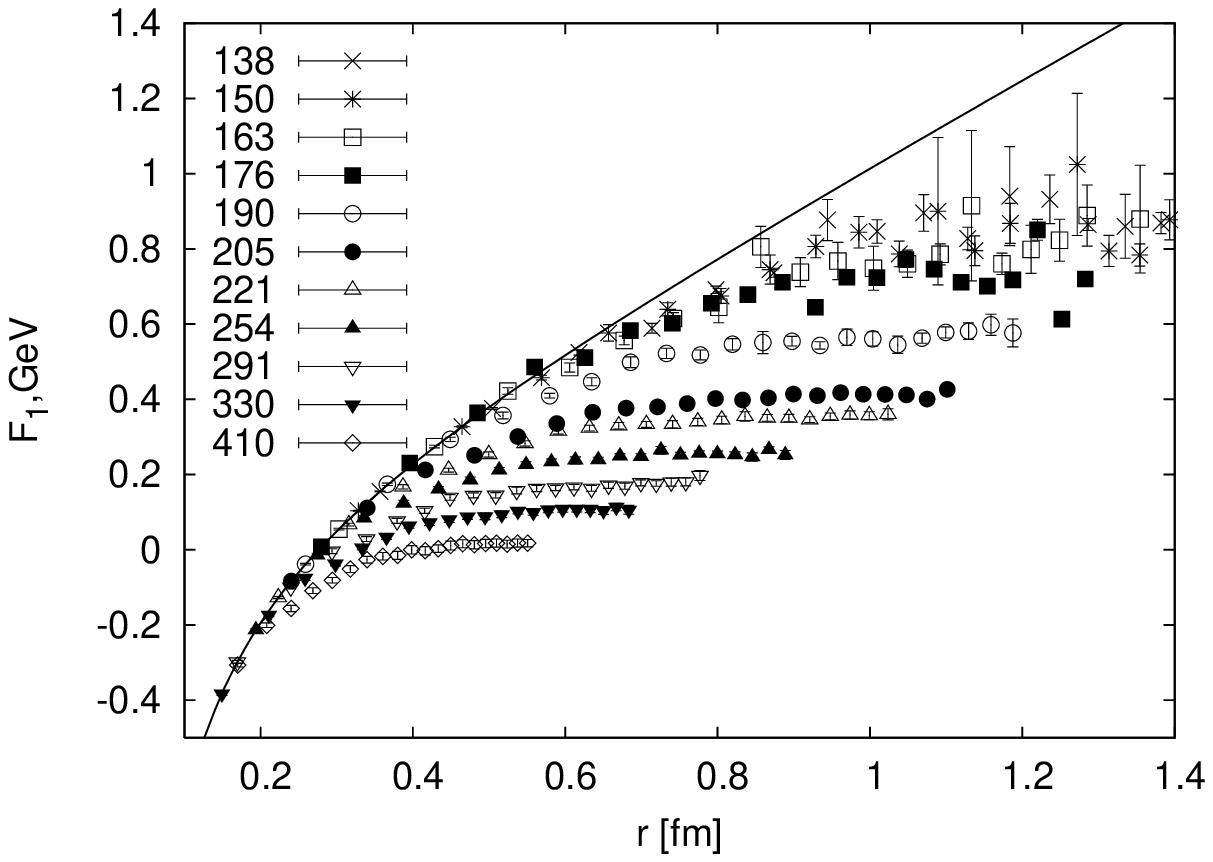}
\end{minipage}
\hspace{0.2cm}
\begin{minipage}{0.48\linewidth}
\includegraphics[width=0.7\linewidth,angle=-90]{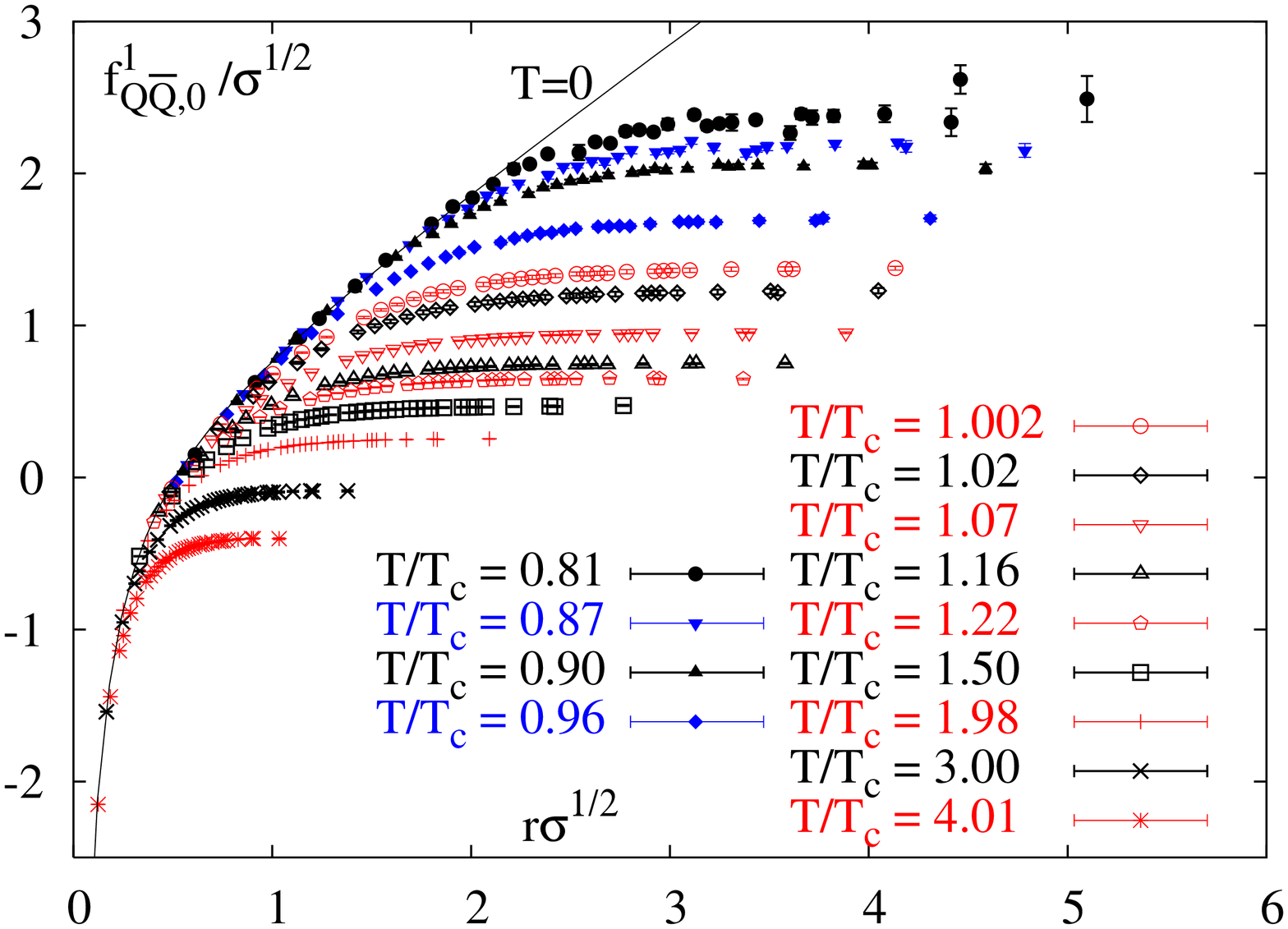}
\end{minipage}
\caption{(Color online) Free energy of a static color-singlet HQ pair as
  computed in lattice QCD for $N_f=3$\protect\cite{Petreczky:2004pz}
  (left) and $N_f$=2\protect\cite{Doring:2005ih} flavors (right). The
  critical temperature is $T_c=193(170)\; \MeV$ for the $N_f=3(2)$
  calculation, and the string tension typically amounts to
  $\sigma^{1/2}\simeq 420\; \MeV=1/(0.45\,\fm)$.  }
\label{fig_FQQ}
\end{figure}
More recently, quantitative lQCD computations of the finite-temperature
color-singlet free energy of a HQ pair, $F_{1}(r;T)$, have
become available, see, e.g., Fig.~\ref{fig_FQQ}. The results nicely
illustrate the color-screening effect and its gradual penetration to
smaller distances. When inserting the in-medium free energy as an
improved estimate of the finite-temperature HQ potential into a
Schr\"odinger equation, the ``melting'' temperature of the $J/\psi$
($\psi'$, $\chi_c$) was found to be just above (below)
$T_c$\cite{Digal:2001ue}.

Further progress in thermal lQCD came with the computation of heavy
quarkonium correlation functions,
\begin{equation}
  G_\alpha (\tau,\bvec r) =
  \langle\langle j_\alpha(\tau,\bvec r) j_\alpha^\dagger(0,\bvec 0)
  \rangle\rangle   
\label{G_x}
\end{equation}
(also referred to as temporal correlators), as a function of Euclidean
time, $\tau$.  $j_\alpha$ represent the creation/annihilation operators
of a hadronic current of given quantum numbers, $\alpha$. In the
pseudoscalar and vector charmonium channels (corresponding to $c$-$\bar
c$ $S$-waves with $\eta_c$ and $J/\psi$ states, respectively), the
Euclidean correlators were found to exhibit a surprisingly weak
temperature dependence up to $\sim$2\,$T_c$, even at large $\tau$,
suggestive for rather stable bound states. The temporal correlators are
related to the physical spectral function, $\sigma_\alpha(E,p;T)$, via
\begin{equation}
G_\alpha(\tau,p;T)=
\int\limits_0^\infty \dd E \ \sigma_\alpha(E,p;T) \
K(E,\tau;T)
\label{G-tau}
\end{equation}
with a thermal integral kernel
\begin{equation}
  K(E,\tau;T) = \frac{\cosh[E(\tau-1/2T)]}{\sinh[E/2T]} \ .
\end{equation}
Eq.~(\ref{G-tau}) implies that the extraction of the spectral function
from the Euclidean correlator requires a nontrivial integral
inversion. Especially at finite $T$, where periodic boundary conditions
limit the information on $G_\alpha(\tau,p;T)$ to a finite interval,
$0\le\tau\le 1/T$, and for a finite number of $\tau$ points, the
unambiguous inversion to obtain $\sigma_\alpha(E,p;T)$ becomes an
ill-defined problem. However, using probabilistic methods (in particular
the so-called maximum entropy method (MEM)), a statistical
reconstruction of $\sigma_\alpha(E,p;T)$ is possible and has been
applied\cite{Asakawa:2003re,Karsch:2003jg}. The approximate constancy of
the temporal correlators lead to spectral functions with rather stable
ground-state peaks corroborating the notion of surviving ground states
well above $T_c$.

\begin{figure}[!t]
\begin{minipage}{0.5\linewidth}
\includegraphics[width=0.99\textwidth]{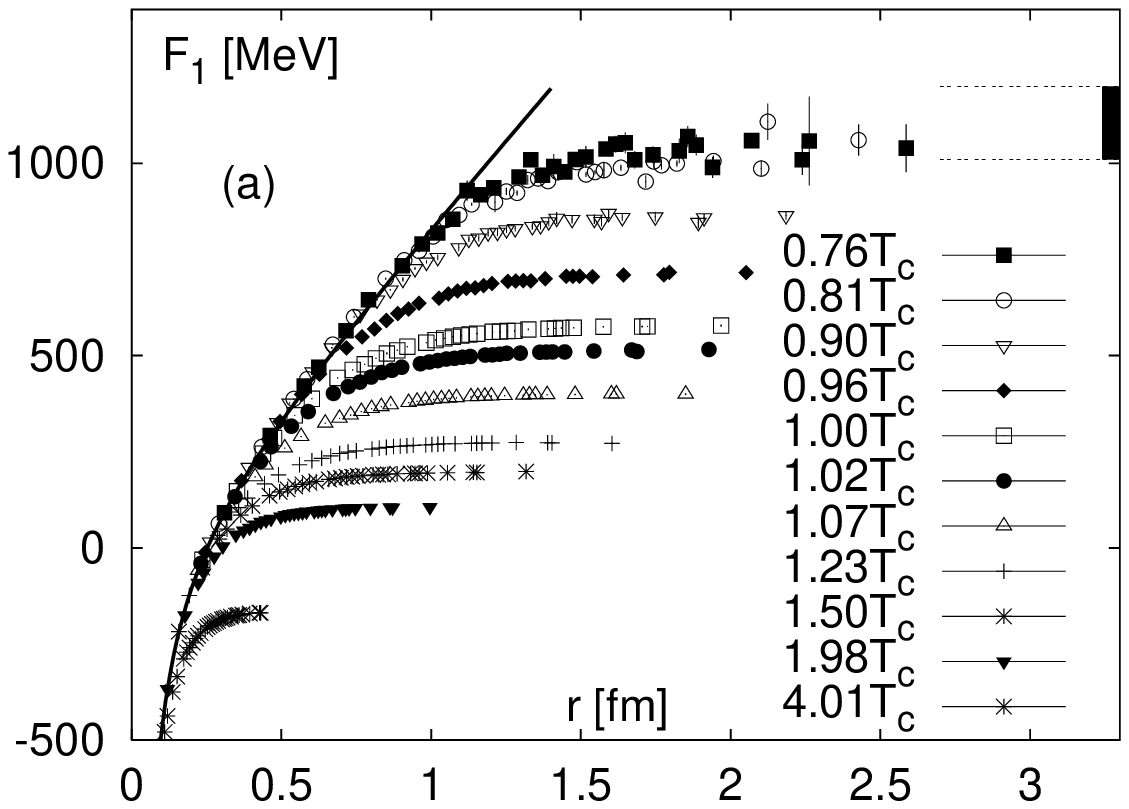}
\end{minipage}
\hspace{-0.3cm}
\begin{minipage}{0.5\linewidth}
\includegraphics[width=0.99\textwidth]{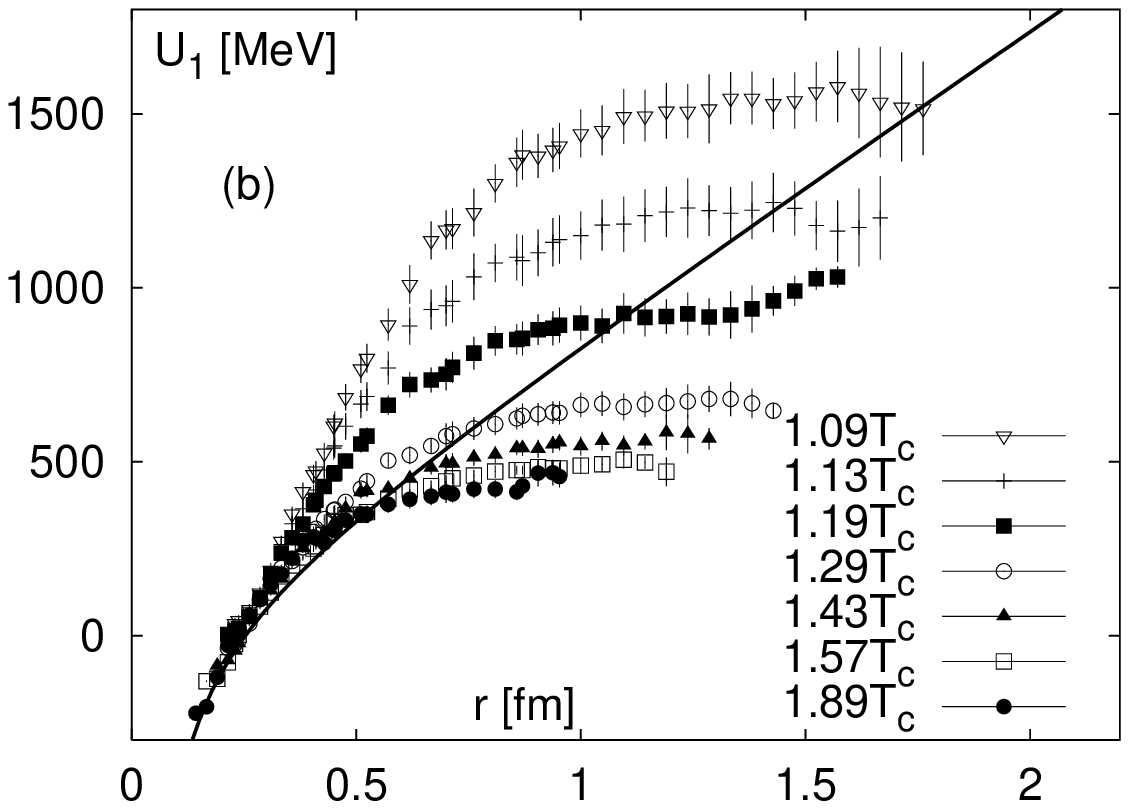}
\end{minipage}
\caption{HQ free energy in the color-singlet channel computed in thermal
  $N_f=2$ lattice QCD (left)\protect\cite{Kaczmarek:2005ui} and
  corresponding numerically extracted internal energy
  (right)\protect\cite{Kaczmarek:2005gi}.  }
\label{fig_FvsU}
\end{figure}
\begin{figure}[!t]
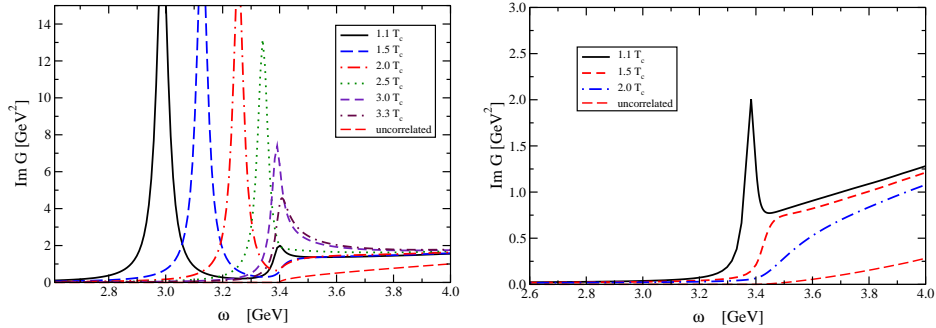

\begin{minipage}{0.49\linewidth}
\includegraphics[width=0.97\textwidth]{ImGccS-tmat0.eps}
\end{minipage}
\begin{minipage}{0.49\linewidth}
\includegraphics[width=0.97\textwidth]{ImGccP-tmat0.eps}
\end{minipage}
\caption{(Color online) Charmonium spectral functions computed in a
  $T$-matrix approach\protect\cite{Cabrera:2006wh} using internal HQ
  free energies extracted from $N_f=3$ thermal lattice
  QCD\protect\cite{Petreczky:2004pz}. For numerical purposes, a small
  charm-quark width of $\gamma_c=20 \; \MeV$ has been implemented into
  the intermediate $Q\bar Q$ propagator in the scattering
  equation~(\protect\ref{Tmat}).}
\label{fig_sf-tmat}
\end{figure}

To resolve the apparent discrepancy with the low dissociation
temperature found in the potential model discussed above, it has been
suggested to employ as potential the internal rather than the free
energy, which are related via
\begin{equation}
F(r;T)= U(r;T) - T S(r;T) \ . 
\label{Fqq-med}
\end{equation}
Especially in the color-singlet channel, the (positive) entropy
contribution rises significantly with $Q$-$\bar Q$ separation, $r$, thus
producing ``deeper'' potentials (cf.~Fig.~\ref{fig_FvsU}) entailing
stronger binding. Consequently, pertinent evaluations of quarkonium
spectra lead to larger dissociation temperatures, which seemingly agree
better with the lQCD spectral functions. These assertions have been made
more
quantitative\cite{Mocsy:2005qw,Wong:2006bx,Cabrera:2006wh,Alberico:2006vw,Laine:2007gj,Mocsy:2007jz,Mocsy:2007yj,Alberico:2007rg}
by employing potential models to calculate in-medium spectral functions,
perform the straightforward integral in Eq.~(\ref{G-tau}) and compare to
the rather precise temporal correlators from lQCD. It is important to
realize that the Euclidean correlators involve the pertinent spectral
function at \emph{all} energies. In Ref.\cite{Mocsy:2005qw} the
in-medium bound-state spectrum obtained from a Schr\"odinger equation
(using either a screened Cornell potential or lQCD internal energies)
has been combined with a perturbative ansatz for the $Q$-$\bar Q$
continuum above threshold. No agreement with lQCD correlators could be
established. In Refs.\cite{Wong:2006bx,Cabrera:2006wh} the importance of
rescattering effects for the interacting $Q$-$\bar Q$ continuum was
emphasized and implemented into the calculations of the $Q$-$\bar Q$
spectral functions. In Ref.\cite{Wong:2006bx} continuum correlations
were implemented via Gamov resonance states in Breit-Wigner
approximation, while in Ref.\cite{Cabrera:2006wh} a thermodynamic
$T$-matrix approach was employed,
\begin{equation}
T_\alpha(E) = V_\alpha  + \int \frac{\dd^3\bvec{k}}{(2 \pi)^3} 
\ V_\alpha \ G_{Q\bar Q}(E;k) \ T_\alpha(E) \ 
[1-f_Q(\omega_k^Q)-f_Q(\omega_k^{\bar Q})] \ ,
\label{Tmat-QQ}
\end{equation}
exactly as introduced in the context of HQ diffusion in
Sec.~\ref{sssec_tmat}, recall Eq.~(\ref{Tmat}). The $T$-matrix approach
enables a consistent treatment of bound and continuum states on equal
footing, as well as the implementation of medium effects (selfenergies)
into the intermediate two-particle propagator, $G_{Q\bar Q}$, recall
Eq.~(\ref{G_qQ}). Pertinent results for $S$- and $P$-wave charmonium
spectral functions, using the internal energy extracted from the $N_f=3$
free energy\cite{Petreczky:2004pz} (left panel of Fig.~\ref{fig_FQQ}),
are shown in Fig.~\ref{fig_sf-tmat} for a constant ($T$-independent)
charm-quark mass of $m_c=1.7 \;\GeV$.
One clearly recognizes the reduction in binding energy as a result of
color screening by the upward moving bound-state peak position with
increasing temperature (the $c\bar c$ threshold is fixed at $2m_c =3.4
\; \GeV$). Also note that nonperturbative rescattering effects close to
and above threshold induce a substantial enhancement in the $Q$-$\bar Q$
spectral function over the non-interacting continuum (indicated by the
red long-dashed lines), an effect which is of prime importance in the
calculation of HQ diffusion in $q$-$Q$ scattering as well. When applied
to the calculation of Euclidean correlators\cite{Cabrera:2006wh} in
Eq.~(\ref{G-tau}), the upward shift of low-energy strength due to the
moving bound states in the $S$-wave spectral function shown in the left
panel of Fig.~\ref{fig_sf-tmat} entails a suppression of $G(\tau)$ with
temperature which disagrees with the weak temperature dependence found
in lQCD.

\begin{figure}[!t]
\begin{minipage}{0.5\linewidth}
\vspace{-0.2cm}
\includegraphics[width=0.99\textwidth]{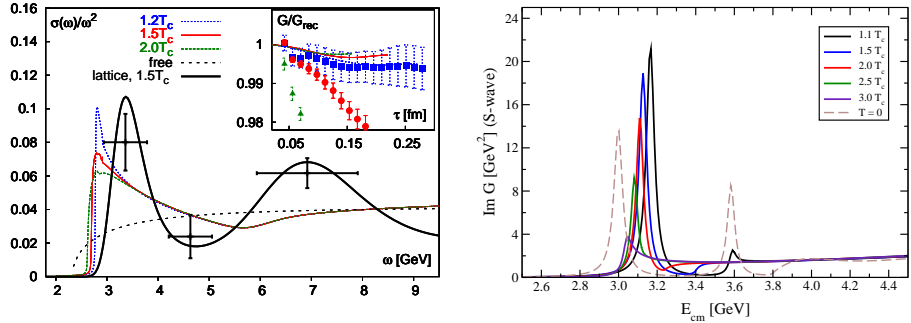}
\end{minipage}
\begin{minipage}{0.48\linewidth}
\includegraphics[width=0.95\textwidth]{ImGccS-tmat-mstar.eps}
\end{minipage}
\caption{(Color online) $S$-wave charmonium spectral functions ($\eta_c$
  or $J/\psi$) computed employing: (i) a screened Cornell potential
  within nonrelativistic Green's function approach
  (left)\protect\cite{Mocsy:2007jz,Mocsy:2007yj}, and (ii) an lQCD-based
  $N_f=3$ internal energy\protect\cite{Petreczky:2004pz} within a
  $T$-matrix approach\protect\cite{Cabrera:2006wh}. Both calculations
  account for in-medium charm-quark masses, and approximately reproduce
  the weak temperature dependence of the temporal correlators computed
  in lQCD up to temperatures of at least $2T_c$ (see, e.g., inset in the
  left panel). However, in calculation (i) the bound-state has
  disappeared at temperatures below $1.5 T_c$, while in calculation (ii)
  it is visible up to $\sim$2.5\,$T_c$.}
\label{fig_sf}
\end{figure}

Another important ingredient to understand the behavior of the
correlators is the temperature dependence of the HQ mass.
Schematically, the, say, $J/\psi$ bound-state mass may be written as
\begin{equation}
m_{J/\psi}=2m_c^*-\epsilon_B \ .
\label{mpsi}
\end{equation}
This illustrates that a small (large) binding energy, $\epsilon_B$, can
be compensated by a small (large) effective quark mass in a way that the
ground-state mass stays approximately constant. Indeed, when
interpreting the asymptotic value of the in-medium potential as an
effective HQ mass correction,
\begin{equation}
m_c^* = m_c^0 + \Delta m_c \quad , \quad
\Delta m_c \equiv X(r=\infty;T)/2 \ ,
\end{equation}
with $X=U$ or $F$ (or an appropriate combination thereof), the use of
$U$ implies strong binding with large effective quark masses while the
use of $F$ leads to weak binding with small $m_c^*$ (recall from
Fig.~\ref{fig_FvsU} that the ``$U$-potential" is deeper than the
``$F$-potential'' but features a larger asymptotic value for
$r\to\infty$). Consequently, it has been found that reasonable agreement
with lQCD correlators can be achieved with different spectral functions,
covering a rather large range of dissociation temperatures, e.g.,
slightly above $T_c$ using screened Cornell potentials similar to the
free energy ($F_1$)\cite{Mocsy:2007jz,Mocsy:2007yj},
$\sim$1.5\,$T_c$ using a linear combination of free and internal
energy\cite{Wong:2006bx,Alberico:2007rg}, or up to $\sim$2.5\,$T_c$ when
using internal energies\cite{Cabrera:2006wh}\footnote{We do not address
  here the issue of so-called zero-mode contributions to quarkonium
  correlators, which arise on the lattice due to the periodic boundary
  conditions in temporal direction\cite{Umeda:2007hy}. These
  contributions are essential to obtain quantitative agreement with the
  lQCD correlators in all mesonic quantum-number channels except the
  pseudoscalar one ($\eta_c$); they are rather straightforward to
  implement in quasi-particle approximation.}, cf.~Fig.~\ref{fig_sf}.

To resolve this redundancy, it will be necessary to develop independent
means of determining the in-medium quark mass and the appropriate
quantity to be identified with the HQ potential. First estimates of the
HQ mass from thermal lQCD have been obtained by approximating the HQ
number susceptibility within a quasiparticle model with effective quark
mass\cite{Petreczky:2008px}. The results suggest a rather moderate
temperature variation of the latter, which deviates significantly from
the perturbative predictions up to $T$$\simeq$3\,$T_c$.  In
Refs.\cite{Laine:2006ns,Laine:2007qy,Brambilla:2008cx} hard-thermal-loop
and HQ effective theory techniques have been applied to derive the
leading terms in a perturbative and HQ mass expansion of a finite
temperature potential. An interesting finding of these investigations is
that the potential develops an imaginary part in the medium which arises
from the Landau damping of the exchanged gluons, representing a decay
channel of the HQ bound state. A more general discussion of the
in-medium decay width of heavy quarkonia, which plays a central role for
phenomenology in heavy-ion collisions, is the subject of the following
section.

The impact of finite-width effects on charmonium correlators has been
studied within the $T$-matrix approach in Ref.\cite{Cabrera:2006wh}, by
implementing an imaginary part into the charm-quark propagators. A
broadening of charmonium spectral functions leads to an enhancement of
the temporal correlators (due to additional strength at lower energies),
which, however, is only a few percent for a charmonium width on the
order of $\Gamma_\Psi$$\simeq$100\,MeV. On the one hand, such a value
for the width is phenomenologically significant, as it implies that
about 60\% of the charmonia decay within a time of $2 \; \fm/c$. On the
other hand, for larger widths, their impact on the correlators should be
accounted for in quantitative comparisons to lQCD ``data''.

\subsubsection{Dissociation Widths}
\label{sssec_diss}
\begin{figure}[!t]
\begin{center}
\includegraphics[width=.33\textwidth]{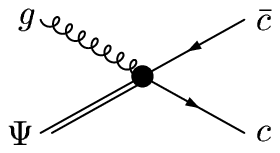}
\hspace{2cm}
\includegraphics[width=.33\textwidth]{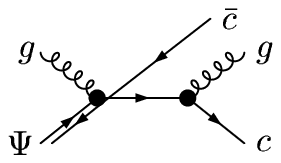}
\end{center}
\caption{Diagrams for quarkonium-dissociation reactions via parton
  impact; left panel:
  gluo-dissociation~\protect\cite{Bhanot:1979vb,Peskin:1979va}; right
  panel: quasifree dissociation~\protect\cite{Grandchamp:2001pf}.  }
\label{fig_psi-dia}
\end{figure}

\begin{figure}[!t]
\begin{minipage}{0.41\linewidth}
\includegraphics[width=1.0\linewidth]{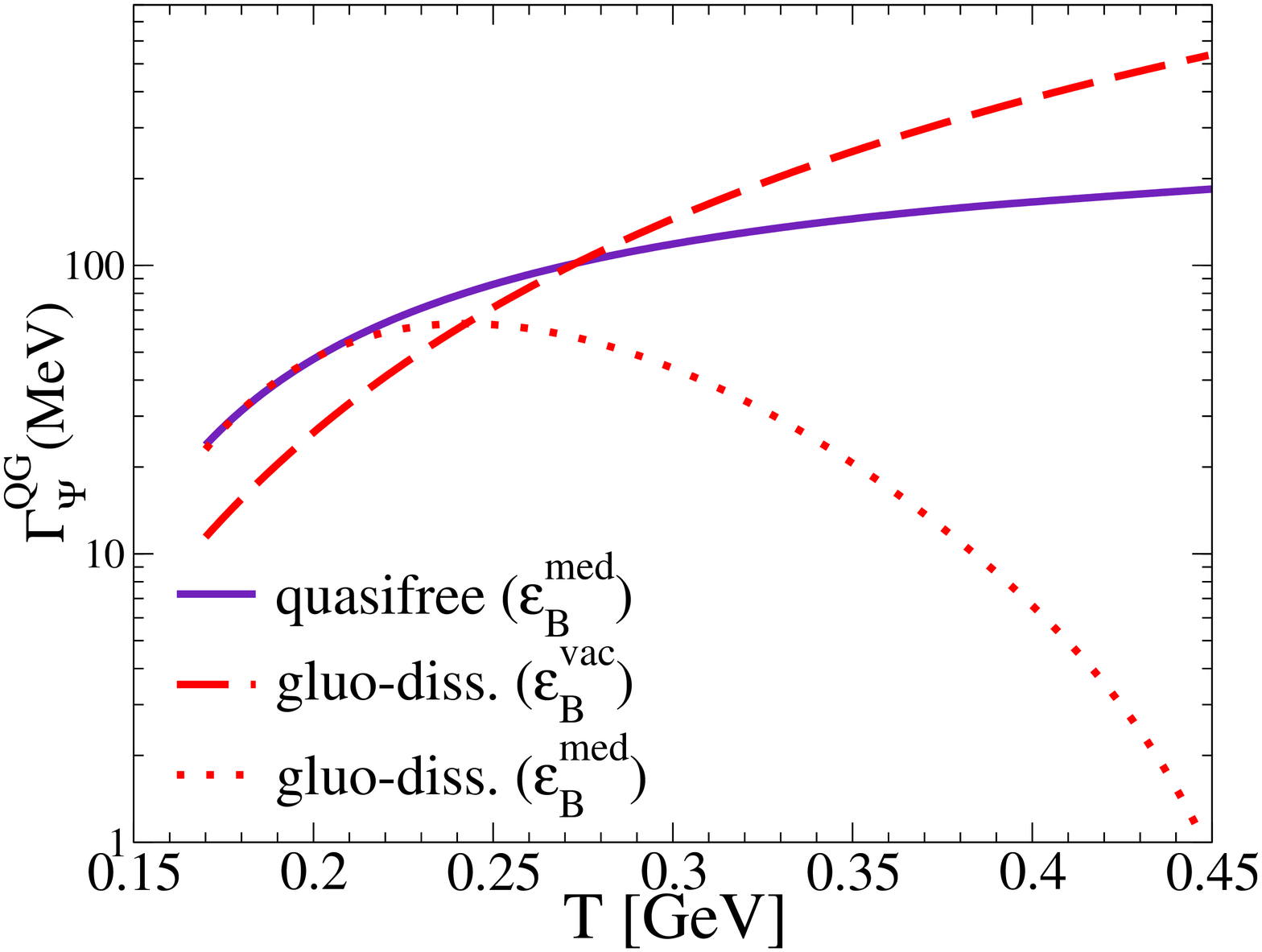}
\end{minipage}
\begin{minipage}{0.58\linewidth}
\includegraphics[width=1.0\linewidth]{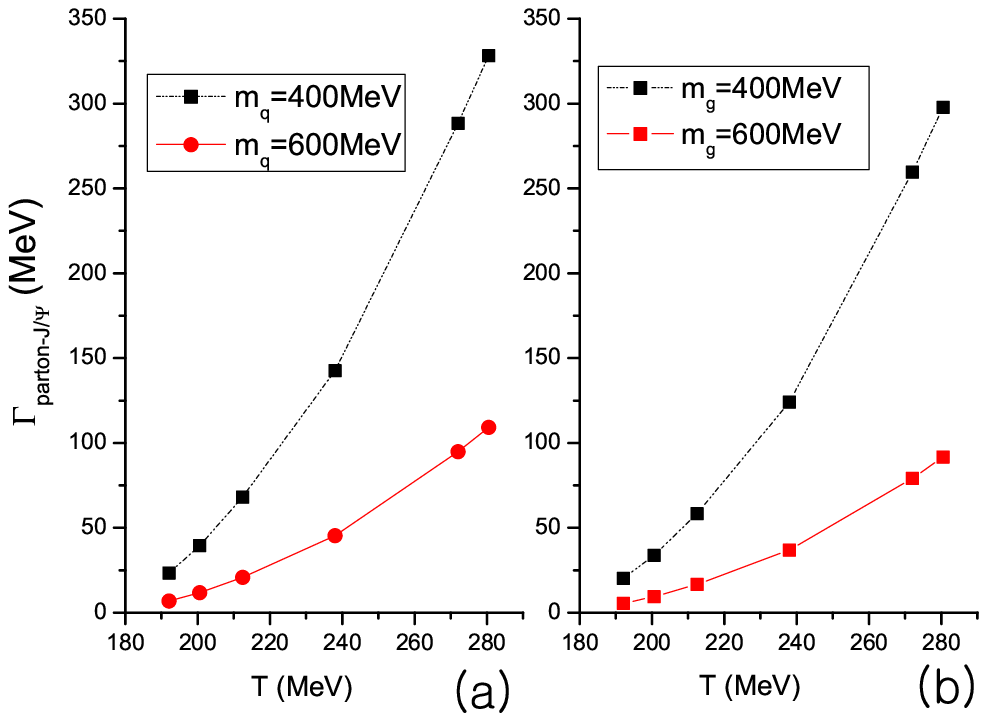}
\end{minipage}
\caption{(Color online) Parton-induced dissociation widths for $J/\psi$
  at rest in a QGP as a function of temperature. Left panel:
  gluo-dissociation width based on the cross section, Eq.~(\ref{gdiss}),
  with free (dashed line) and in-medium decreasing binding energy
  (dotted line), as well as ``quasifree dissociation'' width
  ($J/\psi+p\to c+\bar c +p$) with in-medium decreasing
  $\varepsilon_B$\protect\cite{Grandchamp:2001pf,Rapp:2005rr}.  Right
  panel: full NLO calculation for (a) quark- and (b) gluon-induced
  dissociation for different thermal parton
  masses\protect\cite{Park:2007zza}.}
\label{fig_gamma}
\end{figure}

The spectral width of a quarkonium state propagating through matter can,
in principle, receive contributions from elastic and inelastic reactions
with the medium particles. Elastic scattering affects the momentum
distribution of the quarkonium while inelastic interactions change its
abundance (via dissociation or formation). More formally, the quarkonium
acquires a complex selfenergy which can be expressed via the in-medium
scattering amplitude, ${\cal M}_{\Psi i}$, folded over the (thermal)
distribution, $f_i$, of the medium particles,
\begin{equation}
\Sigma_\Psi (p)= \sum\limits_i\int \frac{\dd^3k}{(2\pi)^3 2\omega_i(k)}
\ f_i(\omega_i(k);T) \ {\cal M}_{\Psi i}(p,k)  \  . 
\end{equation}
The real part of $\Sigma_\Psi$ characterizes in-medium changes of the
quarkonium mass while the imaginary part determines its width,
$\Gamma_\Psi(E)$=-2\,$\im \Sigma_\Psi(E)$\footnote{In addition, mass and
  width changes are induced at the $Q$-$\bar Q$ level via in-medium
  effects on the $Q$-$\bar Q$ potential and direct $\Psi\to Q +\bar Q$
  decays, respectively. These effects can be accounted for, e.g., in
  the underlying $Q$-$\bar Q$ $T$-matrix, Eq.~(\ref{Tmat-QQ}).}.  Most
of the attention thus far has been directed to the inelastic reactions
(rather than elastic scattering). Using the optical theorem to relate
the imaginary part of the forward scattering amplitude to the cross
section, one arrives at the well-known expression
\begin{equation}
\Gamma_\Psi = \sum\limits_i \int 
\frac{\dd^3k}{(2\pi)^3} \ f_i(\omega_k,T)  \
v_{\rm rel} \ \sigma^{\rm diss}_{\Psi i}(s) \ , 
\label{gamma}
\end{equation}
where $v_{\rm rel}$ denotes the relative velocity of the incoming
particles and $s$=$(p+k)^2$ the squared center-of-mass energy of the
$\Psi$-$i$ collision. The first evaluation of the inelastic quarkonium
reaction cross section with gluons was conducted in
Refs.\cite{Peskin:1979va,Bhanot:1979vb}. Employing Coulomb wave
functions for the quarkonium bound state, the leading-order process,
$\Psi + g \to Q +\bar Q$, is the analog of photo-dissociation of
hydrogen, see left panel of Fig.~\ref{fig_psi-dia}.
For an $S$-wave $\Psi$ bound state with binding energy $\varepsilon_B$,
the cross section is given as a function of incoming gluon energy, $k_0$,
by
\begin{equation}
\sigma_{g\Psi}(k_0) = \frac{2\pi}{3} \left(\frac{32}{3}\right)^2
\left(\frac{m_Q}{\varepsilon_B}\right)^{1/2} \frac{1}{m_Q^2}
\frac{(k_0/\varepsilon_B-1)^{3/2}}{(k_0/\varepsilon_B)^5} \ .
\label{gdiss}
\end{equation}
This expression has a rather pronounced maximum structure with the peak
cross section reached for a gluon energy, $k_0^{\rm max}=\frac{7}{5}
\varepsilon_B$. The applicability of the gluo-dissociation formula
should be reasonable for the free bottomonium ground state
($\varepsilon_\Upsilon\simeq 1 \; \GeV$), but borderline for $J/\psi$
($\varepsilon_{J/\psi} \simeq 0.6 \; \GeV$). Taken at face value for QGP
temperatures of $T=300$-$400 \;\MeV$, where the typical thermal energy
of (massless) gluons is around $k_0= 1\; \GeV$, the convolution of the
gluo-dissociation cross section with a thermal gluon distribution
function in Eq.~(\ref{gamma}) results in an inelastic $J/\psi$ width
(lifetime) of $\Gamma_{J/\psi} \simeq 150$-$400\; \MeV$
($\tau_{J/\psi}\simeq 0.5$-$1.3 \,\fm/c$), see the dashed line in the
right panel of Fig.~\ref{fig_gamma}.

The situation changes if the quarkonium binding energy decreases due to
color-screening as discussed in the previous section (or for excited
charmonia which are weakly bound even in vacuum). In this case, the peak
of the gluo-dissociation cross section moves to smaller energies and
becomes rather narrow; the loss of phase space can be basically
understood by the fact that for a loosely bound $\Psi$ state, the
absorption of an on-shell gluon on an (almost) on-shell quark is
kinematically impossible (suppressed). Consequently, with decreasing
binding energy, the cross section has less overlap with the thermal
gluon spectrum~\cite{Grandchamp:2001pf,Rapp:2005rr}, leading to a
decreasing width with temperature (cf.~dotted line in the left panel of
Fig.~\ref{fig_gamma}). This unphysical behavior signals the presence of
other inelastic processes taking over.
In Ref.\cite{Grandchamp:2001pf} the so-called ``quasifree'' dissociation
mechanism has been suggested, $J/\psi+p\to c+\bar c +p$, where a thermal
parton ($p=g,q,\bar q$) scatters ``quasi-elastically'' off an individual
heavy quark in the bound state (see right panel of
Fig.~\ref{fig_psi-dia}). The $p$-$Q$ scattering amplitude has been
evaluated in leading-order (LO) perturbation
theory~\cite{Combridge:1978kx}, including thermal parton and Debye
masses and slightly modified kinematics due to the small but finite
binding energy. While naively of next-to-leading order (NLO) in
$\alpha_s$ compared to gluo-dissociation, the additional outgoing parton
opens a large phase space rendering the quasifree process significantly
more efficient for weakly bound states. Therefore, it readily applies to
excited states as well, thus enabling a treatment of all charmonia on an
equal footing (which is essential even for $J/\psi$ observables, since
the latter receive significant feed-down contributions from $\chi_c$ and
$\psi'$ states, see Sec.~\ref{sssec_trans} below). For a coupling
constant of $\alpha_s\simeq 0.25$, the quasifree dissociation rate
reaches $\Gamma_{J/\psi}=100$-$200\; \MeV$ for temperatures,
$T=300$-$400 \; \MeV$ (cf.~solid line in the left panel of
Fig.~\ref{fig_gamma}).  These values could be substantially enhanced if
non-perturbative $p$-$Q$ scattering mechanisms are operative, much like
the ones discussed in Sec.~\ref{ssec_non-pert}. A complete NLO
calculation for parton-induced charmonium destruction has recently been
carried out in Ref.\cite{Park:2007zza}, including the effects of
in-medium reduced binding energies. The right panels of
Fig.~\ref{fig_gamma} show the results for quark- and gluon-induced
breakup of the $J/\psi$ for $\alpha_s=0.5$ and different (fixed) thermal
parton masses. For temperatures around $250 \; \MeV$, the sum of both
contributions ($\Gamma_{J/\psi}\simeq 350\; \MeV$) is about a factor
$\sim$4 larger than the quasifree results ($\Gamma_{J/\psi}\simeq 80 \;
\MeV$) in the left panel, calculated for $\alpha_s=
0.25$\cite{Grandchamp:2001pf}.  Thus, there is good agreement between
these two calculations, since the rate is basically $\propto \alpha_s^2$
(for $T=250$\MeV the $T$-dependent Debye-mass in
Ref.\cite{Grandchamp:2001pf}, $\mu_D =gT$, amounts to $\mu_D\simeq 440
\; \MeV$).

The three-momentum dependence of the dissociation rate of a moving
quarkonium, $\Gamma_\Psi(p)$, has been calculated for full NLO and the
quasifree rates in Refs.\cite{Song:2007gm,Zhao:2007hh}, respectively. In
both calculations a weak increase of the rate with increasing
three-momentum is found. Since the quasifree cross section is
essentially constant, this increase is caused by the increasing flux of
thermal partons encountered by the moving bound state. A similar result
has been obtained in a calculation employing the AdS/CFT
correspondence\cite{Liu:2006nn} (recall Sec.~\ref{ssec_string} for more
details on this framework and its caveats). On the other hand,
gluo-dissociation leads to a rather pronounced decrease of the
dissociation rate with increasing three-momentum, since the pertinent
cross section is peaked at relatively low energies and falls off rapidly
at large center-of-mass energies, $s=(p+k)^2$. Of course, if
gluo-dissociation becomes ineffective, its three-momentum dependence
becomes immaterial. A decreasing $p$-dependence should also be expected
in quasifree dissociation if non-perturbative (resonance-like) $q$-$Q$
interactions are operative (recall Sec.~\ref{ssec_non-pert}), since the
pertinent cross sections are concentrated at small $\sqrt{s}$ as well.

To utilize quarkonium observables as a probe of QGP formation in URHICs,
it is mandatory to have good control over the modifications of quarkonia
in hadronic matter, in particular their inelastic reaction rates. From
current lattice QCD calculations it is very difficult to extract
information on excited states (say, $\psi'$). In addition, the results
in the $P$-wave (scalar and axialvector) channels (corresponding to
$\chi_{c,0}$ and $\chi_{c,1}$, respectively) are sensitive to the
so-called zero-mode contributions which are not directly related to the
bound-state properties (as briefly mentioned in a previous footnote).
Potential models find that $\psi'$, $\chi_{c,0}$ and $\chi_{c,1}$
``melt'' close to or even below $T_c$, suggesting substantial
modifications in the hadronic phase. Even for the $J/\psi$, hadronic
dissociation may lead to significant suppression (in addition to
suppressed feed down from $\psi'$ and $\chi_c$ states). One main obstacle
in a reliable assessment of these reactions is that low-energy reactions
of the type $h + J/\psi \to D + \bar D +X$ constitute a nonperturbative
problem with little experimental information available to constrain
effective models.
                                                                              
An initial estimate of quarkonium dissociation by light hadrons has been
obtained by using the gluo-dissociation cross section,
Eq.~(\ref{gdiss}), convoluted over the gluon distribution inside
hadrons\cite{Kharzeev:1994pz}. Since the latter is rather soft
($k_0\simeq 0.1 \; \GeV$), the gluon energy is in general not sufficient
to break up the $J/\psi$, leading to a (low-energy) cross section of
order $\sigma_{hJ/\psi}^{\rm inel}\simeq$~0.1\,mb. Quark-exchange
reactions\cite{Martins:1994hd}, e.g., in meson-induced breakup, $h +
J/\psi \to D + \bar{D}$ (including excited $D$ mesons in the
final-state), are presumably more relevant. Effective quark models
predict dissociation cross sections of order 1-2\,mb, see, e.g.,
Refs.\cite{Barnes:2003dg,Ivanov:2003ge}.  An alternative approach is to
construct effective hadronic models, pioneered in
Ref.\cite{Matinyan:1998cb}. Guiding principles are basic symmetries
including gauge invariance for vector mesons ($J/\psi$, $\rho$) as well
as flavor symmetries, most notably $\mathrm{SU}(4)$ (albeit explicitly
broken by the charm-quark mass) and chiral symmetry which is operative
in interactions with (pseudo-) Goldstone bosons ($\pi$ and
$K$)\cite{Lin:2002mp,Haglin:2000ar,Oh:2002vg,Carvalho:2005et,Bourque:2008ta,Blaschke:2008mu}.
The main uncertainty in these models remains a reliable determination of
the cutoff scales figuring into the hadronic vertex form factors. With
cutoff values of around $1 \; \GeV$, the agreement with quark models is
quite reasonable; dissociation reactions induced by $\rho$ mesons appear
to be the most important channel. Their thermal density in hadronic
matter (i.e., for temperatures of $\sim$180\,MeV) is not very large, so
that the total $J/\psi$ width does not exceed a few MeV, and therefore
is substantially smaller than in the QGP. E.g., for a total hadron
density of $\varrho_h=3\varrho_0$ and a thermally averaged cross section
of $\langle \sigma_{hJ/\psi}^{\rm diss}\rangle = 1\; \mathrm{mb}$
(corresponding to significantly larger peak cross sections), a rough
estimate for the dissociation rate gives $\Gamma_{hJ/\psi}^{\rm diss} =
\langle \sigma_{hJ/\psi}^{\rm diss}\,v_{\rm rel}\rangle \, \varrho_h
\simeq 5\; \MeV$. An interesting possibility to constrain effective
hadronic vertices in a rather model-independent way is to use QCD sum
rules\cite{Duraes:2002ux,Carvalho:2005et}. Pertinent estimates yield,
e.g., a thermally averaged $\pi J/\psi$ dissociation cross section of
$\langle \sigma_{\pi J/\psi}^{\rm diss} \ v_{\rm rel} \rangle = 0.3\;
\mathrm{mb}$ at $T=150\;\MeV$, in the same range as the above
estimate. We finally remark that in-medium effects, e.g., modified
spectral distributions of $D$-mesons, can increase the final-state phase
space and lead to an appreciable increase of the dissociation
rate\cite{Blaschke:2002ww}. This is especially pertinent to excited
charmonia like the $\psi'$, whose mass is close to the free $D\bar D$
threshold, so that a slight reduction (or broadening) in the $D$-meson
mass can open the direct decay channel, $\psi'\to D\bar
D$\cite{Friman:2002fs}.
                                      
\subsection{Quarkonium Production in Heavy-Ion Collisions}
\label{ssec_prod}
Similar to the open heavy-flavor sector, a key objective (and challenge)
in the quarkonium sector is to connect their equilibrium properties to
observables in heavy-ion collisions, and eventually deduce more general
insights about basic properties of QCD matter, e.g., color
de-/confinement and Debye screening. Since quarkonium states in
heavy-ion collisions are even more rare than individual heavy quarks, it
is suitable to adopt a transport treatment for their distribution
functions in a realistic ``background medium'' whose evolution is not
affected by the heavy quarks or quarkonia.  The connection between
observables extracted from the distribution function (after its
``transport'' through the medium) and the equilibrium properties
discussed in the previous section is given by the coefficients and
equilibrium limit of the transport equation, as elaborated in the
following section, \ref{sssec_trans}. The current status in comparing
various model implementations to charmonium data at SPS and RHIC will be
assessed in Sec.~\ref{sssec_phen}.

\subsubsection{Quarkonium Transport in Heavy-Ion Collisions}
\label{sssec_trans}
The (classical) Boltzmann equation describing the time-evolution of the
phase-space distribution function, $f_\Psi(\bvec r,\tau;\bvec p)$, of an
(on-shell) quarkonium state, $\Psi$ (with energy $p_0=\omega_p=(\bvec
p^2 +m_\Psi^2)^{1/2}$), may be written as
\begin{equation}
p^\mu \partial_\mu f_\Psi(\bvec r,\tau;\bvec p) =
- \omega_p \ \Gamma_\Psi(\bvec r,\tau;\bvec p) \ f_\Psi(\bvec r,\tau;\bvec p) +
\omega_p \ \beta_\Psi(\bvec r,\tau;\bvec p)  \ 
\label{transport}
\end{equation}
(a mean-field term has been neglected assuming that the real part of the
$\Psi$ selfenergy is small). When focusing on inelastic reactions,
$\Gamma_\Psi(\bvec r,\tau;\bvec p)$ represents the dissociation rate
discussed in Sec.~\ref{sssec_diss} above, which governs the loss term,
i.e., the first term on the right-hand-side (\emph{rhs}) of
Eq.~(\ref{transport}). The $(\bvec r,\tau)$ dependence of $\Gamma_\Psi$
typically converts into a temperature dependence via the fireball
evolution of a heavy-ion reaction for given projectile/target ($A/B$),
collision energy ($\sqrt{s}$) and impact parameter ($b$).  The second
term on the \emph{rhs} of Eq.~(\ref{transport}) is the gain term
accounting for the formation of quarkonia. For a $2 \to 2$ process (as,
e.g., realized via the inverse of gluo-dissociation, $Q +\bar Q \to g
+\Psi$), it takes the form\cite{Yan:2006ve}
\begin{eqnarray}
\beta_\Psi(\bvec p;\bvec r,\tau) &=& \frac{1}{2p_0} \int
\frac{\dd^3k}{(2\pi)^3 2\omega_k} \frac{\dd^3p_Q}{(2\pi)^3 2\omega_{p_Q}}
\frac{\dd^3p_{\bar Q}}{(2\pi)^3 2\omega_{p_{\bar Q}}}
f^Q(\bvec p_Q;\bvec r,\tau) \, f^{\bar Q}(\bvec p_{\bar Q};\bvec r,\tau)
\nonumber \\
&& \times \, W_{Q\bar Q}^{g\Psi}(s) \ 
\Theta[T_{\rm diss}- T(\bvec r,\tau)] \ 
(2\pi)^4 \, \delta^{(4)}(p+k-p_Q-p_{\bar Q}) \  . 
\label{gain}
\end{eqnarray}
To ensure detailed balance, the cross section figuring into the
formation probability, $W_{Q\bar Q}^{g\Psi}(s)=\sigma_{Q\bar Q\to
  g\Psi}^{\rm form} \, v_{\rm rel} \, 4\,\omega_{p_Q}\omega_{p_{\bar
    Q}}$, has to be the same (up to a kinematic and statistical factor)
as the one used in the dissociation rate, Eq.~(\ref{gamma}). For
reactions beyond $2 \leftrightarrow 2$ (such as the quasifree process,
$p+Q +\bar Q \to p +\Psi$), the microscopic evaluation of the gain term
becomes more involved. Note the explicit dependence on the HQ
phase-space distribution functions, $f^{Q,\bar Q}$, in Eq.~(\ref{gain}),
whose modifications in heavy-ion reactions are the central theme in
Sec.~\ref{sec_hq-obs} of this review. The temperature-dependent step
function in Eq.~(\ref{gain}) signifies the limit set by the dissociation
temperature, $T_{\rm diss}$, above which a well-defined $\Psi$ state no
longer exists and thus formation reactions are not meaningful.

It is both instructive and useful for practical applications to simplify
the gain term by integrating out its spatial and three-momentum
dependence. This is possible under the assumption of a homogeneous
medium and thermally equilibrated HQ distribution functions; one
obtains\cite{Grandchamp:2003uw}
\begin{equation}
  \frac{\dd N_\Psi}{\dd \tau} = -\Gamma_\Psi ( N_\Psi - N_\Psi^{\rm eq}) \ ,
\label{rate-eq}
\end{equation}
which now clearly exhibits detailed balance in terms of the approach to
the equilibrium limit, $N_\Psi^{\rm eq}$, of the state $\Psi$. The
latter quantity is given by
\begin{equation}
  N_\Psi^{\rm eq} = V_{\mathrm{FB}} \ n_\Psi^{\rm eq}(m_\Psi; T,\gamma_Q) 
  = d_\Psi \gamma_Q^2 \int \frac{\dd^3p}{(2\pi)^3} f_\Psi(\omega_p;T) \ ,
\label{npsi}
\end{equation}
carrying the explicit dependence on the (in-medium) quarkonium mass,
$m_\Psi$ (which, in turn, depends on a combination of $Q$-$\bar Q$
binding energy and in-medium HQ mass, cf.~Eq.~(\ref{mpsi})); $d_\Psi$
denotes the spin degeneracy of the $\Psi$ state and $V_{\mathrm{FB}}$
the (time-dependent) fireball volume. The appearance of a HQ fugacity,
$\gamma_Q=\gamma_{\bar Q}$, owes its origin to the (theoretically and
experimentally supported) postulate that $Q\bar Q$ production is
restricted to hard $N$-$N$ collisions upon initial nuclear impact. The
number $N_{Q\bar Q}=N_Q = N_{\bar Q}$ of heavy anti-/quarks is then
conserved in the subsequent fireball evolution (separately for charm and
bottom), which is achieved by introducing $\gamma_{Q}$ at given
fireball volume and temperature into the thermal densities of open and
hidden HQ states, i.e.,
\begin{equation}
  N_{Q\bar Q}=\frac{1}{2} N_{\mathrm{op}}\frac{I_1(N_{\mathrm{op}})}{I_0(N_{\mathrm{op}})}+
  V_{\mathrm{FB}} \ \gamma_Q^2\sum\limits_{\Psi} n_\Psi^{\rm eq}(T)
  \ ,
\label{NQQ}
\end{equation}
for either charm ($Q$=$c$) or bottom ($Q$=$b$). The thermal open-charm
(and -bottom) number, $N_{\mathrm{op}}$, depends on whether one is
evaluating it in terms of individual quark states,
$N_{\mathrm{op}}=V_{\mathrm{FB}}\gamma_Q 2 n_Q^{\rm eq}(m_Q^*,T)$
(appropriate for a (weakly interacting) QGP), or in terms of hadronic
states, $N_{\mathrm{op}}=V_{\mathrm{FB}}\gamma_Q \sum_\alpha
n_\alpha^{\rm eq}(T,\mu_B)$. This is, in principle, a nontrivial issue,
since both hadronic and partonic evaluations of $N_{\mathrm{op}}$ can be
subject to corrections, see, e.g.,
Refs.\cite{Grandchamp:2003uw,Grandchamp:2005yw,Andronic:2007zu}. In the
hadronic phase one expects the spectral functions of $D$-mesons,
$\Lambda_c$ baryons, etc. to undergo significant medium effects, e.g.,
reduced masses and/or increased widths. In the partonic phase,
especially close to $T_c$, it is not inconceivable that hadronic bound
states are still present and thus an approximation with weakly
interacting quasiquarks may not be an accurate one. Even within a
quasiquark description, significant uncertainty is associated with the
value of the HQ mass adopted in the calculation of $N_{\mathrm{op}}$ and
thus in the quantitative determination of $\gamma_Q$. The general trend
is that for a given temperature, volume and $N_{Q\bar Q}$, a smaller
value for $m_Q^*$ results in a larger value for $n_Q^{\rm eq}$ and thus
in a smaller value for $\gamma_Q$, which, in turn, reduces $N_\Psi^{\rm
  eq}$ quadratically. The underlying physics is that of relative
chemical equilibrium: for a given number of heavy anti-/quarks, the
latter preferentially occupy the states of the lowest energy. In the
simplest case, where a quasiquark description applies and the $\Psi$
mass is given by the expression (\ref{mpsi}), the $\Psi$ number is
essentially determined by its binding energy (larger $\varepsilon_B$
implying larger $N_\Psi^{\rm eq}$). The gain term as written in
Eq.~(\ref{gain}) is, strictly speaking, only applicable in the
quasiquark approximation. If additional resonances are present in the
medium (e.g., $D$-meson resonances or $cq$ diquark states), additional
reaction channels would have to be included in a coupled rate-equation
framework to account for the competition of these resonances to harbor
$c$ quarks. In the simplified treatment given by Eq.~(\ref{rate-eq}),
this competition is included via the $c$-quark fugacity figuring into
$N_\Psi^{\rm eq}$.

A slightly different view on regeneration and suppression processes in
the QGP is advocated in Ref.\cite{Young:2008he}, based on the strongly
coupled nature of the QGP (sQGP) as produced at SPS and RHIC (i.e., at
not too high temperatures). It is argued that a small charm-quark
diffusion constant (cf.~Secs.~\ref{sec_hq-int} and \ref{sec_hq-obs})
inhibits the separation of the produced $c$ and $\bar c$ pair in the
sQGP. In connection with the survival of $J/\psi$ bound states well
above $T_c$ (as, e.g., in the right panel of Fig.~\ref{fig_sf}), this
enhances the probability for a produced $c\bar c$ pair to bind into a
charmonium state (relative to $p$-$p$ collisions). In particular, this
approach accounts for the possibility that the pairwise produced $c$ and
$\bar c$ quarks do not explore the entire fireball volume as usually
assumed in equilibrium models. Such an effect has also been implemented
in a more simplified manner in the thermal-rate equation approach of
Refs.\cite{Grandchamp:2003uw,Zhao:2007hh} in terms of a time dependent
correlation volume.

Let us briefly discuss the initial conditions for the quarkonium
distribution functions. Starting point are measured quarkonium spectra
in $p$-$p$ collisions. In a heavy-ion collision, these are subject to
modifications before the medium can be approximated with a thermal
evolution. ``Pre-equilibrium'' effects may be distinguished according to
whether they occur before or after the hard $Q\bar Q$-production process
takes place. The former include nuclear modifications of the parton
distribution functions generically denoted as ``shadowing'', as well as
$p_t$ broadening (Cronin effect) attributed to a scattering of the
projectile/target partons on their way through the target/projectile
nucleus prior to the fusion reaction into $Q$-$\bar Q$. In a random-walk
picture, the accumulated transverse momentum is approximated by $\Delta
p_t^2 = a_{gN} \langle l \rangle$, where $\langle l \rangle$ is an
average nuclear path length of both gluons before the hard scattering,
and $a_{gN}$ parameterizes the transverse-momentum kick per path length
in gluon-nucleon scattering. Both ``pre-fusion'' effects are in
principle universal, i.e., not directly linked to the $Q$-$\bar
Q$-production process. In $p$-$p$ collisions, a fraction of $1$-$2 \%$
of $c\bar c$ pairs ($\sim$0.1\% of $b\bar b$ pairs) develop a
correlation that leads to the formation of a charmonium (bottomonium)
state\cite{YELL03}. In nuclear collisions inelastic collisions of the
produced $Q\bar Q$ pair with passing-by nucleons can destroy this
correlation. This so-called nuclear absorption may be parameterized by
an effective absorption-cross section, $\sigma_{\rm abs}^{\Psi
  N}$.\footnote{Nuclear absorption typically occurs at a rather large
  $\Psi$-$N$ center-of-mass energy (comparable to the $\sqrt{s}$ of
  primordial $N$-$N$ collisions) and is therefore in a very different
  energy regime than the low-energy hadronic absorption cross section
  relevant for the later hadron-gas stage of the fireball evolution.}
As is well-known, the finite (and different) formation times of
quarkonia imply that the $Q\bar Q$ pair interacting with a nucleon does,
in general, not represent a fully formed quarkonium, but rather a
pre-resonance state. A microscopic description of nuclear absorption is
therefore a rather challenging task\cite{Cugnon:1993yf,He:1999aj}. At a
minimal level, finite formation times imply that the values for
effective nuclear absorption cross sections should be expected to depend
on collision energy ($\sqrt{s}$), rapidity ($y$) and bound-state quantum
numbers (since different binding energies imply different formation
times). A careful measurement and systematic interpretation of
quarkonium suppression in $p$-$A$ collisions, where the formation of a
thermal medium is not expected (at least at SPS and RHIC), is therefore
an inevitable prerequisite for quantitative interpretations of heavy-ion
data, see, e.g.,
Refs.\cite{Ferreiro:2008wc,Kharzeev:2008cv,Lourenco:2009sk} for recent
work.  Pre-equilibrium effects not only modify the momentum dependence
of the quarkonium distribution functions but also their spatial
dependence.

\subsubsection{Quarkonium Phenomenology in Heavy-Ion Collisions}
\label{sssec_phen}
As discussed in the Introduction, there is ample evidence for both
chemical and thermal equilibration in the low-$p_t$ regime of (bulk)
particle production in ultrarelativistic heavy-ion collisions. A well
defined set of thermodynamic variables characterizing the temperature
evolution and flow fields of the fireball greatly facilitates the
comparison of independent calculations for quarkonium production and
maintains direct contact to their in-medium properties in equilibrated
QCD matter. In this section we therefore focus on rate-equation
approaches implemented into thermal background media.

We recall that the experimental quarkonium yields usually include
feed down contributions due to decays of higher resonances. E.g., for
$J/\psi$ production in $p$-$p$ collisions about $30\%$ ($10\%$) of the
observed number arises from decays of $\chi_c$ ($\psi'$)
states~\cite{Abt:2002vq,Abt:2006va}. The standard assumption in
heavy-ion collisions is that primordial production fractions of excited
states scale as in $p$-$p$ collisions, but subsequent suppression
(and/or regeneration) will change these ratios (due to different
inelastic cross sections at all stages). This needs to be taken into
account for realistic comparisons to heavy-ion data (unless otherwise
stated, it is included in the theoretical models discussed below). After
thermal freezeout, the decay branchings are assumed to be as in vacuum
(since the $J/\psi$ lifetime after freezeout (ca.~$2000\;\mathrm{fm}/c$)
is about a factor of $\sim$200 larger than the fireball lifetime,
in-medium dilepton decays contribute a small fraction to the spectrum
observed in the detectors).

\begin{figure}[!t]
\begin{minipage}{0.48\linewidth}
\includegraphics[width=0.94\textwidth,height=0.77\textwidth]{RAA-j-sps-zhao.eps}
\vspace{-0.6cm}
\end{minipage}
\begin{minipage}{0.48\linewidth}
\hspace{0.1cm}
\includegraphics[width=0.96\textwidth]{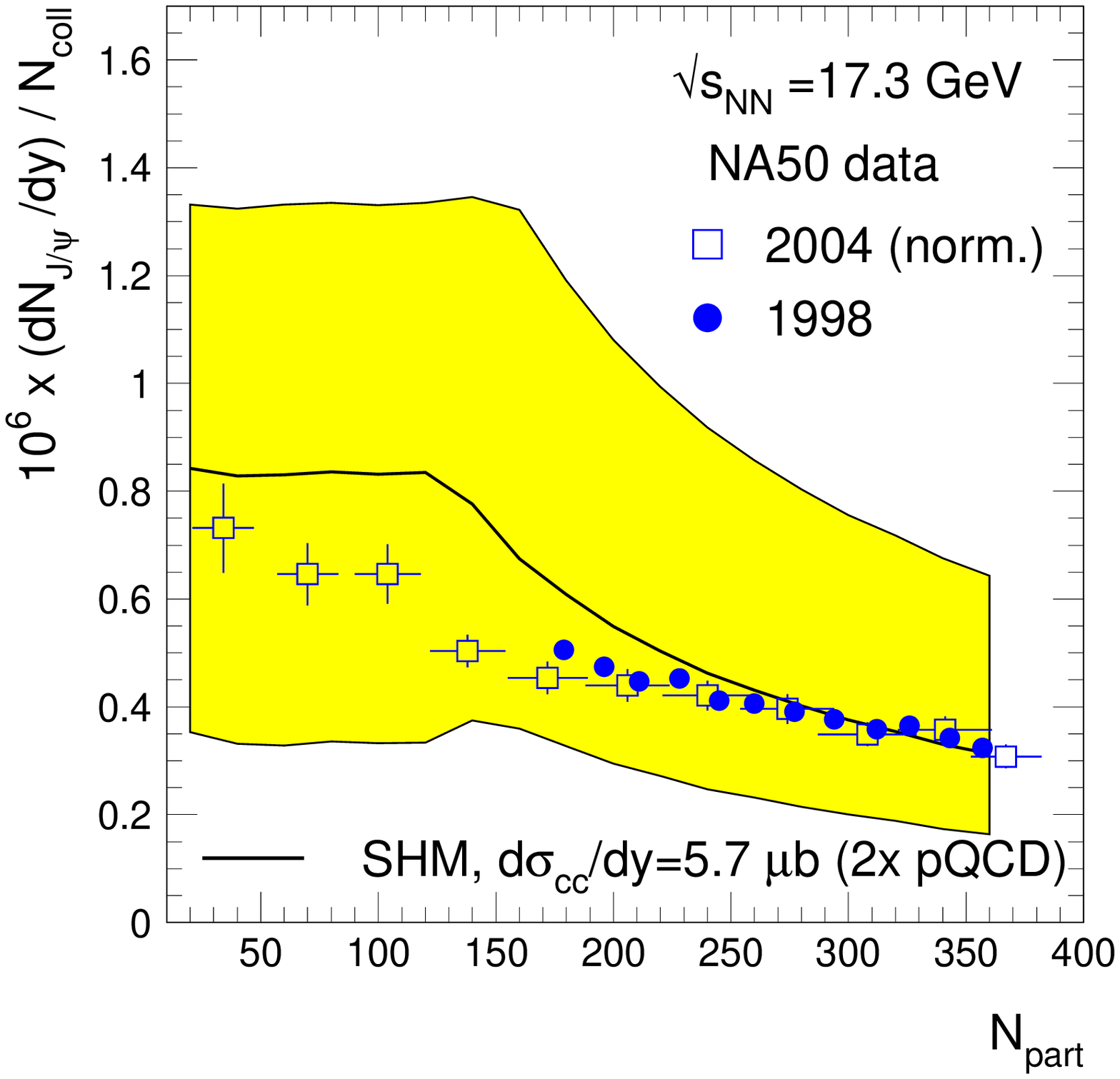}
\end{minipage}
\begin{minipage}{0.48\linewidth}
\includegraphics[width=0.94\textwidth,height=0.79\textwidth]{pt2-j-sps-zhao.eps}
\end{minipage}
\hspace{0.35cm}
\begin{minipage}{0.48\linewidth}
\includegraphics[width=0.9\textwidth]{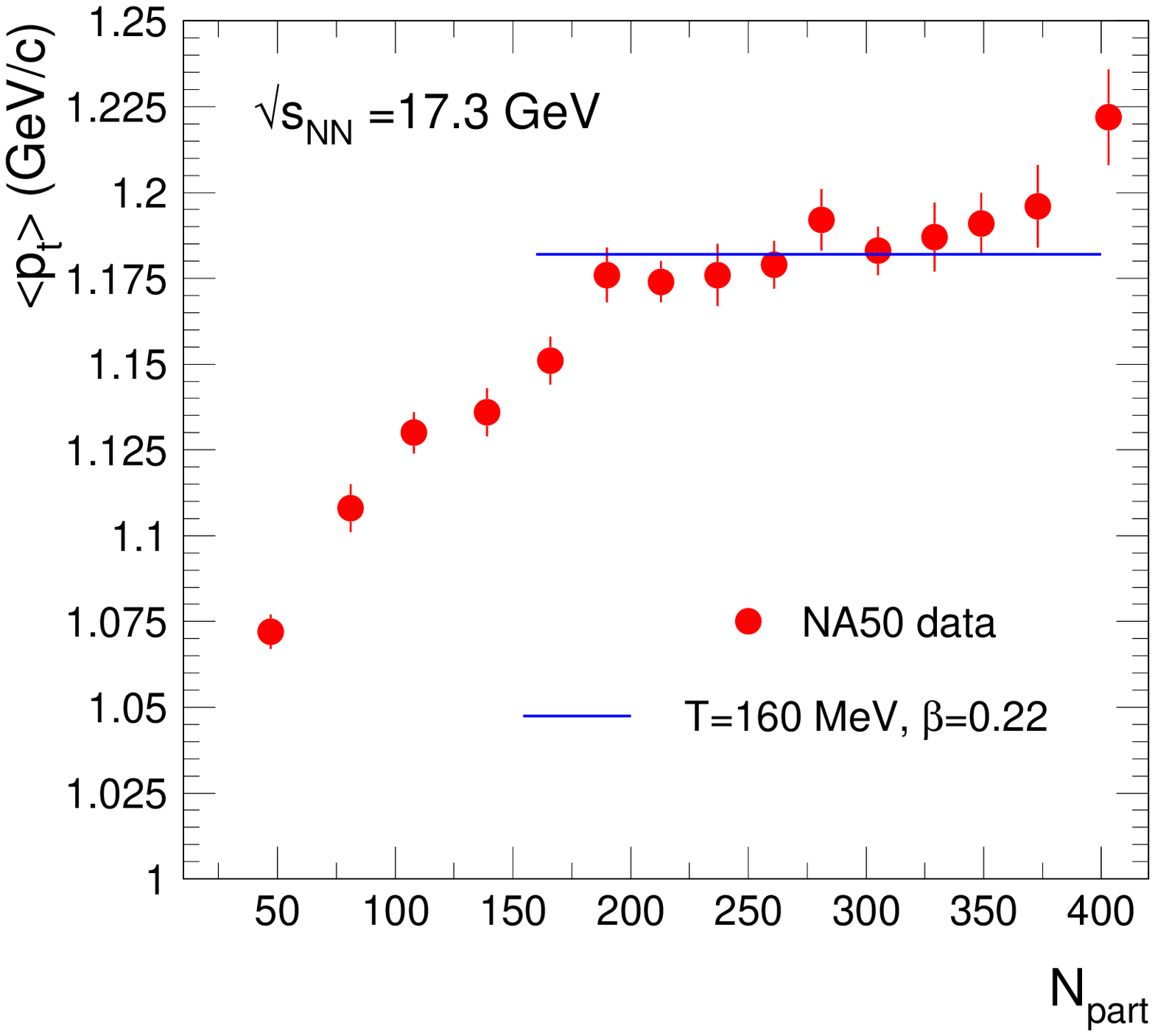}
\end{minipage}
\caption{(Color online) $J/\psi$ production at SPS in the thermal
  rate-equation approach\protect\cite{Grandchamp:2003uw,Zhao:2007hh}
  (left panels) and the statistical hadronization
  model\protect\cite{Andronic:2006ky} (right panels). Upper panels
  display the centrality dependence of the inclusive $J/\psi$ yield
  (normalized to Drell-Yan production or the number of binary $N$-$N$
  collisions), lower panels the average squared $J/\psi$ transverse
  momentum.}
\label{fig_jpsi-na50}
\end{figure}

Let us start by analyzing $J/\psi$ production in Pb-Pb($\sqrt{s} =17.3
\, A\GeV$) collisions at SPS in the context of NA50
data\cite{Abreu:1997ji,Abreu:2000xe,Ramello:2003ig,Alessandro:2004ap},
cf.~Fig.~\ref{fig_jpsi-na50}.
The left panels display the outcome of thermal-rate equation
calculations\cite{Grandchamp:2003uw,Zhao:2007hh}, where quasifree
dissociation rates in the QGP (cf.~left panel of Fig.~\ref{fig_gamma})
and hadronic $\mathrm{SU}(4)$ cross sections for meson-induced
dissociation in the HG are evolved over an expanding fireball model
(adjusted to empirical information on hadron production and flow
velocities). The prevalent effect is identified as suppression in the
QGP, controlled by an effective strong coupling constant, $\alpha_s
\simeq 0.25$, in the quasifree rate. This value is adjusted to reproduce
the suppression level in central collisions (where the average initial
temperature amounts to about $T_0 \simeq 210\;\MeV$). Regeneration is a
rather small effect, based on a $p$-$p$ open-charm cross section of
$\sigma_{\bar cc} =5.5$\,$\mu$b distributed over two fireballs (the
covered rapidity window amounts to $\Delta y = 3.6$, resulting in a
rapidity density of $\dd\sigma_{\bar cc}/\dd y \simeq
1.53\,\mu\mathrm{b}$. A recent compilation\cite{Lourenco:2006vw} of
charm production at fixed-target energies finds a total cross section of
$\sigma_{\bar cc}\simeq 3.6$-$5.2\,\mu\mathrm{b}$; with an experimental
rapidity width of around $\Delta y = 2$\cite{BraunMunzinger:1997xy}, the
resulting rapidity density is approximately $\dd\sigma_{\bar cc}/\dd
y=(2.2\pm0.5)\,\mu\mathrm{b}$).  In addition, a correction for
incomplete charm-quark thermalization has been
implemented\cite{Grandchamp:2003uw} via a kinetic relaxation time
($\tau_c^{\rm eq}$) reducing the equilibrium $J/\psi$ number. The lower
left panel of Fig.~\ref{fig_jpsi-na50} suggests that the centrality
dependence of the average $J/\psi$'s transverse-momentum squared,
$\langle p_t^2\rangle$, is largely governed by the Cronin effect as
extracted from experimental $p$-$A$ data\cite{Abreu:2000xe}. The
quasifree charmonium dissociation rates, which increase with
three-momentum\cite{Zhao:2007hh}, lead to a slight suppression of
$\langle p_t^2\rangle$ for central collisions.

The right panels in Fig.~\ref{fig_jpsi-na50} are calculated within the
statistical hadronization model\cite{Andronic:2006ky}, assuming that all
primordial charmonia are suppressed and production entirely occurs at
the critical temperature for hadronization based on relative chemical
equilibrium of open- and hidden-charm hadrons (with $N_{c\bar c}$ fixed
as in Eq.~(\ref{NQQ})). This also implies that the charm-quark momentum
distributions are kinetically equilibrated. With a rapidity density for
the $p$-$p$ charm cross section of $\dd\sigma_{\bar cc}/\dd
y=5.7\,\mu\mathrm{b}$ the NA50 data can be reproduced reasonably
well. This input $c\bar c$ number is larger by a factor of $\sim$4
compared to the input in the left panels, which accounts for most of the
difference to the regeneration yield in the rate-equation calculation
for central collisions (remaining discrepancies are largely due to the
$c$-quark relaxation correction which becomes more pronounced toward
more peripheral collisions)\footnote{The charm ensemble at SPS is in the
  canonical limit, $N_{\mathrm{op}}\ll 1$, for which
  $I_1(N_{\mathrm{op}})/I_0(N_{\mathrm{op}})\simeq 0.5 N_{\mathrm{op}}$
  in Eq.~(\ref{NQQ}), and thus $N_\psi\propto N_{c\bar c}$.}. The
interpretation of the $J/\psi$'s average $p_t$ is also rather different,
in that it entirely stems from a thermal source (with moderate
collective flow) in the vicinity of $T_c$ (the resulting $\langle
p_t^2\rangle$ is quite consistent with the regeneration component in the
lower left panel of Fig.~\ref{fig_jpsi-na50}).

\begin{figure}[!t]
\begin{center}
\begin{minipage}{0.48\linewidth}
\vspace{0.5cm}
\includegraphics[width=0.95\textwidth]{psip-sps-rate-eq.eps}
\end{minipage}
\begin{minipage}{0.48\linewidth}
\includegraphics[width=1.03\textwidth,height=0.85\textwidth]{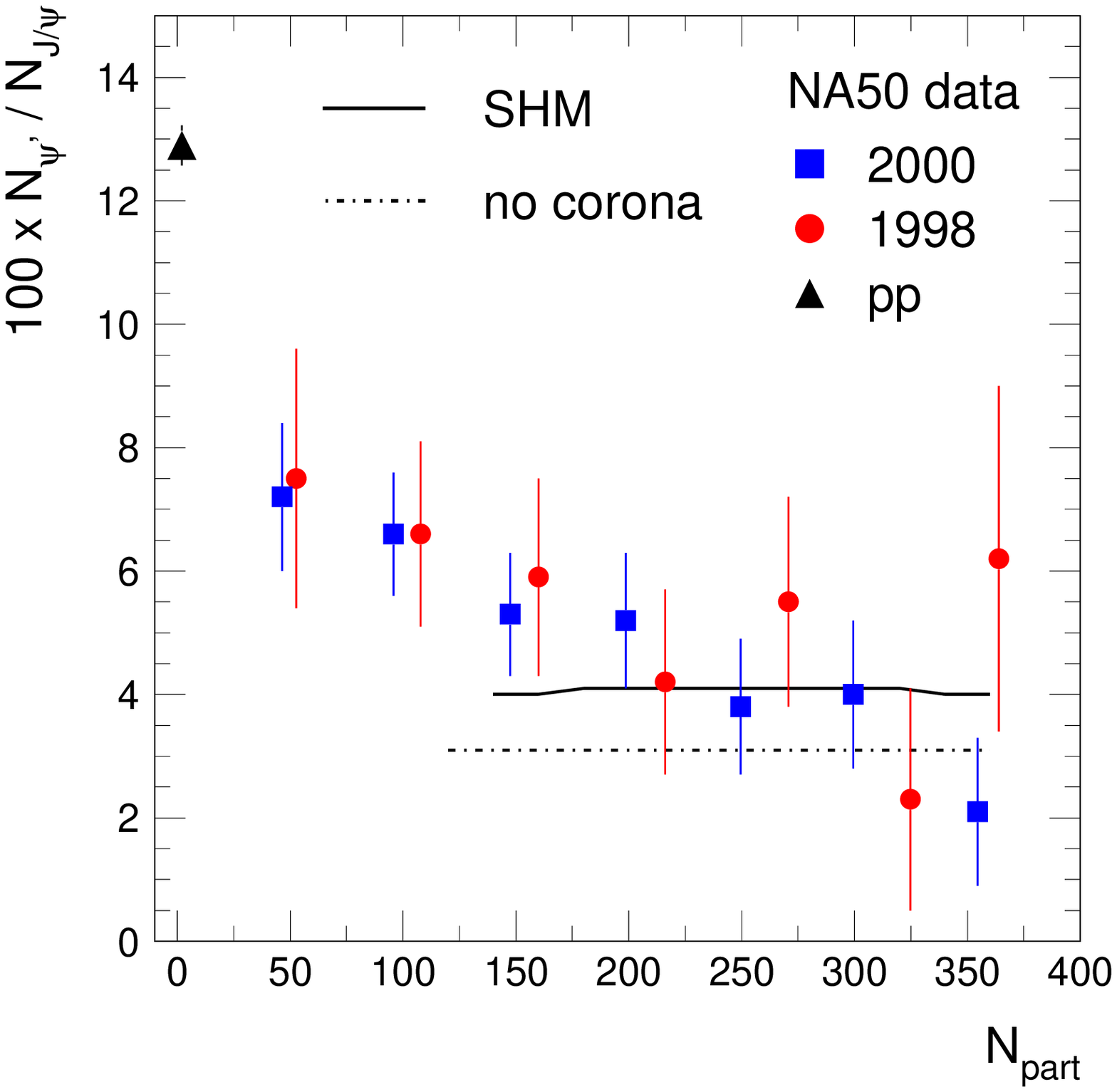}
\end{minipage}
\end{center}
\vspace{-0.3cm}
\caption{(Color online) NA50 data for the $\psi'/(J/\psi)$ ratio in
  Pb-Pb($\sqrt{s}=17.3\;A \GeV$)
  collisions\protect\cite{Abreu:1998vw,Alessandro:2006ju} compared to
  calculations within the thermal rate-equation
  approach\protect\cite{Grandchamp:2003uw,Grandchamp:2004tn} (left
  panel) and the statistical hadronization
  model\protect\cite{Andronic:2006ky} (right panel). In the left panel,
  the dashed (solid) curve is obtained without (with) the inclusion of
  hadronic medium effects (modeled by reduced $D$-meson masses); both
  calculations utilize a $\psi'$ nuclear absorption-cross section of
  $\sigma_{\rm nuc}^{\psi'}=\sigma_{\rm nuc}^{J/\psi} =4.4$\,mb, while
  in the third calculation (dash-dotted line) the value has been updated
  to 7.9\,mb (with the associated band indicating a $\pm$0.6\,mb
  uncertainty). In the right panel, the statistical model yield involves
  either the full hadronization volume (dash-dotted line) or excludes a
  dilute nuclear surface (``corona''), where neither suppression nor
  regeneration is operative.}
\label{fig_psip-na50}
\end{figure}
NA50 has also measured $\psi'$
production\cite{Abreu:1998vw,Alessandro:2006ju,Alessandro:2006jt}.
Using $p$-$A$ collisions, the extracted nuclear absorption cross section
has been updated\cite{Alessandro:2006jt} to $\sigma_{\rm nuc}^{\psi'}
=(7.7\pm0.9)\,\mathrm{mb}$, which is significantly larger than for
$J/\psi$, $\sigma_{\rm nuc}^{J/\psi}=(4.2\pm0.5)\,\mathrm{mb}$. In Pb-Pb
collisions ($\sqrt{s}=17.3\; A\GeV$), the ratio $\psi'/(J/\psi)$ is
suppressed substantially already in rather peripheral collisions,
cf.~Fig.~\ref{fig_psip-na50}\cite{Abreu:1998vw,Alessandro:2006ju}.
Within the thermal rate-equation approach\cite{Grandchamp:2003uw}, this
behavior cannot be explained by inelastic reactions in the QGP alone,
since very little (if any) QGP is formed in peripheral Pb-Pb collisions
at SPS. However, hadronic dissociation of the $\psi'$ can account for
the suppression pattern, but only if in-medium effects are included
(cf. left panel of
Fig.~\ref{fig_psip-na50})\cite{Grandchamp:2003uw,Grandchamp:2004tn},
specifically a reduction of the $D\bar D$ threshold which accelerates
$\psi'$ suppression due to the opening of the direct decay mode, $\psi'
\to D\bar D$ (a similar effect can result from a broadening of the
$D$-meson spectral functions as discussed in Sec.~\ref{ssec_hadronic}).
The updated (larger) $\psi'$ nuclear absorption cross
section\cite{Alessandro:2006jt} also plays a significant role in the
quantitative description of the low-centrality data. The statistical
hadronization model predicts a flat $\psi'/(J/\psi)$ ratio, basically
given by the ratio of thermal densities at the hadronization
temperature. The shape and magnitude of the calculated ratio is rather
consistent with the NA50 data for central and semicentral collisions
where hadronization from a QGP can be expected to be applicable;
deviations occur for more peripheral centralities. Thus, the
$\psi'/(J/\psi)$ ratio does not provide a clear discrimination of
regeneration- and suppression-dominated scenarios at SPS.

\begin{figure}[!t]
\begin{center}
\begin{minipage}{0.5\linewidth}
\includegraphics[width=.9\textwidth]{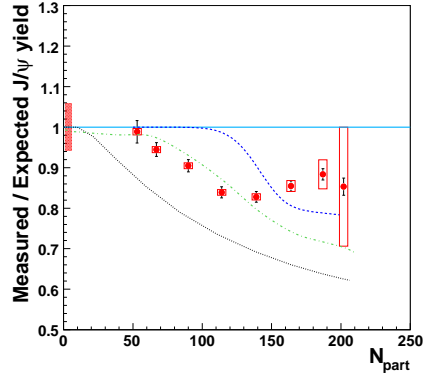}
\end{minipage}
\end{center}
\caption{(Color online) NA60 data for the centrality dependence of
  $J/\psi$ production in In-In ($\sqrt{s} =17.3\; A\GeV$)
  collisions\protect\cite{Arnaldi:2007zz}, compared to theoretical
  predictions based on (from top to bottom): (a) the threshold-melting
  scenario\protect\cite{Digal:2003sg} (dashed line), (b) the thermal
  rate-equation approach\protect\cite{Rapp:2005rr} (dash-dotted line)
  and (c) the comover suppression approach\protect\cite{Capella:2005cn}
  (dotted line). Nuclear absorption effects have been divided out of the
  data and calculations based on measured suppression in $p$-$A$
  collisions.}
\label{fig_jpsi-na60}
\end{figure}
One of the controversies in the interpretation of the NA50 data has been
whether they feature any ``sharp'' drop in their centrality dependence,
e.g., around $E_T\simeq35 \; \GeV$ (or $N_{\rm part}\simeq120)$ in the
upper left (right) panel of Fig.~\ref{fig_jpsi-na50}. Such a drop has
been associated with a threshold behavior for QGP formation resulting in
an abrupt ``melting'' of the $\chi_c$ states due to color
screening\cite{Digal:2003sg} (recall that $\chi_c$ feed down presumably
makes up $\sim$30\% of the inclusive $J/\psi$ yield). The investigation
of this question was one of the main objectives of the successor
experiment of NA50, NA60, where $J/\psi$ production in a medium size
nuclear system (In-In) has been measured\cite{Arnaldi:2007zz}.
Fig.~\ref{fig_jpsi-na60} compares the NA60 $J/\psi$ data as a function
of centrality to three theoretical
predictions\cite{Digal:2003sg,Rapp:2005rr,Capella:2005cn}, all of which
reproduce the NA50 data reasonably well. The predictions of the thermal
rate-equation approach\cite{Grandchamp:2003uw,Rapp:2005rr} roughly
reproduce the onset and magnitude of the suppression (except for the
most central data points); the threshold-melting
scenario\cite{Digal:2003sg} misplaces the onset of the suppression
(which in the data is below $N_{\rm part}=100$, contrary to the Pb-Pb
system) and the comover calculation\cite{Capella:2005cn} overpredicts
the suppression. The leveling-off (or even increasing trend) of the data
for $N_{\rm part}\ge 150$ is somewhat unexpected and deserves further
study.

\begin{figure}[!t]
\begin{minipage}{0.48\linewidth}
\includegraphics[width=0.94\textwidth]{raa-rhic.eps}
\hspace{-0.2cm}
\includegraphics[width=0.98\textwidth]{pt2rhic.eps}
\end{minipage}
\begin{minipage}{0.5\linewidth}
\vspace{-0.6cm}
\includegraphics[width=0.99\textwidth]{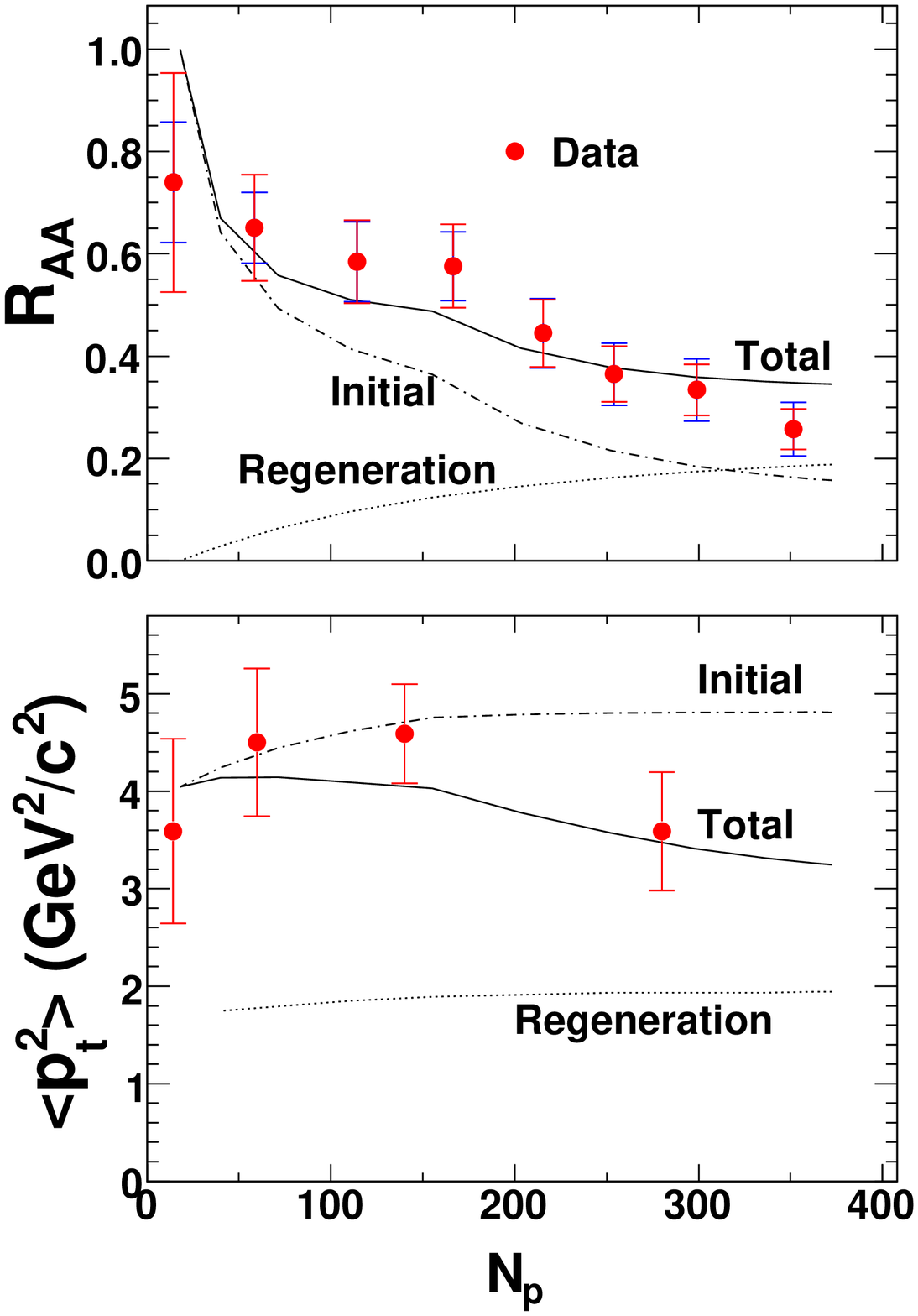}
\end{minipage}
\caption{(Color online) $J/\psi$ production in $200 \; A\GeV$ Au-Au
  collisions at RHIC in the transport approach of
  Ref.\protect\cite{Zhao:2007hh} (left panels, using an expanding
  thermal fireball with reaction rates based on the quasifree
  dissociation process) and Ref.\protect\cite{Liu:2009nb} (right panels,
  using a hydrodynamic evolution with reaction rates based on the
  gluo-dissociation process). The upper (lower) panels display the
  inclusive $J/\psi$ $R_{AA}$ (average $p_t^2$) as a function of
  nucleon-participant number. The data are from the PHENIX
  collaboration\protect\cite{Adare:2006ns}.}
\label{fig_jpsi-rhic}
\end{figure}
The thermal rate-equation framework has been used to predict $J/\psi$
production at RHIC\cite{Grandchamp:2003uw}\footnote{The underlying
  fireball model is the same as used in the open heavy-flavor sector in
  connection with Figs.~\ref{fig_hq-reso}, \ref{fig_hq-tmat}, etc., with
  average initial temperatures of $T_0 \simeq 340$-$370\; \MeV$ for
  semi-/central Au-Au collisions.}. With updates\cite{Zhao:2007hh} for
the experimental input (a smaller nuclear absorption cross
section\cite{Adare:2007gn} and a larger $J/\psi$ number in $p$-$p$
collisions\cite{Adare:2006ns} which figures into the denominator of the
nuclear modification factor and leads to a relative reduction of the
regeneration yield), an approximate agreement with current PHENIX data
on the centrality dependence of inclusive $J/\psi$ production and $p_t$
spectra emerges, see left panels of Fig.~\ref{fig_jpsi-rhic} (the
underlying charm cross section in $p$-$p$, $\sigma_{c\bar c}
=570\,\mu\mathrm{b}$, translates into $\dd\sigma_{\bar cc}/\dd y\simeq
100\,\mu\mathrm{b}$, consistent with PHENIX
measurements\cite{Adare:2006hc,Hornback:2008ur}). The main features of
this interpretation are an about equal share of (suppressed) primordial
and regenerated $J/\psi$'s in central Au-Au collisions (where the
average initial temperature is about $T_0=370 \; \MeV$), as well as a
significant reduction of the average $p_t^2$ due to secondary
production, as compared to primordial production with an estimated
Cronin effect (the latter is not yet accurately determined from $p$-$A$
data).  Consequently, the regeneration component is concentrated in the
low-$p_t$ regime of the spectra. The dependence on the kinetic
relaxation time for $c$ quarks is rather moderate while the inclusion of
the momentum dependence in the quasifree dissociation
rate\cite{Zhao:2007hh} has little effect.

The right panels of Fig.~\ref{fig_jpsi-rhic} show the results of the
rate-equation approach of Refs.\cite{Yan:2006ve,Liu:2009nb}, where
gluo-dissociation rates in the QGP are convoluted over a 2+1-dimensional
hydrodynamic evolution (employing a charm cross section of $\dd
\sigma_{\bar cc}/\dd y=120\; \mu \mathrm{b}$, in line with PHENIX
data\cite{Adare:2006hc}); reactions in the hadronic phase are
neglected. It is very encouraging that the results are in good agreement
with the ones in the left panels\cite{Zhao:2007hh} which are obtained
with similar physics input but in a different realization (e.g.,
fireball vs.  hydro but with comparable initial temperatures ($T_0\simeq
350 \; \MeV$ in central collisions) and charm cross section, etc.). The
centrality dependence of the inclusive $J/\psi$ yield shows a slight
step-structure in the upper right panel\cite{Liu:2009nb}, induced by
different dissociation temperatures for $J/\psi$ ($T_{\rm diss}=320 \;
\MeV$) and $\chi_c$, $\psi'$ ($T_{\rm diss}=T_c=165\;\MeV$) above which
the suppression is assumed to be practically instantaneous (similar
findings have been reported in Ref.\cite{Gunji:2007uy}).  The $\langle
p_t^2\rangle$ of the regenerated component is somewhat smaller in
Ref.\cite{Liu:2009nb} compared to Ref.\cite{Zhao:2007hh}, since in the
former it is computed via continuous regeneration based on
Eq.~(\ref{gain}), while in the latter the $p_t$ spectra of the
regenerated component are approximated with a thermal blast wave.

Recent experimental data on $J/\psi$ production in Cu-Cu collisions
indicate that the nuclear modification factor tends to increase at high
$p_t>5 \; \GeV$\cite{Adare:2008sh,Tang:2008uy}. Such a trend is not
present in QGP suppression calculations based on the quasifree
dissociation rate which increases with the three-momentum of the
$J/\psi$\cite{Zhao:2007hh}. However, the inclusion of
charmonium-formation time
effects\cite{Karsch:1987uk,Blaizot:1987ha,Gavin:1990gm} (which reduce
the dissociation rate due to time dilation in the development of the
hadronic wave packet) and the contributions from bottom
feed-down\cite{Xu:2008} can lead to an increase of $R_{AA}^{J/\psi}$ at
high $p_t$\cite{Zhao:2008vu}.

An initially promising signature to discriminate suppression and
$c$-$\bar c$ coalescence mechanisms is the elliptic flow of charmonia
(which also provides a close connection to, and thus consistency check
with, the collective flow of open charm). If only suppression mechanisms
are operative, the azimuthal asymmetry of the charmonium momentum
distributions entirely develops from the path length differences caused
by the long vs. the short axis of the almond shaped nuclear overlap
zone. As for open-charm (recall middle panel of
Fig.~\ref{fig_elec-rad}), this effect is rather small, generating a
maximal $v_2(p_t)$ of up to $2$-$3\%$\cite{Wang:2002ck}. On the other
hand, for $c$-$\bar c$ coalescence, the charmonium bound state inherits
up to twice the $c$-quark $v_2$\cite{Greco:2003vf,Ravagli:2007xx} at the
time of formation, especially if the $c$-$\bar c$ binding energy is
small (in that case little $v_2$ is carried away by an outgoing light
parton). This effect is further maximized if the coalescence occurs late
in the evolution, e.g., at the hadronization transition (where most of
the elliptic flow is believed to have built up). However, according to
the above discussion, the charmonium regeneration yield is mostly
concentrated at rather low $p_t<3 \; \GeV$, where the magnitude of the
$c$-quark $v_2(p_t/2)$ is not very large (this is a consequence of the
mass ordering of $v_2$, whose rise in $p_t$ is shifted to larger values
for heavier particles), recall, e.g., the right panel of
Fig.~\ref{fig_hq-reso}. Consequently, within the transport models
displayed in Fig.~\ref{fig_jpsi-rhic}, the net $v_2(p_t)$ of regenerated
and primordial $J/\psi$'s combined does not exceed
$2$-$3\%$\cite{Yan:2006ve,Zhao:2008vu,Krieg:2007bc} and would therefore
be difficult to discriminate from primordial production only. An
interesting question, which thus far has received little attention,
concerns elastic interactions of charmonia in the medium and whether
they could contribute to their $v_2$ in heavy-ion collisions.  Elastic
interactions should become more relevant as the binding energy of the
charmonium increases, rendering them more compact objects which are less
likely to break up. Interestingly, the NA60 collaboration has reported a
rather large inclusive ($p_t > 0.5 \; \GeV$) elliptic flow of
$v_2=(6.8\pm4)\,\%$ for $J/\psi$'s in semicentral In-In ($\sqrt{s}
=17.3\;A \GeV$) collisions at the SPS\cite{Arnaldi:2006it}. This
observation will be difficult to explain based on dissociation reactions
alone.

\begin{figure}[!t]
\begin{minipage}{0.54\linewidth}
\includegraphics[width=1.\textwidth]{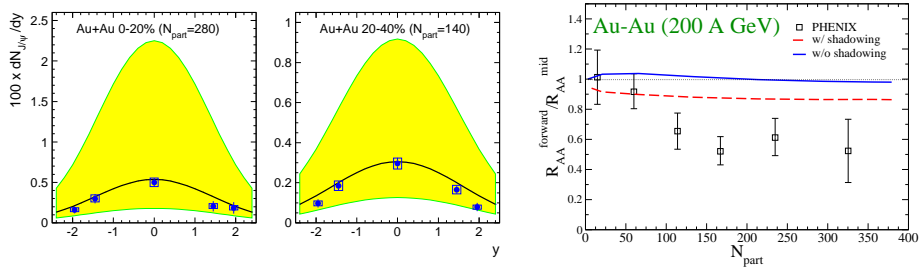}
\end{minipage}
\hspace{0.1cm}
\begin{minipage}{0.4\linewidth}
\includegraphics[width=1.\textwidth]{rhic-y-ratio-zhao.eps}
\end{minipage}
\caption{(Color online) Rapidity dependence of $J/\psi$ production in
  Au-Au ($\sqrt{s}=200\;A\GeV$) collisions at RHIC. Left and middle
  panel: results of the statistical hadronization
  model\protect\cite{Andronic:2006ky} for the $J/\psi$ rapidity density
  in central (left) and semicentral (middle) collisions, compared to
  PHENIX data\protect\cite{Adare:2006ns}. The central lines (shaded
  error bands) correspond to (the uncertainty in) pQCD charm cross
  sections, $\sigma_{pp}^{c\bar
    c}=256^{+400}_{-146}\,\mu\mathrm{b}$\protect\cite{Cacciari:2005rk}.
  Right panel: the ratio of $R_{AA}(N_{\mathrm{part}})$ for forward to
  mid-rapidity $J/\psi$ yields in the thermal rate-equation
  approach\protect\cite{Zhao:2008pp}, compared to PHENIX
  data\protect\cite{Adare:2006ns}.}
\label{fig_rap-rhic}
\end{figure}
Next, we address the rapidity dependence of $J/\psi$ production at RHIC,
where PHENIX measurements in the dielectron channel at central rapidity,
$|y|< 0.35$, and in the dimuon channel, at $|y|=1.2$-$2.2$, indicate a
maximum in $R_{AA}(y)$ around $y=0$ for central and semicentral
collisions. This trend can be nicely reproduced by the statistical
hadronization model as a consequence of the underlying $c$-quark
distributions (cf. left and middle panel in Fig.~\ref{fig_rap-rhic})
which are expected to be narrower than for bulk particle production,
implying higher $c$-quark densities and thus larger charmonium yields at
central rapidity. This dependence is more difficult to explain in the
thermal rate-equation approaches\cite{Zhao:2008pp} where only about
$50\%$ (or less) of the $J/\psi$'s originate from $c$-$\bar c$
regeneration. In addition, the charm ensemble at RHIC is not yet fully
in the grand canonical limit\footnote{In part, this is due to a finite
  correlation volume introduced for $c\bar c$ quarks in the approach of
  Refs.\cite{Grandchamp:2003uw,Zhao:2007hh}.} for which $I_1/I_0\to 1$
in Eq.~(\ref{NQQ}), and thus the sensitivity on the $c$-quark density
(fugacity) is less pronounced. Furthermore, the thermal suppression of
the primordial component exhibits an opposite trend, being slightly less
suppressed at forward $y$ due to reduced light-particle production. It
is quite conceivable that cold-nuclear-matter effects imprint
significant rapidity dependencies on the primordial component (e.g.,
stronger shadowing and/or nuclear absorption at forward
$|y|$)\cite{Kharzeev:2008cv,Ferreiro:2008wc}. In
Ref.\cite{Ferreiro:2008wc}, e.g., stronger nuclear absorption of
$J/\psi$'s at RHIC has been found as a consequence of different
production mechanisms which probe different kinematics in the nuclear
parton distribution functions.

Recent calculations of charmonium production within microscopic
transport models for the bulk-medium evolution can be found, e.g., in
Refs.\cite{Zhang:2002ug,Linnyk:2008uf}. The results are generally quite
reminiscent of the rate-equation calculations discussed above. In
particular, the description of RHIC data requires the inclusion of
regeneration interactions. This is also true for the so-called comover
approach, which has been extended to include charmonium formation
reactions in Ref.\cite{Capella:2007jv}.
     
The interplay of suppression and regeneration should lead to interesting
consequences for the excitation function of charmonium production. The
approximate degeneracy of $J/\psi$ suppression by about a factor of
$\sim$3 in central $A$-$A$ collisions at both SPS and RHIC has been
anticipated in the two-component model of Ref.\cite{Grandchamp:2001pf},
with a rather flat behavior for $\sqrt{s}=17$-$200\,A\GeV$. This
degeneracy is expected to be lifted at higher (LHC, $\sqrt{s}=
5500\,A\GeV$) and lower (FAIR, $\sqrt{s} =8\,A\GeV$) collision
energies. At LHC, the statistical hadronization model predicts the
inclusive $J/\psi$ $R_{AA}$ in central Pb-Pb collisions to recover the
level in $p$-$p$ collisions, i.e., $R_{AA}(N_{\mathrm{part}}=350) \to
1$, based on a $p$-$p$ open-charm cross section of $\dd\sigma_{\bar
  cc}/\dd y=640\,\mu\mathrm{b}$. At FAIR energies, on the other hand,
statistical production is small\cite{Andronic:2007zu}, while
transport\cite{Linnyk:2006ti} and rate-equation approaches predict about
a factor of two suppression, mostly dominated by nuclear absorption. The
effective nuclear absorption cross section will thus be an essential
quantity to be determined in $p$-$A$ reactions at FAIR.

Finally, let us turn to bottomonium production, which adds several new
aspects compared to charmonium production: (i) the binding energies of
bottomonium states are larger by about a factor of $\sim$2 which opens a
wider window to study their dependence on color screening (due to larger
dissociation temperatures) and makes them more robust in the hadronic
phase; (ii) at given collision energy, the number of $b\bar b$ pairs is
substantially smaller than the number of $c\bar c$ ones (e.g., by about
a factor of $\sim$200 at RHIC\cite{Cacciari:2005rk,vanHees:2005wb})
(iii) bottom-quarks are less susceptible to changes in their momentum
distributions due to their factor $\sim$3 larger mass (as discussed in
Secs.~\ref{sec_hq-int} and \ref{sec_hq-obs} of this review). The latter
2 points suggest that regeneration processes play less of a
role\footnote{However, care has to be taken in deducing that this
  renders $\Upsilon$ regeneration irrelevant at RHIC, since (a) the
  bottom ensemble is in the canonical limit, and (b) the regeneration
  yield needs to be compared to the primordial yield: in $p$-$p$
  collisions the ratio $\Upsilon/(b\bar b)\simeq0.1\%$ is about a factor
  of $10$ smaller than in the charm sector where $J/\psi/(c\bar
  c)\simeq1\%$.}.  Early analyses of $\Upsilon$ production in heavy-ion
collisions have focused on the $p_t$-dependence of suppression scenarios
where instantaneous dissociation above a critical energy density has
been combined with formation-time effects, both at
LHC\cite{Karsch:1990wi,Gunion:1996qc,Pal:2000zm} and
RHIC\cite{Pal:2000zm}. The opposite limit of secondary production alone
has been evaluated in the statistical hadronization
model\cite{Andronic:2006ky}. The thermal rate-equation approach,
Eq.~(\ref{rate-eq}), has been applied to $\Upsilon$ production in
Ref.\cite{Grandchamp:2005yw}, in analogy to the charmonium sector as
displayed in the left panels of Figs.~\ref{fig_jpsi-na50} and
\ref{fig_jpsi-rhic}.
\begin{figure}[!t]
\begin{minipage}{0.49\linewidth}
\includegraphics[width=0.95\textwidth]{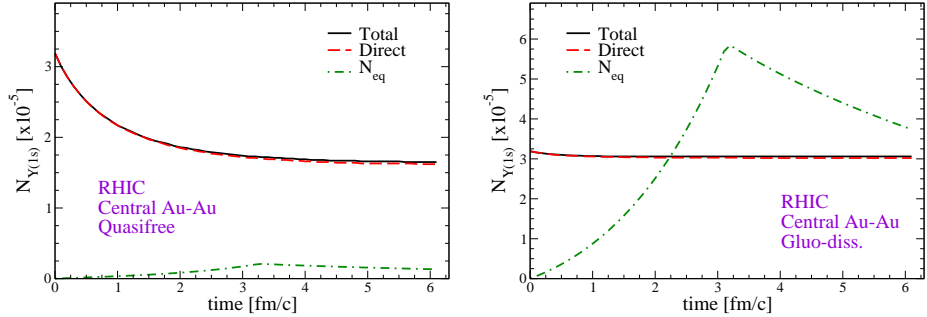}
\end{minipage}
\begin{minipage}{0.49\linewidth}
\includegraphics[width=0.95\textwidth]{ups-time-rhic-gluodiss.eps}
\end{minipage}
\caption{(Color online) Time dependence of $\Upsilon(1S)$ production in
  central Au-Au($\sqrt{s}=200\;A\GeV$) collisions at
  RHIC\protect\cite{Grandchamp:2005yw} using the quasifree
  dissociation-cross section with in-medium reduced binding energies
  (left panel) and the gluo-dissociation cross section with the free
  binding energy (assumed to be $\varepsilon_B^\Upsilon= 1.1\; \GeV$;
  left panel). The time evolution of the background medium is based on
  an expanding fireball as in the left panels of
  Fig.~\ref{fig_jpsi-rhic}.}
\label{fig_ups-rhic}
\end{figure}
The time evolution of the $\Upsilon(1S)$ yield in central Au-Au
collisions at RHIC has been calculated for the following two scenarios
(as shown in Fig.~\ref{fig_ups-rhic}): in the first one (left panel),
reduced in-medium binding energies (according to solutions of a
Schr\"odinger equation with a color-screened Cornell
potential\cite{Karsch:1987pv}) are combined with quasifree dissociation
(and formation) reactions; in the second one, the gluo-dissociation
process is applied to the $\Upsilon(1S)$ with vacuum binding energy
(assuming $m_b=5.28\;\GeV$ in connection with $\varepsilon_B=1.1\;
\GeV$). One finds that color-screening enables a $\sim$40\% suppression
of the $\Upsilon(1S)$ within the first $1$-$2\;\fm/c$ (where the average
medium temperature is above $200\; \MeV$), with insignificant
contributions from regeneration since $N_\Upsilon^{\mathrm{eq}}$ is too
small. On the other hand, with its vacuum binding energy, the
$\Upsilon(1S)$ is basically unaffected at RHIC. This suggests a very
promising sensitivity of $\Upsilon(1S)$ production to color-screening at
RHIC (of course, the observed $\Upsilon(1S)$ yield contains feed down
contributions, amounting to ca.~$50\%$ in $p$-$p$ collisions; this
underlines again the importance of measuring the excited $\Upsilon'$ and
$\chi_b$ states).  Note that $N_\Upsilon^{\mathrm{eq}}$ is quite
different in the two scenarios even though
$m_\Upsilon=2m_b-\varepsilon_B^\Upsilon$ is assumed to be constant in
both cases. The reason is the difference in \emph{relative} chemical
equilibrium: a larger binding energy implies a larger $m_b$, which leads
to a larger $\gamma_b$ via Eq.~(\ref{NQQ}), i.e., it is
thermodynamically more favorable to allocate $b$ and $\bar b$ quarks in
an $\Upsilon$ (relative to a smaller $\varepsilon_B^\Upsilon$ with
smaller $m_b$ and thus smaller $\gamma_b$).


\section{Conclusions}
\label{sec_concl}

Heavy-quark physics is an increasingly useful and adopted tool
in the theoretical analysis of hot and dense QCD matter and its study in
ultrarelativistic collisions of heavy nuclei. Key reasons for this
development lie in a combination of exciting new data becoming available
not only for quarkonia but also for open heavy-flavor observables, 
together with attractive features of charm and bottom quarks from a 
theoretical point of view. These features are, of course, rooted in the 
large scale introduced by the heavy-quark (HQ) mass, which enables the 
use of expansion techniques, most notably HQ effective theories and 
Brownian motion for HQ diffusion in a QGP fluid. Furthermore, the 
production of heavy quarks in nuclear reactions is presumably restricted 
to primary $N$-$N$ collisions which renders HQ spectra a calibrated 
probe of the medium over the \emph{entire range of transverse momentum}.
The latter provides a unique opportunity for a comprehensive 
investigation of the QCD medium at all scales, ranging from diffusion 
physics in the low-$p_t$ limit to a ``standard'' hard probe at 
$p_t\gg m_Q$. Heavy quarks thus connect transport coefficients in the 
QGP and observables in ultrarelativistic heavy-ion collisions in the 
arguably most direct way. Finally, relations between the open and 
hidden heavy-flavor sectors promise valuable mutual 
constraints, both theoretically and phenomenologically. 
In this review we have largely focused on aspects of soft physics for 
HQ propagation and binding in the QGP.
 
Interactions of slowly moving heavy quarks in a QGP are dominated by
elastic scattering off thermal partons. A perturbative expansion of
these interactions, specifically for the HQ diffusion coefficient, shows
poor convergence for coupling constants believed to be relevant for a
QGP as formed in heavy-ion collisions. Several options of amending the
perturbative treatment have been suggested, e.g., a reduced screening
mass or running coupling constant at low momentum transfer. While
increasing the interaction strength, they
inevitably face the problem of little control over higher order
``corrections''. Nonperturbative approaches have been put forward which
can, in principle, overcome this problem by a (partial) resummation of
large contributions. E.g., a potential-based $T$-matrix approach
characterized by a scattering equation becomes particularly promising if
the input interaction can be specified in a model-independent way, i.e.,
from thermal lattice QCD. Currently, open questions remain as to the
validity of the potential approach at finite temperature, and a suitable
definition of the potential from the HQ free energy. Here, a close
connection between the open and hidden heavy-flavor sectors emerges via
the same low-energy interaction operative for HQ diffusion and
quarkonium bound states. Qualitatively, one finds that, if ground-state
quarkonia survive until temperatures well above $T_c$, potential
scattering of heavy quarks in the QGP builds up resonance-like
correlations which are instrumental in obtaining a small HQ diffusion
coefficient close to $T_c$. As an alternative nonperturbative approach, 
HQ diffusion has been estimated in the strong-coupling limit of 
conformal field theory (CFT) using a conjectured correspondence to 
string theory in Anti-de-Sitter (AdS) space. With CFT parameters adapted 
to resemble QCD, the resulting HQ diffusion constant is comparable to 
the $T$-matrix approach close to $T_c$, but is approximately constant 
with increasing temperature while the $T$-matrix interaction and 
ultimately approaches pQCD estimates.

Quantitative applications of the Brownian-motion framework to HQ
observables at RHIC critically hinge on a reliable description of the 
background medium evolution. The latter specifies the ambient conditions 
in the (approximately) thermal bath including its temperature and
collectivity, whose magnitudes directly impact the nuclear
modification and elliptic flow of HQ spectra. A survey of available 
calculations indicates that the translation of a given HQ diffusion 
coefficient into suppression and elliptic flow of HQ spectra is 
currently at the $\sim$$50\%$ accuracy level. This needs to be further
scrutinized and improved. In line with the theoretical expectations 
for HQ diffusion in a QGP at $T$=1-2\,$T_c$, the current data call 
for significantly stronger interactions than provided by LO pQCD. A 
simultaneous and consistent evaluation of spectra ($R_{AA}(p_t)$) 
and elliptic flow ($v_2(p_t)$) is pivotal to this conclusion. A
non-negligible role is played by the hadronization process: heavy-light
quark coalescence processes seem to improve the experimentally observed
correlation of a rather large $v_2$ and a moderately suppressed $R_{AA}$
in the electron spectra at moderate $p_T^e$$\le$3\,GeV. More robust 
conclusions will require a better knowledge of the Cronin effect 
figuring into the initial conditions for the HQ spectra. A theoretically 
appealing aspect of quark coalescence is its close relation to resonance 
correlations in the QGP, which could be at the origin of the 
nonperturbative interaction strength in HQ diffusion. From the 
experimental side, important discrimination power will come with an 
explicit measurement of $D$-mesons, to explicitly separate the bottom 
contribution (present in the electron spectra). Theoretical investigations 
should take advantage of the opportunity to predict angular correlation 
measurements which will become feasible soon. We believe that charm
data at SPS energy would constitute a valuable complement to RHIC data,
which could help in deciding how much of the bulk flow (and suppression)
can be imparted on charm quarks in a medium at smaller temperatures.
 
In the heavy-quarkonium sector we have started with a brief synopsis of
current applications of potential models in medium. At this stage, the
comparison of calculated spectral functions to Euclidean correlation
functions computed in lattice QCD suggests that scenarios with either
strong binding and rather large in-medium HQ mass, or weak binding and
smaller HQ mass, are both viable. On the contrary, inelastic reaction 
rates are rather sensitive to the binding energy and thus to the 
strength of color-Debye screening, especially for bottomonia. This 
translates into a promising discrimination power of bottomonium 
suppression measurements at RHIC and LHC. 
The phenomenology of charmonia is presumably more involved;
e.g., kinetic rate-equation calculations for $J/\psi$ production in
central Au-Au collisions at RHIC indicate that the number of surviving
primordial $J/\psi$'s is comparable to the number of secondary produced
ones via $c$-$\bar c$ coalescence in the QGP (and/or at hadronization). 
Such an interpretation is consistent with $J/\psi$ $p_T$ spectra where 
the coalescence yield is concentrated at low $p_T$, thus reducing the
average $p_T^2$ compared to primordial production. Kinetic approaches
furthermore suggest that regeneration is subleading at the SPS, and that
the observed suppression occurs at energy densities above the critical
one. Deeper insights will follow when advanced theoretical models meet
future precision data, including rapidity dependencies, elliptic flow,
excited charmonia and much needed constraints from d-Au collisions at
RHIC (to pin down cold-nuclear matter effects). An extended excitation 
function via a RHIC energy scan, LHC and FAIR will further disentangle
suppression and regeneration effects. We emphasize
again that the presence of regeneration mechanisms would imply valuable
connections to the open heavy-flavor sector, as coalescing heavy quarks 
necessarily imprint their kinematics on HQ bound states.

In summary, we believe that in-medium HQ physics will continue as a
challenging but rewarding forefront research field for many years to
come, with ample opportunities for surprises, insights and progress.

\vspace{0.5cm}

{\bf Acknowledgments}\\
We are indebted to D.~Cabrera, V.~Greco, C.M.~Ko, M.~Mannarelli,
I.~Vitev and X.~Zhao for productive and enjoyable collaboration on
various aspects of the topics discussed in this review. We furthermore
thank J.~Aichelin, P.~Gossiaux, W.~Horowitz and P.~Petreczky for
illuminating discussions. This work has been supported by a
U.S. National Science Foundation CAREER award under grant
no. PHY-0449489 and by the A.-v.-Humboldt foundation through a Bessel
award (RR) and Feodor-Lynen fellowship (HvH).

\end{document}